\documentclass[11pt]{article}

\usepackage{epsfig}
\usepackage{here}  
\usepackage{cite}  

\begin{document}
\renewcommand{\thefootnote}{\fnsymbol{footnote}}

\begin{center}
{\LARGE  Compton Scattering by Nuclei }\footnote{Supported by Deutsche
Forschungsgemeinschaft (Schu222, SFB201, 436RUS113/510) and DAAD}\\[2ex]

\setcounter{footnote}{0}
\renewcommand{\thefootnote}{\arabic{footnote}}

M.-Th. H\"utt$^{\,a,}$\footnote{e-mail: huett@bio.tu-darmstadt.de},
A.I. L'vov$^{\,b,}$\footnote{e-mail: lvov@x4u.lebedev.ru},
A.I. Milstein$^{\,c,}$\footnote{e-mail: A.I.Milstein@inp.nsk.su},
M. Schumacher$^{\,a,}$\footnote{e-mail:
Martin.Schumacher@phys.uni-goettingen.de}\\[2ex]

\it
$^a$Zweites Physikalisches Institut, Universit\"at G\"ottingen, D-37073
     G\"ottingen, Germany\\
$^b$P.N. Lebedev Physical Institute, Leninsky Prospect 53,  Moscow
117924, Russia\\
$^c$Budker Institute of
Nuclear Physics, 630090 Novosibirsk, Russia
\end{center}

\begin{abstract}
The concept of Compton scattering by even-even nuclei from
giant-resonance to nucleon-re\-so\-nan\-ce energies and the status of experimental and theoretical
researches in this field are outlined.
The description of Compton scattering by nuclei starts from different
complementary approaches, namely from second-order $S$-matrix and
from dispersion theories.
Making use of these, it is possible to incorporate into the
predicted nuclear scattering  amplitudes all the information available
from other channels, {\it viz.} photon-nucleon and photon-meson
channels, and to efficiently make use of models of  the nucleon, the
nucleus and the nucleon-nucleon interaction. The total photoabsorption
cross section constrains the nuclear scattering amplitude in the
forward direction.  The specific information obtained from Compton
scattering therefore stems from the angular dependence of the nuclear
scattering amplitude, providing detailed insight into the dynamics of
the nuclear and nucleon degrees of freedom and into the interplay
between them.  Nuclear Compton scattering in the giant-resonance
energy-region provides information on the dynamical properties of the
in-medium mass of the nucleon.  Most prominently, the electromagnetic
polarizabilities of the nucleon in the nuclear medium can be extracted
from nuclear Compton scattering data obtained in the quasi-deuteron
energy-region. In our description of this latter process special
emphasis is laid upon the exploration of many-body and two-body effects
entering into the nuclear dynamics. Recent results are presented for
two-body effects due to the mesonic seagull amplitude and due to the
excitation of nucleon internal degrees of freedom accompanied by meson
exchanges. Due to these studies the in-medium electromagnetic
polarizabilities are by now well understood, whereas the understanding
of nuclear Compton scattering in the $\Delta$-resonance range is only
at the beginning.  Furthermore, phenomenological methods how to include
retardation effects in the scattering amplitude are discussed and
compared with model predictions.
\end{abstract}

\input pr0.t

\renewcommand{\thefootnote}{\arabic{footnote}}
\setcounter{footnote}{0}

\pagenumbering{roman}
\clearpage
\tableofcontents

\clearpage

\section{Introduction}


\pagenumbering{arabic}

The discovery made by Compton in 1922  \cite{compton23}
that photons are  scattered by free
electrons is one of the mile-stones of modern physics.
In this process an energy shift of the photon was observed which
quantitatively could be explained by assuming an elastic collision
between particles.
Later on, the term Compton scattering has been transferred to photon
scattering by
other objects as there are atoms, nucleons and nuclei.  An
essential difference
between free electrons and these other objects is that the latter have an
internal structure so that excitation and particle emission
processes become possible. In that case the term Compton scattering is
used to denote coherent-elastic photon scattering where the scattering
object coincides in all (internal) quantum numbers in the initial
and final states.
In case of atoms the term Compton scattering occasionally
is replaced by the term Rayleigh scattering \cite{kane86}.
This is of advantage in
cases where the coherent-elastic scattering
process (Rayleigh scattering) is discussed together with the incoherent
scattering process (also called Compton Scattering), where one bound
electron is ejected from the electron cloud
into the continuum. This process then resembles the one discovered by Compton
\cite{compton23} except for the  appearance of binding effects \cite{kane92}.
The term Raman scattering is used in the case where the atom or nucleus
does not return to the initial state but no particle emission
is taking place.
Another related photon scattering process is  coherent-elastic
scattering  in the Coulomb field of nuclei termed
Delbr\"uck scattering \cite{papatzacos75,milstein94}. The three
coherent-elastic photon scattering processes, {\it viz.}\ atomic Rayleigh
scattering, Delbr\"uck scattering and nuclear Compton scattering
are expected to interfere
with each other. This has been observed in several investigations
\cite{milstein94}.

\subsection{Early work on nuclear Compton scattering in the giant-resonance
region}

Scattering of  photons in the region of the ``giant resonance"
of the nucleus was probably first observed by Gaerttner and Yeater
\cite{gaerttner49} and Dressel, Goldhaber, and Hanson \cite{dressel50}.
These works were motivated by Goldhaber and Teller
\cite{goldhaber48} who proposed to understand a pronounced structure in the
photoabsorption cross section between 10 and 30 MeV in terms of a
collective motion of protons against neutrons. The experiments were
made feasible by the advent of betatrons delivering photons
of sufficient energy but suffered from the lack of
efficient photon detectors.
More refined experiments became possible with the advent of NaI(Tl)
scintillation counters which were first applied by Stearns
\cite{stearns52} and Fuller and Hayward \cite{fuller54,fuller56}.
Stearns used monochromatic photons of 17.6 MeV from the $^7$Li(p,$\gamma$)
nuclear reaction whereas Fuller and Hayward used continuous photons
produced by a betatron.

A remarkable progress was made by Hayward et al.\ in 1973 \cite{hayward73}
with the first application of monochromatic linearly polarized photons.
These photons were produced by resonance fluorescence of the well-known
$1^+$ state at 15.1 MeV in $^{12}$C. With these experiments a
discrimination was possible between nuclear Compton scattering
and nuclear Raman scattering,  the latter being  especially
strong for deformed nuclei.

\subsection{Nuclear Compton scattering experiments from the mid-1970th
to the present}

The breakthrough in investigations of nuclear Compton scattering
came with the advent of large-volume  NaI(Tl) detectors and the
availability of quasimonochromatic photons
\cite{D0,H0,L1,D3,H4,W4,W5,A6,N6,L7,
A8,B8,S8,S10,Z10,W10,D11,S11,B12,
pb-lund,D12,L12,M12,W12,W14,F15,
H15,I15,G16,P16,F16,S16,B17,K18,P18,P18a}.
As a first step the technique of positron-annihilation-in-flight has been
applied to produce quasimonochromatic  photons. In early uses
of  bremsstrahlung two different techniques have been developed.
In one of these  the total spectrum of bremsstrahlung photons
was used. By applying very narrow collimators in front of the NaI(Tl)
detectors the low-energy tails of the response functions were largely
suppressed so that large portions of the spectra of scattered photons could
be interpreted in terms of Compton scattering. This method was
efficiently used by B. Ziegler et al.\
\cite{S8,S10,S11} and led to insight into
general properties of nuclear scattering amplitudes
below meson photoproduction threshold, {\it viz.}the
importance of formfactors, the possibility to investigate the electromagnetic
polarizabilities of nucleons in the nuclear medium  ({\it cf.} also
\cite{B8}), {\it etc.}
When using wider collimators in front of the NaI(Tl) detectors the
response functions generate a low-energy tail. In this case only a
narrow energy range at the upper end of the bremsstrahlung spectrum
can be interpreted in terms of Compton scattering.

The techniques described in the foregoing paragraph  have been
more and more replaced by
tagged photons either  from bremsstrahlung or from backscattered laser
light. In this technique the energy of the secondary electron is measured
in a magnetic spectrometer so that the
energy of the primary photon is known  through a coincidence condition
with the Compton scattered  photon or some other reaction product.
This technique was first applied
to nuclear Compton scattering experiments at the University of Illinois
\cite{bowles78}
and  is now standard
in all
laboratories --- {\it viz.} Brookhaven \cite{thorn89}, Lund
\cite{adler90,adler97},
Mainz \cite{anthony91,hall96}, Saskatoon \cite{vogt93}
--- where experiments on Compton scattering by nuclei are
carried out. Back scattering of laser light \cite{thorn89}
and coherent bremsstrahlung
from diamond crystals \cite{lohmann94,kraus97,rambo98}
are also efficient sources of linearly
polarized photons. At low energies (5--100 MeV) also the method of
off-axis tagging \cite{D12,krause91} leads to reasonable degrees of
linear polarization.

An other type of experiments which has to be
mentioned here has achieved
very high energy resolution ($\approx10$ keV) due to the use of
Ge(Li) detectors and due to
the use of photons from nuclear reactions
\cite{kahane74,schumacher81,rullhusen83,
nolte86,fuhrberg89,nolte89,fuhrberg89a}.
These experiments were
restricted to energies from about 5 MeV \cite{schumacher81}
to 17.4 MeV \cite{nolte86}. The high energy-resolution
was of essential help in cases where Compton scattered photons had to be
discriminated from photons stemming from nuclear
resonance fluorescence or nuclear Raman scattering.

\subsection{Motivations for studying nuclear Compton scattering}


Several experiments on Compton scattering by nuclei were concerned with
multipole decompositions of giant resonances
\cite{D0,L1,D3,W4,W5,N6,S8,S10,S11,D12,fuhrberg89}.
The comparatively weak
electric quadrupole amplitude is enhanced through interference
with the dominant electric-dipole amplitude. Therefore, nuclear Compton
scattering became a tool to study the properties of the isovector
electric quadrupole resonance. In fact, this is  the only model
independent method existing for this purpose. The status of this
research will be discussed in section 4.


An interesting feature of nuclear Compton scattering below meson
photoproduction threshold which has been
realized very early is its strong sensitivity to
meson exchange currents (MEC) between nucleons
\cite{L50,christillin4,friarann,
fri76,C75,C78,C80,AR80,A82,ME83,C84,W84,C85,R85,C86,ME86,ME87}.
In this respect
Compton scattering is of essential superiority to electron scattering
at similar momentum transfers
which is to a large extent insensitive to this phenomenon. The study of
MEC  degrees of freedom and their interplay
with nuclear degrees of freedom have
attracted strong interest in many investigations, and different schools have
approached these phenomena in different ways. Meson exchange currents
have long been realized to be the origin for the enhancement of the
integrated dipole strength in comparison with the prediction $\Sigma
_{TRK}$
of the usual Thomas-Reiche-Kuhn (TRK) sum rule \cite{L50}.
The integrated strength of the giant-dipole resonance (GDR) overshoots
the TRK sum rule prediction by the factor (1 + $\kappa^{GDR}$). The
quasi deuteron (QD) photoabsorption process which can be interpreted as
the absorption by correlated proton-neutron pairs inside the nucleus,
provides  an additional strength \cite{ahrens75}
$\kappa^{QD}\Sigma _{TRK}$.
It was a great progress to show \cite{christillin4}
that due to gauge invariance
MEC also have to modify the non-resonant contribution to the scattering
amplitude, i.e.\ the so-called seagull amplitude, which
accompanies  resonant scattering through intermediate
excited nuclear-structure states.
In the absence of MEC the seagull amplitude
would simply be the superposition of the Thomson scattering amplitudes
of  the protons in the nucleus forming the ``kinetic seagull amplitude".
Accordingly it was discussed in what way
the modification
of the total seagull amplitude through an additional ``exchange" or
``mesonic" seagull
amplitude
could possibly be understood as Thomson
scattering by pions in
nuclei \cite{C75,C78,C80,AR80,A82,W84,ME87}. According to
Siegert's theorem \cite{siegert37}
there is no charge-density modification due to mesons in a nucleus
(except for a small relativistic correction)
but  an additional (two-body) current due to the
exchange of charged mesons. Therefore,
the interpretation of the MEC contribution to the  seagull
amplitude requires special care
({\it cf.} sections 3.3 and 7).
It turns out that the giant-resonance part of the
mesonic seagull amplitude is closely related to the in-medium mass of the
nucleon
\cite{nolte86,fuhrberg89a,22,BR80,BWBS87,BR89,
BMP90,BR90,BR91,W93,BRS93,SC94,BR95,FR96}
which is a special case of the general Brown and Rho
scaling \cite{BR90,BR91,W93,BRS93,BR95,FR96,BR96}. One
motivation to study nuclear Compton scattering, therefore, is to get
further insight into   the properties of the in-medium mass of the nucleon.


Below meson photoproduction threshold the internal excitation of the nucleon
enters into the nuclear Compton scattering amplitude via the electromagnetic
polarizabilities of the nucleon
\cite{petrunkin81,lvov93,lvov97}. The electromagnetic
polarizabilities are fundamental constants, characterizing the
response of the nucleon to an external electromagnetic field and, thus, the
dynamical structure
of the nucleon.  The advantage of the electromagnetic polarizabilities
of the nucleon is that they can precisely be measured for the free and
the bound nucleon by carrying out Compton scattering experiments.
Because of this property, the electromagnetic polarizabilities are an
ideal observable for   exploring possible in-medium
modifications of the nucleon structure. Some fundamental theoretical
aspects of the in-medium electromagnetic polarizabilities have been
discussed on the basis of quark models
\cite{drechsel84,lvov92}
and pion-cloud models \cite{R85,ME86,bunatian92}.
In recent time  essential progress has been made in this field of research
both experimentally
and theoretically ({\it cf.} sections 5 and 9).


For energies in the resonance range of the nucleon
above meson photoproduction threshold only first steps have
been made in the investigation of nuclear Compton scattering.
One reason for this delay is that only a very fragmentary set of data
was available, even for the nucleus $^4$He which was the subject of
principal interest for experimental researches
\cite{A6,A8,D11,M12,I15,S16,B17}. On the experimental side considerable
progress has been
made very recently by the G\"ottingen-Mainz collaboration
at MAMI (Mainz)  \cite{S16,K18}
and by the LEGS collaboration at Brookhaven \cite{B17} where
differential cross sections for Compton scattering by $^4$He have been
measured in large angular and energy intervals. These experiments
showed that the widely
accepted theoretical description of Compton scattering in terms of
the $\Delta$-hole model
\cite{os79,os81,koch84,koer94,pas96}
is in serious  disagreement
with experimental data \cite{K18}.  This leads to the conclusion that the
in-medium excitation of the $\Delta$ resonance is not fully understood and
possibly may lead to new insights into the structure of the bound nucleon.

\subsection{Fields related to nuclear Compton scattering}

In addition to nuclear Compton scattering, nuclear Raman scattering
and nuclear resonance fluorescence have remained branches of photon scattering
researches up to  present days. For deformed nuclei with ground-state
spins $I\geq1$ Raman scattering is possible as an incoherent
(non-interfering) elastic
scattering process ending up with a nonexcited  nucleus, but  in a spin-state
differing from that of the initial state. Raman scattering into excited
states
\cite{bowles78,barnoy74,barnoy77,nathan80,bowles81,alarcon89,mondry91}
is  also possible  as an inelastic scattering process.
The theoretical framework
appropriate to describe Raman scattering is the concept of
generalized polarizabilities
\cite{fano60,are67,are67a,aregrei69,are73,arenh85,are86}
leading in a natural way to a discrimination between
scalar (Compton) and tensorial (Raman) scattering  and --- as a third
possibility --- vector scattering.

Nuclear resonance fluorescence  takes place at energies  below
particle threshold of the nucleus as an
elastic photon scattering process through the excitation of
isolated nuclear states
\cite{kneissl90}. The study of this process serves as a tool of
low-energy nuclear spectroscopy which is not considered here in  detail.

\subsection{Organization of the present paper}

The topic of Compton scattering by nuclei cannot be covered completely in
one article. Therefore, this paper is strongly biased by our own
preferences and results.  Historical aspects and
lines of development going back to previous researches have briefly
been covered
in the foregoing parts of this section, or will be mentioned in
introductory parts of  sections to come.
To supplement on this we recommend to take notice also of previous
reports {\cite{arenh85,fuller62,hayward77,molin80,
ziegl85,rull90,lvov90,chris90}.
We believe that the
present report differs from the previous ones by the fact that subnuclear
and retardation phenomena of nuclear Compton scattering have received a
much better understanding due to recent --- partly unpublished ---
researches. Also, due to very recent experimental and theoretical results
Compton scattering in the $\Delta$ resonance range has received a
wider coverage than  before.

The present article is concerned with Compton scattering by
spin-saturated nuclei. This restriction involves that the lightest
nucleus to be discussed here is $^4$He. Other nuclei of experimental
interest are $^{12}$C, $^{16}$O, $^{40}$Ca and $^{208}$Pb. These are the
nuclei where a reasonable amount of experimental data on Compton scattering are
available. In comparison to these
the investigations of the deuteron --- though
being the most fundamental ones --- are at the very beginning
and we give here only  few references to
recent experimental and theoretical works
\cite{lucas94,wilbois95,levchuk95,levchuk98,chen98,karakowski98}.

In {\bf section 2 } we give a definition of the degrees of freedom
of nucleons in a nucleus. Our main goal is to get a clear-cut distinction
between the external (nuclear structure)
and the internal degrees of freedom of the bound
nucleon.
It is pointed  out that the ``classical subnuclear degrees of freedom" viz.
meson exchange currents (MEC) are not specific for the one or the other
part but separately enter into both.

In {\bf section 3 } we formulate an outline of the theoretical aspects of
Compton scattering below meson photoproduction threshold. We start from
relativistic second-order perturbation theory, point out how MEC and
the ``negative energy states" of the nucleus can be taken care of,
and show how dispersion theory can be used to make links to
experimental data.

{\bf Section 4 } is devoted to our present understanding of
giant-resonances. Since previous work has been covered in previous
reviews \cite{arenh85,molin80,ziegl85}
we concentrate on our own regions of researches, as there are
enhancement and retardation effects  and  the dynamics
of the giant-resonance mode.
Furthermore, we
give a detailed description of the retardation problem which is sizable
in nuclear Compton scattering because the wavelengths of the photon
$1/\omega $, where $\omega $ is the photon energy,
and the nuclear radius $R$ are of the same order of
magnitude. The solution of the retardation problem which we present
may appear as a preliminary guess at a first sight.
Nevertheless, the results we obtain are quite precise numerically,
as will be shown in section 10 through a model calculation.

{\bf Section 5 } is devoted to effects  which are observed in the
quasideuteron range of the nucleus. Since the quasideuteron process
itself does not make large contributions to the nuclear scattering
amplitude this section is mainly concerned  with the polarizabilities of
the bound nucleon which belong to the most interesting aspects of nuclear
Compton scattering.

{\bf Section 6 } describes our present knowledge in the
$\Delta$-resonance region.
Unfortunately, the study of nuclear Compton scattering
above meson photoproduction threshold is only at the beginning and very
interesting properties of nuclear Compton scattering in this energy range
have not yet been investigated quantitatively. Accordingly, this section
only provides a brief description of recent points of discussion.

{\bf Section 7 } again is concerned with the giant-resonance region
but describes interpretations of subnuclear effects in terms of
effective masses and effective currents of nucleons in the framework of the
Fermi liquid theory. This section
also refers to modern views, {\it e.g.} the general Brown and Rho scaling
\cite{BR95,FR96}.

The following three sections 8--10 are devoted to recent theoretical
developments. {\bf Section 8 } describes a dispersion theory at fixed momentum
transfer for nuclear Compton scattering.
In {\bf section 9 } a diagrammatic model of meson exchange currents is
discussed, which essentially clarifies the most important properties of
the mesonic seagull amplitude and, in addition, leads to quantitative
results for exchange form factors and meson-induced polarizability
modifications.
In {\bf section 10}
retardation effects are studied within a relativistic oscillator model.

We do not include a discussion of
nuclear Compton scattering in the asymptotic range
beyond 1.5--2.0 GeV into this article, because there has been no recent
research  and because there are comprehensive reviews of
this field  \cite{yennie78}.
Access to literature is given in \cite{rull90} for
diffractive Compton scattering taking place at small scattering angles.
Large-angle Compton scattering takes place via direct photon-quark
coupling which is described in \cite{kroll91,kroll96,ji97,radyushkin97}.


{\it Summary:}
Our main motivation for studying
nuclear Compton
scattering was to investigate in-medium properties of nucleons as there
are\\
{\bf (i) } the in-medium mass of the nucleon,\\
{\bf (ii) } the in-medium electromagnetic polarizabilities,\\
{\bf (iii) } the in-medium $\Delta$ excitation.\\
During  these studies it turned out that some theoretical aspects
which are essential for the data analysis were not understood with the
desirable clarity and, therefore, had to be investigated along with the
in-medium properties.\\ These are\\
{\bf (iv) } the relation between  retardation and higher multipoles,\\ and\\
{\bf (v) } the effects of meson-exchange currents on the in-medium
electromagnetic polarizabilities.\\ Further work has to be done to
understand\\
{\bf (vi) } the effects of meson exchange currents on Compton scattering in the
$\Delta$-range.

\clearpage

\section{The absorption of photons and the degrees of freedom
of the nucleus}

Compton scattering is closely related to the  process of photoabsorption,
since the Compton amplitude for forward scattering can be
expressed via the total absorption cross section with the help of the
optical theorem and a dispersion relation. In this sense the total
photoabsorption cross section is a special case of Compton scattering
and, therefore, a detailed understanding of nuclear
photoabsorption is of essential help for the discussion
of Compton scattering.
Our goal is to investigate the nuclear and subnuclear degrees of freedom
of the nucleus in a consistent way. This means that we first have to
consider the properties of free nucleons and then we have to ask what the
modifications are which enter into the picture as a consequence of nuclear
binding.
As a result  of this discussion we arrive at a clear-cut distinction
between the external (nuclear structure) and internal degrees of freedom
of the bound nucleon.

\subsection{Photoabsorption by  the nucleon}

For  illustration
Fig.\ \ref{nucleon}  shows the photoabsorption cross section of
the proton separated into partial cross sections.
The photoabsorption cross section
of the nucleon starts at the meson photoproduction threshold and continues
through  different regions of photo-excitation mechanisms. In the
region of nucleon resonances  which terminates at  photon energies of about
1.5 to 2 GeV we find three broad lines  corresponding to the
$\Delta$(1232)P$_{33}$, N(1520)D$_{13}$ and N(1680)F$_{15}$ nucleon
resonances located at photon energies of 320 MeV, 740 MeV and 1020 MeV,
respectively.
The excitation of these resonances  proceeds through photons of,
respectively,
$M1$, $E1$ and $E2$ multipolarities. The properties of these resonances and
also those of the weaker ones have successfully been
understood in terms of three constituent quarks located in a binding
potential. In addition we find a nonresonant background corresponding to
one-pion photoproduction up to about 400 MeV and to more complicated
mechanisms above.

\begin{figure}[htb]
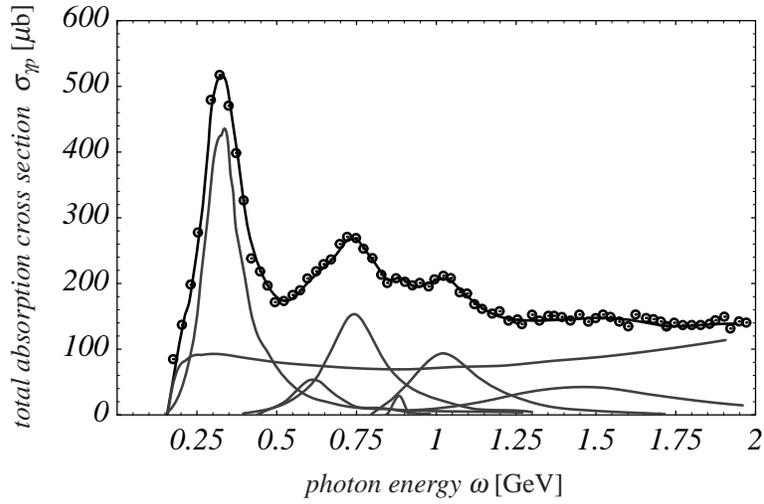

\centering
\caption{\em Photoabsorption cross section of the proton schematically
separated into partial cross sections \cite{armstrong72}  containing
nucleon resonances and a nonresonant component.}
\label{nucleon}
\end{figure}

The range of energies beyond about 2 GeV is frequently called
the asymptotic range  of the photoabsorption cross section. Here we
find  vector mesons, i.e.\ $\omega$,
$\rho$, $\phi$ etc. in the intermediate state. This means that there is
no direct interaction of the photon with the constituents of the nucleon.
In addition, the interaction
of these vector mesons with the nucleon also is not direct but proceeds
peripheral through $t$-channel exchanges.
The main candidates
for these $t$-channel exchanges are the isoscalar tensor meson $f_2(1270)$,
the isovector tensor meson $a_2(1320)$ and the Pomeron ${\cal P}$,
where the Pomeron takes care of the energy independent part of the
photoabsorption cross section, while the two tensor mesons take care of its
energy dependent part.  The isovector meson $a_2(1320)$ has to be introduced
in addition to the isoscalar meson $f_2(1270)$ because proton
and neutron differ slightly in the energy-dependent part of the cross
section  \cite{armstrong72}. The $t$-channel exchanges entering
into the description of forward-direction Compton scattering
have spin 2 in accordance with general conservation rules.

In considering the properties of the nucleon photoabsorption cross section
in the asymptotic range,
we have to be aware that the total photoabsorption
cross section is  subject to inclusive reactions where there is no
detection of the particles in the final state. In this case we predominantly
observe
those processes which make the largest contributions to the
photoabsorption cross section. These are the processes at small transverse
momentum, i.e.\ the peripheral processes mentioned above.
On the other hand, in exclusive reactions where particles in the final
state are detected,
processes with large transverse momenta may be separated
experimentally. Here  we again observe direct couplings of photons with
constituents which now are the individual valence quarks.
Some more detailed discussions of this aspect can be found in articles
contained in  \cite{schumacher90} and in \cite{kroll91,kroll96}.

\subsection{Photoabsorption by the nucleus}

When combining nucleons to form a nucleus three major changes are
observed:

(i) The photoabsorption cross section extends below pion
photoproduction threshold to form two  nuclear-structure modes of
photo excitation, {\it viz.} the giant resonance (GR) mode and the
quasideuteron (QD) mode.  In order to understand these two modes of
excitation, the internal structure of the nucleon probably may  not
be of explicit relevance --- although we do not know this for sure.
To start with, we may use a picture where unmodified nucleons in
their ground states and  mesons being exchanged between them are the
only nuclear constituents to deal with.

Figure \ref{GR-QD-absorption} gives a schematic view of the photoabsorption
cross section  of $^{208}$Pb between the giant resonance range and
the $\Delta$ range.
The  giant resonance range is dominated by the giant-dipole resonance
(GDR). The photoabsorption cross section is well described by a Lorentzian
line which extends below the particle threshold of the nucleus
in the form of narrow lines \cite{zurmuehl83}.
The strengths contained in these individual
narrow lines is given through the interplay  between the Lorentzian
tail of the GDR and the density of levels. This leads to the consequence
that there is a minimum in the individual strengths close to the particle
threshold as indicated in Fig.\ \ref{GR-QD-absorption}. The quasideuteron
cross section starts at the p-n two-particle threshold and extends
into the $\Delta$ range above the pion threshold.\\

\begin{figure}[htb]
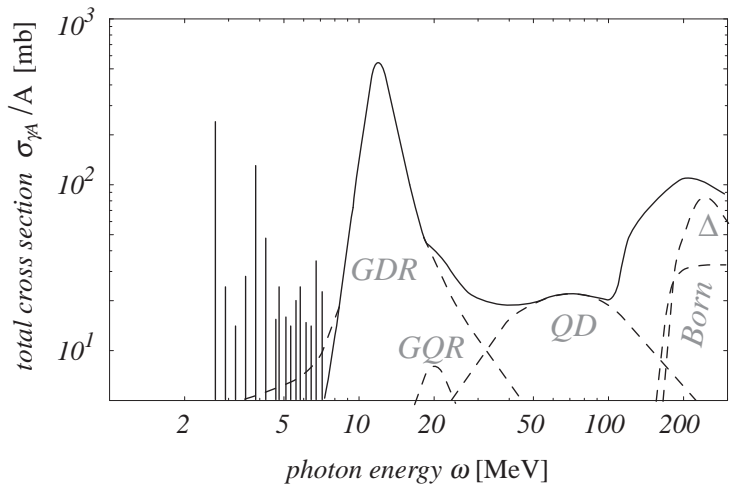
 
\centering
\caption{\em Schematic view of the nuclear photoabsorption cross section
of $^{208}$Pb
from the giant resonance to the $\Delta$ resonance range. GDR:
giant-dipole resonance, GQR: isovector giant-quadrupole resonance,
QD: quasideuteron mode, $\Delta$: $Delta$ resonance of nucleons in the
nucleus, Born: nonresonant photoexcitation of nucleons in the nucleus
through the Born terms of the photo-pion amplitudes  \cite{schumacher88Conf}.}
\label{GR-QD-absorption}
\end{figure}

(ii) Above pion photoproduction threshold in the resonance range of the
nucleon  we find a broadening of the resonances increasing with mass
number $A$. This is shown in Fig.\ \ref{NucleonResonances} where the
normalized photoabsorption cross section for the proton
is compared with those of the three
lightest nuclei.
The solid line represents the normalized photoabsorption cross section
of complex nuclei ($A \simge 4$). In this universal
curve there is no indication left of a $D_{13}$(1520)  resonance.
Irrespective of this, there seems to be  a conservation of the
integrated
photoabsorption strength when going from free to bound nucleons
\cite{ahrens98}.

\begin{figure}[htb]
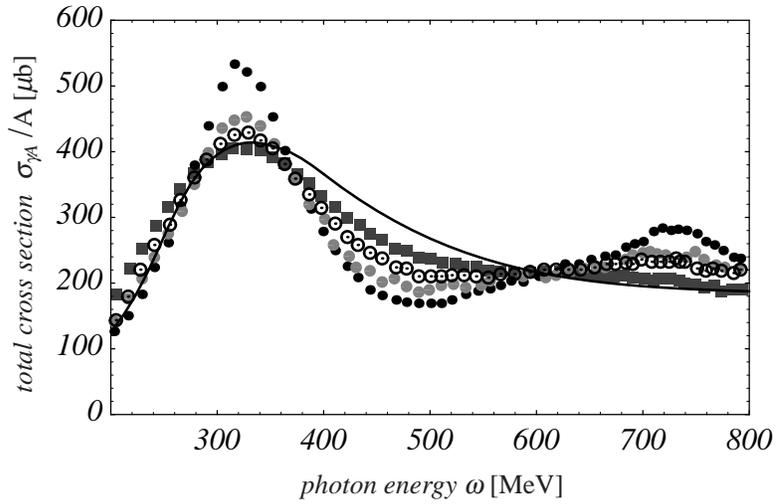
    
\centering
\caption{\em Total photoabsorption cross sections per nucleon. Data are
shown for $^1$H (black circles), $^2$H (gray circles), $^3$He (open circles)
and $^4$He (grey squares) \cite{MacCormic97}.
The solid line represents the universal curve for complex nuclei.}
\label{NucleonResonances}
\end{figure}

(iii) In the asymptotic range of the nucleon photoabsorption cross section
(beyond 1.5 - 2 GeV)
the scaling of the photoabsorption cross section with mass number
$A$ observed through
the universal curve of Fig.\ \ref{NucleonResonances}
is no longer valid. Instead we find a proportionality of the
photoabsorption cross section with $A^{\alpha}$ with $\alpha$
lying somewhere between 1 and 2/3, i.e.\ between a proportionality to the
nuclear volume and the  area of the   nuclear cross-section.
This change in proportionality is
due to the fact that the interaction of the intermediate-state vector
mesons with the nucleus is of very short range.
This property of the photoabsorption cross section is called shadowing
which is illustrated by Fig.\ \ref{shadow}.

\begin{figure}[htb]
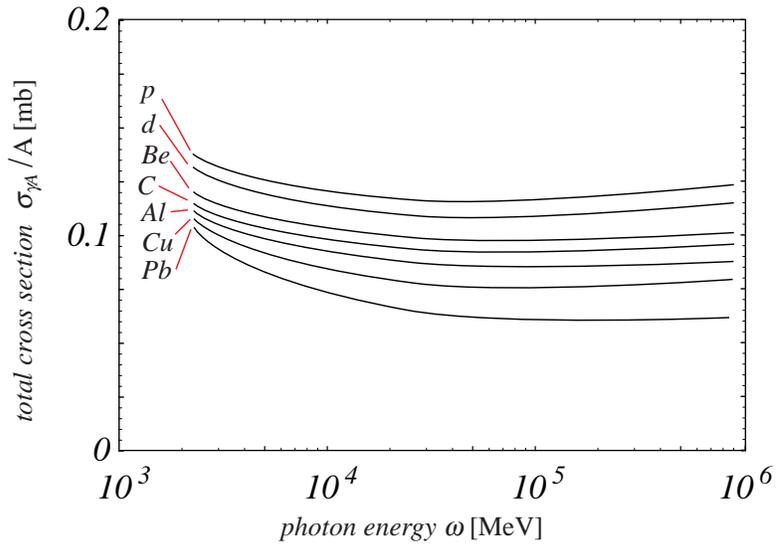
    
\centering
\caption{\em Fits to the total nuclear photoabsorption cross sections
per nucleon as an example of shadowing in the asymptotic region
\cite{ahrens85c}.}
\label{shadow}
\end{figure}

{\it Further properties of giant resonances: } \\
Giant resonances may be viewed as nuclear transitions in the shell model
from occupied states below the Fermi level to unoccupied states  above it.
For a given multipolarity these transitions occur coherently, which
may be visualized in terms of a collective model \cite{steinwedel50}. A further
aspect is the strong interaction between nucleons. This leads to the  need
of replacing particle-hole excitations by quasiparticle-quasihole
excitations, where the latter may be described in terms of the Fermi liquid
theory of finite nuclei. This aspect will be covered in section 7.

As contributions to Compton scattering, the giant resonances can be
characterized by their electromagnetic multipolarity. In this way a
definite dependence on the scattering angle is assigned to each resonance.
The most prominent contribution to  giant  resonances stems from
the isovector giant-dipole resonance (GDR) which is seen in all nuclei
including $^4$He. In heavy nuclei this giant dipole resonance
has  successfully been parameterized in terms of one Lorentzian line
in case of spherical nuclei, or in terms of the superposition of two
Lorentzian lines in case of deformed nuclei
\cite{berman75,dietrich88}.  Axially
deformed means prolate deformation as far as we know. There seems to be
only one case of an investigation of an oblate deformed
nucleus \cite{mondry91}, {\it viz.}\ $^{127}$I. However, in this case
it turned out that the giant resonance  did not behave like a collective
motion with respect to two independent axes.

Most of the available information on photoabsorption cross sections of
medium-mass to heavy nuclei has  been obtained by
photo-neutron experiments \cite{lepretre78}.
For light to medium-mass nuclei
photoabsorption cross sections in the GR and QD ranges have been measured
by photon attenuation experiments \cite{ahrens75}, because for these nuclei
photoneutron experiments do not
lead to information on the total absorption cross section due to sizable
contributions from  proton channels.

It is interesting to note
that the Lorentzian shape of the GDR extends below the
particle threshold where we find individual nuclear levels instead of
a continuum. Some details concerning the properties of
strengths of individual nuclear levels  has  already been
discussed in connection with Fig.\ \ref{GR-QD-absorption}.
In order to prove these expected properties, rather sophisticated  methods
had to be applied \cite{zurmuehl83}. A series of narrow $\gamma$ lines
($\Delta E \approx 10$ eV)     produced
via the neutron capture reaction was used to investigate nuclear
resonance fluorescence due to a random overlap of photons with nuclear
levels. Using statistical methods it was possible to get insight into
the properties of $E1$ excitations in the tail region of giant resonances.
It may be of interest to note that the same statistical ansatz applied
to the giant-resonance tail
also explains
the observed  distributions of electric-dipole strengths in nuclei
\cite{schuma85}.

Figure \ref{Multipoles} gives a schematic  view of the giant-resonance
structure of $^{208}$Pb for the first three multipoles, {\it viz.} E1, M1 and
E2 \cite{SC94}. Each multipole is split up
into an isoscalar and an isovector
component. For the E1 multipole the isoscalar component is identical
with Thomson scattering of the whole nucleus and therefore does not show
up as a separate excitation mode.

\begin{figure}[htb]
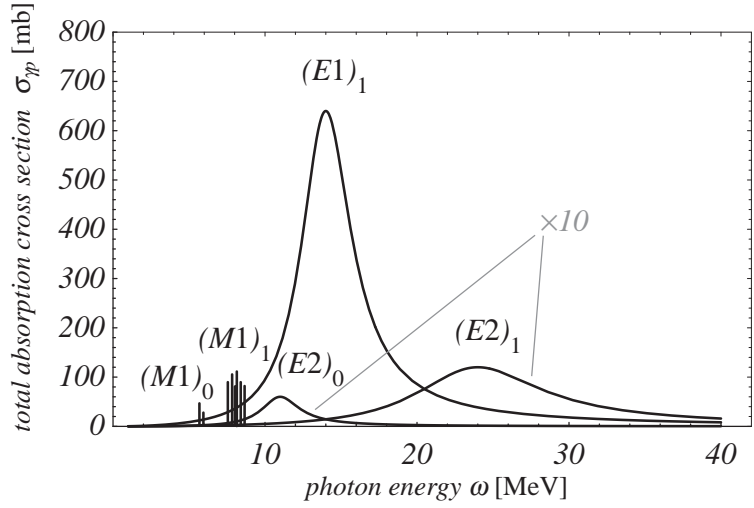
  
\centering
\caption{\it Schematic representation of  giant resonance
multipoles for $^{208}$Pb. (E1)$_1$: isovector giant electric-dipole
resonance (GDR), (M1)$_0$: isoscalar  giant magnetic-dipole
resonance, (M1)$_1$: isovector giant magnetic-dipole
resonance,(E2)$_0$: isoscalar giant  electric-quadrupole
resonance, (E2)$_1$: isovector giant  electric-quadrupole
resonance \cite{SC94}.}
\label{Multipoles}
\end{figure}

To summarize, the GR degree of freedom of the nucleus corresponds
to  the complete
set of one-quasiparticle one-quasihole states in a shell model. This
allows us to introduce a scattering amplitude $T_{GR}$ for the
giant-resonance degree of freedom which separately is gauge invariant,
if the non-resonant (seagull)  contributions are treated properly.

{\it Further properties of quasideuterons:}\\
Quasideuterons (QD) are proton-neutron pairs correlated through the
exchange of charged mesons. Therefore, QD are the smallest cluster
states in nuclei which may show up in the interaction with photons.
There are several approaches to a description of QD excitations of which
we mention the approaches of Levinger \cite{levinger51},
Gottfried \cite{gottfried58} and  Laget \cite{laget81}.
From the  point of view of a proper treatment of degrees of freedom and
sum rules,
the Laget model is the most
appropriate one , because it relates  the total
QD strength to exchange currents via the relation
\begin{equation}
\sigma_{QD}=L\frac{NZ}{A}\sigma^{ex}_D,
\label{lagetQD}
\end{equation}
where $\sigma_{QD}$ is the quasideuteron cross section, $\sigma^{ex}_D$
the exchange part of the deuteron photoabsorption cross section,
$Z$ and $N$ are the charge number and the neutron number, respectively, of
the nucleus and $L$ is  a  parameter.
This means that the total classical or non-mesonic
(Thomas-Reiche-Kuhn)
dipole strength  is contained in  the giant resonances
as is also suggested by shell model considerations. Then meson exchange
currents provide us with an enhancement of the giant resonance strength
and --- in addition --- with the QD strength.  Beyond this favorable
property the Laget formula has not to be taken too literally because
considerable differences are expected between deuterons and
quasi-deuterons.

For the  analysis of Compton scattering data it is most convenient
to use a Lorentzian representation as proposed by
Ziegler \cite{L1}.

Similar to the giant-resonance (GR) degree of freedom the (QD) degree
of freedom may be related to a complete set of isovector 2-particle
2-hole states, allowing us to introduce a separately gauge invariant
amplitude $T_{QD}$, again with the prerequisite that this amplitude is
properly accompanied by non-resonant (seagull) contributions.

{\it Further properties of the  $\Delta$ range of the nucleus:} \\
Figure \ref{NuclearDelta} shows the photoabsorption cross section
of a complex nucleus split up into different contributing components.
The two nonresonant components are due to the quasideuteron effect and
due to the Born term of the photo-pion amplitude. The $\Delta$ resonance
contribution is split up into two components.
The high energy part
with its maximum around 350 MeV corresponds to real-meson photoproduction
whereas the low energy part with its maximum around 260 MeV to
virtual-meson photoproduction.

\begin{figure}[htb]
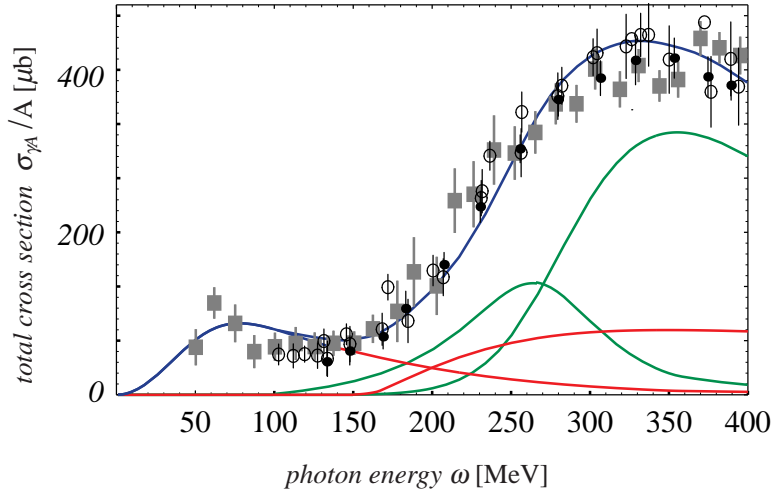
    
\centering
\caption{\em Total nuclear photoabsorption cross sections per nucleon.
Data are shown for nuclei between $^4$He and $^{238}$U
\cite{ahrens85c}. In addition the result of a model
calculation for the universal photoabsorption curve is shown
\cite{HuettMiSchu97}.
The left curve extending underneath the $\Delta$ resonance is the
QD cross section. The $\Delta$-resonance cross section is partitioned
into  two curves, into a non-mesonic part (left resonant curve) and a
mesonic part (right resonant curve). The nonresonant curve
in the $\Delta$ range corresponds to
the Born term of meson photoproduction.}
\label{NuclearDelta}
\end{figure}

This latter phenomenon is well known from photodisintegration experiments
on the deuteron where
a bump-like structure is observed in the non-mesonic cross section
at energies above meson threshold. This bump-like structure occurs through
excitation of the $\Delta$ resonance with subsequent two-nucleon
disintegration of the deuteron. In recent experimental
investigations on $^4$He \cite{wichmann96}  with the  subsequent
theoretical investigations \cite{HuettMiSchu97}
on complex nuclei it was shown that --- after a $\Delta$
excitation --- there is a branching of strength into the non-mesonic channel
and into the mesonic channel with a conservation of strength as compared
to the free nucleon. This consideration shows that the resonance-like
structure in the non-mesonic $pn$ channel is not part of the QD degree of
freedom but of the nucleon internal degrees of freedom which will be
described  by the amplitude $T_N$.
Such a decomposition of the nuclear photoabsorption cross section at
$\Delta $-resonance energies into a mesonic and a non-mesonic channel is
very well understood due to its manifestation in nucleon knock-out
reactions \cite{wakamatsu,oset,ryck1,ryck2,ryck4,ryck5}.

\subsection{Summary on the degrees of freedom of the nucleus}

The most clear-cut distinction
between external and internal degrees of freedom of nucleons in a
nucleus is
contained in the Compton scattering amplitudes of the individual nucleon,
where the
Born terms $T^{Born}$ \cite{lvov97}
correspond to the external degree of freedom and the non-Born terms
$T^{non-Born}$
to the internal ones. The leading Born term is given by the
Thomson scattering amplitude of a point-like particle. This term is
independent  of the photon energy. There are other contributions to the
Born terms which simultaneously depend on the charge and the anomalous magnetic
moment of the nucleon and on the photon energy $\omega $. These
latter terms do not contribute to the nuclear scattering amplitude
in the GR energy region but are in principle observable in the
QD energy range, though the effects are very small (see section 5).

To summarize, we have to discriminate between nuclear  degrees
of freedom (A), which may be
giant resonance states (GR) or quasideuteron states (QD), and nucleon (N)
degrees of freedom,  which may be resonant nucleon states
or asymptotic nucleon states. The correspondence between the
nucleon-structure independent  and nucleon-structure dependent terms
of the nucleon and nuclear Compton scattering amplitudes
is illustrated by Eqs.\ (\ref{ScattAmpN}) and (\ref{ScattAmpA})
\begin{eqnarray}
\label{ScattAmpN}
 T^{nucleon}&=&T^{Born}+T^{non-Born}\\
 \Updownarrow\,\,\,\,\,\,\,\,\,\,\,&&\,\,\,\,\,\, \Updownarrow
 \,\,\,\,\,\,\,\,\,\,\,\,\,\,\,\,\,\,\Updownarrow \nonumber \\
 \label{ScattAmpA}
 T_{tot}\,\,\,\,\,\,\,\,\, & = &\,\,\,\, T_A \,\,\,+\,\,\,\,\,\, T_N
 \end{eqnarray}
with
\begin{equation}
T_A = T_{GR} + T_{QD}.
\end{equation}
The first terms in Eqs.\ (\ref{ScattAmpN}) and (\ref{ScattAmpA})
are similar in the respect that they both show contributions
which are connected to low-energy effective degrees of freedom
of the object involved.

As an overview, the different
degrees of freedom and their main characteristics are summarized in Table
\ref{dofTable}.
From the discussion given above it is clear that the resonant peak at the
high-energy end
of the non-mesonic
$\gamma\, ^A\!X \to \,^{A-2}Y + p + n$
cross section is not part of the QD degree of freedom
but belongs to the nucleon-internal $(N)$ degree of freedom of the nucleus.

  \begin{table}[htb]
 \caption{\em Excitation modes and  degrees of freedom of the nucleus}
 \vspace{3mm}
 \begin{tabular}{|c|c|c|c|}
 \hline
 excitation mode   & corresponding amplitude
          &excitation mode & nuclear objects \\
 of the nucleus    &of the nucleon
          & of the nucleon& and phenomena \\
 \hline
 \hline
 giant &Born terms && quasi-p quasi-h states\\
 resonances (GR)& &   $ -$       &  in  an     average potential\\
 \hline
quasi deuteron&Born terms  &&    proton-neutron pairs\\
excitations (QD)&   &  $-$    &     correlated by charged\\
                   &      &   &meson exchange\\
 \hline
 scaling range&non-Born amplitude&meson-isobar & baryon resonances \\
 $\sigma \propto A$ & in the resonance range &structure &      and mesons in\\
  $(N)$        & & of the nucleon      &    the nuclear medium\\
\hline
 shadowing range&non-Born amplitude& vector dominance& in-medium vector \\
 $\sigma \propto
A^{\alpha}$,\,\, $\alpha < 1$& in the asymptotic range&and $t$-channel
 exchanges &
dominance \& direct \\
 $(N)$ &             &\& direct $\gamma$-quark couplings &
                  $\gamma$-quark couplings\\
\hline
\end{tabular}
 \label{dofTable}
\end{table}

\subsection{The role of Compton scattering in investigations of
the degrees of freedom of the nucleus}

At first sight Compton scattering simply appears to be one further reaction
channel among many others
which --- in addition --- is very difficult to measure. This view would
lead to the conclusion that Compton scattering is not capable of providing
any new information in addition to that obtained from other reaction
channels. This, of course,  is not the case. On the contrary,
Compton scattering is a reaction channel with unique properties
as will be outlined in the following.

Differing from photoparticle physics, Compton scattering is concerned
with the first step of the excitation process only, since the same
matrix element $\langle n| H_{int} |0 \rangle$ of the first-step transition
from the ground state $|0 \rangle$ to the excited state $|n \rangle$
enters into the
absorption and the emission processes. This favorable feature of Compton
scattering makes it possible to study the photoabsorption process without
the disturbing effects of subsequent processes, taking place between
the initial photoabsorption and the final particle emission. For this
reason it is possible to discuss the photoabsorption processes in terms of
the initial photoabsorption process only, and forget about all the other
processes. This property leads to a smaller model dependence of
Compton scattering than observed in photoparticle experiments. The next
favorable property of Compton scattering is its unambiguous relation to
the total photoabsorption cross section via the optical theorem and the
forward-angle dispersion relations. These relations fix the Compton
differential cross section in the forward direction so that additional
properties of the photoexcitation mechanisms and additional genuine
two-photon properties  ({\it e.g.} seagull amplitudes and electromagnetic
polarizabilities) may be studied through measurements of the
angular distribution of Compton scattered photons.

\clearpage
\section{Compton scattering below pion threshold: general aspects}

At present no quantitative consistent description of nuclear Compton
scattering below pion threshold exists which is based on first principles.
Over the years a variety of phenomenological models
has been developed, which provide insight into the physical mechanisms
responsible for contributions to the scattering amplitude.
Many important results were obtained by applying  dispersion relations
to the Compton amplitude (see {\it e.g.} \cite{GGT54,martin1}).
For those parts of the amplitude, which are not easily accessible by
dispersion theory, microscopic calculations have been attempted, {\it e.g.} for
distinct multipole excitations \cite{kurath}, for mesonic effects
\cite{fri76,R85,ME86,ME87,huett96}
and for the contribution from nucleon excitations \cite{drechsel84}.
In this way nuclear Compton scattering can contribute to the
understanding of such subnuclear aspects  as mesonic currents
between the nuclear constituents and virtual excitations of the nucleons.
In the last few years several important pieces of information have been
extracted from the experimental data. At low energies the relative
strengths of electromagnetic multipoles were analyzed
\cite{D0,D3,W5,S8,D12,fuhrberg89,SC94,hay} for
comparison with predictions from multipole sum rules. The interesting
question, whether the electromagnetic polarizabilities of the nucleon
inside the nucleus essentially differ from those of the free nucleon,
has been theoretically addressed \cite{R85,ME86,martin1,bunatian92,huett96}
and experimentally studied
with good accuracy \cite{H15,F16,P18a,martin3}.

In the process of nuclear Compton scattering
the exchange of mesons between nucleons, which represents the
nucleon-nucleon interaction, leads to specific observable phenomena in
the whole energy region considered here.
This can be understood from the fact that mesons, themselves, may be
charged and can, therefore, interact directly with photons.
Best known among these phenomena is the modification of the
Thomas-Reiche-Kuhn (TRK) sum rule \cite{L50}, i.e.\ the appearance
of the so-called enhancement constant $\kappa $.
Originally, the quantity
$\kappa$ was calculated \cite{L50} to be about 0.4
and this value was not in an
obvious disagreement  with the  experimental data then available. These data
\cite{berman75,dietrich88} were
mainly obtained for heavy nuclei and extended up to energies of  about
30 MeV, thus including mainly the giant-resonance region. Later on it was
shown that a considerable amount of photoabsorption strength is located
above the giant resonance region \cite{ahrens75,lepretre78}.
This strength is due to the excitation
of quasideuterons, i.e.\ of clusters consisting of proton-neutron
pairs  correlated through the exchange of charged mesons. Due to this,
the enhancement of giant-multipole strength is increased to $\kappa \sim$
1.0 in total.
A wide variety of model calculations has been carried out for the total
enhancement constant $\kappa $ as well as for the different contributions
to it, which are theoretically or experimentally accessible
\cite{BR80,brown4,arima73}, see also \cite{molin80}.
Parts of $\kappa $ could be related to parameters of Fermi liquid theory
and to the notion of quasi-particle masses \cite{BR95}. A compilation
of results can be found in \cite{EricsonWeise88}.
The contributions to $\kappa $ have also been studied in a
diagrammatic form \cite{R85,ME87},
similar to the approach described in section 9.
In addition, mesonic
exchange currents can imitate a modification of the nucleon
polarizabilities \cite{huett96}.
This contribution may be subtracted in order to
single out a change of the bound nucleon's polarizabilities from its
free values.
The advantage of this subtraction is related to the fact that
meson-exchange contributions are not a local effect
and are accompanied with two-body form factors.
The latter ones are distinctly different from the one-body form factor
which presumably accompanies all single-nucleon contributions, including
polarizabilities of the bound nucleon.
For a more detail discussion see sections 5 and 9.

The effect of mesonic exchange currents on the different electromagnetic
properties of nuclei has been the subject of a large number of research
articles, books and review papers. The reviews by Riska \cite{Riska} and
Arenh\"ovel \cite{arenh85}, as well as the book by Ericson and Weise
\cite{EricsonWeise88} give an excellent introduction to this field.
The description of mesonic exchange effects strongly depends on the
physical process considered. In this review only the scattering of
real photons is discussed. As no complete underlying theory of mesonic
effects in this process exists, most of the results are model-dependent.
Nevertheless, it is possible to obtain a qualitative description of
various features of the Compton amplitude within the approaches,
which will be discussed in this section.

\subsection{The photon-nucleus scattering amplitude in second-order
perturbation theory}

Similar to the case of atomic physics \cite{kane86}
the discussion of nuclear photon scattering may start from second-order
perturbation theory. But dissimilar to atomic physics this approach does
not lead to predictions which directly may be compared with experiments
because of the complexity of the physical situation. Nevertheless
second-order perturbation theory is an essential building block for the
construction of the general phenomenological theory. In principle,
there is no limitation of this approach with respect to the degrees
of freedom taken into account.
However, in the present section we restrict
the application of this method
to the energy region of giant resonances (GR).
Other effects, like the quasideuteron (QD) contribution and the
internal excitation of nucleons,
will be accounted for by alternative approaches.

The
amplitude for the scattering of a photon $\gamma$ by a nucleus $A$ in
its ground state $\gamma$ + $A$ $\to$ $\gamma'$ + $A'$ to the order
$ e^2$ (we set $\hbar =c=1$; $e$ is the charge of the proton and
$e^2=1/137$ is the fine-structure constant) reads
\begin{equation}
\label{sfi}
{\cal T }_{fi}  =
(2\pi)^4\delta^{(4)}(p_i+k_1-p_f-k_2) V \epsilon _1^{\mu }\epsilon _2^{\nu *}
\left(T_{\mu\nu}+S_{\mu\nu}\right)
\end{equation}
where $V$ is the total (infinite) volume, $T^{\mu \nu }$ is given by
\begin{eqnarray}
\label{tfi}
& & (2\pi)^4\delta^{(4)}(p_i+
k_1-p_f-k_2) V \epsilon_1^{\mu } \epsilon_2^{\nu *}
T_{\mu\nu} \nonumber \\
&=& -2\pi
\delta(E_0 + \omega_1 - E_f - \omega_2)  \nonumber
\epsilon_1^{\mu }\epsilon _2^{\nu *} \nonumber
\int d\vx \, d\vy
e^{i\vk_1\cdot\vx}e^{-i\vk_2\cdot\vy}\\
&& {} \times \sum_n\left\{
\frac{\langle f|j_{\nu}(\vy)|n\rangle
\langle n|j_{\mu}(\vx)|i\rangle}{E_0-E_n+\omega_1+i0}
+\frac{\langle f|j_{\mu}(\vx)|n\rangle
\langle n|j_{\nu}(\vy)|i\rangle}{E_0-E_n-\omega_2+i0}
\right\}
\end{eqnarray}
and $S^{\mu\nu}$ is the seagull amplitude \cite{molin80,christillin4} which
provides the gauge invariance for the amplitude ${\cal T }_{fi}$.
Therefore, one has
\begin{equation}
	k_1^{\mu}\,(T_{\mu \nu }+S_{\mu \nu })= (T_{\mu \nu }+S_{\mu \nu })\,
	k_2^{\nu} =0 .
	\label{gauge}
\end{equation}
Representations of $S_{\mu\nu}$ will be discussed below.
In Eq.\ (\ref{tfi}) $\epsilon _1^{\mu }$ and $\epsilon _2^{\nu }$ denote the
polarization vector of the incoming and outgoing photon, respectively.
In the following we will use the Coulomb gauge,
in which the photon polarization vector is
given by $\epsilon _i^{\mu } = (0,\veceps _i)$
with $\veceps _i\cdot \vk _i =0$.
Furthermore, for the sake of simplicity
we consider the case of linearly polarized photons and,
therefore, omit the complex conjugation for the polarization vector
$\veceps _2$.
The four-momenta of the incoming and outgoing photon are
denoted by $k_1 = (\omega _1,\vk _1)$ and $k_2 = (\omega _2,\vk _2)$,
respectively, while the corresponding
quantities for the nucleus are $p_i$ and $p_f$.
The
function $\vj(x)$ is the electromagnetic current associated with the
interaction of the photon with the nucleus, as will be discussed a little
later on.
The summation in (\ref{tfi})
is performed over  a complete set of eigenstates $|n\rangle$  of the nuclear
Hamiltonian with eigenvalues $E_n$.
The state $|i \rangle$
 of the energy $E_0$ stands for the nuclear ground state and
$|f\rangle$ for the final nuclear excited state in the laboratory system.
In the case of Compton
scattering initial and final states are  equal to each other except for the
total kinetic energy of the nucleus. In the laboratory frame
the photon energies $\omega_1$ and
$\omega_2$ are related via the Compton
formula
\begin{equation}
\label{ComptonFormula}
\omega_2=\frac{\omega_1}
   {\displaystyle 1 +
     \frac{\omega_1\rule{0ex}{1.5ex}} {AM} (1-\cos\theta)}
\end{equation}
where $M$ is the nucleon mass, $A$ again is the mass number of the
nucleus and $\theta $ is the scattering angle.

The seagull amplitude
$S^{\mu\nu}$  is identically equal to zero in the case of
a relativistic description of the nucleus and the absence of exchange
currents (or velocity dependent potentials in more general terms
\cite{molin80,christillin4}).
This case
has extensively been investigated for Compton scattering by atoms where
electrons are described as Dirac particles in a self-consistent central
potential \cite{kane86}. These
investigations have provided us with an insight into the general properties
of the amplitude given in Eq.\ (\ref{tfi}).

In the following, a non-vanishing seagull amplitude  enters into
the scattering amplitude for two  reasons. First, we pass to the
nonrelativistic form of the second-order perturbation theory. Then
the negative-energy intermediate states in Eq.\ (\ref{tfi})
give rise to a seagull amplitude $B(\omega,\theta )$.
The other reason are exchange currents which are a consequence of the
isovector nucleon-nucleon interaction. In this case gauge invariance may
be restored by introducing the mesonic seagull amplitude
$S(\omega ,\theta )=\varepsilon _\mu \varepsilon _\nu ^{*'}\,S^{\mu \nu }$.
The appearance of a seagull-like contribution $B$ in a nonrelativistic
description of photon-nucleus scattering can easily be
seen when the photon interaction is introduced in the nonrelativistic
nuclear Hamiltonian via minimal substitution.

\subsection{The nonrelativistic photon-nucleus interaction
and the concept of currents}

In a  nonrelativistic description
\cite{arenh85,chris90,C86,ericson73,baron91}
the interaction between
a nucleon and an electromagnetic field is given by the Hamiltonian
\begin{eqnarray}
\label{Hamiltonian}
H_{int} &=& \sum_{i=1}^A \Bigg[ \frac12 (1+ {\tau}^{(i)}_3)
\left( eA_0(\vr_i)-\frac{e}{2M} [\, \vp_i \cdot \vA(\vr_i) +
  \vA(\vr_i) \cdot \vp_i \, ]
+ \frac{e^2}{2M} \vA(\vr_i) \cdot \vA(\vr_i) \right)
\nonumber \\ && {} \qquad\qquad
 -\frac{e \hbar}{2M} \left( \mu_p \frac12 (1+ {\tau}^{(i)}_3)
  + \mu_n   \frac12 (1-{\tau}^{(i)}_3) \right) \vesigma ^{(i)}\cdot
        (\vecnab \times \vA) \Bigg],
\end{eqnarray}
where $\vesigma ^{(i)}/2$, ${\tau}^{(i)}_3/2$, $\vp_i$ and
$\vr_i$ are the spin operator, the
third component of the isospin operator, the three-momentum and the
coordinate vector of the $i$th nucleon, respectively. The quantities
$\mu _p$ and $\mu _n$ are the magnetic moments of the nucleons.

The term quadratic in $\vA(\vr)$ is related to the
kinetic seagull amplitude $B(\omega,\theta )$. All terms linear
in $\vA(\vr)$ contribute to the one-nucleon electromagnetic
current. The terms depending on the magnetic moments
give rise to spin currents which
are responsible for magnetic spin-flip transitions in a nucleus.
With the help of the usual relation
\begin{equation}
	j_\mu (\vr)=\left. {{{\delta H_{int}}
	\over {\delta A^\mu (\vr)}}} \right|_{A_\nu =0}
	\label{current}
\end{equation}
one obtains the following representation of the
one-body contribution to the nuclear
charge density $\rho (\vr )$ and current $\vj(\vr )$:
\begin{eqnarray}
\label{OneBodyCurrents}
\rho^{[1]}(\vr)  &\equiv &\sum\limits_i {\rho _i^{[1]}(\vr)} =  e
\sum_i \frac{1 + \tau ^{(i)}_3}{2} \delta(\vr - \vr_i)
\nonumber \\
\vj^{[1]}(\vr) &=& e \sum_i  \frac{1 + \tau ^{(i)}_3}{2}
\left( \frac{ \vp_i}{2 M}  \delta(\vr - \vr_i) +
 \delta(\vr - \vr_i)  \frac{ \vp_i}{2 M} \right) \nonumber \\
&-& \frac{e}{2 M} \sum_i \left( \mu_p \frac{ 1+ {\tau}^{(i)}_3}{2}
+  \mu_n   \frac{1-\tau ^{(i)}_3}{2} \right) {\vesigma }_i
\times {\vecnab}\,\,
\delta  (\vr - \vr_i)
\end{eqnarray}

The exchange of charged mesons between proton and neutron leads to an
additional contribution to the electromagnetic current.
The total electromagnetic current fulfills a continuity equation, which
can be decomposed into a relation for the one-body part,
\begin{equation}
\label{ContinuityEq}
\mbox{\boldmath$\nabla $} \cdot \vj^{[1]}(\vr) = -i
\left[ H_0,\rho^{[1]}(\vr)
\right],
\end{equation}
and a similar expression for the two-body current
\begin{equation}
	\vecnab \cdot \vj^{[2]}(\vr)=
	-i[\sum\limits_{i<j} {V_{ij}^{(ex)}},\rho ^{[1]}(\vr)] .
	\label{cont1}
\end{equation}
Here $H_0$ stands for the kinetic and central part of the nuclear
Hamiltonian $H$, while $V_{ij}^{(ex)}$ is the two-nucleon exchange potential.
In Eq.\ (\ref{cont1}) it was taken into account that in the limit of
point-like, static nucleons the
two-body charge density $\rho ^{[2]}$
vanishes since $\rho ^{[2]}(\vr )\propto 1/M$. The vanishing of
$\rho ^{[2]}$ in the static limit is known as Siegert's theorem
\cite{siegert37}.
From $\rho ^{[1]}(\vr )$ one obtains a representation of the operator
$\vD$ for the total dipole moment
\begin{equation}
	\vD=\int {d\vr}\,\rho ^{[1]}(\vr)\,\vr=
	e\sum\limits_i {{{1+\tau _3^{(i)}} \over 2}}\,\vr_i .
	\label{dipole}
\end{equation}
Using this expression, Eq.\ (\ref{cont1}) and representing the exchange
potential as
$V_{ij}^{(ex)}=\vetau ^{(i)}\cdot
\vetau ^{(j)}\;v_{ij}^{(ex)}$,
where $v_{ij}^{(ex)}$ does not depend on the isospin variables,
one finds the following form of the total two-body current
\begin{equation}
	\vJ^{[2]}=\int {d\vr}\;\vj^{[2]}(\vr)=
	i[\sum\limits_{i<j} {V_{ij}^{(ex)}},\;\vD]=-e\sum\limits_{i<j}
	{\left( {\vetau ^{(i)}\times \vetau ^{(j)}}
	\right)_3(\vr_i-\vr_j)v_{ij}^{(ex)}}.
	\label{curr}
\end{equation}

The largest contribution to the two-body current
comes from the $\pi$ and
$\rho$ meson exchange parts of the two-nucleon potential $V_{ij}^{(ex)}$,
which in momentum
space are given by
\begin{eqnarray}
\label{ExchangePotentials}
V^{(\pi )}_{ij}(\vq) &=& - \frac{f^2}{m^2_{\pi}}
(\vetau ^{(i)}\cdot \vetau ^{(j)})
\frac{(\vesigma _i \cdot \vq) (\vesigma _j \cdot \vq) }
{\vq^2 + m^2_{\pi}},
\nonumber \\
V^{(\rho )}_{ij}(\vq) &=& -\frac{f^2_{\rho }}{m^2_{\rho}}
(\vetau ^{(i)}\cdot \vetau ^{(j)})
\frac{(\vesigma _i \times \vq)\cdot (\vesigma _j \times \vq)}{
\vq^2 + m^2_{\rho}}  .
\end{eqnarray}
Here $m_\pi $ and $m_\rho $ is the pion and $\rho $-meson mass,
respectively, $f^2/4 \pi \simeq 0.08$
and $ f^2_{\rho }/4 \pi \sim 4.5-5.5$.

It is convenient to introduce the two-body current in momentum space via
\begin{equation}
	\vj^{[2]}(\vr)={1 \over 2}\sum\limits_{i\ne j} {\int {{{d\vk_1}
	\over {(2\pi )^3}}}}\int {{{d\vk_2} \over {(2\pi )^3}}}\;\exp
        \Big( {i\vk_1\cdot (\vr_i-\vr)+i\vk_2\cdot (\vr_j-\vr)}
        \Big)\,\vj_{ij}^{[2]}(\vk_1,\vk_2)  .
	\label{current2}
\end{equation}
The simplest way to arrive at an explicit form of
the contribution of pion exchange
to the current $\vj_{ij}^{[2]}(\vk_1,\vk_2)$ is
a direct calculation of the Feynman diagrams shown in Figure \ref{pr22}
({\it cf.} section 9).
In these diagrams the vertices are obtained via a minimal substitution
in the corresponding Hamiltonians.
As a result one finds \cite{Riska}:
\begin{eqnarray}
     \vj_{ij}^{[2]}(\vk_1,\vk_2)&=&-ie{{f^2}
	\over {m_\pi ^2}}\left( {\vetau ^{(i)}\times \vetau ^{(j)}}
     \right)_3 \left\{ {{{\vecsigma _i\left( {\vecsigma _j\cdot
	\vk_2} \right)} \over {m_\pi ^2+\vk_2^2}}-
	{{\vecsigma _j\left( {\vecsigma _i\cdot \vk_1} \right)}
	\over {m_\pi ^2+\vk_1^2}}+} \right.
	\nonumber \\
	&&\left. {(\vk_2-\vk_1){{\left( {\vecsigma _i\cdot \vk_1}
	\right)} \over {m_\pi ^2+\vk_1^2}}\,{{\left( {\vecsigma _j\cdot
	\vk_2} \right)} \over {m_\pi ^2+\vk_2^2}}} \right\} .
	\label{current3}
\end{eqnarray}
Similar expressions can be obtained for the contribution of $\rho $-meson
exchange \cite{Riska}.

In a nonrelativistic second-order perturbation theory
the resonance
amplitude $R(\omega,\theta)$ is given by the contribution to
$\epsilon _1^{\mu }\epsilon _2^{\nu } T_{\mu\nu}$ of
only positive-energy intermediate states in the r.h.s.\ of Eq.\ (\ref{tfi}).
This corresponds to using a nonrelativistic Hamiltonian as a starting
point.

In the nonrelativistic approach gauge invariance is provided by
introducing a non-resonant (seagull) amplitude, which is non-zero even
in the absence of mesonic exchange currents.
This part of the seagull amplitude corresponds to the
contribution of negative-energy intermediate states in Eq.\ (\ref{tfi})
in the nonrelativistic approximation.
This so-called kinetic seagull amplitude can be written as
\begin{eqnarray}
	 B(\omega ,\theta )=-{1 \over M}\,\veceps _1\cdot
	 \veceps _2\,\left\langle f \right|\sum\limits_j
	 {\int {d\vx\,d\vy}}\;e^{i\vk_1\cdot \vx}e^{-i\vk_2\cdot
	 \vy}\rho _i^{[1]}(\vx-\vr_i)\,\rho _i^{[1]}(\vy-
	 \vr_i)\left| i \right\rangle
	\nonumber \\
	=-{{e^2} \over M}\,\veceps _1\cdot
	\veceps _2\,\left\langle f \right|\sum\limits_j
	{{{1+\tau _3^{(j)}} \over 2}\exp \left[ {i(\vk_1-\vk_2)\cdot
	\vr_j} \right]}\;\left| i \right\rangle =
	- \frac{Ze^2}{M}\veceps _1 \cdot \veceps _2 F_1(q) ,
	\label{Bkin}
\end{eqnarray}
where $\vq  = \vk_1 -  \vk_2 $ and $F_1(q)$ is the form factor
describing the distribution of protons inside the nucleus.

This representation of the kinetic seagull amplitude is
frequently used in predictions of nuclear photon scattering.
Using (\ref{Bkin}) to represent the
negative-energy states
in (\ref{tfi}) certainly is only an approximation.
However, below meson photoproduction threshold
the possible uncertainty is only of the order
of a few percent as discussed in \cite{SC94}.
Furthermore, the
possibility of using a modified form factor which incorporates
binding effects has been discussed in
\cite{martin1}. This modified
form  is given by \cite{kane86}
\begin{equation}
\label{F1qmodified}
F_1^{(mod )}(q)={1 \over Z}\,\left\langle f
\right|\sum\limits_j {{{1+\tau _3^{(j)}} \over 2}\,\exp
\left[ {i\vq\cdot \vr_j} \right]{M \over {E-V(\vr_j)}}}\;\left| i
\right\rangle
\end{equation}
where $E$ is the total  energy of the nucleon bound in the potential
$V(r)$  which is assumed to be the 4th component of a Lorentz vector.
The modified form factor has been introduced in connection with
atomic Rayleigh scattering \cite{franz35} and later was
investigated and explained   by
Brown et al.\ \cite{brown57}. The predictions of
Brown et al.\ \cite{brown57} concerning the modified form factor
were experimentally confirmed by Schumacher \cite{schumacher69},
who showed that atomic Rayleigh scattering data require the
application of the modified form factor at least for the helicity non-flip
amplitude.

In the hypothetical case of structureless nucleons and the absence
of exchange currents the cross section for nuclear photoabsorption
would  contain  only
giant resonances and, hence, go to zero at about 30 to 40 MeV.
The amplitude $R_{GR}(\omega ,\theta )+B(\omega ,\theta )$ would provide
a complete description of nuclear photon scattering.
Furthermore,
the integral over the total photoabsorption cross section would
fulfill the usual Thomas-Reiche-Kuhn (TRK) sum rule without any
enhancement. Relativistic corrections, which could
in principle destroy this relation, are negligibly small in the giant
resonance region.
If we keep the nucleons structureless
but take into account meson exchange
currents the giant resonance amplitude $R_{GR}$ is modified. Furthermore,
an additional contribution $R_{QD}$ to the resonance amplitude appears,
the so-called quasi-deuteron scattering, which will be discussed in
section 5. This new amplitude $R_{GR}+R_{QD}+B$ is no longer gauge
invariant by itself and a counter term
$S(\omega ,\theta )$, which is
usually called the mesonic seagull amplitude,
is needed to restore gauge invariance.
The total amplitude $T_A$ corresponding to all nuclear degrees of freedom
is then given by
\begin{equation}
	T_A = R_{GR}+R_{QD}+B+S  = R_{GR}+R_{QD}+S_{tot},
	\label{fullamplitude}
\end{equation}
where $S_{tot}$ is the sum of the kinetic and mesonic seagull amplitudes.
The contribution $T_N$ of nucleon-internal degrees of freedom
to the nuclear Compton amplitude, which is not included in
(\ref{fullamplitude}), will be discussed in sections 5 and 6. As
will be seen, below pion threshold the main contribution to this
amplitude is due to the electric and magnetic polarizabilities of the
bound nucleon.

The microscopic consideration of meson exchange has lead to
explicit calculations of the amplitude $S$.
Such calculations have been carried out by
Alberico and Molinari \cite{A82} based on expressions derived by
Christillin \cite{christillin4,C75,C78} and Friar
\cite{friarann}, as well as by H\"utt and Milstein \cite{huett96,huett97}.
In the static limit the mesonic seagull amplitude
at $\omega $=0 is of the following form:
\begin{eqnarray}
	\left. {{\tilde S}(\omega ,\theta )} \right|_{\omega =0}&=&
	{i \over 2}\left\{ {\left\langle f \right|[\veceps _1\cdot
        \vD\,,\;\veceps _2\cdot \vJ^{[2]}]+[\veceps _2\cdot
        \vD\,,\;\veceps _1\cdot \vJ^{[2]}]\left| i \right\rangle }
	\right\}
	\nonumber \\
	&=&{{2e^2} \over 3}\,\veceps _1\cdot
	\veceps _2\;\left\langle f \right|\sum\limits_{j<k}
	{\left( {\tau _+^{(j)}\tau _-^{(k)}+\tau _-^{(j)}\tau _+^{(k)}}
	\right)\;(\vr_j-\vr_k)^2}\;v_{jk}^{(ex)}\left| i \right\rangle ,
	\label{seaglimit}
\end{eqnarray}
where $\tau _\pm ^{(j)}=1/2\;(\tau _1^{(j)}\pm i\,\tau _2^{(j)})$
are the isospin raising and lowering operators.
Here and in the following we use the symbol ${\tilde S}(\omega,\theta)$
to denote a static approximation of $S(\omega,\theta)$.
Due to the isospin structure
of the matrix elements in Eq.\ (\ref{seaglimit}) only
proton-neutron pairs contribute to the mesonic seagull amplitude.
Taking into account the effect of retardation, but still neglecting any
other dependence on the photon energy $\omega $, which is connected with
the $\omega $- and $\vk$-dependence of the mesonic propagators, one has
\begin{equation}
	{\tilde S}(\omega ,\theta ) =
	{{2e^2} \over 3}\,\veceps _1\cdot
	\veceps _2\;\left\langle f \right|\sum\limits_{j<k}
	{\left( {\tau _+^{(j)}\tau _-^{(k)}+\tau _-^{(j)}\tau _+^{(k)}}
	\right)\;(\vr_j-\vr_k)^2}\;v_{jk}^{(ex)}
        \exp \left[ {-i\vq \cdot (\vr_j+\vr_k)/2} \right]
	\left| i \right\rangle
	\label{seag2}
\end{equation}
It is convenient to write this limit of the mesonic
seagull amplitude in the form
\begin{equation}
\label{Sot}
{\tilde S}(\omega,\theta) = - \veceps _1 \cdot \veceps _2
\frac{NZ}{A} \frac{e^2}{M} \kappa F_2(q)
\end{equation}
where $\kappa $ is the enhancement constant and
$F_2(q)$ is the two-body form factor, which characterizes the
distribution of correlated proton-neutron pairs inside the nucleus.
The function $F_2$ is normalized in such a way that $F_2(0)$=1.
A first attempt to qualitatively discuss two-body or exchange form
factors within a microscopic model has been made by Alberico and Molinari
\cite{A82} for the case of $^{208}$Pb. In section 9 the
construction of such form factors is described on more general grounds
within a modified Fermi gas. To this end the full mesonic seagull
amplitude can be represented as a convolution of a spin-isospin
correlation function with matrix elements arising from amputated diagrams
for meson exchange.

\subsection{Low-energy behavior of the scattering amplitude and
dispersion relation}

In this section we discuss another view towards a decomposition of the
Compton scattering amplitude into a sum of resonant and non-resonant parts.
While in the previous section the consideration was based on second-order
perturbation theory and the requirement of gauge invariance, here our
starting point is a dispersion relation for the scattering amplitude
in forward direction.
At the end of the last section the  nuclear
Compton amplitude $T_A$ presented itself as a sum of three
contributions (\ref{fullamplitude}) provided by different physical mechanisms
(see also \cite{ziegl85,martin1,martin3}):
the collective nuclear excitations (giant resonances),
$R_{GR}(\omega ,\theta )$, the scattering by quasi-deuteron clusters,
$R_{QD}(\omega ,\theta )$, and the so-called seagull amplitude,
$S_{tot}(\omega ,\theta )$.
The physical background of this separation is the following:
Via optical theorem and a subtracted dispersion relation the scattering
amplitude in the forward direction is  up to an
additive constant determined by the total photoabsorption cross section.
This cross section contains resonant structures, which at different
energies correspond to the two different excitation mechanisms ({\it cf.} Fig.\
\ref{GR-QD-absorption}) which we call giant-resonance (GR) and quasideuteron
(QD), as was discussed in section 2.

An experimental separation of GR and QD excitations is relatively easy
for heavy nuclei where the giant-dipole resonance (GDR) cross section
containing by far most of the total GR strength is given by
a Lorentzian line in case of spherical nuclei as discussed in section 2.
In case of the validity of a Lorentzian shape rather precise information on the
integrated strength in the giant resonance may be obtained because this
Lorentzian shape may be used to separate the GDR cross section
from the QD cross section.

On the other hand also the shape of the QD
cross section \cite{levinger51,gottfried58,laget81}
is known either in the form of the Levinger parameterization
\cite{levinger51},
the Laget parameterization \cite{laget81} or a (QD) Lorentzian
representation ({\it cf.} sections 2 and 5). Then the shape of the QD cross
section extrapolated to above the pion threshold may be used to
separate the QD excitation from the nucleon internal excitation. In
practice this is achieved with sufficient precision by adjusting a
Lorentzian curve to the QD cross section below pion threshold.

In this sense the nuclear photoabsorption cross section
$\sigma _A(\omega )$ below pion threshold and slightly above
can experimentally be
separated into two parts,
\begin{displaymath}
	\sigma _A(\omega )=\sigma _{GR}(\omega )+
	\sigma _{QD}(\omega ).
\end{displaymath}
This serves as a means to construct the resonance parts
of the scattering amplitude
$R_{GR}$ and $R_{QD}$ as
\begin{equation}
\label{R-GR-disp}
        \mbox{Re} \left( {R_{GR}(\omega ,0)-R_{GR}(0,0)} \right)=
	{{\omega ^2} \over {2\pi ^2}}\,{\cal P}\int\limits_0^\infty
	{{{\sigma _{GR}(\omega ')\,d\omega '} \over {\omega '^2-\omega ^2}}}
	\label{mesonGR1}
\end{equation}
and
\begin{equation}
        \mbox{Im} R_{GR}(\omega ,0)=
  {\omega  \over {4\pi }}\,\sigma _{GR}(\omega ) .
	\label{mesonGR2}
\end{equation}
The same is valid for $R_{QD}$. This widely used procedure corresponds to the
application to the once-subtracted forward-angle dispersion relation
and of the optical theorem to partial cross sections. This procedure
is legitimate because the GR and QD excitations correspond to separate
degrees of freedom.

The application of dispersion relations to the total
nuclear Compton cross section
dates back to the early work of Gell-Mann, Goldberger and Thirring
\cite{GGT54}.
There it was shown that for the total scattering
amplitude $T_{tot}(\omega,\theta)$, which
--- in our language --- includes all degrees of freedom of the nucleus
a subtracted dispersion relation may be formulated:
\begin{equation}
\label{DispRel}
\mbox{Re}\left( T_{tot}(\omega,0) - T_{tot}(0,0) \right) =
\frac{\omega^2}{2 \pi^2} {\cal P} \int_0^{\infty}
\frac{\sigma_{tot}(\omega')}{{\omega'}^2-{\omega}^2}d\omega' ,
\end{equation}
where $T_{tot}(0,0)=-Z^2e^2/AM$.
For convenience in the further discussion we introduce a ``free-nucleon
photoabsorption cross section" via
\begin{equation}
\label{FreeNuclAbs}
A \sigma_N^{free}(\omega) = Z \sigma_p({\omega}) + N \sigma_n(\omega) ,
\end{equation}
where $\sigma_p(\omega)$ and $\sigma_n(\omega)$ are the free proton and
neutron photoabsorption cross sections, respectively. The cross section
of Eq.\ (\ref{FreeNuclAbs}) defines
a ``free-nucleon scattering amplitude"
for the ensemble of nucleons in a nucleus through
\begin{equation}
\label{FreeNuclAmp}
\mbox{Re}\left( T_N^{free}(\omega,0) -  T_N^{free}(0,0) \right)=
\frac{\omega^2}{2 \pi^2} {\cal P} \int_0^{\infty}
\frac{A\sigma_N^{free}(\omega')}{{\omega'}^2-{\omega}^2}d\omega'
\end{equation}
with $T_N^{free}(0,0)= -Ze^2/M$.
Similarly, one may think of the difference between the total
photoabsorption cross section $\sigma_{tot}$
and the one $\sigma_A$, where only nuclear degrees of
freedom contribute, as a bound-nucleon cross section:
\begin{equation}
\label{BoundNuclCross}
\sigma_N^{bound}(\omega) = \frac1A (\sigma_{tot}(\omega) -
\sigma_A(\omega)).
\end{equation}
From (\ref{DispRel}), (\ref{FreeNuclAmp}) and (\ref{BoundNuclCross})
we obtain
\begin{eqnarray}
\label{GGTrel}
\mbox{Re}\left( T_{tot}(\omega,0) -  T_N^{free}(\omega,0) \right) &=&
\frac{NZ}{A}\frac{e^2}{M}+
\frac{\omega^2}{2 \pi^2} {\cal P} \int_0^{\infty}
\frac{\sigma_A(\omega')}
{{\omega'}^2-{\omega}^2}d\omega'\nonumber \\  &+&
\frac{\omega^2}{2 \pi^2} {\cal P} \int_0^{\infty}
\frac{A \left(  \sigma_N^{bound}(\omega') - \sigma_N^{free}(\omega')
\right)}
{{\omega'}^2-{\omega}^2}d\omega'
\end{eqnarray}
which is the basis of our further conclusions.

For $\omega \to \infty$ Eq.\ (\ref{GGTrel}) provides us with the
Gell-Mann-Goldberger-Thirring (GGT) sum rule\footnote{In the original
work the nuclear (A) and nucleon internal (N) d.o.f.\ were simply
separated through a cut at pion threshold.}, {\it viz.}
\begin{eqnarray}
\label{GGTsumrule}
\int_0^{\infty} \sigma_A(\omega)d\omega =2\pi^2\frac{NZ}{A}\frac{e^2}{M}
&+&2\pi^2 \mbox{Re}\left( T_{tot}(\infty,0) -  T_N^{free}(\infty,0) \right)
\nonumber \\
&+&\int_0^{\infty}A(\sigma_N^{free}(\omega)-\sigma_N^{bound}(\omega))d\omega.
\end{eqnarray}
In the original paper \cite{GGT54}
it was assumed that the
asymptotic term in  Eq.\ (\ref{GGTsumrule}) is equal to
zero. Then Eq.\ (\ref{GGTsumrule}) has a very attractive interpretation
by stating that the  enhancement of strength contained in the integral
over the nuclear photoabsorption cross section $\sigma_A(\omega)$ is
equal to the reduction of strength in the nucleon photoabsorption cross
section due to nuclear binding. Later on it was realized
\cite{weiseGGT1,weiseGGT2}
that the basic assumption of GGT
about the properties of the asymptotic term in (\ref{GGTsumrule})
is not correct because of vector dominance phenomena. Due to this fact
the GGT sum rule needs refinements. The refinement  discussed by
Ahrens et al.\ \cite{weiseGGT3}
fixes the dispersion integral in the second
line of (\ref{GGTrel}) at $\omega=314$  MeV instead of $\omega \to \infty$
thus making use of  zero crossings of the real parts of scattering
amplitudes. Although in this way reasonable numbers are obtained for the
enhancement constant $\kappa$ the interesting physical content of the
original GGT sum rule  \cite{GGT54} is not preserved. Furthermore,
there seem to be no indications for a reduction of photoabsorption strength
above pion threshold which might show up as an enhancement of
photoabsorption strength below pion threshold  \cite{ahrens98}.

Some new aspects of the GGT-sum rule may be obtained by showing
that this sum rule is not a property of the photoabsorption
cross section but rather a property of the mesonic seagull-amplitude
$S(\omega,\theta)$. The arguments leading to this conclusion are as
follows.
The total seagull amplitude $S_{tot}$, which is the sum of the kinetic
and mesonic seagull amplitudes,
has an imaginary part only above pion threshold.
From Eq.\ (\ref{GGTrel}), together with (\ref{mesonGR1}) one obtains
a dispersion representation of the mesonic seagull amplitude in the form
\cite{R85,martin1}
\begin{equation}
\label{PolCorr}
\mbox{Re} {S(\omega,0)} -S(0,0) + \mbox{Re} {T_N(\omega,0)} =
\frac{\omega^2}{2\pi^2} {\cal P}
\int_{m_\pi }^{\infty} \frac{A \sigma_N^{bound}(\omega')}
{\omega'^2-\omega^2}d\omega'  ,
\end{equation}
where on the l.h.s.\ the energy is restricted to the range below meson
threshold where the amplitudes are real. This relation has the
following interpretation. The difference $S(\omega,0)-S(0,0)$ is the
``dynamic" part of the mesonic seagull amplitude at zero angle, i.e.\ that
part which is energy dependent in the forward direction, corresponding
to meson exchanges beyond the static limit of the meson exchange potential.
In the order $\omega^2$ the l.h.s.\ of Eq.\ (\ref{PolCorr})
corresponds to the electromagnetic polarizabilities of the nucleon and
their modification due to binding.
From (\ref{PolCorr}) we are led to the supposition
that the quantity $\sigma_N^{bound}(\omega)$
should  consist of two different parts one,
${\sigma'_N}^{bound}(\omega)$, related to $T_N$
and an other, $\Delta \sigma^{(S)}(\omega)$, related to
$S(\omega,0)-S(0,0)$. Below meson photoproduction threshold the amplitude
$T_N$ may be expressed in terms of the in-medium electromagnetic
polarizabilities  ${\tilde \alpha}_N$ and ${\tilde \beta}_N$
(see section 5.2 for more details)  via
\begin{equation}
T_N(\omega,\theta)= A \left( {\tilde \alpha}_N g_{E1}(\theta)
+ {\tilde \beta}_N g_{M1}(\theta) \right) \omega^2 F_1(q)+{\cal
O}(\omega^4).
\label{in-mediumPol}
\end{equation}
Then the supposition implies that only the part
${\sigma'_N}^{bound}(\omega)$ is
responsible for  the $genuine$ in-medium electromagnetic
polarizabilities ${\tilde \alpha}_N$ and ${\tilde \beta}_N$
whereas the rest, $\Delta \sigma^{(S)}(\omega)$, leads to the
relations
\begin{eqnarray}
\mbox{Re} {S(\omega,0)} - S(0,0)&=&\frac{\omega^2}{2\pi^2}{\cal P}
\int_{m_{\pi}}^{\infty}\frac{-
\Delta\sigma^{(S)}(\omega')}{{\omega'}^2-\omega^2}d\omega' \nonumber\\
\mbox{Im} {S(\omega,0)}&=&-\frac{\omega}{4\pi}\Delta\sigma^{(S)}(\omega).
\label{MesonRelations}
\end{eqnarray}
The sign of $\Delta \sigma^{(S)}(\omega)$ in (\ref{MesonRelations})
has been chosen negative in order
to make this quantity itself positive. The proof of this statement
is given by Eq.\ (\ref{S-rule}) where the positive right side of the
equation is consistent with a negative signs in (\ref{MesonRelations}).
In the  case that
\begin{equation}
\lim_{\omega \to \infty} S(\omega,0) =0
\label{Sinfty}
\end{equation}
which appears to us plausible,
we are lead to a sum rule in the following form
\begin{equation}
\int_{m_{\pi}}^{\infty} \Delta \sigma^{(S)} (\omega')d\omega'
     =2\pi^2 \frac{NZ}{A} \frac{e^2}{M} \kappa
\label{S-rule}
\end{equation}
 where use has been made of Eq.\ (\ref{Sot}).
In a formal sense the sum rule (\ref{S-rule}) is exactly identical
to what Gell-Mann Goldberger and Thirring \cite{GGT54} had in mind,
but the physical interpretation is different in two essential points:\\
(i) The sum rule (\ref{S-rule}) does not describe a property
of the photo\-absorption cross section but rather a property of the
mesonic seagull amplitude $S(\omega,\theta)$. This is an important
difference, because due to this, $\Delta \sigma^{(S)}(\omega)$ may not  be
interpreted  in terms of a reduction of photoabsorption
strength of the individual nucleons through a transfer to energies
below $\pi$-threshold by some mechanism.
Instead, $\Delta \sigma^{(S)}(\omega)$
has to be understood as being due to an   additional two-body effect. \\
(ii) The condition for the validity of the sum rule (\ref{S-rule})
is not the equality of absorption cross-sections for the free and bound
nucleon in the high-energy limit, but instead is given by the
property (\ref{Sinfty}) of the mesonic seagull amplitude which should
already be fulfilled in the resonance region of the nucleon photoabsorption
cross-section. 
This point has been cleared up by Schumacher et al. \cite{martin1}.
A favorable consequence of this finding is that the sum
rule (\ref{S-rule}) does not interfere with the shadowing effect as
suspected in \cite{weiseGGT1,weiseGGT2,weiseGGT3}.



In addition to the fulfillment of a dispersion relation, the main
constraint on the nuclear Compton amplitude is the low-energy
theorem \cite{friarann,silbar,thirring}.
It states that the Compton scattering amplitude at
$\omega =0$ is equal to the (coherent) Thomson limit
\begin{equation}
	T_A(0,\theta )=-\,\veceps _1\cdot \veceps
	_2\;{{Z^2e^2}
	\over {AM}}    .
	\label{thomson}
\end{equation}
Let us represent the low-energy limit of the
resonance part of the Compton amplitude in a form which is convenient
for the further discussion:
\begin{equation}
	R_{GR}(0,\theta )=\,\veceps _1\cdot \veceps _2\;{{e^2}
        \over M}\,\,{ZN \over A}\,(1+\k)    ,
	\label{resonance}
\end{equation}
where $\k$ is a constant which, as it will become clear later,
is the essential parameter characterizing all phenomena connected with the
enhancement of sum rules due to mesonic effects in the giant resonance
region.
Taking into account the fact that $R_{GR}(\omega ,0)$
vanishes as $\omega $ tends to infinity,
with the help of the dispersion relation
(\ref{mesonGR1}) one has
\begin{equation}
\frac{1}{2 \pi^{2} }\, \int\limits_{0}^{ \infty }
\sigma _{GR}(\omega ) d\omega = R_{GR}(0,0) =
    \frac{NZ}{A}  \frac{e^2}{M} (1+ \k).
\label{mesonGR3}
\end{equation}

In the limit of $\omega=0$, one could expect that only
the electric dipole operator contributes to scattering and
that only unretarded quantities should appear.
Therefore, one might think that the unretarded $E1$ resonance amplitude
\begin{equation}
   \Ro_{GDR}(0,\theta) = \veceps _1\cdot \veceps _2
    \;\frac{e^2}{M} \; \frac{ZN}{A} \,(1+\ko),
\label{R-GDR-unretarded}
\end{equation}
which introduces the unretarded enhancement parameter $\ko$
given by the unretarded electric-dipole photoabsorption cross section
$\so_{GDR}$ (again $GDR$ stands for ``giant dipole resonance"),
\begin{equation}
  \frac{1}{2 \pi^{2} }\, \int\limits_{0}^{ \infty }
   \so_{GDR}(\omega ) d\omega = \Ro_{GDR}(0,0) =
 \frac{NZ}{A}
       \frac{e^2}{M} (1+ \ko),
\end{equation}
coincides with the dispersion amplitude (\ref{resonance}).
The appropriate equation
\begin{equation}
   R_{GR}(0,\theta) = \Ro_{GDR}(0,\theta)
\label{R-gerasimov}
\end{equation}
and its consequence
\begin{equation}
   \k=\ko
\label{kappa-gerasimov}
\end{equation}
make the essence of Gerasimov's argument \cite{gerasimov64}, which
is one of the most frequently cited (and critisized
\cite{matsuura73,friar75,arenh77,arenh79,schmitt85}) properties
of the multipole contributions to giant resonance excitations.
It states that the corrections due to retardation and the effects of
multipoles higher than electric dipole cancel in the integrated total
photoabsorption cross section.
Gerasimov's argument is not precisely valid in nonrelativistic
models because they do not maintain the strict validity
of dispersion relations \cite{matsuura73,friar75,arenh77}.
Even in relativistic models which obey the constraints of dispersion
relations (see \cite{matsuura73,arenh77,arenh79} and section 10),
Eq.\ (\ref{R-gerasimov}) does not hold as an identity (see section 10.4).
Nevertheless, expected deviations are rather small,
as will be discussed in some more detail in section 4.4.
Data analyses given in section 4 supports the assumption
that the difference between $\k$ and $\ko$ can be neglected
in practical applications.
In the following discussion in section 3 we disregard the difference
between $\k$ and $\ko$.

Returning to the remaining low-energy contributions
to the amplitude $T_A$ we now define
the quasideuteron amplitude at zero energies, $R_{QD}(0,\theta )$,
in a similar way to
Eq.\ (\ref{mesonGR3}) as
\begin{equation}
	R_{QD}(0,\theta )=\veceps _1\cdot
	\veceps _2\;\,{{e^2} \over M}\,{{ZN} \over A}\kappa ^{QD}  .
	\label{mesonQD2}
\end{equation}
Therefore, one has
\begin{equation}
\frac{1}{2 \pi^{2} }\, \int\limits_{0}^{ \infty }
\sigma _{QD}(\omega ) d\omega =
 \frac{NZ}{A}
       \frac{e^2}{M} \kappa ^{QD} .        \label{mesonQD3}
\end{equation}
From comparing (\ref{thomson}) with (\ref{resonance}) and (\ref{mesonQD2})
one obtains
\begin{equation}
	S_{tot}(0,\theta )=-\veceps _1\cdot \veceps _2\;{{Ze^2}
	\over M}\,\left( {1+{N \over A}\kappa } \right)
	\label{mesonS2}
\end{equation}
for the low-energy limit of the seagull amplitude, where
$\kappa = \k + \kappa ^{QD}$.

Early attempts
\cite{christillin4,C75,C78,AR80,A82,ME87} to understand the term
proportional to $\kappa $ in (\ref{mesonS2})
were based on the idea that this term may be interpreted as
Thomson scattering by charged pions in the nucleus.
This effect cannot be seen in electron scattering
experiments, because this is a one-photon process which averages over the
positive and negative charges of the pions. Thomson
scattering on the other hand
is related to the  squares of pion charges and, therefore,
there is no cancellation.
In consequence it should be possible give a
number for the pion excess per nucleon which was derived \cite{ME87}
to be $n_{\pi} \approx 0.07$.
However, this idea dating back to a work of Christillin and Rosa-Clot
\cite{C78} implies a special interpretation of Siegert's theorem
\cite{siegert37} in the sense that there is a number density
of pions without a charge density. In some sense this may be in line with
Siegert's theorem
because there are positive and negative pions so that the two charges may
cancel to zero. By following Siegert's arguments  literally
also number densities should be excluded. This latter point of view
has been substantiated by Weyrauch and Arenh\"ovel \cite{W84}
who realized that pion charges only occur as ``transition charges".
In other words, the term
proportional to $\kappa $ does not appear due to a modification of the
number of particles in the nucleus  but due to a modification
of the electromagnetic current.

\subsection{Multipole expansion of the scattering amplitude for
giant resonances}

Via a dispersion relation and the optical theorem the resonance
structure of the Compton scattering amplitude in forward direction is
completely determined by nuclear photoabsorption. However, in order to
obtain the angular dependence of the scattering amplitude, it is
necessary to consider the multipolarity of the resonances.
The multipole-dependent transition operators are obtained by expanding
the photon field $\vA$ into multipole components.
For this purpose let us consider a circularly polarized photon
with a polarization vector $\vecxi _\mu $ equal to
\begin{equation}
	{\vecxi}_{\pm 1} = \mp \frac{1}{\sqrt{2}}({\bf e}_x \pm i {\bf e}_y)
	\label{PolVector}
\end{equation}
where ${\bf e}_x$, ${\bf e}_y$ are two orthogonal unit vectors in the plane
perpendicular to the photon momentum $\vk$, which is
directed along the axis $z$, $\vk = \omega {\bf e}_z$, and
$\mu  = \pm $1 denotes the helicity of the photon.
The multipole decomposition of the photon wave function
for a fixed helicity $ \mu $ is
then given by
\begin{eqnarray}
\label{MultipoleDecomp}
	\vecxi _\mu \,e^{i\,\vk\cdot \vr}=
	\sum\limits_{L=1}^{\infty} {i^L\,\sqrt {2\pi (2L+1)}\,\left\{
        {\mu \vA_{L\mu }(\vr,M)+
	\vA_{L\mu }(\vr,E)} \right\}}   .
\end{eqnarray}
Here, at any angular momentum projection $\mu$, $|\mu| \le L$,
\begin{eqnarray}
\label{VectorSpherical}
     \vA_{L\mu }(\vr,M) &=& j_L(kr)\,\vY_{LL\mu }(\hat \vr) \, ,
\nonumber \\
     \vA_{L\mu }(\vr,E) &=&
     \frac{1}{\omega }\mbox{\boldmath$\nabla $}
	\times \left( {j_L(kr)\,\vY_{LL\mu }(\hat \vr)} \right)
\\
   &=& i\; \sqrt{\frac{L  }{2L+1}} j_{L+1}(kr) \vY_{L~L+1~\mu}
     - i\; \sqrt{\frac{L+1}{2L+1}} j_{L-1}(kr) \vY_{L~L-1~\mu}
\nonumber
\end{eqnarray}
are the vector potentials for the magnetic and electric multipoles,
respectively, $j_L(kr)$ is the spherical Bessel function.
The vector spherical harmonic
$\vY_{JlM}(\hat \vr)$ is related
to the usual spherical harmonics $Y_{lm}(\hat \vr)$ via
\begin{displaymath}
        \vY_{JlM}(\hat \vr)=\sum\limits_{m \mu}
     {C_{1\mu ,lm}^{JM}\,Y_{lm}(\hat \vr)}\,\vecxi _\mu
\end{displaymath}
with $\vecxi_0 = {\bf e}_z$ and
with $C_{j_1  m_1 ,j_2  m_2}^{J  M}$ denoting the
Clebsch-Gordan coefficient for the coupling of angular momenta
$\vj_1$ and $\vj_2$ to $\vJ$.
The quantity $\hat \vr $ is
the unit vector in the direction of $\vr$.

Choosing the axis $z$ along the momentum $\vk_1$ of the initial photon
in Compton scattering, we can use the above expressions
(with $\vk = \vk_1$) to write the multipole transition operators as
\begin{equation}
   Q_{\lambda L}
  = \sum_\mu  \left( \vecxi _\mu ^*\cdot \veceps _1 \right) \,
   Q_{\lambda L}^{(\mu )} \, ,
\end{equation}
where
\begin{equation}
\label{TransitionOp}
   Q^{(\mu)}_{\lambda L} = (2 \pi)^{1/2} (2L+1)^{1/2} i^L
     \int d \vx\,\vj(\vr)\cdot
      \vA_{L \mu}(\vr, \lambda) , \quad \lambda=E,M.
\end{equation}
Strictly speaking, such a multipole decomposition makes a sense
in the center-of-mass of the nucleus. In a more general situation
the operators (\ref{TransitionOp}) carry both convectional (center-of-mass)
and intrinsic electromagnetic currents. Since
the nucleus enters to the initial, final and
intermediate states in the direct and crossed term
of the Compton scattering amplitude with different
velocities, the convectional current cannot be completely excluded.
However, for low energies and heavy nuclei which we only consider here
such complications are irrelevant, and we can safely assume in most
situations that the transition operators
always transform the ground state of the nucleus,
$|0\rangle$, into an excited state $|\nu\rangle$ with $\nu \ne 0$.

In Eq.\ (\ref{TransitionOp}) each operator contains all orders in $\omega
$ starting from $\omega ^{L}$ for the magnetic multipoles and from
$\omega ^{L-1}$ for the electric ones.
When in Eqs.\ (\ref{VectorSpherical}), (\ref{TransitionOp}) the substitution
$j_L(\omega r)\to (\omega r)^L/(2L+1)!!$ is made, one obtains the
corresponding transition operators in the long-wavelength limit
which we call unretarded and denote
by  $\unret{Q}{(\mu)}_{\lambda L}$
and $\unret{Q}{}_{\lambda L}$.
In particular,
\begin{equation}
\label{Q-E1-unretarded}
   \unret{Q}{}_{E1} =
   \int d \vx\,\vj(\vr) \cdot \veceps_1
   = i[H, \veceps_1\cdot \vD ].
\end{equation}
The amplitude resulting from the unretarded operators
will be called  ``unretarded", while the exact
operators are understood to lead to the ``retarded" amplitude.
If the wavelength of the incoming photon is of the same order as the
nuclear radius, it is not effective to expand the photon plane wave
with respect to $\omega r$
since many powers of $\omega r$ and many multipoles contribute.
At these photon energies the effect of retardation is essential. The problem
of retardation is one of the main topics of our review and will be
discussed in sections 4 and 10.

For the final photon which is not directed along the axis $z$,
similar expressions can be used with
a rotation $R: \vk_1 \to \vk_2$
applied to the polarizations $\vecxi_\mu$
and multipoles $\vA_{L \mu}(\vr, \lambda)$:
\begin{equation}
   \tilde Q_{\lambda L}^R
  = \sum_\mu  \left( \vecxi _\mu ^{R\, *} \cdot \veceps _2 \right) \,
    Q_{\lambda L}^{(\mu )\,R} ,
\end{equation}
where
\begin{equation}
    Q^{(\mu)\,R}_{\lambda L} = (2 \pi)^{1/2} (2L+1)^{1/2} i^L
     \int d \vx\,\vj(\vr)\cdot
      \vA^R_{L \mu}(\vr, \lambda) .
\end{equation}
The rotated and unrotated operators are
related by ordinary $D$-functions of Wigner,
\begin{equation}
\label{Q-rotated}
   Q^{(\mu)\,R}_{\lambda L} =
  \sum_{\mu'} Q^{(\mu')}_{\lambda L} D^L_{\mu\mu'}(R).
\end{equation}

Following the general idea that it is possible to represent the total
scattering amplitude as a sum, where each contribution corresponds to a
different physical mechanism of excitation, we now consider the
scattering by giant resonances only, together with the corresponding
counterparts coming from the seagull amplitude.
Therefore, $T_{GR}=R_{GR}+B+S_{GR}$ with
$S_{GR}= S(\kappa \to \k)$ and $S$ from Eq.\ (\ref{Sot}).
One then arrives at the following decomposition of the giant resonance
scattering amplitude:
\begin{eqnarray}
\label{IntCoordinates2}
T_{GR}(\omega,\theta)&=& -  \veceps _1 \cdot
\veceps _2\frac{Z e^2}{M} F_1(q) -    \veceps _1 \cdot
\veceps _2\frac{NZ}{A} \frac{e^2}{M}\k F_2(q)
\nonumber \\ && \quad
  -\sum_{\lambda L}   \sum_{\nu} g^{\lambda L}(\theta) \left\{
\frac{ | \langle \nu|Q_{\lambda L} |0 \rangle ^2 |}
{E_0 - E_{\nu}^{\lambda L} + \omega  + i {\Gamma}^{\lambda L}_{\nu}/2} +
\frac{ | \langle \nu|Q_{\lambda L} |0 \rangle ^2 |}
{E_0 - E_{\nu}^{\lambda L} - \omega  - i {\Gamma}^{\lambda L}_{\nu}/2}
\right\} ,
\end{eqnarray}
where unrotated transition operators are used
(with any polarization which is irrelevant)
and all the angular and polarization dependence
of the resonance amplitude is translated into
the angular functions $g^{\lambda L}(\theta)$ which
incorporate the $D$-function in (\ref{Q-rotated})
and the factors of $\mu$ in (\ref{MultipoleDecomp}) for each photon.
If both photons are circularly polarized and have the
helicities $\mu_1=\pm 1$ and $\mu_2=\pm 1$, then
\begin{equation}
\label{g-lambda-L}
    g^{EL}(\theta)  = \left[ D^L_{\mu_1 \mu_2}(R) \right]^*,
\quad
    g^{ML}(\theta) = \mu_1\mu_2 \left[ D^L_{\mu_1 \mu_2}(R) \right]^*.
\end{equation}
Since we consider Compton scattering, where the final and initial
internal nuclear states are equal, and restrict ourselves to spin-0 nuclei,
it follows from angular momentum and parity
conservation that only products of multipole components with the same
$\lambda L$ contribute to Eq.\ (\ref{IntCoordinates2}).
For simplicity, we have skipped recoil corrections
in the energy denominator, which are negligibly small in the giant
resonance region. We also neglected the additional $\omega $-dependent
terms in the kinetic seagull amplitude  (relativistic corrections, which are
related to negative-energy intermediate states) and in the mesonic seagull
amplitude (modifications of the electromagnetic polarizabilities).
Such corrections are small in the giant resonance region.
The separation of the scattering amplitude into multipoles and its
application to giant resonances
has been discussed by Arenh\"ovel, Danos and Greiner \cite{are67a}.
If the transition operators are used in the long-wavelength approximation
and, at the same time, the dependence of the resonance denominators on
the photon energy is neglected, one observes a strict cancellation for
each electric multipole $EL \ne E1$
between the contributions to the resonance amplitude and those coming
from the multipole expansion of the seagull amplitude. This general
property
of scattering amplitudes is a consequence of low-energy theorems
\cite{friarann} and may be traced back to low-energy expansions of
spherical Bessel functions.
The terms from the electric dipole give rise to the Thomson limit,
in accordance with the low-energy theorem.

The interpretation of Eq.\ (\ref{IntCoordinates2})
in terms of individual nuclear levels is inappropriate
in the nuclear continuum because of the formation
of giant resonances \cite{BrownBolsterli}, where the strength of
a large number of individual nuclear states $n$
(including the continuum)
is concentrated into one GR nuclear state $\nu$.
That is why we use in Eq.\ (\ref{IntCoordinates2}) the sum over
$\nu \lambda L$ rather than a sum over $n$
standing in the first-principle Eq.\ (\ref{tfi}).
In contrast to the continuum states $n$, the collective states
$\nu$ have finite widths $\Gamma^{\lambda L}_\nu$
which give an imaginary part to the resonance amplitude $R_{GR}$
as is also prescribed by Eq.\ (\ref{tfi}).
Note that the presence of the widths in the crossed term in
(\ref{IntCoordinates2}) is not supported by Eq.\
(\ref{tfi}), in which the crossed term has no imaginary
part at positive energies $\omega$ at all.
However, the Lorentzian form of (\ref{IntCoordinates2}) is a rather
popular approximation often used to describe
dispersion effects in dissipative systems like an oscillator with a
friction.
The form given in Eq.\ (\ref{IntCoordinates2}) could arise
if one uses the widths $\Gamma_\nu^{\lambda L}$
depending on the photon energy, like, for example, the radiative widths
$\Gamma_\gamma \sim (\hbar \omega)^3$ \cite{BohrMottelson},
and naively extrapolates them to negative $\omega$
(where the widths, strictly speaking, must be zero).
A mathematical advantage of the Lorentzian approximation to
partial resonance amplitudes is that the Lorentzian function
\begin{equation}
\label{RL}
   R_L(\omega) =
    \frac{1}{\omega _r^2 - \omega ^2 - i\omega \Gamma_0}
  = -\frac{1}{2\omega_0} \left(
     \frac{1}{\omega - \omega_0 + i\Gamma_0/2} -
     \frac{1}{\omega + \omega_0 + i\Gamma_0/2} \right),
\end{equation}
with $\omega_0=\sqrt{\omega_r^2 - \Gamma_0^2/4}$,
has no singularities in the upper semiplane ${\rm Im}\,\omega >0$
of the complex energy and thus obeys the causality condition.
It exactly satisfies the dispersion relation
\begin{displaymath}
  {\rm Re} R_L(\omega) = \frac{2}{\pi} {\mathcal P} \int_0^\infty
   {\rm Im} R_L(\omega')
   \frac{\omega' \,d\omega'}{{\omega'}^2 - \omega^2} ,
\end{displaymath}
what makes its use especially convenient in dispersion calculations.
A disadvantage is that the crossing symmetry
$R_L(\omega )=R_L(-\omega)$ is not
fulfilled for the imaginary part of $R_L$.
However, this does not create problems as long as
the Lorentzian approximation for absorption cross sections
is used for positive $\omega$ only.
On the experimental side it is well known that heavy nuclei show GDR
photoabsorption cross-sections in the form of Lorentzian curves for
spherical nuclei. These Lorentzian photoabsorption cross sections can
be generated in a formal way from the sum in Eq.\
(\ref{IntCoordinates2}) by keeping only one level.

The coupling of giant resonance modes and collective
nuclear surface modes has been discussed by Arenh\"ovel et al.\
\cite{are67a,aregrei69}.
In this case the terms in
(\ref{IntCoordinates2}) are the components into which the giant resonances
are fragmented due to  the coupling of collective surface and
volume modes. In this case a relation between widths $\Gamma_\nu$ and
energies $E_\nu$ of the components of the form
$\Gamma_\nu \sim E_\nu^{3/2}$ is assumed.

In a collective model, giant resonances which we are dealing with here
are due to collective motions of nucleons.  In case of the giant-dipole
resonance (GDR) this is the isovector electric-dipole
motion of protons against neutrons for which the dynamic model of
Steinwedel and Jenssen \cite{steinwedel50} has
been developed.
In a shell model this same GDR phenomenon corresponds to $1p{-}1h$
excitations of nucleons between neighboring oscillator shells ($1 \hbar
\omega$ transitions)  ({\it cf.} {\it e.g.} \cite{molin80}).
The most elegant way to incorporate two-body currents into this
one-body picture is provided by the Fermi liquid theory. In this model
the strong interaction between nucleons leads to an
effective mass $M^*$ of quasiparticles and a modification of the
isovector current.
Because of the special importance
of this aspect for the understanding of subnuclear degrees of freedom
in isovector giant-resonances we devote a special section to this
problem (section 7).

Other important GR modes are the isoscalar and isovector giant-quadrupole
resonances. Enhancement effects on these GR modes have not been discussed
except for a first approach in one of our previous works \cite{SC94}.
Following the essential ideas of
this approach we will discuss the enhancement of higher multipoles in the
present work on more general grounds.

As pointed out above, giant resonances may be considered
as nuclear  states which
themselves are coupled to a very large number of strongly overlapping
individual nuclear states.
These GR states may be viewed as coherent superpositions of transitions
of a given multipolarity between occupied and unoccupied shell model
states located below and above the Fermi level, respectively.
In the following we slightly rearrange the resonance part of
the GR amplitude, Eq.\ (\ref{IntCoordinates2}).
We introduce the resonance energy $(E^{\lambda L}_{\nu})_r$
of the GR states instead of $E^{\lambda L}_{\nu}$ via
$(E^{\lambda L}_{\nu})_r^2=(E^{\lambda L}_{\nu}-E_0)^2
+ ({\Gamma}^{\lambda L}_{\nu})^2/4$.
Furthermore, we separate the $\omega$-independent contribution
coming from the $E1$ transition.
Using Eqs.\ (\ref{resonance}) and (\ref{Q-E1-unretarded}),
we may transform the resonance part
in such a way that the low-energy limit appears explicitly:
\begin{eqnarray}
\label{Rforward}
R_{GR}(\omega,\theta) &=& \frac{e^2}{M}
   \frac{NZ}{A}(1+\k) \,g^{E1}(\theta)
\nonumber \\ && {}
+ \sum_{\lambda L} \sum_{\nu}
 2 \, (E^{\lambda L}_{\nu}-E_0)
\frac{(E^{\lambda L}_{\nu})^2_r-\omega ^2 + i \omega
\Gamma^{\lambda L}_{\nu}}{\left((E^{\lambda L}_{\nu})^2_r -
\omega ^2\right)^2 + ( \omega
\Gamma^{\lambda L}_{\nu})^2} \,
\Theta _\nu ^{\lambda L}(\omega) \,g^{\lambda L}(\theta),
\end{eqnarray}
where
\begin{eqnarray}
         \Theta _\nu ^{\lambda L}&=&\left| {\left\langle \nu
         \right|Q_{\lambda L}\left| 0 \right\rangle }
         \right|^2\;,\quad \lambda L\ne E1
        \label{modify} \\
        \Theta _\nu ^{E1}&=&\left| {\left\langle \nu
        \right|Q_{E1}\left| 0 \right\rangle }
        \right|^2-\left| {\left\langle \nu  \right|
        \unret{Q}{}_{E1}\left| 0
        \right\rangle } \right|^2 + {{\omega ^2} \over 3}
       |\langle \nu | \vD | 0 \rangle |^2
        \nonumber
\end{eqnarray}
are energy-dependent coefficients and
the operator $\vD$ is the dipole moment.

In the following the subscript $r$ of $(E^{\lambda L}_{\nu})_r$ will be
omitted for convenience.
Due to the transformation of the electric dipole contribution, the lowest
multipoles in the sum, $E1$, $M1$ and $E2$, are all proportional to
$\omega ^2$ in the low-energy limit. Close to resonance energies
a good approximation for the
angular dependence of the amplitude $R_{GR}$ can be obtained by
multiplying each
term in the sum with respect to $\lambda L$ in Eq.\ (\ref{Rforward})
by the multipole angular functions
$g^{\lambda L}(\theta)$.
Using (\ref{g-lambda-L}) and (\ref{dfun-Legendre}), one can obtain that
\begin{equation}
   g^{EL}(\theta) = A_L(z)\, \veceps_1 \cdot \veceps_2
                  + B_L(z)\, \vs_1 \cdot \vs_2,
\quad
   g^{ML}(\theta) = A_L(z)\, \vs_1 \cdot \vs_2
                  + B_L(z)\, \veceps_1 \cdot \veceps_2,
\end{equation}
where $\vs_i  = \hat \vk_i \times \veceps_i$,
$\hat \vk_i  = \vk_i /\omega $, $z= \hat \vk_1\cdot \hat \vk_2$ and
\begin{equation}
    A_L(z) = \frac{2}{L(L+1)}
          (P'_L(z) + zP^{\prime\prime}_L(z)),
\quad
    B_L(z) = -\frac{2}{L(L+1)} P^{\prime\prime}_L(z).
\end{equation}
Here $P_L(z)$ are Legendre polynomials.
For the lowest multipoles the functions $g^{\lambda L}(\theta)$
are given by:
\begin{eqnarray}
\label{gangle}
g^{E1}(\theta) & = & \veceps _1 \cdot \veceps _2  \nonumber\\
g^{M1}(\theta) & = & \vs_1 \cdot \vs_2 \nonumber \\
g^{E2}(\theta) & = & 2 (\veceps _1 \cdot \veceps _2)\,
(\hat \vk_1 \cdot \hat \vk_2)
              - \vs_1 \cdot \vs_2 \nonumber \\
g^{M2}(\theta) & = & 2(\vs_1 \cdot \vs_2) \,
(\hat \vk_1 \cdot  \hat \vk_2)
-\veceps _1 \cdot \veceps _2   \nonumber\\
	g^{E3}(\theta ) & = & {1 \over 4}\left[ {15\left( {\hat \vk_1
	\cdot \hat \vk_2}
	\right)^2-1} \right]\,\veceps_1\cdot \veceps_2-{5
	\over 2}\left( {\hat \vk_1\cdot \hat \vk_2} \right)\left( {\vs_1\cdot
        \vs_2} \right).
          \end{eqnarray}
In order to obtain the angular distribution at energies outside the
resonance region, it is necessary to use further properties
of the scattering amplitude, which are closely related to the problem of
retardation.
In other words, it is necessary to discuss the energy dependence of
the quantity $\Theta _\nu ^{\lambda L}$ entering into (\ref{Rforward}).
This energy dependence is quite different at small energies,
in the center of the GR region and in the high-energy limit.

\subsection{Multipole and isospin decomposition of the seagull amplitude}

For the further consideration it is useful to understand the
isospin structure of the seagull amplitude, as well as its multipole
decomposition.
The underlying idea has been adopted from the hydrodynamic model
of  giant resonances, introduced by Steinwedel and Jensen \cite{steinwedel50}
which describes the isovector motion of protons against neutrons in a
giant-dipole resonance excitation. In this model the relevant dynamic
quantity is the square of the difference of proton and neutron
densities  $(\rho_p-\rho_n)^2$ leading to an increase in potential energy
in case of a local deviation from the average. In a similar way also the
in-phase, or isoscalar,  variation of the local density leads to giant
resonance modes with the isoscalar giant-dipole mode being equivalent
to Thomson scattering of the whole  nucleus.

In accordance with Eq.\ (\ref{seaglimit}) the total seagull amplitude
$S_{tot}$ in the static limit is equal to
\begin{equation}
	S_{tot} =
	{i \over 2}\left\{ {\left\langle f \right|[\veceps _1\cdot
	\vD\,,\;\veceps _2\cdot (\vJ^{[1]}+\vJ^{[2]})]+[\veceps _2\cdot
	\vD\,,\;\veceps _1\cdot (\vJ^{[1]}+\vJ^{[2]})]\left| i \right\rangle }
	\right\}
	\label{seaglimit2}
\end{equation}
with
\begin{displaymath}
   \vJ^{[1]}={e \over M}\sum\limits_i {{{1+\tau _3^{(i)}} \over 2}\;\vp_i}
\end{displaymath}
and $\vJ^{[2]}$ from Eq.\ (\ref{curr}).
In order to single out the isoscalar and isovector contributions to
Eq.\ (\ref{seaglimit2}) we represent the dipole moment operator
in the following form:
\begin{equation}
	\vD=e\sum\limits_i {{{1+\tau _3^{(i)}}
	\over 2}\;\vr_i}=\left( {e\,{N \over A}\sum\limits_i
	{{{1+\tau _3^{(i)}} \over 2}\;\vr_i}-e\,{Z \over A}\sum\limits_i
	{{{1-\tau _3^{(i)}} \over 2}\;\vr_i}} \right)+eZ\,{1
	\over A}\sum\limits_i {\;\vr_i} .
	\label{dipole2}
\end{equation}
The term in round brackets in Eq.\ (\ref{dipole2}) is the intrinsic
dipole moment operator.  It depends on relative coordinates of
protons and neutrons and hence has an isovector structure.
The remaining term in (\ref{dipole2}) represents the dipole moment of the
nucleus as a whole. It does not distinguish coordinates of protons
and neutron and, therefore, is an isoscalar operator.%
\footnote
{Using these terms in such a context we ignore
that the electric charge of the nucleus itself, $Z$,
has an isovector component.}
It is evident from Eq.\ (\ref{curr}) that the commutator of $\vJ^{[2]}$ with
the isoscalar part of the dipole moment vanishes. Therefore, exchange
currents do not contribute to the isoscalar part of the seagull amplitude.
The commutator of $\vJ^{[1]}$ with the isoscalar part of the dipole operator
is nonzero and leads to the isoscalar part of the seagull amplitude:
\begin{equation}
	S_{tot}^{(is)}=-{{Z^2} \over A}\;{{e^2} \over M}\veceps _1
	\cdot \veceps _2  .
	\label{is}
\end{equation}
The isovector part of the dipole moment has a nonzero commutator with both
$\vJ^{[1]}$ and $\vJ^{[2]}$ and, as a result, the isovector seagull amplitude
in the static limit is given by
\begin{equation}
	S_{tot}^{(iv)}=-{{ZN} \over A}{{e^2} \over M}\;\left(
	{1+\kappa } \right)\;\veceps _1\cdot \veceps _2 .
	\label{iv}
\end{equation}
In Eq.\ (\ref{iv}) we use the quantity $\kappa$ instead of $\k$ for the
formal reason that
the current operators introduced in Eq.\ (\ref{seaglimit2}) do not
discriminate between GR and QD excitations. Nevertheless, the application
of the results obtained here may be restricted giant resonances, as
will be done in the next section when discussing GR sum rules.

As soon as retardation corrections are included via the photon plane wave
the corresponding form factors appear in the seagull amplitude with $F_1$
being related to the current $\vJ^{[1]}$ and $F_2$ to $\vJ^{[2]}$ ({\it cf.}  Eqs.\
(\ref{Bkin}) and (\ref{seag2})):
\begin{equation}
	S_{tot}^{(is)}=-{{Z^2} \over A}\;{{e^2} \over M}F_1(q)\;\veceps _1
	\cdot \veceps _2\;,\quad S_{tot}^{(iv)}=-{{ZN}
	\over A}{{e^2} \over M}\;\left( {F_1(q)+\kappa \,F_2(q)} \right)\;
	\veceps _1\cdot \veceps _2  .
	\label{isiv}
\end{equation}
In order to obtain a multipole decomposition of the seagull amplitude
$S_{tot}$, we introduce one- and two-body densities $\rho _{1,2}$,
which are related to the form factors $F_{1,2}$ via
\begin{equation}
	\rho _i(r)=\int {{{d\vq} \over {(2\pi )^3}}}\,F_i(q)\,\exp
	\left( {i\vq \cdot \vr} \right)  .
	\label{rho12}
\end{equation}
Then
\begin{equation}
\label{FormMultipoles}
\veceps _1 \cdot \veceps _2 F_i(q) = \sum_{l=0}^\infty
f_l^{(i)}(\omega ) G_l(\theta)
\end{equation}
with
\begin{equation}
\label{BesselIntegral}
f_l^{(i)}(\omega ) = 4 \pi \int_0^\infty dr r^2 \rho_{i}(r) j^2_l(\omega
r) ,
\end{equation}
$G_0(\theta )=g^{E1}(\theta )$ and, for $l>0$,
\begin{equation}
G_l(\theta) =  \frac12 (l+2) g^{E(l+1)}(\theta) +\frac12 (2l+1)
g^{Ml}(\theta) + \frac12 (l-1) g^{E(l-1)}(\theta) .
\label{GlMultipoles}
\end{equation}
Here $g^{\lambda L}(\theta)$ is the angular distribution function for the
multipole $\lambda L$, as before. A derivation of the expansion given in
Eqs.\ (\ref{FormMultipoles}) and (\ref{GlMultipoles})
can be found in appendix C.
Substitution of these relations into Eq.\ (\ref{isiv}) leads to the multipole
decomposition of the seagull amplitude
$S_{tot}=S_{tot}^{(is)}+S_{tot}^{(iv)}$:
\begin{eqnarray}
	&&S_{tot}=
	\sum\limits_{\lambda L} {S_{tot}^{\lambda L}
	(\omega )\,g^{\lambda L}(\theta )}\;,\quad S_{tot}^{EL}(\omega )=
        -{{Ze^2} \over M}\left(
   {{{(L+1)f_{L-1}(\omega)+Lf_{L+1}(\omega)} \over 2}\,} \right)\;,
	\label{seagulldecom} \\
	&&S_{tot}^{ML}(\omega )=-{{Ze^2} \over M}\left(
	{{{2L+1} \over 2}} \right) f_L(\omega)\;,\quad
	f_l(\omega)=f_l^{(1)}(\omega)+
	{N \over A} \kappa f_l^{(2)}(\omega)  .
	\nonumber
\end{eqnarray}
When one uses only the lowest-order terms with respect to $\omega $ for
each multipole, one obtains expressions for the unretarded multipole
amplitudes entering into $S_{tot}$:
\begin{eqnarray}
        &&\unret{S}{EL}_{tot}(\omega )=-{{Ze^2}
	\over M}\,{{(L+1)\,\omega ^{2L-2}}
	\over {2\,\left[ {(2L-1)!!} \right]^2}}
	\left( {\left\langle {r^{2L-2}} \right\rangle _1+
        {N \over A}\,\kappa \,
       \left\langle {r^{2L-2}} \right\rangle _2} \right)  ,
		\label{seagulldecom2} \\
        &&\unret{S}{ML}_{tot}(\omega )=
	-{{Ze^2} \over M}\,{{(2L+1)\,\omega ^{2L} }
	\over {2\,\left[ {(2L+1)!!} \right]^2}}
	\left( {\left\langle {r^{2L}} \right\rangle _1+
	{N \over A}\,\kappa \,\left\langle {r^{2L}} \right\rangle _2} \right)
	\;,\quad
	\left\langle {r^{2l}} \right\rangle _{1,2}=4\pi
	\int\limits_0^\infty  {dr\,r^{2l+2} \rho _{1,2}}  .
	\nonumber
\end{eqnarray}
As was already pointed out all contributions from electric multipoles in Eq.\
(\ref{seagulldecom2}) except for $E1$ cancel the corresponding
low-energy expansion of the resonance amplitude. In the case of magnetic
multipoles such cancellations do not appear.
This asymmetry between electric and magnetic multipoles
becomes  important when we discuss sum rules
in section 4. There it will become evident that sum rules in a strict sense
are only possible for electric multipoles.

\clearpage

\section{Compton scattering by giant resonances}

In the present section we discuss the status of Compton scattering
by giant resonances. The usual way of constructing the
resonance (positive energy) scattering amplitude of Compton scattering by giant
resonances is given by the Lorentzian representation of photoabsorption
cross sections and the application of the optical theorem and
the once-subtracted dispersion relation. The amplitudes
obtained in this way are valid close to the peaks of the
resonances but quite obviously fail in the low-energy and high-energy
limits. These deficiencies can be cured by introducing nonretarded
amplitudes which are valid in the low-energy limit and by introducing
retardation formfactors to take into account retardation effects
\cite{SC94,baron91}.
This procedure is in accordance with the nonrelativistic Hamiltonian
but violates the requirements dispersion relations which  may be
understood as a violation of causality \cite{GGT54}.
When correcting for this effect
a satisfactory representation of the resonance (positive energy)
amplitude including the effects of retardation
is obtained as will be shown in section 10 through a model
calculation. On the basis of this theoretical approach we discuss
recent theoretical and experimental results obtained for the giant
resonances.

Before proceeding further it is worth to briefly explain why
the second-order perturbation theory, which results in
Eq.\ (\ref{Rforward}) for the resonance amplitude $R_{GR}$, is not
fully suitable for data analyses and why dispersion relations
are invoked for an alternative construction for $R_{GR}$.
The main practical problem with a use of Eq.\ (\ref{Rforward}) is
that an evaluation of this equation involves both the spectrum of
excited states and the energy dependences of the transition
matrix elements $\langle \nu | Q_{\lambda L} |0\rangle$.
These matrix elements are not known with the necessary precision
when $\omega R \simeq 1$ (here $R$ is the nuclear radius),
{\it i.e.} when the energy $\omega$ lies well above
the region of the giant dipole and giant quadrupole resonances
and when retardation effects come fully into play.
At such energies matrix elements between the ground state $|0\rangle$ and
excited states $|\nu\rangle$ with $L \sim \omega R$
are no longer suppressed by threshold factors
like $(\omega R)^L$ or $(\omega R)^{L-1}$. Therefore,
these excited states, being fully negligible
in Eq.\ (\ref{Rforward}) in the low-energy region
$\omega R \ll 1$, do give an essential contribution
to the real part of $R_{GR}$ at high energies.
We refer to section 10 for a model example.

On the other hand, the dispersion relation (\ref{R-GR-disp})
suggests that the high-energy behavior of $R_{GR}(\omega,0)$
is controlled by photoabsorption in the low energy region at $\omega R \ll1$,
where the  giant dipole and quadrupole resonances are located,
so that the knowledge of details of the high-energy behavior
of the transition matrix elements should be unnecessary.
From this reason it is  attractive to try to formulate the resonance
amplitude $R_{GR}$ by using dispersion relations and to extend
Eq.\ (\ref{R-GR-disp}) to nonzero angles.
One possible extension based on fixed-$t$ dispersion relations
is considered in section 8.
As we will see, it can be developed for low $t$ (or small angles)
but also runs into trouble when the region of high $t$, {\it i.e.}
$-t R^2 \simge 1$, is considered.

In this situation a phenomenological approach was developed \cite{SC94}
which combines the advantages of the two approaches and
is apparently applicable at all energies of practical interest ---
up to the pion photoproduction threshold ---
as confirmed by a model study in section 10.
To help with the understanding of the different  steps
and to provide a dictionary of the notation used,
we briefly summarize the procedure of finding the partial resonance
amplitudes $R^{\lambda L}_{GR}$.
To avoid confusion note that all the partial amplitudes
used in this section do not include the constant $R_{GR}(0,0)$,
Eq.\ (\ref{resonance}), so that all of them vanish at $\omega=0$.

At the first stage naive once-subtracted dispersion relations
for partial amplitudes are postulated
and the corresponding ``dispersion" amplitudes $\tilde R^{\lambda L}_{GR}$
are obtained as a sum of  few Lorentzians with parameters given by
the structure of the total photoabsorption cross section.
These ``dispersion" amplitudes are not assumed to be identical
with the proper amplitudes $R^{\lambda L}_{GR}$
because they have either a wrong
threshold (at $\omega\to 0$) or a wrong asymptotic (at $\omega\to\infty$)
behavior, or both. They are only used to construct
unretarded,  $\unret{R}{\lambda L}_{GR}$, 
and retarded, $\hat{R}^{\lambda L}_{GR}$, amplitudes
and to fix their scale.
It is assumed that the (total) retarded amplitude $\hat{R}_{GR}$
is nothing but the full unretarded amplitude $\unret{R}{}_{GR}$
multiplied by retardation form factor squared, $F_{(ret)}^2(\omega)$.
The matching of the ``dispersion" amplitude $\tilde {R}_{GR}$
and the retarded amplitude $\hat {R}_{GR}$ is done in such a way that
these two amplitudes yield the same absorption
cross sections at the giant-resonance peaks.
The retarded amplitudes have theoretically sound behavior both at
low and high energies
as long as one stays on the real axis of complex $\omega$-plane.
Being also correct at the resonance peaks,
they are assumed to be in general very close to
the proper amplitudes $R^{\lambda L}_{GR}$.
However, a comparison with the requirement of the forward-angle
dispersion relation reveals a mismatch $\Delta R(\omega)$ between
the total ``dispersion" amplitude $\tilde R_{GR}(\omega,0)$
and the total retarded amplitude $\hat R_{GR}(\omega,0)$.
In order to remove this mismatch, the electric dipole
amplitude $R_{GR}^{E1}$ is appropriately
corrected with $\Delta R$, and this gives the final resonance amplitude
$R_{GR}$. The whole construction is chosen to ensure vanishing of
the resulting resonance amplitude $R_{GR}$
in the high-energy limit at any angle.

The described procedure cannot be strictly derived from
first principles. Moreover, some features of thus obtained resonance
amplitude $R_{GR}$ are in contrast to those found in relativistic models;
among them is
the absence of a $q$-dependent form factor $F(q)$ in the asymptotics
of $R_{GR}(\omega,\theta)$ (see section 10).
Nevertheless, that is the best what was proposed to describe
Compton scattering through the giant resonance region,
and this  Ansatz finds a numerical confirmation
both experimentally and through model studies.

Now we proceed with giving details of the described approach.

\subsection{The giant resonance amplitude in the Lorentzian representation}

The general form of the resonance amplitude
for the forward direction
given in (\ref{Rforward})
consists of a Lorentzian part and a coefficient $\Theta _\nu ^{\lambda L}$,
which is energy dependent and is expected
to be different at low energies, in the center of the
giant-resonance region and in the high-energy limit, respectively.
In the following we, therefore, will discuss three different regimes
of the resonance amplitude
corresponding to the three energy regions.

The great advantage of Compton scattering experiments is the fact
that in the
forward direction the amplitude is fixed through the optical theorem,
the dispersion relation and the low-energy theorem.
Then the specific information of Compton scattering
is contained in the angular distribution of the differential cross
section. For applications this means that the resonance
amplitudes as formulated in Eq.\ (\ref{Rforward}) has to be replaced
by a more practical version.
In the vicinity of the giant resonances
this is obtained through the following procedure.
We consider the  total photoabsorption
cross section as a superposition of Lorentzian lines
\begin{equation}
\label{xsect}
\sigma(\omega) =  \sum_{\lambda L} \sigma^{\lambda L}(\omega) =
\sum_{\lambda L} \sum_{\nu} \sig
\frac{ (\omega \Gam)^2}{{\left[ (\Enu)^2 - \omega ^2 \right]}^2 +
\omega ^2 (\Gam)^2}
\end{equation}
where the sum with respect to $\nu $ is taken
over all Lorentzians contributing to the multipoles
$\lambda L$. The quantities $\Enu$, $\Gam$ and $\sig$ denote the resonance
energies, widths and peak cross sections of the Lorentzians, respectively.
A resonance  amplitude $\tilde{R}{}_{GR}$, which is applicable in the
resonance region and corresponds to the photoabsorption cross section
of (\ref{xsect}), then has the form
\begin{eqnarray}
\label{ResGR}
&&\tilde{R}{}_{GR}(\omega,\theta) = \frac{NZ}{A} \frac{e^2}{M} (1+\k )
g^{E1}(\theta ) +
\sum_{\lambda L}
\tilde{R}{}^{\lambda L}(\omega , \theta)  \\ \nonumber
&& = \frac{NZ}{A} \frac{e^2}{M} (1+\k ) g^{E1}(\theta ) +
\frac{\omega ^2}{4\pi}
\sum_{\lambda L, \nu} \sig \Gam
\frac{(\Enu)^2 - \omega ^2 + i\omega \Gam}
{ \left[ (\Enu)^2 - \omega ^2 \right]^2 + \omega ^2(\Gam)^2}g^{\lambda
L}(\theta) .
\end{eqnarray}
Here and below $\tilde R^{\lambda L}(\omega,\theta)$ means the product
$\tilde R^{\lambda L}(\omega) \,g^{\lambda L}(\theta)$.
In the low and high energy limit a special discussion of
the amplitude's properties is required.

The energy-dependent multipole amplitudes $\tilde{R}{}^{\lambda L}$
in (\ref{ResGR}) are built in such a way that they satisfy
the optical theorem,
\begin{equation}
\label{OpticalTheorem}
\mbox{Im} \tilde{R}{}^{\lambda L}(\omega,0)
=\frac{\omega}{4 \pi } \sigma^{\lambda L}(\omega) ,
\end{equation}
and obey the naive subtracted dispersion relations
\begin{equation}
\label{DispersionRelation}
\mbox{Re} \tilde{R}{}^{\lambda L}(\omega,0)
=\frac{\omega ^2}{2 \pi^2 } {\cal P}\int_0^{\infty}
\frac{\sigma^{\lambda L}(\omega')}{(\omega')^2 - \omega^2}
\,d{\omega'}.
\end{equation}
Here the subtraction is replaced by zero, what is allowed because the
subtractions for all multipoles are contained in the first term of the
r.h.s.\ of (\ref{ResGR}) and, furthermore, are of electric-dipole
multipolarity.

In the high-energy limit $\omega \to \infty$,
the dispersion resonance amplitudes (\ref{DispersionRelation})
are given by
\begin{equation}
\label{HighLimit}
\mbox{Re} \tilde{R}{}^{\lambda L}(\infty,0)
=  - \frac{1}{2 \pi^2} \int_0^{\infty}
{\sigma^{\lambda L}(\omega')} \,d{\omega'} = - \sum_{\nu}
\frac{\sigma_{\nu}^{\lambda L} \Gamma_{\nu}^{\lambda L}}{4\pi } ,
\end{equation}
where the explicit form of the partial absorption cross sections,
Eq.\ (\ref{xsect}), was used.
Since the giant resonance amplitude has to vanish in the high-energy limit,
we obtain from (\ref{ResGR}) and (\ref{HighLimit}) the GGT sum rule:
\begin{equation}
\label{IntegratedXsect}
 \sum_{\lambda L} \int_0^{\infty} \sigma^{\lambda L}(\omega') d\omega'
= \sum_{\lambda L ,\nu} \frac{\pi}{2}
  \sigma_{\nu}^{\lambda L} \Gamma_{\nu}^{\lambda L}
= 2 {\pi}^2 \frac{NZ}{A} \frac{e^2}{M} (1 + \k).
\end{equation}

A further useful property of the scattering amplitude in the Lorentzian
form concerns the second negative momentum of the integrated cross section.
We discuss this property here as a reference for further applications.
One has
\begin{equation}
\label{SecondMomentum}
\lim_{\omega \to 0} \frac{\mbox{Re} \tilde R{}^{\lambda L}(\omega,0)}{\omega^2}
= \frac{1}{2 \pi^2 } \int_0^{\infty} \frac{\sigma^{\lambda L}
(\omega')}{(\omega')^2} d\omega'
\equiv  \frac{1}{2 \pi^2} \sigma^{\lambda L}_{-2}
= \frac{1}{4 \pi }\sum_{\nu} \frac{\sig \Gam}{(\Enu)^2} ,
\end{equation}
where in the last step the explicit (Lorentzian) form of the amplitudes
$\tilde R{}^{\lambda L}$ has been used again.

It is possible to relate the resonance amplitude, Eq.\ (\ref{ResGR}), to
the forward-direction form, Eq.\ (\ref{Rforward}), which yields
a representation of the matrix elements
$\Theta _\nu ^{\lambda L}$ applicable {\em in the resonance region}:
\begin{equation}
        \Theta _\nu ^{\lambda L}(\omega)={{\omega ^2\sigma _\nu ^{\lambda L}
	\Gamma _\nu ^{\lambda L}} \over {8\pi \,E_\nu ^{\lambda L}}} .
	\label{rel1}
\end{equation}
Note that in this form of the matrix elements
$\Theta _\nu ^{\lambda L}$ all effects of retardation are included. Such
effects are important, when the product of the photon energy $\omega $
with the nuclear radius $R$ is of the order of unity,
which is the case in an actual experiment.

\subsection{The unretarded resonance amplitude}

The dispersion amplitude (\ref{ResGR}) is built in such a way that
it vanishes at high energies in the forward direction.
Such vanishing appears as a result of the exact balance
between the first term in (\ref{ResGR})
and the partial amplitudes $\tilde R{}^{\lambda L}(\infty,0)$,
as is stated by Eq.\ (\ref{IntegratedXsect}).
However, this balance is destroyed at angles $\theta \ne 0$, when
the angular coefficients $g^{\lambda L}(\theta)$ of different multipoles
become different.
Accordingly, Eq.\ (\ref{ResGR}) does not vanish in
the high-energy limit.
Another deficiency of Eq.\ (\ref{ResGR}) which has to be cured
is that the multipole amplitudes $\tilde R{}^{\lambda L}(\omega,0)$
do not have the proper $\omega$ dependence at low energies 
except for the lowest
multipoles $\lambda L = E1, E2, M1$.

As the first step to remove these deficiencies
we introduce now an unretarded form $\Ro_{GR}$ of the giant resonance
amplitude $R_{GR}$, where each multipole transition operator
is taken into account only in its lowest order in $\omega $
and manifestly results in the correct low-energy behavior.
Our construction is as follows.
In the low-energy limit  each multipole (electric and magnetic) in the
total nuclear scattering amplitude  $T_A(\omega,\theta)$ is
proportional to ${\omega}^{2L}$
(except for the Thomson $E1$ amplitude).
This property can be obtained from general
considerations based on analytical properties of the amplitude,
as will be seen in section 8.
The amplitude $R_{GR}(\omega,\theta)$
in the form of
Eq.\ (\ref{Rforward}), together with (\ref{TransitionOp})
teaches us that in the low-energy limit
each electric multipole component is proportional to ${\omega}^{2L-2}$, while
the magnetic multipole starts from ${\omega}^{2L}$. As was pointed out
above the $EL$ multipole contributions
of the order $\omega^{2L-2}$ in the seagull
amplitude cancel the corresponding terms in the resonance
amplitude, which leads to the correct low-energy behavior for the sum.
Taking into account these properties we may write
\begin{eqnarray}
\label{LowGR}
&&\Ro_{GR}(\omega,\theta) =  \frac{NZ}{A}\frac{e^2}{M}(1 +
\ko )g^{E1}(\theta) +
\sum_{\lambda L}
\unret{R}{\lambda L}(\omega, \theta)= \\ \nonumber
&& \frac{NZ}{A}\frac{e^2}{M}(1 + \ko )g^{E1}(\theta)
+ \frac{\omega ^2}{4\pi} \sum_{\lambda L, \nu}
\unret{\sigma}{\lambda L}_{\nu } \Gam
\frac{(\Enu)^2 - \omega ^2 + i\omega \Gam}
{ \left[ (\Enu)^2 - \omega ^2 \right]^2 + \omega ^2(\Gam)^2}
\Omega^{\lambda L}_\nu (\omega)
g^{\lambda L}(\theta) ,
\end{eqnarray}
where
\begin{eqnarray}
\label{Omega}
\Omega^{E1}_\nu (\omega)&=& 1  \nonumber\\
\Omega^{EL}_\nu (\omega) &=& (\omega / E_\nu^{EL})^{2L-4}
   \quad \mbox{for $L \geq 2$} \nonumber\\
\Omega^{ML}_\nu (\omega) &=& (\omega / E_\nu^{ML})^{2L-2}
   \quad \mbox{for $L \geq 1$}.
\end{eqnarray}
In accordance with Eq.\ (\ref{R-GDR-unretarded}),
the unretarded enhancement parameter $\ko$ stands in the unretarded
amplitude (\ref{LowGR}). Assuming that the $E1$ component
of Eq.\ (\ref{LowGR}) vanishes at high energies,
we find the (modified) TRK sum rule
\begin{equation}
\label{IntegratedXsect0}
 \int_0^{\infty} \unret{\sigma}{E1}(\omega') d\omega'
= \sum_\nu \frac{\pi}{2} \unret{\sigma}{E1}_{\nu} \Gamma_{\nu}^{E1}
= 2 {\pi}^2 \frac{NZ}{A} \frac{e^2}{M} (1 + \ko).
\end{equation}
Here the corresponding unretarded absorption cross section is
\begin{equation}
\label{xosect}
\so(\omega ) =  \sum_{\lambda L} \unret{\sigma}{\lambda L}(\omega) =
\sum_{\lambda L, \nu }\unret{\sigma}{\lambda L}_{\nu }
\frac{ (\omega \Gam)^2}{{\left[ (\Enu)^2 - \omega ^2 \right]}^2 +
\omega ^2 (\Gam)^2}
\Omega^{\lambda L}_\nu (\omega)  .
\end{equation}
The difference between this expression for $\so (\omega )$ and the
corresponding quantity $\sigma (\omega )$ from Eq.\ (\ref{xsect}) lies in the
appearance of the $\omega $-dependent factor $\Omega^{\lambda L}_\nu $ and in
the unretarded peak cross sections $\unret{\sigma}{\lambda L}_\nu$.
A comparison of
$\unret{\sigma}{\lambda L}(\omega)$ and $\sigma^{\lambda L}(\omega)$ will be
given in section 10 in the framework of a specific model.
It should be noted that the energy-dependent factor
$\Omega^{\lambda L}_\nu$ becomes
effective for electric octupoles, magnetic quadrupoles and higher
multipoles.
At energies much larger
than the resonance energies this form of the unretarded multipole cross
sections cannot be applied.
Nevertheless, in the case of electric multipoles
energy-weighted sum rules can be derived, which serve as a guideline in
the interpretation of experimental data.

The cancellation between the leading terms in the
unretarded resonance amplitudes $\unret{R}{\lambda L}$
and in the expansion of the seagull amplitude leads to
sum rules for electric multipoles. Since different mechanisms of excitation
are involved, it is possible to formulate such sum rules independently
for the isoscalar ($is$) and the isovector ($iv$) contributions.
Due to the fact that we are restricting ourselves to giant resonance
phenomena, it is necessary to understand which part of the total seagull
amplitude $S_{tot}$ corresponds to this degree of freedom. Clearly, the
kinetic seagull amplitude $B$ has to be included. Therefore, as
mesonic effects appear in the isovector part only ({\it cf.} Eq.\ (\ref{iv})),
the isoscalar sum rules can straightforwardly be written down:
\begin{equation}
\label{IsoscalarSumRule}
\unret{\sigma }{EL,(is)}_{-(2L-2)}
   = 2\pi^2\lim_{\omega \to 0} \frac{\unret{R}{EL,(is)}
(\omega ,0)}{\omega ^{2L-2}}
= {\pi}^2 \frac{Z^2 e^2}{AM}
\frac{L+1}{[(2L-1)!!]^2} {\left\langle {r^{2L-2}} \right\rangle}_1 \; ,
\quad L>1  .
\end{equation}
The r.h.s.\ of Eq.\ (\ref{IsoscalarSumRule}) follows from the decomposition of
the seagull amplitude, Eq.\ (\ref{seagulldecom2}), together with Eq.\
(\ref{is}). Note that no such sum rule exists for $E1$, as the observed
electric dipole resonance is an isovector excitation.
In the case of isovector sum rules the situation is more complicated.
In section 3.2 we discussed the full mesonic seagull amplitude obtained
in the static limit. Clearly, only a part of this amplitude, previously
denoted by $S_{GR}$, enters into the giant resonance
amplitude $T_{GR}$, which we discuss now. However, it is impossible to
extract $S_{GR}$ from $S$ without further assumptions, as long as we
basically argue  in the framework of the  Fermi gas model which
cannot discriminate between
GR and QD modes of excitation. Strictly speaking,
it is not sufficient to only pass from $\kappa $ to $\ko $ in the
amplitude $S$, but also the dependence on momentum transfer is modified.
This means that the form factor $F_2^{GR}$ entering into $S_{GR}$ is
different from the two-body form factor $F_2$ appearing in the amplitude
$S$. Such a modification can only be discussed in a model-dependent way,
{\it e.g.} by following the frequently applied 
(see {\it e.g.} \cite{EricsonWeise88})
line of thought that the main
contribution to $S_{GR}$ comes from the central part of the exchange
potential. We discuss the quantitative
consequences of this assumption in appendix B. Here we denote the
normalized density, which corresponds to the form factor $F_2^{GR}$ by
$\rho _2^{GR}$ and write the isovector sum rule as
\begin{eqnarray}
\label{IsovectorSumRule}
\unret{\sigma }{EL,(iv)}_{-(2L-2)}&=&2\pi ^2\lim_{\omega \to 0}
\left( {\frac{\unret{R}{EL,(is)}
(\omega ,0)}{\omega ^{2L-2}}  + \frac{NZ}{A}\frac{e^2}{M}(1 + \ko )
\delta _{L,1}
} \right)    \\
&=& {\pi}^2 \frac{NZ}{A}\frac{e^2}{M}
\frac{L+1}{[(2L-1)!!]^2}
\left( {\left\langle {r^{2L-2}} \right\rangle _1+
\ko \,\left\langle {r^{2L-2}} \right\rangle _2^{GR}} \right) .
\nonumber
\end{eqnarray}
The label $GR$ given here to the two-body formfactor $F_2^{GR}$, the
two-body density $\rho _2^{GR}$ and the two-body averaged  radius
$\left\langle {r^{2L-2}} \right\rangle _2^{GR}$
indicates that these quantities now are
restricted to the GR d.o.f.\ and, therefore, may be slightly different
from the corresponding quantities introduced in section 3.5 where no
distiction between the GR and QD d.o.f.\ has been made.
The upper indices ($is$) and ($iv$) for the absorption cross section
correspond to restricting the sum with respect to $\nu $ in
Eq.\ (\ref{xosect})
to isoscalar and isovector giant resonance excitations, respectively.
It should be stressed that these sum rules
are valid for the unretarded photoabsorption cross sections and not
immediately for the experimental (retarded) photoabsorption cross
sections. Sum rules for the latter case will be discussed below in
section 4.5 and in section 8.
A useful relation for the purpose of data analysis as well as for the
following discussion is the sum of Eqs.\ (\ref{IsoscalarSumRule})
and (\ref{IsovectorSumRule}) for the
case of the electric quadrupole. One has
\begin{equation}
      \unret{\sigma }{E2}_{-2} =
      \unret{\sigma }{E2,(is)}_{-2}  + \unret{\sigma }{E2,(iv)}_{-2} =
	{{\pi ^2} \over 3}\,{{Ze^2} \over M}\,\left( {\left\langle {r^2}
        \right\rangle _1+{N \over A}\ko \left\langle {r^2}
  \right\rangle _2^{GR}}
	\right)  .
	\label{quadSR}
\end{equation}
In the particular case $L=1$, Eq.\ (\ref{IsovectorSumRule})
gives the sum rule (\ref{IntegratedXsect0}).

\subsection{The retardation problem}

The principal goal of a theoretical
investigation of nuclear Compton scattering is
to obtain a consistent description of the energy and angular dependence
of the scattering amplitude in a wide energy region. In the previous
section different quantities have been introduced.
The ``dispersion" amplitudes
$\tilde R^{\lambda L}$ provide a formulation of Compton scattering
in the resonance region only. Since their parameters are taken from
experiment, these amplitudes are retarded quantities.
The corresponding unretarded amplitudes
$\unret{R}{\lambda L}$ were designed to correctly describe the low-energy
Compton scattering, but they were expressed in terms of unretarded
quantities. Therefore, in order to obtain in a wide energy region
(also including high energies) a representation of the Compton scattering
amplitude expressed only via observable quantities, it is necessary to
establish a relation between unretarded and retarded quantities.
A formulation of this retardation problem in terms of the Bessel functions
entering into the multipole transition operators has been given in
section 3.4 in the following way:
In an expansion \cite{forest}
of the plane wave for the photon into
terms with definite parity and total angular momentum $L$,
Eq.\ (\ref{MultipoleDecomp}),
each term contains a spherical Bessel function $j_{L}(\omega r)$.
An amplitude obtained using the exact multipole operators is called
``retarded", while in the case where  the substitution
$j_L(\omega r)\to (\omega r)^L/(2L+1)!!$ is made, the
resulting amplitude is called ``unretarded".
The importance of retardation effects is different in the different
energy regions, i.e.\ at low energies, in the resonance region and in the
high-energy limit, which for our present purpose can be identified with
the pion threshold.
Only at very low energies the parameter of expansion $\omega R$, with the
nuclear radius $R$, is small. In the resonance region this parameter is
still smaller than unity, but no longer negligible, while at high
energies one has $\omega R \sim 1$.

Let us consider again the representation of the (retarded) scattering amplitude
$T_{GR}$ coming from second-order perturbation theory.
In Eq.\ (\ref{IntCoordinates2}) the seagull part of the scattering amplitude
includes the effects of retardation through the form factors.
A further
important property of (\ref{IntCoordinates2}) is that in the high-energy
limit the last term, i.e.\ the resonance amplitude, vanishes and,
therefore, the high-energy
limit of the scattering amplitude is represented by the first two (seagull)
terms.
Our solution of (i.e.\ the Ansatz for) the retardation problem
with respect to the resonance
amplitude consists of two parts.  We have to fix
(i)  how the low-energy form $\Ro _{GR}(\omega,\theta)$
of Eq.\ (\ref{LowGR}) and the retarded Lorentzian form
$\tilde R _{GR}(\omega,\theta)$ of Eq.\ (\ref{ResGR})
are related to each other, and
(ii) through which mechanism the high-energy form
of the resonance amplitude tends to zero.
We follow the ideas of our previous
work \cite{SC94} and then --- in later sections --- discuss the predictions of
fixed-$t$ dispersion relations and calculations within a relativistic
harmonic-oscillator model.

The interpolation between $\Ro _{GR}(\omega,\theta)$ and
$\tilde R _{GR}(\omega,\theta)$
starts with the introduction of a ``retarded`` resonance amplitude
\begin{equation}
\label{Retardation1}
\hat R _{GR}(\omega,\theta) = \Ro _{GR}(\omega,\theta) \, F^2_{(ret)}(\omega ).
\end{equation}
This retarded resonance amplitude $\hat{R} _{GR}(\omega,\theta)$
reflects our
``classical" expectation of the retardation effect. The
quantity $F_{(ret)}(\omega)$ is a form factor due to the spatial
distribution of quasiparticles. It takes into account the different
phases of the electromagnetic wave  with respect to the location of these
quasiparticles. In a classical picture one form factor in Eq.\
(\ref{Retardation1}) appears as a result of calculating the total force
acting on an extended charged object. The second form factor results
from the calculation of the photon radiation off this extended object
due to acceleration. This picture is confirmed by the consideration of
the relativistic oscillator model ({\it cf.} section 10).
As a
consequence (\ref{Retardation1}) is the logical form of a retarded
amplitude in a classical picture. This type of retardation has already been
discussed by Barashenkov and Kaiser \cite{barashenkov62}
in the case of the nucleon
and by Petrun'kin \cite{petrunkin64} and
Ericson and H\"ufner \cite{ericson73} in a quantum-mechanical treatment
of Thomson scattering. More recently it has been derived by Baron et al.\
\cite{baron91} in a nonrelativistic model calculation
of retardation effects on
the resonance part of the giant dipole resonance.

Strictly speaking, due to the close structural relation between the
resonance and seagull parts of the amplitude,
it is not obvious, what type of form factor should be
used for $F_{(ret)}(\omega )$ in Eq.\ (\ref{Retardation1}). One may
expect that different types of excitations
lead to different form factors. For example, while for isoscalar
excitations the one-body form factor $F_1$ seems appropriate,
isovector excitations should
be related to an isovector form factor $F_{(iv)}^2=
(F_1^2+\ko (F_2^{GR})^2)/(1+\ko )$.

In order to fix the coefficients
$\unret{\sigma}{\lambda L}_\nu$  contained
in the low-energy form, Eq.\ (\ref{LowGR}), of the resonance amplitude
$\Ro _{GR}(\omega,\theta)$, we use the condition that
the retarded amplitude yields through the optical theorem the correct
magnitude of the absorption cross section at the resonance peaks:
\begin{equation}
\label{hat-tilde-relation1}
{\rm Im}\,\hat{R}_{GR}^{\lambda L}(\Enu,\theta) =
{\rm Im}\,\tilde R _{GR}^{\lambda L}(\Enu,\theta).
\end{equation}
This gives relations
\begin{equation}
\label{hat-tilde-relation2}
   \unret{\sigma}{\lambda L}_\nu \,  F^2_{(ret)}(\Enu) =\sig
\end{equation}
between the unretarded $\unret{\sigma}{\lambda L}_\nu$
and the experimental peak cross sections $\sig$.

Thus defined, the retarded resonance amplitude $\hat{R} _{GR}(\omega,\theta)$
fulfills one of the most essential conditions we are looking for
since it apparently vanishes
in the high-energy limit owing to the factor $F_{(ret)}(\omega)$.
In spite of this and in spite of its
classical plausibility, the resonance amplitude
$\hat{R} _{GR}(\omega,\theta)$ in the
form of (\ref{Retardation1}) is not fully acceptable because it violates
causality. This can easily be seen if for example the proton density is
approximated by a Gaussian distribution, i.e.\ $F_{(ret)}(\omega )=
\exp(-\omega^2/(4\lambda ))$
where the parameter $\lambda $ is fixed by demanding that it
reproduces the usual mean square nuclear charge radius \cite{baron91}.
In this particular form one can easily derive that
$\hat{R}(\omega,\theta)$ of Eq.\ (\ref{Retardation1}) tends to infinity
in part of the upper complex-$\omega$ half-plane for $|\omega| \to
\infty$. Consequently,
this form of the resonance  amplitude violates
causality \cite{GGT54}.
Similar difficulties ({\it e.g.} additional poles in the upper half-plane of
$\omega $) may appear for other distributions of quasiparticles.
However, as it will be seen in section 10 within the relativistic
oscillator model, it is possible that such energy-dependent form factors
are compensated by the sum over all intermediate states, if all matrix
elements are taken into account exactly. In this way
the correct analytical properties of the total amplitude may in principle
be restored.
Since it is necessary to approximate these matrix elements in all
applications of the general formalism, a different approach for
overcoming the difficulties of an incorrect analytical behavior has been
suggested in \cite{SC94}.

The incorrect analytical behavior of the ``retarded" amplitude
(\ref{Retardation1}) manifests itself in that it does not obey the
GGT dispersion relation at the forward angle. That means that there is
a mismatch
\begin{equation}
 \Delta R(\omega) = \tilde R _{GR}(\omega,0) - \hat{R}_{GR}(\omega,0)
\end{equation}
between the exact (dispersion) amplitude $\tilde R _{GR}(\omega,0)$
and the ``retarded" amplitude (\ref{Retardation1}).
It was proposed in \cite{SC94} to consider this difference as an electric
dipole correction to the ``retarded" amplitude and to finally write
the resonance amplitude $R _{GR}(\omega,\theta)$ including the
effects of retardation in the following form:
\begin{equation}
\label{Retardation2}
R_{GR}(\omega,\theta)=\hat{R}_{GR}(\omega,\theta) + \veceps _1 \cdot
\veceps _2 \, \Delta R(\omega).
\end{equation}
By construction, the amplitude (\ref{Retardation2}) is in
agreement with forward-direction dispersion theory, just
because of the above specific choice of $\Delta R(\omega)$.
Simultaneously, the amplitude (\ref{Retardation2}) preserves
all the favorable properties of the
retarded amplitude $\hat{R}(\omega,\theta)$.
The  high-energy limit of the resonance
amplitude (\ref{Retardation2}) is determined by $\tilde R _{GR}(\omega,0)$
and has the form
\begin{eqnarray}
 R_{GR}(\omega ,\theta ) &\approx&
      \tilde R _{GR}(\omega,0) \,g^{E1}(\theta) \nonumber \\
  &=& \left( {{{NZ} \over A}{{e^2}
         \over M}\,(1+\k )+{{\omega ^2} \over {4\pi }}
	 \sum\limits_{\lambda L,\nu } {\sigma _\nu ^{\lambda L}
	 \Gamma _\nu ^{\lambda L}{{(E_\nu ^{\lambda L})^2-\omega ^2+
	 i\omega \Gamma _\nu ^{\lambda L}} \over {((E_\nu ^{\lambda L})^2-
	 \omega ^2)^2+\omega ^2(\Gamma _\nu ^{\lambda L})^2}}}}
	 \right)g^{E1}(\theta )
        \nonumber \\ && \qquad
        \to {1 \over {4\pi \omega ^2}}\sum\limits_{\lambda L,\nu }
	{\sigma _\nu ^{\lambda L}\Gamma _\nu ^{\lambda L}\left(
	{(\Gamma _\nu ^{\lambda L})^2-(E_\nu ^{\lambda L})^2+
	i\omega \Gamma _\nu ^{\lambda L}} \right)}\,g^{E1}(\theta )  .
	\label{Rinfty}
\end{eqnarray}
For the last line in Eq.\ (\ref{Rinfty}) the sum rule
(\ref{IntegratedXsect}) has been used. The Ansatz
(\ref{Retardation2}) implies that we attributed an
electric dipole characteristic to this high-energy limit.%
\footnote{\label{footnote3}
Within the relativistic model considered in section 10,
the asymptotics of the amplitude $R_{GR}(\omega,\theta)$ is more
complicated and contains also a retardation form factor dependent
on the momentum transfer $q$. Numerically, however, such a form factor
is not very important, because it becomes only effective when the
asymptotic amplitude is already small (see section 10.5).}
In section 8 the properties of Eq.\ (\ref{Retardation2}) are discussed
within the framework of fixed-$t$ dispersion relations.

The adopted procedure is equivalent to using the dispersion
relation for the dipole $E1$ amplitude (however
not with only the $E1$ cross section but rather with a modified one)
and to introducing threshold ($\Omega^{\lambda L}_\nu$)
and retardation ($F_{(ret)}$) factors to other ``dispersion" multipoles.
This is explicitly shown by another form of Eq.\ (\ref{Retardation2})
which summarizes all different peaces of the amplitude:
\begin{eqnarray}
\label{Retardation3}
    R_{GR}(\omega,\theta) &=&
         \frac{NZ}{A} \frac{e^2}{M} (1+\k ) g^{E1}(\theta )
\nonumber \\ &&
   {} + \frac{\omega ^2}{4\pi} \sum_{\lambda L,\nu}
       \sig \Gam \frac{(\Enu)^2 - \omega ^2 + i\omega \Gam}
        { \left[ (\Enu)^2 - \omega ^2 \right]^2  + \omega ^2(\Gam)^2}
  \left[\Omega^{\lambda L}_\nu (\omega)
   g^{\lambda L}(\theta) \right]_{(ret)} ,
\end{eqnarray}
where
\begin{equation}
\label{Omega-retarded}
   \left[ \Omega^{\lambda L}_\nu (\omega) g^{\lambda L}(\theta)
  \right]_{(ret)}
     = g^{\lambda L}(\theta)
      + \left[  \frac{ F^2_{(ret)}(\omega)} {F^2_{(ret)}(\Enu)}
     \Omega^{\lambda L}_\nu (\omega) - 1 \right]
       \left( g^{\lambda L}(\theta) - g^{E1}(\theta) \right).
\end{equation}
We may note that the unretarded enhancement parameter $\ko$
standing in the unretarded amplitude $\Ro_{GR}(\omega,\theta)$
and in the retarded amplitude $\hat R_{GR}(\omega,\theta)$
does not appear in the final amplitude (\ref{Retardation3}).
Instead, only the retarded quantity $\k$ appears as it should,
owing to the constraint of the GGT sum rule. So, the parameter
$\ko$ cannot be {\em directly} determined from Compton scattering
data. Given, however, the absorption cross section and
the form factor, one can calculate $\ko$
through Eq.\ (\ref{IntegratedXsect0}) and compare it with $\k$.
This will be done in the following sections.

The above-mentioned replacement of the unretarded enhancement parameter
$\ko$ by the retarded one, $\k$, in Eq.\ (\ref{Retardation3})
takes place only, because the correction $\Delta R(\omega)$ in
Eq.\ (\ref{Retardation2}) was attributed to the $E1$ multipolarity.
Therefore, one can conclude that the electric dipole factor appearing
in Eq.\ (\ref{Retardation2}) and persisting up to the asymptotics
(\ref{Rinfty}) is obligatory at low energies.
At high energies it might be modified, and this is exactly what
is observed in the model considered in section 10, where an additional
form factor appears in the asymptotics of $R_{GR}$.
However, we have to repeat again (see footnote $^{\ref{footnote3}}$)
that the same model suggests that such modifications
are not important  numerically.  (See section 10.5 for more detail.)

The resonance amplitude $R_{GR}$, Eqs.\ (\ref{Retardation2})
or (\ref{Retardation3}),
together with the total giant-resonance seagull contribution
\begin{equation}
        B(\omega,\theta) + S_{GR}(\omega,\theta) =
    -\frac{e^2}{M} \Big( Z \,F_1(q)
     + \frac{NZ}{A}\,\k F_2^{GR}(q) \Big) \;g^{E1}(\theta),
\end{equation}
composes the part of the total amplitude $T_A$, Eq.\ (\ref{fullamplitude}),
related to the giant-resonance degrees of freedom.
The remaining contributions in Eq.\ (\ref{fullamplitude})
are discussed in section 5.

The effects of retardation on the giant-multipole amplitudes are
illustrated in Fig.\ref{pr21pr20}.
The imaginary parts are not noticeably affected by the retardation factor
$F^2_{(ret)}(\omega)$ whereas  the effect on the real parts is very large,
as can be seen from the differences between the solid and the dotted
curves. However, it is important to realize that the damping of the
scattering amplitude through $F^2_{(ret)}(\omega)$ is a physical effect
only for multipoles other than electric dipole. For the electric dipole
this damping (solid curve in the upper figure) is an artifact of a
nonrelativistic theory which violates the requirements of dispersion
theory. The corresponding correction given by $\Delta R(\omega)$  of
(\ref{Retardation2}) not only restores the nondamped dotted curve of the
upper figure but --- even more --- enhances this amplitude by adding
constructively those parts of the real amplitudes of the higher multipoles
which have been damped away by  $F^2_{(ret)}(\omega)$.   As a consequence,
retardation does not lead to a decrease of the real amplitude
as a whole but to a change in its multipolarity. The real amplitudes
of the higher multipoles vanish in the high-energy limit and the
electric-dipole amplitude becomes larger by the same amount. The total
resonance amplitude $R_{GR}(\omega,\theta)$ vanishes in the high-energy
linit due to a cancellation of the two terms on the r.h.s.\
of the first line of
(\ref{Rinfty}).

\begin{figure}[htb]
\centering
\caption{\it Scattering amplitudes for the giant-dipole resonances
(upper figure) and isovector giant-quadrupole resonance (lower figure)
of $^{208}$Pb in the forward direction: Im~$R^{\lambda L}(\omega,\theta=0)$
(dashed),  Re~$\tilde R^{\lambda L}(\omega,\theta=0)$ (dotted) and
Re~${\hat  R}^{\lambda L}(\omega,\theta=0)$ (solid).
}
\label{pr21pr20}
\end{figure}

\subsection{ Gerasimov's argument}

The developed model for nuclear Compton scattering in the
giant-resonance region allows one to determine the parameter $\ko$ which is
of great importance for understanding the effective mass of the nucleon
in a nucleus.
As was emphasized above,  the integrated strength of the unretarded $E1$
absorption cross section is determined through experimental
data on photon absorption indirectly, {\it viz.} by virtue of
an extrapolation based on the specific form of retardation corrections,
Eq.\ (\ref{hat-tilde-relation2}), implemented in the model.
In the next subsections we will see that such an extrapolation gives
a result very close to that predicted by Gerasimov's argument
\cite{gerasimov64} which is therefore considered here.

Gerasimov's argument \cite{gerasimov64}
says that the photoabsorption cross section integrated over
all multipole components is equal to the integrated unretarded dipole
strength and that, therefore, retardation corrections pertinent to
the $E1$ absorption of real photons are compensated in the integral
by other multipoles.
This argument has been criticized in a number of papers
\cite{matsuura73,friar75,arenh77,arenh79,schmitt85} (see also section 10)
for not being valid in general. In those papers it was explicitly shown
that the alleged compensation does not happen for a particle bound
in a potential (without meson-exchange forces) and that relativistic
corrections destroy Gerasimov's argument.

It is instructive to see where exactly a flaw is in the
Gerasimov's original derivation. It is in an illegal combination of the
dispersion relation (\ref{R-GR-disp}) and the time-ordered perturbation
theory (\ref{tfi}) which give seemingly similar but nevertheless
different resonance amplitudes when applied in the usual context of
low energies $\omega \ll M$.
The same total scattering amplitude $T$ is represented differently
in the dispersion and perturbation theory:
\begin{equation}
   T(\omega,0) = S^{disp}(\omega,0) + R^{disp}(\omega,0)
  = S^{p.t.}(\omega,0) + R^{p.t.}(\omega,0)
\end{equation}
(in this discussion we omit subscripts like GR or GDR and assume that
$S$ means the total seagull amplitude including the kinetic part $B$;
for the sake of clarity, we denote the ``dispersion" amplitude
$\tilde R(\omega,0)$ by $R^{disp}(\omega,0)$ here).
Considering as an example Compton scattering on a Dirac particle
of mass $M$ bound in a potential and restricting oneself
to energies $\omega \ll M$, one includes all negative-energy
intermediate states in Eq.\ (\ref{tfi}) to the (effective) seagull
$S^{p.t.}$, so that $R^{p.t.}$ is given by positive-energy components
of the total amplitude $T$:
\begin{equation}
\label{R-unret-through-T+}
   R^{p.t.}(\omega,0) = T_+(\omega,0).
\end{equation}
When the retardation is turned off and only electric dipole transitions
are included (this limit corresponds to a formal use of $\vk=0$
but $\omega \ne 0$ in Eq.\ (\ref{tfi})), the corresponding
unretarded $E1$ resonance scattering amplitude 
$\unret{R}{p.t.}_{E1}(\omega,0) = \unret{T}{}_{+,E1}(\omega,0)$
vanishes at ``high" energies $\omega \sim \omega_m$,
where $\omega_m$ is an arbitrary energy which is much smaller than $M$
but much larger than the binding energy of the particle involved.
Therefore the unretarded resonance amplitude 
satisfies an {\em unsubtracted} dispersion relation
\begin{equation}
     {\rm Re}\,\unret{R}{p.t.}_{E1}(\omega,0) =
      \frac{1}{2\pi^2} {\cal P} \int_0^{\omega_m}
     \unret{\sigma}{}_{E1}(\omega') \,
     \frac{{\omega'}^2 \,d\omega'} {{\omega'}^2 - \omega^2},
\end{equation}
which gives
\begin{equation}
\label{R-unret-through-sigma}
     R^{p.t.}(0,0)  \equiv  \unret{R}{p.t.}_{E1}(0,0)  =
      \frac{1}{2\pi^2} \int_0^{\omega_m}
     \unret{\sigma}{}_{E1}(\omega) \,d\omega.
\end{equation}

As a simple analysis shows
(see section 10.3 and Eq.\ (\ref{T+andT-results}) below),
the positive-energy part alone of the total amplitude $T$ does not satisfy
the unsubtracted dispersion relation, because $T_+$ does not
vanish at ``high" $\omega \sim \omega_m$.
Therefore, in order to have the unsubtracted dispersion
relation valid for the ``dispersion" resonance amplitude $R^{disp}$,
\begin{equation}
     {\rm Re}\,R^{disp}(\omega,0) =
      \frac{1}{2\pi^2} {\cal P} \int_0^{\omega_m}  \sigma(\omega') \,
     \frac{{\omega'}^2 \,d\omega'} {{\omega'}^2 - \omega^2}
\end{equation}
(note that the ``dispersion" amplitude was {\it defined} through
this dispersion integral)
and accordingly to have the relation
\begin{equation}
\label{R-disp-through-sigma}
     R^{disp}(0,0) =
      \frac{1}{2\pi^2} \int_0^{\omega_m}  \sigma(\omega) \, d\omega,
\end{equation}
the amplitude $R^{disp}$ has to include, at least partly,
negative-energy contributions of $T$. Since the total amplitude
satisfies the GGT dispersion relation, we conclude that
\begin{equation}
\label{R-disp-through-T+}
   R^{disp}(\omega,0) = T(\omega,0) - T(\omega_m,0),
\end{equation}
where the constant $T(\omega_m,0)$ ensures vanishing of the
``dispersion" resonance amplitude at ``high" energies.
In this equation the amplitude $T(\omega,0)$ includes both
positive and negative energy contributions and therefore
$R^{disp}(\omega,0)$ does so.%
\footnote{Since the Lorentzian parameterization of the ``dispersion"
amplitude $R^{disp}$, Eq.\ (\ref{ResGR}) at $\theta=0$, is constructed
exactly in a way to ensure its vanishing at high energies,
this parameterization is an approximation
to Eq.\ (\ref{R-disp-through-T+}).
Therefore it intrinsically includes a part of the negative-energy
contribution $T_-$.}
Comparing Eqs.\ (\ref{R-unret-through-T+}) and (\ref{R-disp-through-T+}),
we conclude that their difference at zero energy,
\begin{eqnarray}
\label{deltaR-gerasimov}
   \Delta R^{Gerasimov}(0,0) &=& R^{disp}(0,0) - R^{p.t.}(0,0)
\nonumber \\ &=&
   T_-(0,0) - T(\omega_m,0) =
   T_-(0,0) - T_-(\omega_m,0) - T_+(\omega_m,0),
\end{eqnarray}
is determined by the energy dependence of the negative-energy
contribution (if any) and by the asymptotics of the positive-energy one.
Just this difference prevents the integrals (\ref{R-unret-through-sigma})
and (\ref{R-disp-through-sigma}) to be identical
and generally leads to a violation of Gerasimov's argument.

As an illustration of what happens, let us consider positive and
negative energy parts of the amplitude $T(\omega,0)$ in the case of a free
Dirac particle having a momentum $\vp$ \cite{SC94}.
Such a case is easy for a treatment, and it is relevant because
it shows the behavior of the scattering amplitude
at energies much higher than the binding energy.
It is clear from scale arguments that the ``high"-energy
($\omega \sim \omega_m$) asymptotics of the scattering amplitude
is determined by a coherent sum of local scattering contributions
found for different points in the potential well with
a constant local potential. Therefore it is sufficient to consider
scattering off free particles (with a local mass $M(r)=M+V_s(r)$)
and then to take a proper average.
We can disregard a possible Lorentz-vector potential $V_v$, because
it makes only local shifts of energies $E \to E+V_v(r)$ which
do not change any local observables including
the scattering amplitude ({\it cf.} \cite{friar75,goldberger68}).

Considering the case of forward scattering on the free Dirac particle,
we have
\begin{eqnarray}
\label{T+andT-}
   T_+ &=& e^2 U^+ \left[
   \frac{(\vecal\cdot\veceps_2) \,\Lambda^+_{\vp+\vk} \,(\vecal\cdot\veceps_1)}
          {E_{\vp+\vk} - E_\vp - \omega}  +
   \frac{(\vecal\cdot\veceps_1) \,\Lambda^+_{\vp-\vk} \,(\vecal\cdot\veceps_2)}
          {E_{\vp-\vk} - E_\vp + \omega}  \right] U,
\nonumber \\
   T_- &=& e^2 U^+ \left[
   \frac{(\vecal\cdot\veceps_2) \,\Lambda^-_{\vp+\vk} \,(\vecal\cdot\veceps_1)}
          {-E_{\vp+\vk} - E_\vp - \omega} +
   \frac{(\vecal\cdot\veceps_1) \,\Lambda^-_{\vp-\vk} \,(\vecal\cdot\veceps_2)}
          {-E_{\vp-\vk} - E_\vp + \omega}  \right] U,
\end{eqnarray}
where $U$ is the Dirac spinor ($U^+U=1$) and $\Lambda^\pm_\vp$ are the
projectors onto the positive and negative energy states:
\begin{equation}
  \Lambda^\pm_\vp = \frac12 \left( 1 \pm
     \frac{\vecal\cdot\vp + \beta M}{E_\vp} \right),
   \quad E_\vp = \sqrt{M^2 + \vp^2}.
\end{equation}
Keeping only the spin-independent part of the amplitude and the leading
relativistic correction with respect to $1/M$, we find
in the semirelativistic case of $p \ll M$, $\omega \ll M$:
\begin{eqnarray}
\label{T+andT-results}
   T_+ &=& \frac{e^2}{M} (\veceps_1\cdot\veceps_2)
      \left( -\frac{\omega^2}{4M^2} \right)
     -\frac{e^2}{M^3} (\veceps_1\cdot\vp) \, (\veceps_2\cdot\vp),
\nonumber \\
   T_- &=& \frac{e^2}{M} (\veceps_1\cdot\veceps_2)
      \left(-1 + \frac{p^2}{2M^2} + \frac{V_s}{M}
     +\frac{\omega^2}{4M^2} \right)
     +\frac{e^2}{M^3} (\veceps_1\cdot\vp) \, (\veceps_2\cdot\vp),
\end{eqnarray}
where we have explicitly restored the Lorentz-scalar potential $V_s$
hidden in the particle mass $M$.
Terms with $\omega^2$ in Eq.\ (\ref{T+andT-results})
are quantum corrections $\sim \hbar^2$ which are
specific for the spin 1/2 particle having a magnetic moment $e\hbar/2M$; 
they are absent in the spinless case (see section 10.3).
Moreover, the $\omega^2$ terms cancel in the
sum $T=T_+ + T_-$ and therefore do not appear in the resonance amplitude
(\ref{R-disp-through-T+}).
The term with the energy $V_s + p^2/(2M)$ in $T_-$ describes
a physical effect, which is a modification of
the high-energy behavior of the total amplitude $T$,
in exact accordance with the findings
of Goldberger and Low \cite{friar75,goldberger68}:
\begin{equation}
\label{Ttotal}
    T(\omega,0) = -(\veceps_1\cdot\veceps_2)
   \left[\frac{e^2}{E_\vp}\right]_{M \to M+V_s} \simeq 
    \frac{e^2}{M} (\veceps_1\cdot\veceps_2)
    \left(-1 + \frac{p^2}{2M^2} + \frac{V_s}{M} \right) .
\end{equation}
Using Eq.\ (\ref{Ttotal}) for the evaluation of $T(\omega_m,0)$
and Eq.\ (\ref{T+andT-results}) for the evaluation of $T_-(0,0)$,
we find that the difference (\ref{deltaR-gerasimov})
of the resonance amplitudes 
and therefore the difference of the GGT and TRK integrals
is determined solely by the term with
$(\veceps_1\cdot\vp) \, (\veceps_2\cdot\vp)$ in $T_-$, Eq,\
(\ref{T+andT-results}). Specifically,
\begin{equation}
   \Delta R^{Gerasimov}(0,\theta) =
      \frac{e^2}{M^3} (\veceps_1\cdot\vp) \, (\veceps_2\cdot\vp).
\end{equation}
This perfectly agrees with a direct evaluation of the GGT integral
done by Friar and Fallieros \cite{friar75}.
See section 10 and, in particular, Eq.\ (\ref{Rminus})
for an alternative derivation of the 
$(\veceps_1\cdot\vp) \, (\veceps_2\cdot\vp)$ component of $T_-$.

Applying these results to nucleons in the nucleus,
taking a sum over $Z$ protons, averaging over nucleon momenta $\vp$
and subtracting the center-of-mass contribution
(which is assumably unaffected by relativistic corrections),
we conclude that the GGT sum rule including relativistic effects but not
including meson exchange currents is given by
\begin{equation}
\label{GGT-relativ}
   \frac{1}{2\pi^2} \int_0^{\omega_m}  \sigma(\omega) \, d\omega
    = \frac{Ze^2}{M} \left( 1- \frac{1}{M}
       \left\langle \frac{p^2}{2M} + V_s \right\rangle \right)
      - \frac{Z^2e^2}{M_A},
\end{equation}
where $M_A \simeq AM$ is the mass of the nucleus.
The relativistic correction arising in the r.h.s.\ of (\ref{GGT-relativ})
cannot be found without further assumptions about a splitting of
the mean-field potential into the Lorentz-scalar and Lorentz-vector parts
$V_s$ and $V_v$, respectively.
We suggest that the net result might be very similar to that
contained in the last term in (\ref{GGT-relativ}). That is, it
might be roughly equivalent to the replacement of the free nucleon mass
$M=938.9$ MeV by the mass $\overline{M} = M_A / A \simeq 931.2$ MeV
(as found for $^{12}$C).
If so, this would have only a tiny effect of about 0.8\% on the sum rule
and would give a small negative contribution $-0.008$
to the extracted value of $\k$ and $\ko$.

Similarly, the difference of the GGT and TRK sum rules
including relativistic effects but not including meson exchange currents
is given by
\begin{equation}
\label{delta-sigma-gerasimov}
   \frac{1}{2\pi^2} \int_0^{\omega_m} \Big( \sigma(\omega)
    - \unret{\sigma}{}_{E1}(\omega) \Big)\, d\omega
    = \frac{Ze^2}{M} \left\langle \frac{p^2}{3M^2} \right\rangle.
\end{equation}
This equation may suggest that the value of the unretarded enhancement
parameter $\unret{\kappa}{}$ has to be smaller than the value of
the retarded one, $\kappa$, by
\begin{equation}
   \unret{\kappa}{} - \kappa \simeq
   -\frac{A}{N} \langle \frac{p^2}{3M^2} \rangle  \sim  -0.05.
\end{equation}
Here the Fermi-gas estimate of the nucleon average momentum squared
was used at the standard nuclear density $\rho_0 = 0.17~\rm fm^{-3}$
and $A/N$ was set to be 2.
Such a large difference between $\unret{\kappa}{}$ and $\kappa$
would, of course, be important.
However, within the simple potential model used it is not clear
whether such an expected difference is concentrated in
the giant resonance region or in the quasi-deuteron one.
As we will see, the retardation model developed in the previous
sections, leads to a close agreement between
the enhancement parameters $\ko$ and $\k$ relevant to the giant resonances.

\subsection{The effects of enhancement and retardation in
electric multipole sum rules}

The general phenomenology of giant resonances gives the possibility
to predict the integrated strengths of electric giant
multipoles including the effects of retardation and enhancement.
As each multipole component in the photoabsorption cross section may in
principle consist of several Lorentzians, it is convenient for the
further discussion to introduce a quantity
$\overline {\left( {F^2} \right)^{(\lambda L)}}$
which is the
squared retardation form factor at resonance energy averaged over
the different multipole components:
\begin{equation}
        \overline {\left( {F^2} \right)^{(\lambda L)}}
    =\left( {\sum\limits_\nu  {{{
        \unret{\sigma}{\lambda L}_{\nu} \Gamma _\nu ^{\lambda L}}
    \over {(E_\nu ^{\lambda L})^{2L-2}}}}\,F_{(ret)}^2(E_\nu ^{\lambda L})}
        \right)\left( {\sum\limits_\nu  {{{\unret{\sigma}{\lambda L}_{\nu}
	\Gamma _\nu ^{\lambda L}} \over {(E_\nu ^{\lambda L})^{2L-2}}}}\,}
	\right)^{-1}  .
	\label{averform}
\end{equation}
In case of only one Lorentzian line contributing to the multipole
$\lambda L$ Eq.\ (\ref{averform}) reduces to
$\overline {\left( {F^2} \right)^{(\lambda L)}}=F_{(ret)}^2(E_\nu
^{\lambda L})$ which will be used in the following applications.
In principle it could be necessary to distinguish between isoscalar and
isovector averages of the retardation form factor. As in the case of the
retardation form factor itself we will neglect such differences.

From Eqs.\ (\ref{IntegratedXsect}), (\ref{IntegratedXsect0})
and (\ref{hat-tilde-relation2}),
together with Eq.\ (\ref{averform}) it follows that
the relevant sum rule for the giant-dipole resonance (GDR) is given by
\begin{eqnarray}
	\label{GDRsumRule}
       \int\limits_0^\infty
     {\sigma ^{GDR}(\omega )\,d\omega } &=& {\pi  \over 2}
	\sum\limits_\nu  {\sigma _\nu ^{E1}\Gamma _\nu ^{E1}}= {\pi  \over 2}
        \sum\limits_\nu  {\unret{\sigma}{E1}_{\nu}\Gamma _\nu ^{E1}}\,
	F_{(ret)}^2(E_\nu ^{E1})
\nonumber \\
        &=&  2\pi ^2{{NZ} \over A}{{e^2} \over M}(1+\ko)\,
	\overline {\left( {F^2} \right)^{(E1)}}  .
\end{eqnarray}
In (\ref{GDRsumRule}) $\sigma^{GDR}(\omega)$  is the giant-dipole resonance
component of the experimental (retarded) photoabsorption
cross section and $\sigma^{E1}_\nu $, $\Gamma^{E1}_\nu $ and  $E^{E1}_\nu $
are the Lorentz
parameters of the $\nu $th component of the giant-dipole resonance,
denoting the peak cross section,
the width and the peak position, respectively. The quantity
$\overline {\left( {F^2} \right)^{(E1)}}$
on the r.h.s.\ of Eq.\ (\ref{GDRsumRule}) characterizes the
modification of the Thomas-Reiche-Kuhn (TRK) sum rule due to the effect
of retardation.


In the same way one obtains sum rules
for the electric giant-quadrupole resonances,
following from Eqs.\ (\ref{SecondMomentum}), (\ref{IsoscalarSumRule})
and (\ref{IsovectorSumRule}):
\begin{eqnarray}
	&&\int\limits_0^\infty  {{{\sigma _{E2}^{(is)}(\omega )}
	\over {\omega ^2}}d\omega }={\pi  \over 2}\sum\limits_\nu
	{\matrix{{^{(is)}}\cr
{}\cr
}}\frac{\sigma _\nu ^{E2}\Gamma _\nu ^{E2}}{(E_\nu ^{\lambda L})^{2}}
={{\pi ^2} \over 3}{{Z^2}
\over A}{{e^2} \over M}\left\langle {r^2} \right\rangle _1\,
\overline {\left( {F^2} \right)^{(E2)}}
	\label{GQRsumRule1} \\
	&&\int\limits_0^\infty  {{{\sigma _{E2}^{(iv)}(\omega )}
	\over {\omega ^2}}d\omega }={\pi  \over 2}\sum\limits_\nu
	{\matrix{{^{(iv)}}\cr
{}\cr
}}\frac{\sigma _\nu ^{E2}\Gamma _\nu ^{E2}}{(E_\nu ^{\lambda L})^{2}}
={{\pi ^2} \over 3}{{NZ}
\over A}{{e^2} \over M}\left( {\left\langle {r^2} \right\rangle _1+
\ko \left\langle {r^2} \right\rangle _2^{GR}} \right)\,
\overline {\left( {F^2} \right)^{(E2)}} ,
	\label{GQRsumRule2}
\end{eqnarray}
which differ from the similar relations (\ref{IsoscalarSumRule})
and (\ref{IsovectorSumRule}) by the fact that here retardation has been
taken into account explicitly by the averaged form factor appearing on
the r.h.s.

From Eqs.\ (\ref{GDRsumRule})--(\ref{GQRsumRule2}) it is possible to make
predictions for the properties of the three most important
giant resonances, the giant-dipole resonance (GDR), the isovector (IVGQR) and
isoscalar (ISGQR) giant-quadrupole resonances,
if we also take into account experimental results, as far as they are
known.
In order to clarify the quantitative influence
of  mesonic currents, retardation and the relative strength of the
different multipoles, it appears useful to introduce the following
quantity:
\begin{equation}
	\gamma^{\lambda L}=\left( {\int\limits_0^\infty
	{\sigma _{\lambda L}(\omega )\,d\omega }} \right)\,\left( {2\pi ^2{{NZ}
	\over A}\,{{e^2} \over M}} \right)^{-1}  ,
	\label{sigzero}
\end{equation}
which is the total multipole absorption strength given in units
of the (unmodified) TRK sum rule. The cross section $\sigma _{\lambda
L}(\omega )$ in Eq.\ (\ref{sigzero}) includes the effects of enhancement
and retardation and may in principle be the result of an experiment.
In the case, where only one Lorentzian contributes,
Eq.\ (\ref{sigzero}) can be written for the three lowest electric multipoles,
i.e.\ GDR, IVGQR and ISGQR,
in the form
\begin{eqnarray}
	\gamma ^{E1}&=&\left( {1+\ko } \right)\,F_{(ret)}^2(E_{E1})
        \label{sigzerolow} \nonumber \\
	\gamma _{(iv)}^{E2}&=&{1 \over 6}(E_{(iv)}^{E2})^2\left(
	{\left\langle {r^2} \right\rangle _1+\ko \left\langle {r^2}
	\right\rangle _2^{GR}} \right)\,F_{(ret)}^2(E_{(iv)}^{E2})
	\nonumber \\
	\gamma _{(is)}^{E2}&=&{Z \over 6N}(E_{(is)}^{E2})^2\left\langle {r^2}
        \right\rangle _1\,F_{(ret)}^2(E_{(is)}^{E2}).
\end{eqnarray}
Replacing in Eq.\ (\ref{sigzerolow}) the form factor by $1 $
and $\ko $  by $0$
one obtains expressions for the integrated
multipole cross sections $\gamma _{0}^{\lambda L}$ without any
effects of mesonic currents (enhancement) and retardation.
These quantities will be called the unmodified integrated cross sections
which are theoretical quantities to be obtained from sum rules not
containing enhancement and retardation effects.
The inclusion of the
form factor corresponds to taking into account retardation , while
$\ko$ is responsible for mesonic contributions (enhancement).
For the case of $^{208}$Pb these and other
quantities are listed in Table \ref{PbGiantMultipoles}.
The widths and peak positions of the giant resonances
are taken from the following
references: GDR \cite{veyssiere70},
IVGQR  \cite{S8} and
ISGQR \cite{bertrand80}.
The unmodified and the modified integrated cross sections are related
to each other through the relation
\begin{equation}
\gamma ^{\lambda L}=\gamma ^{\lambda L}_0
   \times {\rm MEF} \times F^2_{(ret)}(E_{\lambda L}),
\label{definition}
\end{equation}
where MEF depends on $\lambda L$ and means
a meson enhancement factor (any $\rm MEF=1$ when $\ko=0$).
Using this equation and the definitions of $\gamma ^{\lambda L}$ and
$\gamma ^{\lambda L}_0$, the quantities MEF
may easily be obtained from Eqs.\ (\ref{sigzerolow}).
For the
calculation of  $F^2(E_{\lambda L})$ the same prescriptions have been used
as in our previous paper \cite{SC94}. For the one-body radius
${\left\langle {r^2} \right\rangle _1}^{1/2}$ the charge radius as
determined by electron scattering has been used, i.e.\ for $^{208}$Pb
${\left\langle {r^2} \right\rangle _1}^{1/2}$= 5.51 fm.
For a big nucleus like
$^{208}$Pb this quantity should be
a good measure for the square-averaged radius
defined through the distribution of (pointlike) protons,
{\it viz.}\ ${\left\langle {r^2} \right\rangle _p}^{1/2}$ and,
therefore, also  for the isoscalar radius,
but  differences may be expected \cite{SC94} for the
isovector radius $({\left\langle {r^2} \right\rangle _1}+\ko
{\left\langle {r^2} \right\rangle _2^{GR}})/(1+\ko)$. The two-body
radius entering into this latter quantity was determined
\cite{SC94} through experimental and theoretical arguments
and a numerical value of $\left( \left\langle {r^2} \right\rangle _2^{GR}
\right)^{1/2} \equiv \left\langle {r^2} \right\rangle_{pn}^{1/2}
=(4.3 \pm 0.8)$ fm has been obtained.
The value of $\ko$ has been chosen in such a way that the experimental
integrated
GDR cross section of $^{208}$Pb, $\gamma ^{E1}$=1.38, is reproduced.
This number has been obtained through
photo-neutron
experiments    \cite{veyssiere70}
on $^{208}$Pb. Exactly the same result is obtained as an
average over a larger
number of nuclei in the $^{208}$Pb mass range, where the peak cross sections
have partly  been redetermined through precise Compton scattering
experiments. This will be described in section 4.6. The
importance of the content of Table \ref{PbGiantMultipoles} lies in the
fact,   that a prediction for the ``experimental" enhancement,
i.e.\ the combination
of retardation and enhancement is obtained for the IVGQR which
is an  unknown quantity otherwise. The
mesonic enhancement factor of ${\rm MEF} =1.28$ calculated from $\ko$
and the  radii ${\left\langle {r^2} \right\rangle}_1$ and
${\left\langle {r^2} \right\rangle}_2^{GR}$
combines with the
retardation factor $F_{(ret)}^2(E_{(iv)}^{E2})=0.88$ to give an
``experimental" enhancement of
${\rm MEF} \times F_{(ret)}^2(E_{(iv)}^{E2})=1.13$. This result of $13\%$
is the only information we have on the ``experimental" enhancement
because experiments are by far not precise enough to measure this
quantity. Details will be given in section 4.7.

\begin{table}[htb]
\centering
\caption{\it Giant multipoles in $^{208}$Pb: Peak energy
$E_{\lambda L}$, width $\Gamma$, unmodified integrated cross section
$\gamma ^{\lambda L}_0$ in units of the (unmodified) TRK sum
rule, retardation factor $F^2(E_{\lambda L})$, mesonic enhancement factor
MEF and experimental or modified integrated cross section
$\gamma ^{\lambda L}$.}
\vspace{5mm}
\label{PbGiantMultipoles}
\begin{tabular}[ht]{cccc}
\hline
Parameter&GDR&IVGQR&ISGQR\\
\hline
$E_{\lambda L}$&13.42 MeV& 22.5 MeV& 10.6 MeV\\
$\Gamma$&4.1 MeV&9.0 MeV&3.0 MeV\\
$\gamma ^{\lambda L}_0$  &1&0.066&0.007\\
$F^2_{(ret)}(E_{\lambda L})$ &0.945&0.88&0.98\\
MEF&1.46&1.28&1\\
$\gamma ^{\lambda L}$&1.38&0.074&0.007\\
\hline
\end{tabular}
\end{table}

It is interesting to note that the numbers contained in Table
\ref{PbGiantMultipoles} are in line with the predictions of Gerasimov's
argument. The sum of the integrated strengths of the three multipoles
is $1+\k = 1.38 + 0.074 + 0.007 =1.46$
whereas the corresponding non-retarded dipole strength is
$1+\ko=1.46$ in units of the unmodified TRK sum rule.
This certainly is an important confirmation of Gerasimov's argument
since  the two numbers compared here are  (almost) completely
independent of each other.

\subsection{Scaling of giant resonance parameters via Compton
scattering}

The quantity we wish to determine is the unretarded enhancement constant
for the giant-dipole resonance, {\it viz.}\ $\ko $,
which is the {\it  universal  parameter}
of giant resonance enhancement phenomena. In principle this quantity can
be determined from photoabsorption experiments. In a first place
these are photoneutron  experiments \cite{berman75,dietrich88,berman87}
for heavy nuclei and total photoabsorption
experiments  \cite{ahrens75}
for light nuclei. The reason for this difference in methods
is that for heavy nuclei
the photoneutron channel largely dominates the photoproton channel
and   total photoabsorption measurements are inaccurate because of
the $e^+\,e^-$ pair production cross section which is much larger than the
nuclear photoabsorption cross section.
On the other hand, at low mass numbers the photoproton and
the photoneutron channels are about equal in strength. Since
low-energy  protons are difficult to detect, especially for solid
targets the total photoabsorption cross section measurements are
preferable.

In case of photoneutron
experiments there may be one or more neutrons in the final state per
photoabsorption process. Furthermore, the neutrons may be emitted after a
fission process. This means that photoneutron experiments require a precise
counting technique for neutrons and a method to determine the neutron
multiplicity. These experiments, therefore, gave an excellent overview of the
properties of giant resonances \cite{berman75,dietrich88,berman87},
but suffer from systematic uncertainties because
of the fact that the intensity of the photon beam and the detection
efficiency of the neutron detector have to be known on an absolute scale.
As a consequence, the integrated cross sections measured in different
laboratories showed large deviations from each other
\cite{berman75,dietrich88}. On the
other hand, the relative cross sections measured in one experiment for
different photon energies, i.e.\ the shape of the cross-section curve,
turned out to be much more reliable. Therefore,
general scaling factors of the cross sections were required in order to
improve on the integrated cross sections.

The amplitude for forward-angle Compton scattering is related to the total
photoabsorption cross section via optical theorem and dispersion relation.
Therefore, the total photoabsorption cross section may also be determined
via Compton scattering provided the angular distribution of Compton
scattered photons is well enough known so that an extrapolation to zero
angle becomes possible and provided there is enough information to carry
out the dispersion integral. Given these premises, Compton scattering
provides a rather precise information because there are no principal
difficulties  in arriving at absolute numbers. The reason for this is
that the projectiles  and the reaction products are photons of (almost)
the same energy. This makes it possible to measure the rates of incident
photons and scattered photons with the same detector and thus to avoid
the determination of detection efficiencies on an absolute scale. A
further advantage of Compton scattering is that it is possible to very
largely reduce the effects of $e^+\,e^-$ pair production. This latter
process enters into the Compton scattering amplitude  via  Delbr\"uck
\cite{milstein94}
scattering. Since Delbr\"uck scattering takes place in the Coulomb field
surrounding the nucleus this process is much more peaked
to the forward direction than
nuclear Compton scattering and, therefore, can be avoided except for small
scattering angles or photon energies below 10 MeV where sizable
contributions of Delbr\"uck scattering are also observed at
large angles \cite{milstein94}. As a result the
determination of the  Compton differential cross section only depends on the
relative rates of incoming and scattered photons and makes
this method   a favorable
tool to scale the photon-neutron data. One
experiment of this type has been carried for $^{209}$Bi \cite{nolte86},
being an immediate neighbor of $^{208}$Pb. The advantage
of investigating $^{209}$Bi instead of     $^{208}$Pb  is, that because
of the much higher
level density the cross section is a smooth function  of energy whereas that
of  $^{208}$Pb is rather fragmented \cite{fuhrberg89}.
Furthermore, information on magnetic moments to compare with when
studying the Fujita-Hirata relation (see section 7) is most
clear-cut for $^{209}$Bi which has one valence nucleon out side a closed
core. The enhancement constant obtained
in this way was rather large, amounting to $\gamma ^{E1} = 1.46 \pm 0.05$.
Later on it was realized that in spite of the smoothness of the cross
section the exact value for  $\ko$ depends
on very fine details of the shape of the measured cross sections
\cite{fuhrberg89a,nolte89,dale88} so that from  one single experiment alone
the desired systematic precision can hardly be achieved.
This point has been investigated by
Fuhrberg et al.\ \cite{fuhrberg89a}, showing
that in  spite of precise scaling through a Compton scattering experiment
minor differences in available photoneutron cross sections lead to
sizable differences in $\gamma^{E1}$. Details of this investigation
are shown in Figs.\ \ref{scale1}--\ref{scale3}.
The results of the two different Compton scattering experiments
\cite{nolte86,dale88}
are shown
in Fig.\ \ref{scale1}. The data points at 9.0, 11.4 and 17.74 MeV
depicted by closed circles have been measured using photons from nuclear
reactions \cite{nolte86}, the other data points by tagged photons
\cite{dale88}.
It is interesting to note that the data from the two experiments coincide
on a percent level of pecision at the high-energy side of the spectrum
at 17.4 Mev
and that there are no data from tagged photons below 12.5 MeV. There
are two different photoabsorption experiments shown in Figs.\ \ref{scale2}
and \ref{scale3}. The photoabsorption data of Fig.\ \ref{scale2} are in
agreement with the solid curve after applying a scaling factor of 1.35
whereas the data of Fig.\ \ref{scale3} are partly in favour of the solid
curve and partly in favour of the dashed curve after multiplying with a
scaling factor of 1.09.
As a whole the solid curve appears to be more
justified than the dashed curve but on the other hand the difference
--- at least at a first sight --- appears to be not dramatic. But
nevertheless, when drawing conclusions with repsect to the retarded
enhancement constant $\kappa^{GDR} = \gamma^{E1}-1$
the difference between curves become
essential. For the solid curve we get $\kappa^{GDR}=0.46 \pm 0.05$ whereas
for the dashed curve $\kappa^{GDR}=0.29 \pm 0.05$. Summarizing we can say
that two different Compton scattering experiments lead to the same
large scaling factors for the available photoabsrption data. In spite of
this, a slight difference in the shapes of the photoabsorption cross sections
on the low-energy sides leads to a sizable ($3 \sigma$) difference in
$\kappa^{GDR}$. In view of the rather precise Compton cross sections shown in
Fig.\ \ref{scale1} at 9.0. and 11.4 MeV the larger of the two
$\kappa^{GDR}$ results appears more likely to be true.

\begin{figure}[htb] 
\centering
\caption{\em Differential cross sections for Compton scattering by
$^{209}$Bi. Scattering angle $\theta$ = 135$^{\circ}$. Data points at
9.0, 11.4 and 17.74 MeV are
measured by \cite{nolte86}. Other data  points are measured by  \cite{dale88}.
Solid curve: calculated using the GDR parameters of \cite{nolte86}.
Dashed curve: calculated using the GDR parameters of \cite{dale88}.}
\label{scale1}
\end{figure}

\begin{figure}[htb]   
\centering
\caption{\em Photoabsorption cross sections for  $^{209}$Bi
measured by \cite{harvey64}  multiplied by a scaling factor of 1.35.
Solid curve:
calculated using the GDR parameters of \cite{nolte86}. Dashed curve: calculated
using the GDR parameters of \cite{dale88}.}
\label{scale2}
\end{figure}

\begin{figure}[htb]  
\centering
\caption{\em Photoabsorption cross sections for $^{209}$Bi
measured by \cite{young72} multiplied by a scaling factor of 1.09. Solid curve:
calculated using the GDR parameters of \cite{nolte86}.
Dashed curve: calculated using the GDR parameters of \cite{dale88}.}
\label{scale3}
\end{figure}

The conclusion from the foregoing is, that
it has advantages to rely on averages over a larger number of nuclei
rather than on the result obtained for one single nucleus.
On the theoretical side
it was realized \cite{SC94} that  instead
of the {\em experimental} quantity $(\gamma ^{E1} -1)=\kappa^{GDR}$
the  {\em unretarded} quantity $\ko$
is of interest for a comparison with models, as outlined in the previous
section.

An extensive investigation of Compton scattering in the giant resonance
region in addition to total photoabsortion measurements has been
summarized in our previous paper \cite{SC94}.
The result of this investigation is
contained in  Table \ref{EffectiveMasses}.
For nuclei in the spherical region around $^{208}$Pb, i.e.\ for mass numbers
$A=197$--209, the average integrated  cross section amounted to
$\gamma ^{E1} = 1.38 \pm 0.05$ leading to
\begin{equation}
\label{NumericalKappa}
\ko = 0.46 \pm 0.05
\end{equation}
as a best value for the {\em unretarded} GDR enhancement constant in the
$^{208}$Pb range \cite{SC94}. By change this number coincides with the one
given in our original experiment \cite{nolte86} as documented in Figs.\
\ref{scale1}--\ref{scale3}. For the neighbouring mass regions
of deformed nuclei the quantities $\gamma ^{E1}$ proved to be smaller
than for the spherical region as also shown in Table \ref{EffectiveMasses}.
This observation is rather interesting and will be interpreted in
connection with the
effective mass $M^*$ of the Fermi liquid theory ({\it cf.} section 7)
quoted in column 4 of Table \ref{EffectiveMasses}.

\begin{table}[htb]
\centering
\caption{\em  Average experimental (retarded) integrated cross section
$\gamma^{E1}$ in units of the (unmodified) TRK value,
average unretarded enhancement constant
$\protect\ko$ and effective mass $M^*$ with
$M^*/M=1/(1+ 3{\protect\ko} /4)$ following from
Fermi liquid theory ({\it cf.} section 7, Eq.\ (\protect\ref{mstar}))
for three different nuclear mass ranges. }
\label{EffectiveMasses}
\vspace{5mm}
\begin{tabular}{|c|c|c|c|}
\hline
A& $\gamma ^{E1}$ & $\ko$ & $M^*/M$\\
\hline
181       & 1.27  & 0.34 & 0.80\\
197--209  & 1.38  & 0.46 & 0.75\\
232--238  & 1.27  & 0.34 & 0.80\\
\hline
\end{tabular}
\end{table}

In case of the total photoabsorption measurements which are the favorable
method at low and intermediate mass numbers
there are
also two major difficulties and, thus, require a rescaling through Compton
scattering. The first is that the measurement of the total
photoabsorption cross section can only be done as a measurement of the
target-in target-out difference. The second is that even for light nuclei
$e^+\,e^-$ pair production is the dominating process so that this process
has to be calculated precisely and subtracted from the data. The
precise calculation of the $e^+\,e^-$ pair production cross section is by
far not without problems because of quantum-electrodynamic corrections
entering into the calculation. Therefore, also in this case a scaling via
Compton scattering leads to improvements on the photoabsorption cross
section in general and to tests on the structure of  the
photoabsorption cross section \cite{P16,P18a}.

\subsection{The isovector giant-quadrupole resonance and higher multipoles}

Giant multipole resonances other than the giant-dipole resonance are an
important
subject of  photonuclear physics. In the previous sections we have mainly
covered the theoretical aspects of these investigations. An overview over
the status of research both experimental and theoretical up to the late
1970s is given in a conference proceedings edited by Bertrand
\cite{bertrand79}.

The isoscalar and isovector magnetic-dipole resonances have been
identified below particle emission threshold as isolated levels or
sequences  of isolated levels \cite{laszewski85,wienhard82}.
The favorable tool for these investigations is nuclear resonance fluorescence.

The isoscalar electric giant-quadrupole resonance
has been discovered \cite{pitthan71}  by electron scattering experiments
at energies below the peak of the giant-dipole resonance and investigated
through different
nuclear reactions \cite{VanDerWoude79}. Because of this location at the
low energy side of the giant dipole resonance and
because of the small strength located in this resonance is not possible
to observe the isoscalar quadrupole excitation through Compton scattering.
Most of the information regarding the location of the corresponding
isovector excitation has also been
obtained from inelastic electron scattering experiments \cite{pitthan79},
which show a concentration of quadrupole strength at an excitation energy
of $\sim 130\times A^{-(1/3)}$ MeV. However, the interpretation of the results
in terms of excitation
strength remains uncertain. Other methods of determining location and
strength  mainly rely on the
interference  of the quadrupole resonance  with the predominant electric
dipole resonance,
leading to for-aft asymmetries. Among these are  photon-nucleon
reactions, the  radiative capture of nucleons and Compton scattering
experiments.

The IVGQR in  $^{208}$Pb has been investigated via Compton scattering of
unpolarized \cite{S8}  and polarized \cite{D12} photons
where in the latter case linear polarization has been produced through
off-axis tagging.
These experiments confirm that at least one isovector quadrupole sum rule
is exhausted by these resonances though the experimental errors forbid
precise conclusions on the overfulfilment of the sum rule. One example
of Compton scattering experiments  identifying the IVGQR is discussed
in Fig.\ \ref{IVGQR}.
The quadrupole strength shows up as an interference of $E2$ and $E1$
resonance amplitudes. A further property of the IVGQR of $^{208}$Pb which
only can be seen through Compton scattering at high energy-resolution
by using photons from nuclear reactions and Ge(Li) detectors
\cite{fuhrberg89} is its partial fragmentation. This
experiment \cite{fuhrberg89} showed that there is E2 strength located at
about 17.6 MeV.

\begin{figure}[htb]  
\centering
\caption{\em  Cross-section ration
$\sigma(\theta=150^{\circ})/\sigma(\theta=60^{\circ})$
for Compton scattering of unpolarized photons by $^{208}$Pb \cite{S8}.
Solid line: Calculated including the IVGQR. Dashed line: Calculated not
including the IVGQR.
The IVGQR shows up as
interference of the $E2$ amplitude with the predominant $E1$ amplitude from
the GDR.}
\label{IVGQR}
\end{figure}

Studies of $(n,\gamma)$ reactions have provided valuable data on the IVGQR
in nuclei ranging from $^{40}$Ca to $^{208}Pb$
\cite{drake81,zorro87,bergqvist84,laymon92}. However, neutron capture
experiments are limited by the difficulty of producing a sufficiently
intense  flux of high-energy monochromatic neutrons.

Data on the isovector quadrupole resonance of $^{40}$Ca have been
obtained via the photo-neutron reaction $(\gamma,n)$ and the $(n,\gamma)$
neutron capture reaction \cite{sims97}. Data analysis in terms of
models estimate the isovector quadrupole resonance to be at an energy
of 31.0 MeV with a width of 16.0 MeV, and exhausting most of the energy
weighted IVGQR sum rule.

A study of Compton scattering by $^{12}$C and $^{40}$Ca
carried out by Wright et al.\ \cite{W4,W5} did not lead to
the identification of localized IVGQR strength.

\subsection{The electromagnetic polarizabilities of the nucleus}

After first estimates of the electric polarizabilities of
nuclei due to Migdal \cite{migdal44} (see also \cite{levinger60}),
the electromagnetic polarizabilities of the nucleus were addressed by
Ericson and H\"ufner \cite{ericson73}
for the case of no enhancement of the giant multipole strength
and by Friar \cite{friarann} on a more general ground. In this section
we want to use our formalism to include the effects of enhancement
and to give a transparent formulation of the retardation effects.
For this purpose we start from the amplitude for Compton scattering by
giant resonances, {\it viz.}
\begin{equation}
\label{TotalAmplitude}
T_{GR}(\omega ,\theta) = B(\omega ,\theta) + S_{GR}(\omega ,\theta) +
R_{GR}(\omega ,\theta)
\end{equation}
where $B(\omega ,\theta)$ is the kinetic seagull amplitude,
$S_{GR}(\omega ,\theta)$ is the GR part of the mesonic seagull amplitude,
which does not contain any energy dependence except of the form factor, as
introduced before, and $R_{GR}(\omega ,\theta)$ is
the resonance amplitude.
The electromagnetic polarizabilities are obtained by
expanding the scattering amplitude up to the quadratic order of
$\omega $, i.e.\ at small $\omega $ the amplitude $T_{GR}$ can be written as
\begin{equation}
	T_{GR}(\omega ,\theta )=-{{Z^2e^2} \over {AM}}\,\veceps _1\cdot
	\veceps _2+\omega ^2\left( {{\bar \alpha}_{GR}\,\veceps _1\cdot
	\veceps _2+ {\bar \beta}_{GR} \,\vs_1\cdot \vs_2} \right)  .
	\label{defpol}
\end{equation}
This expansion is first given for the forward direction,
making use of the optical theorem and the once-subtracted dispersion
relation ({\it cf.} Eq.\ (\ref{mesonGR1})). One obtains a form valid for $\omega $
much below giant resonance energies:
\begin{equation}
\label{OptTheorem}
\mbox{Re}T_{GR}(\omega ,0)  =   -\frac{Z^2 e^2}{A M} +
 \frac{\omega ^2}{2 \pi^2} \int_0^{\infty}\frac{\sigma^{GR}
 (\omega ')}{\omega '^2}
d\omega ' + {\cal O}(\omega ^4).
\end{equation}
From (\ref{OptTheorem}) we read off the Baldin--Lapidus sum rule
\cite{baldin60} for the
sum of electric and magnetic polarizabilities due to giant resonance
excitations in the form
\begin{equation}
\label{AlphaPlusBeta}
\bar{\alpha}_{GR} + \bar{\beta}_{GR} = \frac{1}{2 \pi^2}\int_0^{\infty}
\frac{\sigma^{GR}(\omega )}{\omega ^2} d\omega
= {1 \over {4\pi }}\sum\limits_{\lambda L,\,\nu }
{{{\sigma _\nu ^{\lambda L}\Gamma _\nu ^{\lambda L}}
\over {(E_\nu ^{\lambda L})^2}}}   .
\end{equation}
In order to arrive at the electromagnetic polarizabilities of the nucleus
we have to take into account that in addition to giant resonances the
quasideuteron and nucleon-internal degrees of freedom lead to
contributions. Applying
the Baldin--Lapidus sum rule  to all three contributions we arrive
at the relation
\begin{equation}
\label{AlphaPlusBeta1}
\bar{\alpha}{}_{tot} + \bar{\beta}_{tot} = (\bar{\alpha}{}_{GR} +
\bar{\beta}{}_{GR}) + (\bar{\alpha}{}_{QD} +  \bar{\beta}{}_{QD}) +
A (\tilde{\alpha}_N + \tilde{\beta}_N),
\end{equation}
where $\bar{\alpha}{}_{QD}$ and   $\bar{\beta}{}_{QD}$ are the QD electric
and magnetic polarizabilities, while again
$\tilde{\alpha}_N$  and  $\tilde{\beta}_N$ are the in-medium electromagnetic
polarizabilities of the nucleon.
A rough estimate shows that the main contribution to
electromagnetic polarizabilities of the nucleus stems from the giant
resonances with the QD degree of freedom contributing additional 3 per cent
and the nucleon internal excitations giving  0.3 per cent.
Therefore, our present discussion can be restricted to the GR degree of
freedom.

The Baldin--Lapidus sum rule cannot simply be
separated into an electric part and
a magnetic part. In order to nevertheless obtain explicit expressions for
$\bar{\alpha}_{GR}$ and $\bar{\beta}_{GR}$ one can
either make use of dispersion sum rules
written separately for the electric and
magnetic polarizabilities (see \cite{petrunkin81,lvov79} and section 8)
or make use of second-order perturbation theory.
The latter procedure is as follows: We
expand the representation (\ref{Rforward}) of the giant resonance
amplitude $R_{GR}$ with respect to $\omega $, as well as the
corresponding seagull amplitude $B + S_{GR}$ and take into account the
fact that the resonant E2 contribution proportional to $\omega ^2 $
cancels explicitly with the part coming from the seagull amplitude.
As a result we get
\begin{equation}
\label{AlphaBeta}
\bar{\alpha}_{GR} = \alpha_0 + \Delta \alpha \;,\quad
\bar{\beta}_{GR}  =  \beta_{para} + \beta_{dia}  ,
\end{equation}
where
\begin{equation}
\label{Alpha0Beta0}
\alpha_0  =  \frac23 \sum_{n\ne 0} \frac{|\left\langle {0|\vD|n}
\right\rangle |^2}{E_n - E_0}        \;,\quad
\beta_{para}  =  \frac23 \sum_{n\ne 0} \frac{|\left\langle {0|{\bf M}|n}
\right\rangle |^2}{E_n - E_0}  ,
\end{equation}
are the so-called ``proper" electric and paramagnetic polarizabilities,
respectively. The sums in Eq.\ (\ref{Alpha0Beta0}) are taken over all
intermediate states corresponding to an internal excitation and $\bf M$
is the magnetic dipole operator.
The quantity $\Delta \alpha$ is the retardation correction
of the electric polarizability  and $\beta_{dia}$ the diamagnetic
polarizability.
For giant resonances these quantities are
given by
\begin{equation}
\label{PolCorr1}
\Delta \alpha  =  \frac13 \frac{Z^2 e^2}{A M}
\left\langle {r^2} \right\rangle   _1   \;,\quad
\beta_{dia} = - \frac16 \frac{Ze^2}{M}
\left( {\left\langle {r^2} \right\rangle _1+\ko {N
\over A}\left\langle {r^2} \right\rangle^{GR}_2} \right)
 - \frac{\left\langle { {\bf D}^2} \right\rangle }{2 A M } \, .
\end{equation}
The contribution $\alpha _0$ is obtained from the part of
$\Theta _\nu ^{E1}$ in Eq.\ (\ref{modify}) containing the
electric dipole moment operator. The retardation correction
$\Delta \alpha $ has two contributions, one from the expansion of the
form factor in the seagull amplitude and the other from the expansion of
the difference
$$\left| {\left\langle \nu
        \right|Q_{E1}\left| 0 \right\rangle }
        \right|^2-\left| {\left\langle \nu  \right|
        \unret{Q}{}_{E1}\left| 0
        \right\rangle } \right|^2$$
in Eq.\ (\ref{modify}).
Similarly, $\beta _{para}$ corresponds to the contribution of
$\Theta _\nu ^{M1}$ in the resonance amplitude, Eq.\ (\ref{Rforward}). The
first term in Eq.\ (\ref{PolCorr1}) for $\beta _{dia}$  as in the case of
$\Delta \alpha $ is due to the expansion of the form factors in the
seagull amplitude, but in contrast to the expression for $\Delta \alpha $
the mesonic contribution (proportional to $\ko$) appears explicitly. The
second part of $\beta _{dia}$ is a recoil correction, which is small
as it contains an
additional factor $1/A$ in comparison to the first term. In our review we
do not consider any recoil corrections, but write this term only for the
sake of completeness.
The two quantities $\Delta \alpha$ and $ \beta_{dia}$ are common
ingredients  to all
quantum mechanical treatments of the electromagnetic polarizabilities,
either of atoms \cite{friar81} or of nuclei \cite{ericson73,friarann}
and hadrons \cite{petrunkin81,lvov92,petrunkin64}. An extensive
investigation of the two quantities
in the framework of a nonrelativistic theory
is given by L'vov and Schumacher \cite{lvov92}.
According to all these investigations the
quantity $\Delta \alpha$ has its origin in the form factor squared of the
charged composite object at the momentum of the incoming or outgoing photon.
The formula (\ref{PolCorr1}) for $\Delta \alpha$,
first established by Petrun'kin (see in \cite{petrunkin81,petrunkin64}),
is independent of the model used to describe the hadron or nucleus
and also valid in the relativistic case,
what was recently re-emphasized in \cite{lvov98}.

A discussion of the electromagnetic polarizabilities of the nucleus
including the effects of enhancement has been given in \cite{SC94}
in the framework of the phenomenology outlined in section 3.
These polarizabilities have two contributions, one from the expansion of
the form factors in the seagull amplitude $B + S_{GR}$, the other from
the resonance amplitude $R_{GR}$. Using Eqs.\ (\ref{seagulldecom}),
(\ref{seagulldecom2}) and (\ref{Retardation1})
we arrive at the multipole decompositions
\begin{eqnarray}
\nonumber
&&\left( {B+S_{GR}} \right)^{(E1)}=-{{Ze^2} \over M}\left[ {\left( {1+
{N \over A}\ko } \right)-{{\omega ^2} \over 3}\left( {\left\langle {r^2}
\right\rangle _1+{N \over A}\ko
\left\langle {r^2} \right\rangle _2^{GR}} \right)}
\right]\,g^{E1}(\theta )     \\
\nonumber
&&\left( {B+S_{GR}} \right)^{(M1)}=-{{Ze^2} \over M}{{\omega ^2}
\over 6}\left( {\left\langle {r^2} \right\rangle _1+{N \over A}\ko
\left\langle {r^2} \right\rangle _2^{GR}} \right)\,g^{M1}(\theta )     \\
\nonumber
&&\left( {B+S_{GR}} \right)^{(E2)}=-{{Ze^2} \over M}{{\omega ^2}
\over 6}\left( {\left\langle {r^2} \right\rangle _1+{N \over A}
\ko \left\langle {r^2} \right\rangle _2^{GR}} \right)\,g^{E2}(\theta )    \\
\label{MultDecomp4}
&&\hat R_{GR}={{NZ} \over A}{{e^2} \over M}\left[ {\left( {1+\ko }
\right)-{{\omega ^2} \over 3}\left( {\left\langle {r^2} \right\rangle _1+
\ko \left\langle {r^2} \right\rangle _2^{GR}} \right)} \right]\,
g^{E1}(\theta )  \nonumber \\ && \qquad\qquad
  + \frac{\omega ^2}{4 \pi } \sum_{E1,M1,E2,\, \nu}
\frac{\sigma_{\nu}^{\lambda L}}{F^2_{(ret)}(E_{\nu}^{\lambda L})}
\frac{\Gamma_{\nu}^{\lambda L}}{(E_{\nu}^{\lambda L})^2} g^{\lambda
L}(\theta) ,
\end{eqnarray}
where the sum for the resonance amplitude is written with taking into
account only the first three
multipoles ($E1$, $M1$ and $E2$), as in the order $\omega ^2$
higher multipoles do not contribute.
Here we used the retarded form $\hat R_{GR}$ of the resonance amplitude,
as it is most appropriate at energies below the resonance region. In
addition, we took into account the fact that different retardation form
factors should appear for the one-nucleon and the two-nucleon
contribution.
Inserting Eq.\ (\ref{MultDecomp4}) into the definitions, Eq.\
(\ref{Alpha0Beta0}), we find
\begin{eqnarray}
\nonumber
\alpha_0 &=& \frac{1}{2 \pi^2} \int_0^\infty
\frac{\unret{\sigma}{E1}(\omega )}{\omega ^2}
  d\omega  = \frac{1}{4 \pi } \sum_{\nu}
\frac{\sigma_{\nu}^{E1}}{F^2_{(ret)}(E_{\nu}^{E1})}
\frac{\Gamma_{\nu}^{E1}}{(E_{\nu}^{E1})^2} \\
\label{alpha0beta02}
\beta_{para} &=& \frac{1}{2 \pi^2} \int_0^\infty
\frac{\unret{\sigma}{M1}(\omega )}{\omega ^2} d\omega  =
\frac{1}{4 \pi } \sum_{\nu}
\frac{\sigma_{\nu}^{M1}}{F^2_{(ret)}(E_{\nu}^{M1})}
\frac{\Gamma_{\nu}^{M1}}{(E_{\nu}^{M1})^2}   .
\end{eqnarray}
In the same way we may also obtain expressions for
$\Delta \alpha$ and ${\beta }_{para}$ from Eq.\ (\ref{MultDecomp4}),
together with (\ref{PolCorr1}).
The quantity  $\Delta \alpha$
has two terms, $\Delta \alpha_{s}$  and $\Delta \alpha_{r}$
stemming from the seagull amplitude $\left( {B+S_{GR}} \right)^{(E1)}$
and from the $E1$ part of the resonance amplitude
$\hat R_{GR}$, respectively, whereas ${\beta}_{para}$
has only one term coming from
$\left( {B+S_{GR}} \right)^{(M1)}$. These terms are
\begin{eqnarray}
\nonumber
\Delta \alpha _s &=& \frac13 \frac{Z e^2}{M } \left( \left\langle {r^2}
\right\rangle _1 + \frac{N}{A} \ko
\left\langle {r^2} \right\rangle _{2}^{GR} \right) \\
\nonumber
\Delta \alpha_r &=& - \frac13 \frac{e^2}{M} \frac{NZ}{A}
\left( \left\langle {r^2}
\right\rangle _1 + \ko
\left\langle {r^2} \right\rangle _{2}^{GR} \right) \\
\nonumber
\Delta \alpha &=& \Delta \alpha_s + \Delta \alpha_r =
\frac13 \frac{Z^2 e^2}{A M} \left\langle {r^2} \right\rangle _1   \\
\label{DeltaPara4}
\beta_{dia} &=& - \frac16  \frac{Z e^2}{M} \left( \left\langle {r^2}
\right\rangle _1 + \frac{N}{A} \ko
\left\langle {r^2} \right\rangle _2^{GR} \right)  .
\end{eqnarray}
It is seen that a sizable dependence of the
diamagnetic polarizability $\beta_{dia}$ on the enhancement constant
$\ko $ exists.

The main assumption in putting together these expressions is the following:
In Eq.\ (\ref{MultDecomp4}) the retarded resonance amplitude $\hat R_{GR}$
has been used. Clearly, the influence of the function $\Delta R$
introduced in section 4.4 should be small in the low-energy expansion.
Nevertheless, it may be useful to understand
what contributions are coming from this correction.
The angular dependence in Eq.\ (\ref{Retardation2})
shows that the function $\Delta R$
only influences the electric polarizability, but not the magnetic one.
Substituting the definitions of $\tilde R_{GR}$ and $R_{GR}$ in Eq.\
(\ref{Retardation2}) and extracting the term proportional to $\omega ^2$
we obtain an
expression for the $\Delta R$-contribution
$\alpha (\Delta R)$ to the electric polarizability:
\begin{equation}
	\alpha (\Delta R)={1 \over {4\pi }}\sum\limits_{\lambda L,\,\nu }
	{{{\sigma _\nu ^{\lambda L}\Gamma _\nu ^{\lambda L}}
	\over {(E_\nu ^{\lambda L})^2}}-{1 \over {4\pi }}}
        \sum_{E1,M1,E2,\, \nu }
        {{{\unret{\sigma }{\lambda L}_\nu
       \Gamma _\nu ^{\lambda L}} \over {(E_\nu ^{\lambda L})^2}}}
         +{1 \over 3}\,{{NZ}
        \over A}\,{{e^2} \over M}\left(
   {\left\langle {r^2} \right\rangle _1+\ko
	\left\langle {r^2} \right\rangle _2^{GR}} \right) .
	\label{alphaDR1}
\end{equation}
Making use of the relation (\ref{hat-tilde-relation2}) and noting again
that
\begin{displaymath}
        F_{(ret)}^2(\omega )\approx 1
    -{1 \over 3}\,{{\left\langle {r^2}
	\right\rangle _1+\ko \left\langle {r^2} \right\rangle _2^{GR}}
        \over {1+\ko }} \, \omega^2,
\end{displaymath}
we find a simple expression for $\alpha (\Delta R)$:
\begin{equation}
	\alpha (\Delta R)={1 \over {4\pi }}\sum\limits_{M2,E3,...}
         {{{\unret{\sigma }{\lambda L}_\nu \Gamma _\nu ^{\lambda L}}
	 \over {(E_\nu ^{\lambda L})^2}}}  .
	\label{alphaDR2}
\end{equation}
The most important property of Eq.\ (\ref{alphaDR2}) is the fact that only
multipoles higher than the electric quadrupole contribute to the
correction $\alpha (\Delta R)$.
This statement can independently be
confirmed via the Baldin--Lapidus sum rule, Eq.\ (\ref{AlphaPlusBeta}).
Subtracting all polarizability contributions,
Eqs.\ (\ref{alpha0beta02}) and (\ref{DeltaPara4}),
from this relation, one easily comes to the same conclusion.
Evidently, $\alpha (\Delta R)$ is highly suppressed.
Similar effects will be discussed on the basis of fixed-$t$ dispersion
relations in section 8.

\begin{table}[htb]
\centering
\caption{\em Electromagnetic polarizabilities of the nucleus $^{208}$Pb
and nucleon in
units of $10^{-4} fm^3$. The nuclear polarizabilities are devided by the
number $A$ of nucleons in the nucleus. The nucleon polarizabilities
are proton-neutron averages
${\bar \alpha}_N=\frac12 ({\bar \alpha}_p+{\bar \alpha}_n$) and
${\bar \beta}_N=\frac12 ({\bar \beta}_p+{\bar \beta}_n$)
of the free nucleons.}
\vspace{5mm}
\begin{tabular}{|c|c|}
\hline
$\alpha_0$/A& 1202\\
$\Delta\alpha_s$/A  &   68    \\
$\Delta\alpha_r$/A  & $-54$   \\
${\bar \alpha}_{GR}$/A & 1216 \\
$\beta_{dia}$/A     & $-36$   \\
\hline
${\bar \alpha}_N$& 11.3 $\pm$ 1.5 \\
${\bar \beta}_N$ &  3.7 $\mp$ 1.5 \\
\hline
\end{tabular}
\label{pol}
\end{table}

For the nucleus $^{208}$Pb  numerical values
for the electromagnetic polarizabilities are obtained using the
parameters given in section 4.5. The results are given in Table \ref{pol}
together with the proton-neutron averages of the free-nucleon,
the latter taken from section 5.3. For the magnetic
polarizability of the nucleus we only quote the diamagnetic component
because the paramagnetic component is expected to be considerably smaller.
A fraction of 85\% of the nuclear diamagnetism is due to kinetic currents
and the remaining 15\% due to exchange currents.

\clearpage

\section{Compton scattering in the quasideuteron range}


In framework of our concept to describe nuclear Compton scattering
in terms of three different degrees of freedom,
{\it viz.}\ giant-resonance (GR),
quasi-deuteron  (QD) and nucleon-internal (N), we now consider the
second of these contributions, i.e.\ the amplitude $T_{QD}(\omega,\theta)$.
We may write the total nuclear
scattering amplitude in the from
\begin{equation}
\label{d.o.f.}
T_{tot}(\omega,\theta)= T_{GR}(\omega,\theta) + T_{QD}(\omega,\theta)
+T_N(\omega,\theta) = T_A(\omega,\theta) + T_N(\omega,\theta)
\end{equation}
where $T_A(\omega,\theta)$ collects the two amplitudes,
$T_{GR}(\omega,\theta)$ and $T_{QD}(\omega,\theta)$, which are
related to the external degrees of freedom of the nucleon.
In the QD energy region we observe a superposition of contributions
from all three degrees of freedom
with the QD part being the least important one. The reason for
this surprising property  may be understood as follows.
In the QD range the scattering
amplitude $T_{tot}(\omega,\theta)$ consists of contributions
from the imaginary
part of $T_{QD}(\omega,\theta)$ and the superposition of the real parts
of all three amplitudes. The real part of the
QD amplitude is suppressed in middle of the QD range because of its
sign change at the peak of the QD cross section.
As a consequence the resulting scattering amplitude
is mainly due to the high-energy tail of the GR amplitude,
$T_{GR}(\omega,\theta)$, and the
low-energy tail of the amplitude
$T_N(\omega,\theta)$
which is related to  the in-medium electromagnetic
polarizabilities $\tilde{\alpha}_N$
and $\tilde{\beta}_N $ of the nucleons. The most interesting topic of
Compton scattering experiments in the QD range  is
the experimental determination of the in-medium polarizabilities and their
comparison with the values for the free nucleon.

In spite of the small size of the QD contribution to the Compton
differential cross sections we have to discuss this amplitude
with  care  in order to take it into account with the highest possible
precision.
This will be done in the following.

\subsection{The quasideuteron  amplitudes}

The quasideuteron amplitude has two parts, {\it viz.}
\begin{equation}
\label{QDamplitude}
T_{QD}(\omega,\theta) = S_{QD}(\omega,\theta) + R_{QD}(\omega,\theta).
\end{equation}
The seagull amplitude $S_{QD}(\omega,\theta)$
has a static part $\tilde{S}_{QD}(\omega,\theta)$
which is energy-independent in the forward direction and a dynamic,
energy-dependent part
which can
be  parameterized in terms of mesonic corrections $\delta \alpha $
and $\delta \beta$ to the in-medium electromagnetic
polarizabilities $\tilde{\alpha}_N$ and $\tilde{\beta}_N$ of the
nucleon. By definition the quantities  $\tilde{\alpha}_N$ and
$\tilde{\beta}_N$
are one-body effects which are related to the nucleon internal
coordinates only, whereas the quantities  $\delta \alpha $
and $\delta \beta$ are related to two-body effects.
The important field of our current research concerning two-body effects
will be discussed in section 9. The static part of the mesonic seagull
amplitude is of the same structure as the corresponding
GR part. This is
\begin{equation}
\label{SeagullQD}
\tilde{S}_{QD}(\omega,\theta) = - \frac{NZ}{A} \frac{e^2}{M}
\kappa ^{QD}g^{E1}(\theta) F_2(q)
\end{equation}
where again $\kappa ^{QD}$ is the quasideuteron part of the enhancement
constant $\kappa = \k + \kappa^{QD}$
and $F_2(q)$ the two-body form factor.
In principle the exact form factor $F_2^{(QD)}(q)$ for the quasideuteron
contribution to the static mesonic seagull-amplitude
differs from the (total) two-body form factor $F_2(q)$ introduced
in section 3.2 and the following
relation holds: $\k F_2^{(GR)}(q) + \kappa ^{QD} F_2^{(QD)}(q) =
\kappa F_2(q)$, where the
function $F_2^{(GR)}(q)$ appears in the $GR$ part of the seagull amplitude as
discussed in sections 4 and 9 and in
appendix B. However, the difference between $F_2^{(QD)}(q)$ and $F_2(q)$
is much smaller than between $F_2^{(GR)}(q)$ and
$F_2(q)$ and may be neglected, as
will be shown in section 9.
In the following discussion we will not
distinguish between $F_2^{(QD)}(q)$ and $F_2(q)$.
The meson exchange corrections of the electromagnetic polarizabilities
enter into the scattering amplitude in the form
\begin{equation}
\label{meson-corr}
S_{QD}(\omega,\theta)-\tilde{S}_{QD}(\omega,\theta) =
\omega ^2 \, A \,\left( \delta \alpha\,
\veceps _1 \cdot \veceps _2 +\delta \beta \, \vs_1
\cdot \vs_2 \right) F_2(q) +{\cal O}(\omega^4).
\end{equation}
Recently it was found \cite{huett97} that the form factor $F_2(q)$ in Eq.\
(\ref{meson-corr}) is not identical with the two-body form factor
of Eq.\ (\ref{SeagullQD}). This difference, which is not negligible in the case
of relatively light nuclei, is due to the complicated structure of the
two-nucleon correlation function and will be discussed in section 9
within a modified Fermi gas model, together with the polarizability
modifications $\delta \alpha $ and $\delta \beta $.
For convenience all such energy dependences of the (total) mesonic seagull
amplitude $S(\omega,\theta)$ is put into $S_{QD}(\omega,\theta)$,
so that $S_{GR}(\omega,\theta)$ only consists of the
term proportional to $\k $. This is possible, because any model
investigation within the Fermi gas model may only address the total mesonic
seagull amplitude $S(\omega,\theta)$
with no strict possibility to separate it into the GR and QD parts.
The reason for this is that the Fermi gas model does not take into account
the specific differences between GR and QD excitations.
Therefore, this separation into GR and QD parts can only be understood
in the discussion of properties of the two resonance amplitudes
$R_{GR}(\omega,\theta)$
and $R_{QD}(\omega,\theta)$ if we go beyond the Fermi
gas model, by explicitly taking
into account the different nuclear excitation mechanisms corresponding to
the two processes. A possibility to extend the distinction between the GR and
QD parts to the mesonic seagull amplitude is provided by the Fermi liquid
theory as discussed in section 7.

If we suggest that the resonance part of the
quasi-deuteron contribution does not involve the nucleus as a whole, but
rather is connected with deuteron-like subsystems, its dependence on
momentum transfer should be given by the form factor $F_2(q)$, which is
in contrast to the giant resonance case, where no such form factor appears.
The energy dependence may be represented in a Lorentzian
form, which leads to
\begin{equation}
\label{ResQD}
R_{QD}(\omega,\theta) =  \veceps _1 \cdot \veceps _2 F_2(q)
\bigg\{ \frac{NZ}{A}\frac{e^2}{M}\kappa^{QD}
+ \frac{\omega ^2}{4\pi} \sigma _{QD} \Gamma _{QD}
\frac{E _{QD}^2 - \omega ^2 + i\omega \Gamma _{QD}}
{ ( E _{QD}^2 - \omega ^2 )^2 + \omega ^2\Gamma _{QD}^2} \bigg\}.
\end{equation}
Equation (\ref{ResQD}) expresses
our supposition that the scattering amplitude of each of the contributing
quasi\-deuterons follows an electric-dipole characteristic which should be
the case. For the energy independent term the factor $F_2(q)$ is necessary
to ensure vanishing of the resonance amplitude $R_{QD}(\omega,\theta)$
in the high-energy  limit for all angles.
The assumption that the amplitude $R_{QD}(\omega,\theta)$
vanishes in the high-energy
limit leads again to the sum rule (\ref{mesonQD3}) for the integral over the
absorption cross section $\sigma _{QD}$. By explicitly using the
Lorentzian form for $\sigma _{QD}$ one finds
\begin{displaymath}
	{{1} \over {4\pi }} \sigma _{QD}\Gamma _{QD} ={{NZ}
	\over A}\,{{e^2} \over M}\,\kappa ^{QD}
\end{displaymath}
which is a useful relation between Lorentzian parameters and enhancement
constant  $\kappa ^{QD}$ .
It is interesting
to note that as a result the seagull amplitude
$\tilde{S}{}_{QD}(\omega,\theta)$ and the
energy-independent term of the resonance amplitude $R_{QD}(\omega,\theta)$
cancel exactly.

In photon-nucleon investigations another phenomenological
description other than a purely
Lorentz\-ian form exists for $\sigma _{QD}(\omega )$. It is known as
the Levinger representation (\cite{lev1}, see also \cite{lag1}) and has
the following form:
\begin{equation}
	\sigma _{QD}(\omega )=
	L{{NZ} \over A}\,\sigma _D(\omega )\,\exp (-D/\omega )
	\label{levinger}
\end{equation}
where $L$ and $D$ are some parameters and $\sigma _D$ is the experimental
photoabsorption cross section for the deuteron.
The exponential function
takes into account that due to Pauli blocking not the total deuteron
absorption strength should appear in  $\sigma _{QD}$. However, for the
analysis of Compton scattering data  it is convenient and sufficient
to use Lorentzian fits to the expression given in (\ref{levinger}).

\subsection{The single-nucleon contribution to Compton scattering up
to quadratic order in the photon energy}

We now want to introduce the amplitude $T_N(\omega,\theta)$, which is
connected with the third degree of freedom, i.e.\ with the contribution
from individual nucleons to the nuclear Compton scattering amplitude.
Our consideration starts from the spin-independent part of the scattering
amplitudes $T_p$ and $T_n$ for the free proton and neutron,
respectively,  \cite{petrunkin81} in the laboratory system,
\begin{eqnarray}
\label{impulsp}
T_p & = &  \frac{e^2}{M}
\Bigg( -\veceps _1 \cdot \veceps _2
+  {\omega_1}{\omega_2} \frac{2 {\kappa_p} +  {\kappa_p}^2}{4M^2}
\veceps _1 \cdot \veceps _2  \nonumber
\hspace{40mm}\\
& & {}  - {\omega_1}{\omega_2} \frac{(1 + \kappa_p)^2}{4M^2}
\vs_1 \cdot \vs_2 \, \hat \vk_1 \cdot  \hat \vk_2
+ {\omega_1}{\omega_2} \frac{1}{4 M^2}
\vs_1 \cdot \vs_2 \Bigg)   \nonumber \\
& & {}  + \omega_1 \omega_2 {\bar \alpha}_p
\,\veceps _1 \cdot \veceps _2
+ \omega_1 \omega_2  {\bar \beta}_p
\,\vs_1 \cdot \vs_2 + {\cal O}({\omega}^4),
\end{eqnarray}
\begin{eqnarray}
\label{impulsn}
T_n & = & \frac{e^2}{M}
\Bigg(
{\omega_1}{\omega_2} \frac{ {\kappa_n}^2}{4M^2}
\veceps _1 \cdot \veceps _2
 - {\omega_1}{\omega_2} \frac{{\kappa_n}^2}{4M^2}
\vs_1 \cdot \vs_2 \,\hat \vk_1 \cdot  \hat \vk_2
\Bigg)  \hspace{10mm} \nonumber\\
& & {}  + \omega_1 \omega_2 {\bar \alpha}_n
\,\veceps _1 \cdot \veceps _2
+ \omega_1 \omega_2 {\bar \beta}_n
\,\vs_1 \cdot \vs_2 + {\cal O}({\omega}^4),
\end{eqnarray}
The quantities
$\kappa_p$ and $\kappa_n$ are the anomalous magnetic moments of the
proton and neutron, respectively, and ${\bar \alpha}_p$, ${\bar \beta}_p$,
$\bar\alpha_n$ and $\bar\beta_n$ are the electromagnetic polarizabilities
of the free proton and neutron, respectively.
Both amplitudes, Eqs.\ (\ref{impulsp}) and (\ref{impulsn}),
can be divided into a
non-Born part (containing the electromagnetic polarizabilities) and a
Born part (containing all other terms).
When the nucleons are embedded in a nucleus, recoil effects may be
disregarded so that ${\omega}_2 =  {\omega}_1$ to a good approximation.
Furthermore, form factors have to be introduced in order to take into
account the spatial distributions of nucleons in a nucleus. Clearly, it
is also necessary to account for the Fermi motion of the nucleons.
In the energy region considered here this contribution
is of the same order as relativistic corrections and may thus be skipped
for our present purpose.

The only problem in considering these
single-nucleon contributions is to avoid any double-counting with respect
to the other parts of the nuclear scattering amplitude.
It can directly be seen that (i) the non-Born contributions have not yet
been included in any other part of the amplitude $T_{tot}$ and (ii) the
first term in Eq.\ (\ref{impulsp}) is evidently related to the kinetic
seagull amplitude $B(\omega,\theta)$. The remaining terms in Eqs.\
(\ref{impulsp}) and  (\ref{impulsn})      are
corrections proportional to $\omega ^2$ and
depend on the anomalous magnetic moments of the nucleons.
If one compares the values of $\bar{\alpha}_N$ and $\bar{\beta}_N$
given in section  5.3 with the coefficients at $\omega ^2$ in the nucleon
Born terms one can see that the latter are essentially smaller
(of the order of $0.4\times10^{-4} ~\mbox{fm}^3$) and, therefore, may be
neglected.
Moreover, the contribution of the magnetic moment is partly
included into the resonance amplitudes $R_{GR}$ and $R_{QD}$.
Thus, we have to keep only the nucleon non-Born contributions
to the nuclear Compton amplitude which read
\begin{eqnarray}
\label{nonBorn}
T^{non-Born}(\omega,\theta) & = &
 \omega ^2   \Big[ Z \Big( {\bar \alpha}_p
\,\veceps _1 \cdot \veceps _2
+ {\bar \beta}_p
\,\vs_1 \cdot \vs_2 \Big) \hspace{50mm} \nonumber\\
& &  {} + N \Big( {\bar \alpha}_n
\,\veceps _1 \cdot \veceps _2
+ {\bar \beta}_n
\,\vs_1 \cdot \vs_2 \Big) + {\cal O}({\omega}^4)  \Big] F_1(q).
\end{eqnarray}
In this representation, Eq.\ (\ref{nonBorn}),
use has been made of the
assumption that protons and neutrons
follow the same one-body form factor $F_1(q)$.
From $T^{non-Born}(\omega,\theta)$ of Eq.\ (\ref{nonBorn}) we may construct
the amplitude $T_N(\omega,\theta)$ of (\ref{d.o.f.}) by introducing
the in-medium electromagnetic polarizabilities of the nucleon
$\tilde{\alpha}_N$ and $\tilde{\beta}_N$ leading to
\begin{equation}
T_N(\omega,\theta)=\omega ^2\, A \,\left[ \left( \tilde{\alpha}_N
\,\veceps _1 \cdot\veceps _2  + \tilde{\beta}_N\, \vs_1
\cdot \vs_2 \right) + {\cal O}(\omega^4) \right] F_1(q).
\label{T-N}
\end{equation}
For the special case of no modification of the electromagnetic
polarizabilities in the nuclear medium we have
\begin{equation}
\tilde{\alpha}_N \equiv
\bar{\alpha}_N= \frac{Z}{A} \bar{\alpha}_p + \frac{N}{A} \bar{\alpha}_n
\hspace{20mm}
\tilde{\beta}_N
\equiv \bar{\beta}_N = \frac{Z}{A} \bar{\beta}_p + \frac{N}{A} \bar{\beta}_n .
\label{Pol}
\end{equation}

\subsection{Status of the free polarizabilities of the nucleon}

Though electromagnetic polarizabilities of hadrons have been discussed
already for a long time, some clarifying remarks concerning these quantities
are advisable at the beginning. The electromagnetic
polarizabilities ${\bar \alpha}_p$ and ${\bar \beta}_p$ of the free proton
appearing in the Baldin-Lapidus sum rule or in the differential
cross section for proton Compton scattering of real photons (RCS)
\cite{petrunkin81} are most natural characteristics of the
second-order proton response to external electromagnetic fields.
Details of this view are discussed below.
The often used expressions like
$\bar \alpha = \alpha_0 + \Delta \alpha$ and
$\bar \beta = \beta_{para} + \beta_{dia}$
introduce quantities (they are normally termed
the static electric polarizabilty $\alpha_0$, the retardation
correction of the electric polarizability $\Delta\alpha$,
the paramagnetic polarizability $\beta_{para}$ and
the diamagnetic polarizability $\beta_{dia}$)
which have only a theoretical significance but cannot be
measured in real (or even gedanken) experiments.

There is a persisting prejudice that $\bar\alpha$ has not its own fundamental
sense and rather appears in Compton scattering as an artificial
combination of the genuine polarizability $\alpha_0$ and the
retardation correction $\Delta \alpha$. The latter is thought
to be caused by non-static effects pertinent to real
photons.  They think that the static polarizability $\alpha_0$
determines a coupling of the particle to a static electric field and
therefore $\alpha_0$ alone is measured under static conditions which allow
to expell the retardation effects.  Such a view point is wrong, what
is easily demonstrated by a general consideration based on effective
Lagrangians \cite{lvov93}.

A rigorous theorem, which, for the sake of simplicity, we formulate here
for a {\em spinless} composed hadron or nucleus of a known mass $m$
and electric charge $e$, is the following.
Whenever the particle stays or slowly moves in the region
outside external electromagnetic charges or currents
$j_\mu^{ext}(\vr,t)$ which create the probing electric $\vE$ (or
magnetic $\vH$) field, the internal structure of the particle and its
low-energy long-wavelength response to the field
is characterized by the only additional parameter
$\bar\alpha$ through an effective potential
$-\frac12\bar\alpha \vE^2$
(or by $\bar\beta$ through $-\frac12\bar\beta \vH^2$, respectively)
\cite{lvov93}. This implies that
the so-called static polarizability is irrelevant
whenever the particle has a charge and an internal size
$\langle r^2 \rangle \ne 0$,  thus leading to
$\bar\alpha = \alpha_0 + e^2\langle r^2 \rangle / (3m) \ne \alpha_0$.
This theorem signifies that all standard tools considered in textbooks
as methods for measuring the electric polarizability
(like placing the particle into the field of an electric capacitor
and looking at its energy shift or at an induced dipole moment)
give $\bar\alpha$ rather than $\alpha_0$ \cite{lvov87}.
There is no way to measure $\alpha_0$ instead of $\bar\alpha$
but putting the probing external charges $j_0^{ext}(\vr,t)$
{\em inside} the particle, {\it e.g.} in an electron scattering
experiment with observation of a secondary real or virtual photon.
In an experiment like this one could simultaneously measure
$e^2\langle r^2 \rangle$ and $\bar\alpha$ and, therefore,
determine $\alpha_0$.
This means that the polarizabilities $\bar\alpha$ and $\bar\beta$
introduced in Compton scattering studies are structure parameters which have
a more general sense than sometimes assumed.
In essense, nothing changes when spin of the particle is included.

Since we are dealing here with Compton scattering itself, the relevance
of the free-nucleon polarizabilities $\bar\alpha_N$ and $\bar\beta_N$
needs no further explanations. However we have to explain how
the in-medium polarizabilities of the nucleon,
$\tilde\alpha_N$ and $\tilde\beta_N$,
enter to nuclear Compton scattering.
For a clear-cut definition of the in-medium polarizabilities
\cite{R85,martin1} we have to distinguish between
(i) effects of rebuiding the internal ({\it e.g.}, quark) structure of the nucleon
entering into $\tilde\alpha_N$ and $\tilde\beta_N$
as one-body quantities and
(ii) two-body effects of meson exchange currents between $pn$ pairs
which may be denoted by $\delta\alpha$ and $\delta\beta$.
From a phenomenological point of view developed in sections 3 and 4
these latter quantities are not part of the in-medium
electromagnetic polarizabilities but are rather used to parametrize
that part of the ``mesonic seagull amplitude"
(Thomson scattering by correlated $pn$ pairs)
which is energy dependent in the forward direction
({\it cf.} Eq.\ (\ref{meson-corr}); for more details see section  9).
In a first step of the  analysis of experimental data the convolution
of the true in-medium electromagnetic polarizabilities and the meson
exchange corrections are  determined. This leads to the effective
in-medium  electromagnetic polarizabilities
\begin{equation}
\label{polariz-effective}
   {\tilde \alpha}^{eff}_N = {\tilde \alpha}_N + {\delta \alpha},
\quad
   {\tilde \beta}^{eff}_N = {\tilde \beta}_N + {\delta \beta}.
\end{equation}
The quantities $\delta \alpha$ and $\delta \beta$ have to be
calculated and, therefore, introduce some model dependence into the
determination of ${\tilde \alpha}_N $ and ${\tilde \beta}_N$.
However, as will be
shown later, the respective model dependent uncertainties are not large.

We have to repeat our words said in the beginning of section 3
that it would be also possible to accept the effective quantities
(\ref{polariz-effective}) as an alternative (and different)
definition of the in-medium polarizabilities. Since the free-nucleon
polarizabilities (at least, the electric one)
are essentially determined by the pion cloud of the nucleon
\cite{weiner85,bernard95}, the in-medium modifications of the
pion contribution to $\bar\alpha_N$ and $\bar\beta_N$
like a Pauli blocking \cite{ME86,bunatian92} have also a right
to be attributed to the in-medium polarizabilities of the nucleon.
However, in no way such effects should be added to
the mesonic seagull contribution because this leads to the double counting.
Considering in the following the in-medium polarizabilities
we always mean the quantities ${\tilde \alpha}_N $ and ${\tilde \beta}_N$
which are free from the mesonic seagull correction.

In nuclei with $Z=N$ which are investigated in the following the arithmetic
averages of the proton and neutron electromagnetic polarizabilities are
observed. From the Baldin-Lapidus sum rule applied to experimental
photoabsorption cross sections the sum
$({\bar \alpha} + {\bar \beta})_N = 14.5 \pm 0.5$
(in units of $10^{-4}~\rm fm^3$) is very well known
\cite{damashek70,lvov79,babusci98,babusci98a},
where later estimates \cite{babusci98,babusci98a}
give values by a few per cent lower that the older ones
\cite{damashek70,lvov79}.
Among the reasons for that is that
experiments on Compton scattering by the proton in the $\Delta$-resonance
range \cite{huenger97} and photoproduction experiments
(like \cite{MacCormic97}) carried out at MAMI (Mainz)
lead to a more precise determination of the resonant $M_{1+}$ amplitude
of the $\Delta$ photoexcitation. Now \cite{arndt97,hanstein98}
this amplitude is by about 3\% lower than it was assumed
a few years ago in the SAID (SM95)
parametrization of photo-meson amplitudes \cite{arndt96}.

Though, apparently there is some room for discussion,
the overall precision of $({\bar \alpha} +{\bar \beta})_N$ is by far
good enough for the purpose of our present data analysis. Furthermore,
adjustments in the predicted differential cross sections are possible at
small angles within the limits given by the errors of the nuclear
photoabsorption cross sections ({\it cf.} Fig.\ \ref{pr31pr28}),
making small differences
in $({\bar \alpha} +{\bar \beta})_N$ unobservable. Note that the free
parameter of our data analysis is the difference of in-medium
polarizabilities  $({\tilde \alpha} - {\tilde \beta})_N$ which has no
influence on the differential cross section at zero angle.

Separately, polarizabilities of the free proton have been
measured in proton Compton scattering experiments performed
at photon energies below photo-meson threshold
\cite{goldanski60,baranov74,federspiel91,zieger92,macgibbon95}.
Their results are partly summarized
in the review paper \cite{petrunkin81} and
in the Review of Particle Physics 1998 \cite{PDG98}
(in fact, the latter reference relies on
a ``global average" over experiments of 90's
derived in \cite{macgibbon95}
on the base of the Baldin--Lapidus sum rule and experimental
data on the differential cross section of proton Compton scattering).
It was found that
$\bar\alpha_p = \rm 12.1 \pm 0.8~(stat+syst) \pm 0.5~(theor)$ and
$\bar\beta_p  = \rm  2.1 \pm 0.8~(stat+syst) \pm 0.5~(theor)$.

The knowledge of polarizabilities of the neutron is less certain.
Using model-dependent dispersion relations at fixed $t$,
the differences
$\bar\alpha_p - \bar\alpha_n \simeq  -1.4$ and
$\bar\beta_p  - \bar\beta_n  \simeq   0.1$
are theoretically expected \cite{lvov79,babusci98a}.
On the experimental side the first meaningful number
for the electric polarizability of the neutron
was measured by Rose et al.\ \cite{rose90}
through quasifree Compton scattering on neutrons
bound in the deuteron. The precision of this experiment was surpassed
by an experiment on scattering of neutrons in the  Coulomb field of Pb
nuclei, enriched in $^{208}$Pb \cite{schmied91}.
The reported number%
\footnote{We give that number and a similar number below
with a small correction of 0.62
added which takes care of relativistic effects lost in the fully
nonrelativistic treatment of the Coulomb scattering experiments
\cite{lvov93,lvov87,bawin97}.}
${\bar \alpha}_n = \rm 12.6 \pm 1.5 ~(stat) \pm 2.0 ~(syst)$
perfectly agrees with the above theoretical expectation.
However, more recently a new experiment on Coulomb scattering of the
neutron \cite{koester95} gave a rather different polarizability
$\bar\alpha_n = 0.6 \pm 5$.  It was also claimed that the high accuracy of
the former Coulomb scattering result \cite{schmied91} may possibly be
grossly overestimated \cite{enik97}.
A detailed discussion of these points and a possible experimental way
out of this problem has recently been addressed by Wissmann et al.\
\cite{wissmann98}.

Given the experimental results for proton
polarizabilities and the theoretical evaluations
of the proton--neutron differences, the works on nuclear
Compton scattering discussed below
use the following values for averaged free-nucleon polarizabilities:
${\bar \alpha}_N = 11.3 \pm 1.5$ and ${\bar \beta}_N=3.7 \mp 1.5$.
These numbers should be compared with what is inferred from
nuclear Compton scattering experiments themselves.
The free-nucleon value of the sum is slightly shifted down
in the nuclear medium to about
$({\tilde \alpha} + {\tilde \beta})_N = 14.0$ \cite{L12}. This
observation of an approximately constant sum of electromagnetic
polarizabilities does not exclude that the relative sizes of the in-medium
electric ${\tilde \alpha}_N$ and magnetic ${\tilde \beta}_N$ polarizabilities
may be considerably different from the corresponding free-nucleon values
due to meson exchange-currents and/or modifications of the internal
structure of the nucleon. In this connection it is of interest that
in-medium shifts as large as ${\Delta \alpha}^{eff}_{exp} = - 8$ and
${\Delta \beta}^{eff}_{exp} = + 8$ have been reported on the basis of
analyses of Compton scattering experiments on $^{16}$O \cite{F16}.

\subsection{Meson exchange currents}

In complex nuclei meson exchange currents are a consequence of the
interaction between proton-neutron
pairs (quasideuterons). Following the notation of our previvous work
\cite{H15} and the discussion of section  5.1,
the  corresponding modification of the nuclear scattering
amplitude  up to the order $\omega^4$ may be written in the form
\begin{equation}
\label{DeltaM}
S(\omega,\theta) - {\tilde S}(\omega,\theta)
 = \Big(\delta \alpha \, g_{E1}(\theta) +
\delta \beta \, g_{M1}(\theta)  \Big) \omega^2 F_2(q).
\end{equation}
The r.h.s.\ of  (\ref{DeltaM}) takes care of the fact that
the meson exchange corrections  $\delta\alpha$ and
$\delta\beta$ are related to a two-body effect and, therefore, go along
with the formfactor $F_2(q)$.
The approximative definition in the following Eq.\ (\ref{DeltaM1})
is more convenient,
because in this case $\delta\alpha$ and $\delta\beta$ can be directly
compared with the  electromagnetic polarizabilities of the free  nucleon.
\begin{equation}
\label{DeltaM1}
S(\omega,\theta) - {\tilde S}(\omega,\theta)\approx
\left(\delta \alpha^{(1)}\, g_{E1}(\theta) +
\delta \beta^{(1)} \, g_{M1}(\theta) \right)  \omega^2  F_1(q).
\end{equation}
A calculation of  $\delta\alpha$ and  $\delta\beta$       has been
carried out by H\"utt  and Milstein \cite{huett96} for nuclear matter.
The result is  $\delta{\alpha}^{(1)}(A = \infty)= - 3.4$
and $\delta{\beta}^{(1)}(A = \infty)= +
2.4$ ({\it cf.} Table \ref{MesonCorr}).
For finite nuclei the same    authors \cite{huett97}
find the results also listed in Table \ref{MesonCorr}.
We see that the meson exchange corrections of the electromagnetic
polarizabilities become smaller with decreasing mass number. The reason
for this is that the possibility for two-body or --- even more ---
three-body effects to take
place becomes smaller for decreasing mass number. Three-body
effects are essential because they enter into the
process through correlations.
In $^4$He the possibility for two-body effects is largely reduced
and that for three-body effects almost absent.

\begin{table}[htb]
\caption{\it Meson exchange corrections of the in-medium electromagnetic
polarizabilities in units of \protect{$10^{-4}fm^3$}}
\vspace{3mm}
\begin{center}
\begin{tabular}{|c|c|c|}
\hline
A&\protect{$\delta{\alpha}^{(1)}$}&\protect{$\delta{\beta}^{(1)}$}\\
\hline
$\infty$& $-3.4$ & $+2.4$ \\
  40    & $-2.4$ & $+2.1$ \\
  16    & $-1.3$ & $+1.0$ \\
   4    & $\approx 0$ & $\approx 0$ \\
\hline
\end{tabular}
\end{center}
\label{MesonCorr}
\end{table}

\subsection{Experiments on the bound-nucleon electromagnetic polarizabilities}

Figure \ref{pr19} shows a typical experimental arrangement as used
at the MAX laboratory of the university of Lund (Sweden)
\cite{adler90,adler97}.
The photon beam having an energy of about 100 MeV hits a thin metal foil
serving as a bremsstrahlung radiator. Quasi monochromatic photons are
obtained through a coincidence condition between an event in one of the
NaI(Tl) detectors (A, B, C) and an event in the tagger. The tagger
consists of a magnetic spectrometer which directs the unused electrons
into the beam dump, and  electrons which have lost energy through
bremsstrahlung production on an array of plastic scintillators. These
plastic scintillators provide energy channels for the coincident
bremsstrahlung photons.

\begin{figure}[htb]
\centering
\caption{\em Typical experimental arrangement as used at the MAX laboratory
of the university of Lund (Sweden) \cite{adler90,adler97}.
The photon beam having
an energy of about 100 MeV hits a thin metal foil serving as a
bremsstrahlung radiator. Quasi monochromatic photons are produced
through a coincidence condition  between an event in one of the
NaI(Tl) detectors (A, B, C) and an event in the tagger.}
\label{pr19}
\end{figure}

For the data interpretation the resonance amplitude $R(\omega,\theta)$
and the static part of the mesonic seagull amplitude ${\tilde
S}(\omega,\theta)$
have to be calculated in addition to the kinetic seagull amplitude
$B(\omega,\theta)$. This is possible with very good precision for nuclei
where the total photoabsorption cross section is known. To illustrate this,
the total photoabsorption cross sections of $^{16}$O and $^{40}$Ca measured
by Ziegler et al.\ \cite{ahrens75}
are shown in Fig.\ \ref{pr31pr28}.
The experimental photoabsorption cross section as fitted by the solid
line is partitioned into Lorentzian lines. Except for the sum of all these
Lorentzian lines, the only constraint is that the QD cross section as given by
the dotted line resembles the Levinger representation of Eq.\
(\ref{levinger}) as close as possible. All the other Lorentzian lines may
be interpreted in terms of GR components.
The experiments carried out to measure
the in-medium electromagnetic polarizabilities use photons of 50 to 80
MeV. At these energies only the tails of the GR components are of
relevance since the real part of the QD amplitude has a zero crossing
in this energy range.
This makes the determination of the amplitude $R(\omega,\theta)$
quite unambiguous. For the  determination of the static part
of the mesonic seagull amplitude  ${\tilde S}(\omega,\theta)$ only
the integrated GR and QD cross sections have to be known in order to
determine $\k $ and $\kappa^{QD}$. The formfactor $F_2(q)$ is taken
from model calculations ({\it cf.} section 9).

\begin{figure}[t]
\centering
\caption{\em
Total photoabsorption cross sections \cite{ahrens75}
for $^{40}Ca$ (upper figure) and
$^{16}O$ (lower figure)
partitioned into components having Lorentzian shapes. Dashed curves: GR
components. Dotted curves: QD components.
}
\label{pr31pr28}
\end{figure}

On the basis of the prerequisites discussed in the preceding paragraph,
we now discuss the present status of our knowledge about
in-medium electromagnetic polarizabilities based on recent results
\cite{P18a}   obtained
for $^{40}$Ca and $^{16}$O. Throughout this section we will
calculate the predictions
for different ``effective" in-medium electromagnetic
polarizabilities ${\tilde \alpha}^{eff}_N$ and ${\tilde \beta}^{eff}_N$.
These quantities  are introduced
such that they  may be inserted into Eq.\ (\ref{T-N}) in order to take care of
different tentative choices for the in-medium electromagnetic
polarizabilities including the effects of meson-exchange currents and of
modifications of the internal structure of the nucleon, while simultaneously
the mesonic seagull-amplitude
$S(\omega,\theta)$ is represented through its static approximation
${\tilde S}(\omega,\theta)$.
These choices are\\
(i) the free-nucleon electromagnetic polarizabilities,   \\
(ii) the free-nucleon electromagnetic polarizabilities supplemented by
meson exchange corrections predicted for the finite nucleus,\\
(iii) the free-nucleon electromagnetic polarizabilities supplemented
by meson exchange corrections predicted for nuclear matter.\\


{\bf $^{40}$Ca}:
Figure \ref{CaAngDep}  shows the energy
distributions of the
differential cross sections
obtained in  recent experiments \cite{P16,P18,P18a}
together with predictions.
The nuclear part of the amplitudes, i.e.
$B(\omega,\theta)+{\tilde S}(E,\theta)+R(E,\theta)$, is fixed through
the experimental
total photoabsorption cross section
so that the only  parameters are  the effective electromagnetic
polarizabilities ${\tilde \alpha}^{eff}_N$ and  ${\tilde \beta}^{eff}_N$.
All three curves are in good agreement with each other and with
the data points at $\theta =45^{\circ}$. This shows that the nuclear
part of the scattering amplitudes has been calculated with sufficient
precision, so that the angular dependence of the differential cross section
may be interpreted in terms of
properties of the effective electromagnetic polarizabilities. The only
remaining precaution is concerned with the isovector giant-quadrupole resonance
(IVGQR) which also has an effect on the angular distribution
of the differential
cross section. After taking the IVGQR into account using experimental
photoabsorption data \cite{bergqvist84,sims97}, the uncertainty  of
the strength and location of the IVGQR
leads to modifications of the angular
distribution, being essentially smaller than
the differences between the lower (dashed) curves
and the (solid) curves  in the middle and, thus, is irrelevant for our
conclusion that there is no noticeable modification of the in-medium
electromagnetic polarizabilities compared to the free ones.
Indeed, there is apparent  preference of the experimental data for
the solid curves i.e.\ the predictions including the free-nucleon
electromagnetic polarizabilities supplemented by the
meson-exchange corrections  ${\delta \alpha}^{(1)}$ and ${\delta \beta}^{(1)}$
predicted for the finite nucleus. However, for $^{40}$Ca it apparently makes no
major difference whether we make use of the  nuclear
matter (A = $\infty$, dotted curves) predictions or the predictions for
the finite nucleus (A = 40, solid curves). In the next paragraph we will
see that such differences become visible for the smaller nucleus $^{16}$O.

\begin{figure}[htb]
\caption{ \em Experimental elastic differential cross
sections \cite{P16,P18,P18a}  for $^{40}$Ca
versus scattering angle compared with predictions.
E$_{\gamma}$ = 58 MeV (upper Figure),
E$_{\gamma}$ = 75 MeV (lower Figure).
The curves are calculated for  (i) the free-nucleon electromagnetic
polarizabilities (dashed), (ii) the free-nucleon electromagnetic
polarizabilities supplemented by meson exchange corrections predicted for
the finite nucleus (solid), and (iii) the free-nucleon electromagnetic
polarizabilities supplemented by meson exchange corrections predicted for
nuclear matter (dotted).
}
\label{CaAngDep}
\end{figure}

{\bf $^{16}$O:}
Figure \ref{O16DiffAngle} shows data obtained for $^{16}$O versus
scattering angle.
For this smaller nucleus the meson exchange corrections
$\delta \alpha^{(1)}$ and $\delta \beta^{(1)}$ of the
electromagnetic polarizabilities predicted for the finite nucleus (A =
$\infty$, solid curve) shows a sizable difference from the corresponding
prediction for nuclear matter (A = 16, dotted curve). The lower curves
(dashed) have
been calculated with the meson-exchange corrections set to zero.
It is apparent that the solid curves
provide the optimum fit to the majority of the
experimental data --- if we tentatively exclude the two data points for
E$_{\gamma}$ = 75 MeV and about $\theta$ = 60$^{\circ}$ from the present
consideration. This finding
leads to the conclusion that the predicted meson-exchange corrections are
in line with our experiments and that there are no indications of an
additional in-medium modification of the electromagnetic polarizabilities.

\begin{figure}[htb]
\centering
\caption{\em Differential cross sections \cite{H15,P18,P18a}
for Compton scattering by $^{16}O$
versus scattering angle compared with predictions. $E_{\gamma}$ = 58 MeV
(upper figure); $E_{\gamma}$ = 75 MeV (lower figure).
The curves are calculated for (i) the free-nucleon electromagnetic
polarizabilities (dashed), (ii) the free-nucleon electromagnetic
polarizabilities supplemented by meson exchange corrections predicted for
the finite nucleus (solid), and (iii) the free-nucleon electromagnetic
polarizabilities supplemented by meson exchange corrections predicted for
nuclear matter (dotted).
}
\label{O16DiffAngle}
\end{figure}

\subsection{Summary and Conclusions}

The data presented above have led to a satisfactory
consistency in our conclusion concerning
the in-medium electromagnetic polarizabilities. This conclusion is that
the proton-neutron averages of in-medium  electromagnetic polarizabilities
${\tilde \alpha_N}$ and ${\tilde \beta_N}$  are the same as the
corresponding quantities
${\bar \alpha_N}$ and ${\bar \beta_N}$ for the free nucleon within a
precision of the order of  \,$\pm 2.5\times10^{-4} ~\rm fm^3$,
if we do not put too
much weight on some points discussed in the next paragraph.
Furthermore, there are strong indications that the data obtained for
$^{40}$Ca and $^{16}$O are in favour of the predicted  \cite{huett97}
meson-exchange corrections $\delta \alpha$ and $\delta \beta$
of the electromagnetic polarizabilities.
It was known before
that the sum of electromagnetic polarizabilities
${\bar \alpha_N}+{\bar \beta_N}$ is not modified through binding.
The essential additional message
of our result is that the difference of electromagnetic polarizabilities
${\bar \alpha_N}-{\bar \beta_N}$ also remains the same. In the dispersion
theory of nucleon Compton scattering \cite{lvov97} the difference
${\bar \alpha_N}-{\bar \beta_N}$  is an independent input parameter
which may be related to the exchange of the scalar $\sigma$ meson in the
$t$-channel. On the other hand the scalar $\sigma$  meson is believed to be
responsible for the largest part of the binding potential between nucleons
in a complex nucleus.
This observation suggests that there might be a relation between
the quantity ${\tilde \alpha_N}-{\tilde \beta_N}$
measured in nuclear Compton scattering and the nucleon
binding potential. The tentative
conclusion from our experimental result, therefore, may be that the
role  the $\sigma$ meson plays for the structure of the nucleon
is not modified when the nucleons are imbedded in nuclear matter.

The conclusions drawn in the foregoing section are weakened to some extent
by some unsolved problems which should be cleared up in further studies:
(i) The free-nucleon polarizabilities of the neutron are not known with
the desirable reliability. (ii) The angular distribution of
Compton-scattering differential cross-sections for $^{12}$C and $^{16}$O
at an energy of 75 MeV deviates from the predictions at $\theta$ =
60$^{\circ}$ by 20--30 \% \cite{H15}. These deviations have not been
removed by the recent theoretical studies of H\"utt and Milstein
\cite{huett96,huett97}.

\clearpage

\section{Compton scattering above $\pi$ meson  threshold}



Nuclear Compton scattering above $\pi$ meson threshold has three
regions of current interest where experimental work has been carried out:\\
(i) Compton scattering in  resonance range where the experiments were
restricted to the energy region of the $\Delta$ resonance.\\
(ii) Compton scattering in the asymptotic range at small transverse
momenta $p_T$, where shadowing and vector-meson dominance phenomena
may be studied (see {\it e.g.} \cite{rull90}).\\
(iii) Compton scattering in the asymptotic range at large transverse
momenta $p_T$, where the direct coupling of photons to valence quarks may be
studied. These experiments are of interest because the wave function
of quarks in the nucleon may be investigated. Data are available only for the
proton (see {\it e.g.} Kroll et al.\ \cite{kroll91,kroll96}).

In principle there is also a fourth type of Compton scattering
experiments at high $p_T$, {\it viz.} the deep inelastic scattering
of photons by quarks \cite{wormser90}. These deep inelastic Compton
processes are not coherent-elastic with respect to the nucleon or
nucleus but resemble Compton scattering by single quarks with
subsequent hadronization. As a consequence a jet of hadrons is
leaving the nucleon after  Compton scattering has taken place.

Our discussion here is rather brief and restricted to the 
$\Delta$-resonance region of nuclei, i.e. to case (i).  
The latter choice is made
because this is the only region where recent experimental data are
available and because the theoretical ideas which are tested through the
experiments have a large overlap with those considered in the rest of
the present paper.

\subsection{Pioneering work on Compton scattering in the resonance range}

The study of nuclear Compton scattering above $\pi$ meson threshold
started on the experimental side with the work of Hayward and Ziegler
\cite{H4}.  The experiment has been carried out with a low-duty
factor linear accelerator then available in Mainz and with a
comparatively small (25 cm$\oslash \times 36$ cm) NaI(Tl) detector
positioned at one scattering angle of $\theta = 115^\circ$ and in an
energy range from 180 to 375 MeV.
One difficulty in these experiments was the separation of elastic
from inelastic photon scattering processes. Therefore, the main
interest in subsequent theoretical work  \cite{arenhoevel85,vesper85}
was directed to this problem. These investigations were based on
$\Delta$-hole model approaches
\cite{koch84,lenz75,os79,klingenbeck80} including inelastic processes
and led to some  agreement with the experimental data.  These first
investigations showed that the proper experimental separation of
elastic and inelastic components is essential for this type of
experiments. This was one of the major motivations for purchasing the
large Mainz 48 cm$\oslash \times 64$ cm NaI(Tl) detector \cite{W14}.

\subsection{Compton scattering by  $^{12}$C in the $\Delta$ range}

In a first application of the large Mainz 48 cm$\oslash \times 64$ cm
NaI(Tl) detector elastic and inelastic photon scattering from
$^{12}$C was investigated \cite{W14} at $\theta_{LAB}= 40^\circ$
using the tagged-photon beam at the 855 MeV microtron MAMI (Mainz).
With a width of tagger intervals of $\Delta E$= 2 MeV and an energy
resolution of the NaI(Tl) detector of 1.5~\% the separation of the
elastic and the inelastic lines was sufficient to obtain  elastic
(Compton) and inelastic differential cross sections in an energy
interval from 200 MeV to 500 MeV.
The differential cross sections for Compton scattering were compared
with calculations carried out in the $\Delta$-hole model
\cite{koer94} leading to a remarkably good agreement between theory
and experiment at photon energies above 250 MeV and a drastic
disagreement at photon energies below 250 MeV.  The occurance of this
discrepancy was very surprising because $\theta_{LAB}=40^\circ$ is
close enough to the forward direction so that any theory which
reproduces the photoabsorption cross section and has a correct
low-energy limit should also reproduce the Compton scattering
differential cross section.
It is interesting to note that the same observation was made in first
interpretations of Compton differential cross sections measured for
$^4$He at the same angle  $\theta_{LAB}=40^\circ$ \cite{S16,pas96}
This observation seemed to suggest that  some omission should be in the
$\Delta$-hole approach as formulated in Refs.\ \cite{koch84,koer94}.
In the meantime this omission was identified to
be the neglect of the  seagull amplitudes $B(\omega,\theta)+{\tilde 
S}(\omega,\theta)$
which were discussed in section 3.
Indeed after including the seagull amplitudes $B(\omega,\theta)+{\tilde 
S}(\omega,\theta)$
the agreement between theory and the $\theta_{LAB}=40^\circ$ data
for $^{12}$C and $^4$He became much better, as is further discussed below.

\subsection{Compton scattering by $^4$He in the $\Delta$ range}

Among the nuclei to be studied by Compton scattering $^4$He is of special
interest, because this nucleus combines the well-known structure of a
few-body system with the binding energy of a complex nucleus. Furthermore,
the first excited state is at 20 MeV, thus making the separation of
elastic and inelastic processes easy.
Previous experiments on Compton scattering by $^4$He in the $\Delta$
range \cite{A6,A8,D11,M12,I15} were carried out using the bremsstrahlung
method and, therefore, may have suffered from the well-known systematic
errors contained in that method. In spite of that, the use of large NaI(Tl)
detectors made it possible to get good data on the elastic reaction
at energies close to the end points of the bremsstrahlung beams.

The first fully high-resolution Compton scattering experiments on
$^4$He were carried out only recently with a remarkable progress.  At
LEGS (BNL) \cite{B17}, the first data were obtained with a polarized
and tagged photon beam ($\Delta E\approx$ 5 MeV) produced by
backscattering of laser light on 2.5 GeV electrons.  Photons were
detected with a 48 cm$\oslash\times 48$ cm NaI(Tl) detector.
Measurements were performed at six angles between $31^\circ$ and
$130^\circ$ which produced angular dependences of the differential
cross sections and beam asymmetries at 5 energies between 206 and
310 MeV.
At MAMI (Mainz), for the first time Selke et al.\ \cite{S16}
performed measurements at a scattering angle of $\theta_{LAB} =
37^\circ$ through the whole $\Delta$ region from 120 to 510 MeV.
The tagged-photon beam with $\Delta E\approx 2$ MeV and the large
Mainz 48 cm$\oslash\times 64$ cm NaI(Tl) detector were used in that
experiment.  The latter data were supplemented by Kraus et al.\
\cite{K18} leading to additional data at $\theta_{LAB}=37^\circ$
taken with 48 cm$\oslash\times 64$ cm NaI(Tl) detector and to
additional data at energies between 150 and 300 MeV for the two angles of
$\theta_{LAB} = 93^\circ$ and $\theta_{LAB} =  137^\circ$ covered by
smaller 25.4 cm$\oslash\times 35.6$ cm NaI(Tl) detectors.  In this
work also the beam asymmetry was determined between 230 and 270 MeV
using linear polarization produced by the coherent bremsstrahlung in
a diamond crystal.

The results of the recent Mainz experiments \cite{S16,K18}
for the differential cross section are shown in Fig.\ \ref{HeliumDelta}
together with theoretical predictions to be
discussed below.
Not shown in that Figure are the LEGS data \cite{B17} which, for example,
lie in between the data points of
Mainz \cite{K18} and Saskatoon \cite{D11} at $90^\circ$
and agree with  both, the Mainz and Saskatoon data, at $130^\circ$.
A sample of the LEGS data can be seen in Fig.\ \ref{He4-results} below.
As was already found in the very first measurements
at Bates (MIT) \cite{A6,A8}, any version of the $\Delta$-hole model ---
the original version of Ref.\ \cite{koch84},
a simplified version \cite{pas96} based on a local-density
approximation, and a version \cite{koer94} corrected \cite{K18}
for the seagull contributions to the elementary $\gamma N$ scattering
amplitude  ---
fails to explain the large size of the differential cross section
observed at angles $\theta_{LAB} \simge 90^\circ$.
This is in contrast to the fact that  a nice agreement exists between the
experimental data and the modern predictions \cite{K18}
at a  small scattering angle of $\theta_{LAB}$ = 37$^{\circ}$.
Therefore, something fundamental must be  wrong with our 
present understanding of
Compton scattering in the $\Delta$ range.

\begin{figure}[htb]
\caption{\em Differential cross sections $(d\sigma/d\Omega)_{LAB}$
for Compton scattering by  $^4$He \cite{K18} compared with predictions.
The full circles are calculated within the $\Delta$-hole model
including the seagull amplitudes \cite{K18}.
The results of the schematic model (dashed)
and the extended schematic model (solid) are also shown \cite{K18}.}
\label{HeliumDelta}
\end{figure}

\subsection{Calculations of Compton scattering in the  $\Delta$ range}

The discrepancies  found between
the experimental $^4$He data and the theory at large angles
demonstrate the  necessity of comparing the main components of
the existing different approaches with each other 
and to search for possible lines of developments which may solve
the problems, apparently existing in all them. We
will not discuss here whether the solution of these
problems may be possible in a genuine isobar-meson model or whether
it might be necessary to introduce different degrees of freedom
like those of constituent quarks.
Some ideas going into the latter direction have been published
in recent literature \cite{huber90}.

{\it The schematic model:}\\
A zero-order approach to a theoretical interpretation of Compton
scattering in the $\Delta$ range has been attempted through the
``schematic model".  It is based on the suppositions (i) that
the  absorption and emission of photons takes place locally on one nucleon,
(ii) that different nucleons contribute equally and coherently, thus
introducing into the nuclear Compton scattering amplitude
$T_{tot}(\omega,\theta)$ the one-body nuclear form factor, and
(iii) that the local spin-averaged $\gamma N$ scattering amplitude
has a pure magnetic dipole, 
$M1$, characteristic  (or optionally $E1$ but not both).
In the schematic model the energy dependence is not taken from the
elementary $\gamma N$ scattering amplitude but rather from the full
nuclear amplitude at zero angle, $T_{tot}(\omega,0)$.  The latter
quantity can be derived from  the total photoabsorption cross section
$\sigma_{tot}(\omega)$ through the optical theorem and the GGT dispersion
relation, thus giving the dispersion-theoretical differential cross
section ${d\sigma_{GGT} /d\Omega}$.
One then arrives at the simple formula \cite{K18,H4}
usually used in the CM frame:
\begin{equation}
   \frac{d\sigma}{d\Omega}(\omega,\theta) =
   \frac{d\sigma_{GGT}}{d\Omega}(\omega,0)\,
          |F_1(q)|^2\,\,\,\frac{1+\cos^2\theta}{2}.
\label{schematic}
\end{equation}
Here $q$ is the momentum transfer and $F_1(q)$ the one-body
formfactor. It is not expected that the schematic model
(\ref{schematic}) makes a very realistic prediction except for very forward
directions, because {\it e.g.} of the neglect of any multipolarity
except for $M1$, which is an oversimplification for the $\Delta$
energy-range.

{\it The extended schematic model:}\\
The nuclear scattering amplitude in the ``extended schematic model"
\cite{K18} is given by a combination of several pieces which are
results of different  reaction mechanisms. In  close
analogy with the considerations given in sections 3 to 5, we write
\begin{eqnarray}
     T_{tot}(\omega,\theta) &=& T_A(\omega,\theta)
               + T_N(\omega,\theta),
\nonumber \\
     T_N(\omega,\theta) &=& T^{E1}(\omega,0) \,F_1(q) \,g^{E1}(\theta)
             + T^{nm}_{\Delta}(\omega,0) \,F_2(q) \,g^{M1}(\theta)
\nonumber \\
  &&   {} \qquad
             + T^{m}_{\Delta}(\omega,0) \,F_1(q) \,g^{M1}(\theta).
\label{TotalAmpl}
\end{eqnarray}
In (\ref{TotalAmpl}) the amplitude $T_A(\omega,\theta)$ contains all
the relevant nuclear contributions as there are the kinetic and
mesonic seagull amplitudes and some high-energy tail from the
positive-energy part of the quasideuteron amplitude.
These pieces were discussed in sections 3 and 5.
The amplitude $T_N(\omega,\theta)$ is related to nucleon excitations
including the  $\pi N$ continuum.
It contains one term of electric-dipole ($E1$)
multipolarity and two terms of magnetic-dipole ($M1$) multipolarity.
The term $T^{E1}(\omega,0)$ is calculated  via optical theorem and
dispersion relation from the total photoabsorption cross section of
the nucleon excluding the $\Delta$-resonance contribution. This is
justified because the nonresonant one-pion channel and the
$D_{13}(1520)$ resonance are of $E1$ multipolarity and also the
nonresonant two-pion channel is predominantly of $E1$ multipolarity.
At low energies, a large part of $T^{E1}$ comes from the so-called
Kroll-Ruderman mechanism of pion photoproduction.
The $\Delta$-resonance itself appears in two terms,
$T^{nm}_{\Delta}(\omega,0)$ and $T^{m}_{\Delta}(\omega,0)$,
corresponding to the nonmesonic ($nm$) and the mesonic ($m$)
photoabsorption cross section of the nuclear $\Delta$ resonance
(see Fig.\ \ref{NuclearDelta}).
Of the three terms on the r.h.s.\ of (\ref{TotalAmpl}) the nonmesonic
$\Delta$ term goes along with the formfactor $F_2(q)$ because this
term corresponds to a two-body process.
At  forward angle the total amplitude (\ref{TotalAmpl})
necessarily coincides with the dispersion-theoretical amplitude
$T_{GGT}(\omega,0)$ provided the sum of partial cross-sections used to
calculate the different pieces in (\ref{TotalAmpl}) give 
$\sigma_{tot}(\omega)$, i.e. the total photoabsorption cross section of
the nucleus. Therefore, it is not  surprising that the nuclear Compton
scattering data at small angles find an explanation in the framework
of such a model. Really important is that by making use of both 
multipolarities, $M1$ and
$E1$, and by including the seagull contributions
improvements at large angles are obtained.

{\it The $\Delta$-hole model:}\\
Recent calculations of Compton scattering in the $\Delta$-hole
approach \cite{pas96,koer94,K18} originate from the treatment of the
problem given by Koch, Moniz and Ohtsuka \cite{koch84}.  Due to
simplifications introduced by the local density approximation, which
was used in the later works, many parallels with the two  models
described above 
become transparent.  The scattering amplitude schematically reads
\begin{equation}
\label{T-Delta-hole-general}
   T_{tot}(\omega,\theta) = A T^{\gamma}_{\Delta}(\omega)
     g^{M1}(\theta) F_\Delta(q)
     + \Big( Z T^{\gamma}_T(\omega)
    + A T^{\gamma}_{KR}(\omega) \Big) g^{E1}(\theta) F_1(q).
\end{equation}
Here the first term gives the $\Delta$-resonance part of the
scattering amplitude.  It appears from
the model as local and has a pure $M1$ multipolarity.  However, the form
factor $F_\Delta(q)$ it is accompanied with is not identical with the
one-body form factor $F_1(q)$, because $F_\Delta(q)$ counts both, the
nucleon distribution in the nucleus and a nuclear-density dependence
of elementary Compton scattering on elements of the nuclear volume
through the $\Delta$-hole excitation,
which appears due to the density-dependent width and self-energy
of the dressed $\Delta$ in the nuclear medium.
The second term is of  $E1$ multipolarity. It represents the Thomson part
of the $\gamma p$ scattering amplitude (which was omitted in
\cite{koch84,koer94}) and the Kroll-Ruderman part which appears due
to an intermediate isovector excitation of the nucleon to the
$s$-wave $\pi N$ state.  It is assumed that the $E1$ part is not
density dependent and, therefore,  is accompanied with the one-body form
factor $F_1(q)$.

Most recently such an approach was used in Ref.\ \cite{pas96},
in which the $\Delta$ resonance amplitude takes the form%
\footnote{Strictly speaking, the form factor $F_\Delta(q)$
in Eq.\ (\ref{T-Delta-hole}) depends on  both, the momentum transfer $q$ and
the photon energy $\omega$.}
\begin{equation}
\label{T-Delta-hole}
  A T^{\gamma}_{\Delta}(\omega) F_\Delta(q) =
  \frac49 \,\frac{f^2_\gamma}{m_\pi^2} \omega_c^2
  \int d\vr \, \rho(\vr) \, e^{i\vq\cdot\vr} \, \Big[
  G_{\Delta h}(\rho(\vr),s_+) + G_{\Delta h}(\rho(\vr),s_-)  \Big].
\end{equation}
Here $f_\gamma$ is the $\gamma  N\Delta$ coupling, $\rho(\vr)$
the nuclear density, $\omega_c$  the photon energy in the rest frame
of the  produced $\Delta$ , and $G_{\Delta h}$  the local density
approximation to the $\Delta$-hole Green function:
\begin{equation}
  G_{\Delta h}(\rho,s) = \Big[ \sqrt{s} - M_\Delta
   + \frac{i}{2} \tilde\Gamma(\rho,s) - \Sigma_\Delta(\rho,s) \Big]^{-1},
\end{equation}
where $\tilde\Gamma$ and $\Sigma_\Delta$ are the Pauli-blocked
width and self-energy of the $\Delta$ in the nuclear matter
\cite{koch84,oset87,oset91}. The effective energies $s_\pm$ in the $s$-
and $u$-channels of the $\Delta$-hole excitation include small shifts
caused by  Fermi motion of the nucleons
and read in the Fermi gas approximation:
\begin{equation}
\label{s-eff}
   s_\pm = M^2 \pm 2\omega \Big(M + \frac35\,\frac{k_F^2}{2M} \Big),
\end{equation}
where $k_F$ is the (local) Fermi momentum,
$k_F^3 = \frac32\pi^2\rho$.

In the microscopic calculations of the $\Delta$-hole model the
constraint of the GGT dispersion relation is not used.  Therefore, an
agreement of the calculated differential cross section with
the experimental data at small angles
supports the validity of the model for nuclear Compton scattering.
The agreement at energies below 350 MeV is largly improved when the
kinetic (Thomson) seagull contribution is included \cite{pas96,K18}.
At energies above the $\Delta$ peak the results of different
implementations \cite{pas96,K18} of the $\Delta$-hole model are
different.  A nice agreement with the experimental small-angle data
\cite{S16} was found in Ref.\ \cite{K18}, in which the $\Delta$
contribution was calculated with the parameters borrowed from
\cite{koer94,uda90}.

{\it Missing effects at large angles -- nonlocality:}\\
However, all the considered models fail to correctly describe the 
experimental results at large angles, where the $^4$He data
\cite{A6,A8,D11,B17,K18} largely exceed the predictions.  Comparing
the different calculations, one may conclude that the
density-dependent self-energy of the $\Delta$ which results in a
shift and broadening of the $\Delta$ peak
is not the only effect which has to be taken into
account to explain nuclear Compton scattering in the $\Delta$ range.
At high momentum transfers the self-energy cannot be considered as
a local operator and therefore nonlocal effects should be taken into account.
Nonlocal two- or many-body $E1$ contributions might be important as well.
This view was supported by a calculation done by L'vov and Petrun'kin
\cite{lvov90}, in which large two-body $E1$ contibutions were
obtained as a result of effective contact interactions arising in a
gauge-invariant description of Compton scattering on a bound nucleon
--- see \cite{levchuk95} for a more detailed motivation and
derivation of such an approach and its relation with the standard
language.  In spite of a phenomenological success of the
proposed scheme \cite{lvov90,B17}, it cannot be fully applicable in
the $\Delta$ range, because {\it e.g.} the contact $E1$
contributions arise there as purely real, what cannot be true at
energies exceeding the pion threshold.

The following simple model provides  strong evidence that nonlocal
effects missing in the relations (\ref{T-Delta-hole-general})
and (\ref{T-Delta-hole}) may be responsible for the long-standing
discrepancy between the $\Delta$-hole model and the $^4$He
Compton scattering data at large angles.
The model we consider includes only one- and two-body mechanisms of
photon scattering by $^4$He mediated by  one-pion exchange.
They are shown in Fig.\ \ref{He4-diagrams}.
We disregard the $\rho$-meson exchange.
Pion absorption or rescattering by intermediate nucleons is disregarded too
what makes this model inapplicable close to the $\Delta$-resonance
peak where the absorption is very strong.
In spite of all these oversimplifications, the model is not
fully unreasonable at energies like 200 MeV,
{\it i.e.} well below the $\Delta$ peak.

Diagram {\it a} in Fig.\ \ref{He4-diagrams} gives the
impulse approximation. The corresponding amplitude is equal to
\begin{equation}
\label{IA}
    T_{IA}(\omega,\theta) =  A\, F_1(q) \,
     \langle\, T_{\gamma N}(\omega_{eff},\theta) \,\rangle
\end{equation}
and is given by the spin-isospin average of the elementary
$\gamma N$ scattering amplitude on the free nucleon.
The effective energy in Eq.\ (\ref{IA}),
\begin{equation}
\label{omega-eff}
    \omega_{eff} =
    \omega + \frac{q^2}{8M} \Big(1-\frac{1}{A} \Big),
\end{equation}
is obtained by assuming that the active nucleon has the momentum
\begin{equation}
  \vp_{eff} = -\frac{\vq}{2} \Big(1-\frac{1}{A}\Big)
\end{equation}
in the Lab frame and that the rest of the nucleus is on shell
(see {\it e.g.} \cite{landau78}). The elementary $\gamma N$ scattering
amplitude could  easily be approximated by the $\Delta$-pole
diagram plus a Thomson and Kroll-Ruderman-like background
as this was done {\it e.g.} in Refs.\ \cite{koch84,pas96}.
However, in the present context we use instead numerical results
obtained in the framework of the dispersion theory \cite{lvov97}.
The differential cross section for Compton scattering  by $^4$He
obtained within the impulse approximation (\ref{IA})
at the energy $\omega = 206$ MeV is compared with predictions of
the $\Delta$-hole model \cite{pas96}
and with available data \cite{B17,K18} in Fig.\ \ref{He4-results}.
Though the $\Delta$-hole model includes the density-dependent
modification of the $\Delta$-resonance contribution, this modification
is numerically not a big effect at 206 MeV and  the predictions
of the IA and the $\Delta$-hole model are quite close to each other.
The only exception is the region of large angles where
the difference is big. However, this difference is not related
with the IA or the $\Delta$-hole model itself and is mostly
caused by our choice for the effective energy (\ref{omega-eff})
which includes a $q$-dependent correction absent in
the Ansatz (\ref{s-eff}).

\begin{figure}[htb]
\caption{\em
One- and two-body contributions to nuclear Compton scattering.
Diagram {\it a} is the impulse approximation.
Diagram {\it b} describes real pion production followed by
pion absorption.
Diagrams {\it c} are required by the gauge invariance; in particular,
the last diagram contains the contact $\gamma\pi N\Delta$ vertex.}
\label{He4-diagrams}
\end{figure}

\begin{figure}[htb]
\caption{\em Differential cross sections $(d\sigma/d\Omega)_{CM}$
for Compton scattering by $^4$He at $E_\gamma = 206$ MeV.
Data are from \protect\cite{B17} (solid circles) and from
\protect\cite{K18} (open circles). Shown are predictions
of the $\Delta$-hole model in the local-density approximation
\protect\cite{pas96}, impulse approximation, Eq.\ (\protect\ref{IA}),
and the total contribution including the two-body MEC,
Eq.\ (\protect\ref{MEC}).}
\label{He4-results}
\end{figure}

The two-body contributions are given by diagrams {\it b} and {\it c}
in Fig.\ \ref{He4-diagrams}. Diagram {\it b} is determined by
amplitudes of pion photoproduction which we take in the tree
approximation with fixed nucleons, {\it i.e.} including
the Kroll-Ruderman $\gamma\pi NN$ vertex,
the pion-in-flight current, and $\Delta$ excitations in the
direct and crossed channels.
In the diagram {\it b} the intermediate pion has the energy $\omega$
and, therefore, can be real (not only virtual).
Accordingly, even without $\Delta$ contributions
the diagram {\it b} contributes to the imaginary part of the
nuclear Compton scattering amplitude.
In contrast to this, the diagrams {\it c} do not contain real-pion propagators.
Actually, the diagrams {\it c} do not represent a self-dependent contribution
but they rather ensure the gauge invariance
of the total two-body amplitude.
In particular, the last of the diagrams {\it c}
contains the contact $\gamma\pi N\Delta$ vertex
which is labeled as the $\Delta$--Kroll--Ruderman vertex
in Fig.\ \ref{He4-diagrams}.
Explicit formulas for the kernel amplitude
$\langle\, T_{b+c}(\vk_1,\vk_2,\vQ) \,\rangle$
which corresponds to the diagrams {\it b} and {\it c} with
omitted wave functions of external photons and nucleons
and spin-isospin averaged over nucleons in
the nucleus are very similar to those given in Ref.\ \cite{huett96}
(see also section 9, Eqs.\ (\ref{b1}) and (\ref{deltatij})).
The kernel amplitude depends  on the
photon momenta $\vk_1$, $\vk_2$ and on the internal momentum $\vQ$ which
defines the sharing of the total momentum transfer $\vq$
between the momenta $\frac12\vq -\vQ$, $\frac12\vq + \vQ$
transferred to each of the two interacting nucleons.
Assuming  oscillator wave function for $^4$He,
\begin{equation}
   \Psi(\vp_1,\ldots\vp_A) =
    {\rm const} \times \exp\Big\{-\frac{C_p}{2}
        \Big(\vp_1^2 + \ldots + \vp_A^2 - \frac{1}{A}\vP^2 \Big) \Big\},
    \quad \vP = \vp_1 + \ldots + \vp_A,
\end{equation}
we fix the oscillator parameter $C_p$ through the one-body formfactor:
\begin{equation}
\label{F-one-body}
   F_1(q) = \exp\Big(-\frac{C_p}{4}\,\frac{A-1}{A}\,q^2 \Big)
       =\exp\Big( -\frac{q^2}{4\alpha^2} \Big),
\end{equation}
where $\alpha=165$ MeV is determined from the electric radius of $^4$He and
that of the proton.Then the two-body form factor is equal to%
\footnote{Note that we use the oscillator wave function
only in order to evaluate the two-body form factor.
Finding the amplitude (\ref{IA}), we use a more accurate form factor
found through experimental data on electron scattering
by $^4$He \cite{datatab}.}
\begin{equation}
\label{F-two-body}
   F_2(q) = \exp\Big(-\frac{C_p}{8}\,\frac{A-2}{A}\,q^2 \Big).
\end{equation}
For $A=4$ we have $F_2(q) = [F_1(q)]^{1/3}$.
Then the total MEC amplitude corresponding to diagrams {\it b} and {\it c}
reads
\begin{equation}
\label{MEC}
    T_{MEC}(\omega,\theta) = A(A-1) \,F_2(q) \int
     \frac{d\vQ}{(2\pi)^3} \exp\Big(-\frac{C_p}{2}\vQ^2 \Big) \,
       \langle\, T_{b+c}(\vk_1,\vk_2,\vQ) \,\rangle.
\end{equation}

In the limit $\omega=0$ the mesonic amplitude (\ref{MEC})
determines the (unretarded) enhancement parameter $\unret{\kappa}{}$
(see section 9 for more detail).
Actually the value of $\unret{\kappa}{}$ obtained through
the two-body diagrams of the lowest order in the $\pi NN$ coupling
is rather small \cite{huett96}, so that $\unret{\kappa}{}$
is dominated by two-pion exchanges and three-body interactions.
In the chiral limit of $m_\pi \to 0$ the amplitude
(\ref{MEC}) vanishes at $\omega=0$ due to the exact
compensation between the diagrams {\it b} and {\it c}.
However, when the energy $\omega$ becomes higher than the pion threshold,
such compensation is destroyed. In particular,
the diagram {\it b} gets an imaginary part at $\omega > m_\pi$,
whereas the diagram {\it c} without $\Delta$ contributions
remains real. At energies above pion threshold the amplitude (\ref{MEC})
is not suppressed any more by the small pion mass and is numerically large.
Adding the meson-exchange amplitude $T_{MEC}$ to the amplitude
of the impulse approximation $T_{IA}$, we find a strong enhancement
in the differential cross section at backward angles
shown in Fig.\ \ref{He4-results} at the energy 206 MeV which brings the 
predictions of the model into a qualitative agreement with the data.
At the energy 253 MeV the agreement with available data \cite{B17,K18}
is almost perfect but this may be accidental
because the model is not expected to work well close to the $\Delta$ peak.
On the other hand, the IA and the $\Delta$-hole model grossly underestimate
the differential cross section at backward angles at these energies.

The increase  obtained at backward angles is caused by mainly three reasons.
First, the two-body formfactor does not suppress the MEC contribution
as much as the one-body formfactor does the IA contribution;
for instance, $F_2(q) / F_1(q) \simeq 2.5$ at 206 MeV and  backward angles.
Second, the medium correction to photon scattering through
$E1$ photoproduction which is described by diagrams {\it b} and {\it c}
with only electric couplings (these are diagrams without $\Delta$)
is quite large and predominately imaginary.
It explains about 50\% of the obtained enhancement (in the amplitude)
at 206 MeV and backward angles. Such a correction was not included into 
the 
$\Delta$-hole calculations of nuclear Compton scattering.
Third, equally large and also predominately imaginary
is the nonlocal correction due to propagation of the $\Delta$-excitation
from one nucleon to an other one through intermediate pion propagation.

The  model considered here  clearly is too simple to provide a perfect
description of the $^4$He Compton-scattering data, especially
close to the $\Delta$ peak where
multiple processes including pion absorption and rescattering
are very important. In addition to pion exchange,  $\rho$-meson
exchange has also to be included.
Moreover, nucleons should not to be considered fixed and also
couplings to $NN$ continuum-states should be incorporated;
in the $\Delta$-hole model these couplings are a large part
of the so-called spreading potential of the $\Delta$-hole  states.
It might also be important to calculate two-body contributions
with a more accurate wave function of $^4$He including the $D$-wave.
Despite all these omissions, we believe that we can conclude
that realistic calculations of nuclear Compton scattering
at high momentum transfers
should properly take into account the nonlocal effects.

\clearpage

\section{Fermi liquid theory and nuclear Compton scattering}

An appropriate method to describe enhancement phenomena of the giant
dipole resonance is within Landau's Fermi liquid   theory
\cite{20,21}
of finite systems as first applied  by Migdal \cite{22,23}.
The Fermi liquid theory  is concerned with strongly interacting particles,
which are replaced by weakly interacting
quasiparticles analogous to particles
in an ideal Fermi gas with a residual interaction.

In a finite system, the problem has to be
treated in close analogy to the nuclear shell model, with the main
difference that particle-hole
excitations are replaced by quasiparticle-quasihole excitations.
Giant-dipole excitations involve transitions between neighboring
oscillator shells from below the Fermi energy-level to above it. In case
of particle-hole excitations these transitions exhaust exactly one TRK sum
rule so that the transition to the quasiparticle picture should contain
the overfulfilment of the TRK sum rule as represented by $\ko $. This
will be shown in this section.

The spectrum of collective excitations ({\it e.g.} sound and spin waves of
giant resonances)  and the response of the system to an
external (electromagnetic) field are determined by the quasi-particle
scattering amplitude near the Fermi surface for zero scattering angle.
This amplitude is a function of the angle between the initial
quasi-particle momenta and is satisfactorily described by the first two
Legendre polynomials.
The coefficients of these polynomials
are the constants introduced into the theory and are the main parameters
of Landau's theory of Fermi liquids \cite{20}.

In the Fermi liquid theory the system of particles with the spin 1/2
is replaced by a system of quasiparticles with the same spin. This
theory is aimed at the  description of low-lying excitations. Each
quasiparticle is characterized  by the momentum $\bf p$ and the quasiparticle
number $n({\bf p})$. Strictly speaking, this quasiparticle
number is a matrix $\hat n({\bf p})\equiv n_{\alpha\,\beta}({\bf p})$
in spin and isospin spaces
(for the sake of simplicity we assume that each index
simultaneously corresponds to spin and isospin variables).
The total number of particles $N$ is equal to the
total number of quasiparticles:
\begin{equation}
\mbox{Tr}\int \hat n({\bf p})\,d\tau =\frac{N}{V}\, ,\quad
d\tau=\frac{d{\bf p}}{(2\pi)^3}\, ,
\end{equation}
where $V$ is the volume of system. In the ground state
one has $n_{\alpha\,\beta}^{(0)}({\bf p})=
\delta_{\alpha\,\beta}\theta (p_F-p)$ as
in a Fermi gas. Due to the interaction between quasiparticles, the
total energy of the system is not the sum of quasiparticle energies. The
energy matrix $\hat{\epsilon}\equiv\epsilon_{\alpha\,\beta}({\bf p})$ of
the (individual) quasiparticle is defined
as a variational derivative of the total energy $E$ :
\begin{equation}\label{fermi3}
\frac{\delta E}{V}=
\mbox{Tr}\int\hat\epsilon({\bf p})\delta\hat n({\bf p})\,d\tau\quad .
\end{equation}
In the ground state $\epsilon_{\alpha\,\beta}({\bf p})=
\epsilon^{(0)}({p})\delta_{\alpha\,\beta}$ and
\begin{equation}\label{mass}
\epsilon_F=\epsilon^{(0)}(p_F)\quad ,\quad
\left.\frac{\partial\epsilon^{(0)}(p)}{\partial p}\right |_{p=p_F}=v_F=
{p_F\over M^*}\quad ,
\end{equation}
where  $M^*$  is the effective mass of the quasiparticle.
Giant-dipole excitations involve transitions between neighboring
oscillator shells from below the  Fermi energy to above  it. Therefore,
the relevant nucleons have momenta close to the Fermi momentum.
Near the Fermi surface $\epsilon^{(0)}({p})=\epsilon_F+v_F(p-p_F)$.
Including quasiparticle excitations the energy matrix
$\epsilon_{\alpha\,\beta}({\bf p})$  depends on
the quasiparticle numbers $n_{\alpha\,\beta}({\bf p})$ :
\begin{equation}\label{fermi1}
\epsilon_{\alpha\,\beta}({\bf p})=
\epsilon^{(0)}({p})\delta_{\alpha\,\beta}+
\int f_{\alpha\gamma,\beta\delta}({\bf p},{\bf p}^{\prime})
\delta n_{\delta\,\gamma}({\bf p}^{\prime})d\tau^{\prime}\, ,
\end{equation}
where $\delta n_{\delta\,\gamma}({\bf p})$ is the difference between
$n_{\delta\,\gamma}({\bf p})$ and $n^{(0)}_{\delta\,\gamma}({\bf p})$
and $f_{\alpha\gamma,\beta\delta}({\bf p},{\bf p}^{\prime})$
describes the  quasiparticle-quasiparticle interaction. In this function
one can put $p=p^{\prime}=p_F$. Therefore, it depends only on the directions
of the vectors ${\bf p}$ and ${\bf p}^{\prime}$.
In matrix notations one can write  Eq.\ (\ref{fermi1}) as
\begin{equation}\label{fermi2}
\hat\epsilon({\bf p})=\hat\epsilon^{(0)}({\bf p})+
\mbox{Tr}^{\prime}\int\hat f({\bf p},{\bf p}^{\prime})
\delta\hat n({\bf p}^{\prime})d\tau^{\prime}\quad ,
\end{equation}
where $\mbox{Tr}^{\prime}$ denote the trace with respect to indices
corresponding to momentum ${\bf p}^{\prime}$.
Using (\ref{fermi3}) and (\ref{fermi2}) one can write an
expression for the excitation energy:
\begin{equation}\label{fermi4}
\frac{\delta E}{V}=
\mbox{Tr}\int\hat\epsilon^{(0)}({\bf p})\delta
\hat n({\bf p})\,d\tau +
{1\over 2}\mbox{Tr}\,\mbox{Tr}^{\prime}
\int\!\int\hat f({\bf p},{\bf p}^{\prime})
\delta\hat n({\bf p}^{\prime})
\delta\hat n({\bf p})d\tau d\tau^{\prime}\quad .
\end{equation}
Usually the following representation is used for the operator
$\hat f({\bf p},{\bf p}^{\prime})$ of the interaction between
quasiparticles:
\begin{equation}\label{fermi5}
N_0\hat f({\bf p},{\bf p}^{\prime})=
F+\vetau\cdot\vetau^{\prime}\,F^{\prime}+
\vesigma\cdot\vesigma^{\prime}\,G+
\vetau\cdot\vetau^{\prime}
\vesigma\cdot\vesigma^{\prime}\,G^{\prime}\, .
\end{equation}
Here
\begin{equation}
N_0=4\int\delta(\epsilon^{(0)}(p)-\epsilon_F)\,d\tau
=\frac{2p_FM^*}{\pi^2}
\end{equation}
is the density of states on the Fermi surface (the factor four
corresponds to two possible states in both, spin space and isospin space),
$\vetau$
and $\vesigma$ are Pauli matrices for isospin and spin, respectively.
The functions $F,\, F^{\prime},\, G$ and $G^{\prime}$ depend on the angle
$\theta$ between ${\bf p}$ and ${\bf p}^{\prime}$ and can be expanded
with respect to Legendre polynomials:
\begin{equation}\label{expen}
F=\sum F_lP_l(\cos \theta) \quad ,
\end{equation}
and the same for the other functions. The coefficients in this series are
some phenomenological constants which should be extracted from
experiment. Usually only the zeroth and first harmonics are essential.
It possible to describe a big variety of processes inside the nucleus
with only a few coefficients $F_l,\, F^{\prime}_l,\, G_l$ and
$G^{\prime}_l$ (see \cite{23}).

It follows from Galilei invariance that the momentum density is equal to
the density of mass flow.
Since the number of particles is equal to the number of
quasiparticles, one has
\begin{equation}
\mbox{Tr}\int {\bf p}\;\hat n({\bf p})\,d\tau =
M\,\mbox{Tr}\int \frac{\partial\hat\epsilon}{\partial {\bf p}}
\hat n({\bf p})\,d\tau \, ,
\end{equation}
where $M$ is the nucleon mass. Substituting
$\hat n({\bf p})=\hat n^{(0)}({\bf p})+
\delta\hat n({\bf p})$ and using  (\ref{fermi2}), we get for the
isoscalar current:
\begin{equation}\label{fermi6a}
{\bf j}_0=\mbox{Tr}\int \frac{{\bf p}}{M}\,\delta\hat n({\bf p})\,d\tau =
\mbox{Tr}\int \left[
\frac{\partial\epsilon^{(0)}}{\partial {\bf p}}
-\mbox{Tr}^{\prime}\int
\frac{\partial n^{(0)}({\bf p}^{\prime})}
{\partial {\bf p}^{\prime}}
\hat f({\bf p}^{\prime},{\bf p})d\tau^{\prime}\right]
\delta\hat n({\bf p})d\tau\quad .
\end{equation}
Taking into account the relation
\begin{equation}
\frac{\partial n^{(0)}({\bf p})}{\partial {\bf p}}=
-\delta(\epsilon^{(0)}(p)-\epsilon_F)
\frac{\partial\epsilon^{(0)}(p)}{\partial {\bf p}}
\end{equation}
and also the representation (\ref{fermi5}), we finally obtain
\begin{equation}\label{fermi6b}
{\bf j}_0=\mbox{Tr}\int \frac{{\bf p}}{M}\,\delta\hat n({\bf p})\,d\tau =
\mbox{Tr}\int
\frac{\partial\epsilon^{(0)}}{\partial {\bf p}}
\left[1+\frac{F_1}{3}\right]\delta\hat n({\bf p})d\tau\quad ,
\end{equation}
where $F_1=(F_1^{pp}+F_1^{pn})/2$ with
$ F_1^{pp}= F_1^{nn}\not= F_1^{pn}=F_1^{np}$ being the coefficients
at  $l=1$ for the interaction of $pp$, $nn$ and $np$ quasiparticles
(see  Eqs.\ (\ref{fermi5}) and (\ref{expen})).
It is seen that due to the interaction between  quasiparticles
any motion of particles is accompanied by a backflow of quasiparticles
(corresponding to the term proportional to $F_1$).
From Eqs.\ (\ref{mass}) and(\ref{fermi6b}) it follows that the
effective mass $M^*$ is equal to:
\begin{equation}\label{mass1}
M^*=M\left[1+\frac{F_1}{3}\right]\, .
\end{equation}
For the isovector current ${\bf j}_1$ one obtains in the same way:
\begin{equation}
{\bf j}_1=\mbox{Tr}\int \frac{\partial\hat\epsilon}{\partial {\bf p}}
\tau_3\hat n({\bf p})\,d\tau \,=
\mbox{Tr}\int
\frac{{\bf p}}{M}\frac{1+F_1^{\prime}/3}{1+F_1/3}\tau_3\,
\delta\hat n({\bf p})d\tau\quad .
\end{equation}
The electromagnetic current is given by
${\bf j}_{em}=(e/2)({\bf j}_0+{\bf j}_1)$. Therefore,
\begin{equation}\label{emcurrent}
{\bf j}_{em}={e\over 2}
\mbox{Tr}\int
\frac{{\bf p}}{M}\left[\,1+\tau_3 +
{1\over 3}\,\frac{F_1^{\prime}-F_1}{1+F_1/3}\,\tau_3\,\right]
\delta\hat n({\bf p})d\tau\quad .
\end{equation}
Note that $F_1^{\prime}=( F_1^{pp}-F_1^{pn})/2$. The application of a
relativistic approach \cite{BaymChin76,BE85,BWBS87}leads to
some modification of the mentioned results.
A review of the numerous articles devoted to this subject can be found in
\cite{BWBS87}.

In order to obtain an information on the parameters of Fermi liquid theory
for nuclei it is possible to start from a pion-exchange
and $\rho$-meson-exchange interaction in the Born approximation. Then one
can apply the Brueckner approach to take into account higher orders
of perturbation theory with respect to the interaction between particles (see
\cite{BR80,25,27,28} and the review \cite{29}).
The operator for the one-pion-exchange potential is of the form
\begin{eqnarray}\label{vpi}
&\displaystyle
V_{\pi}(\vp)=-\frac{f^2}{m^2_\pi(m^2_\pi+p^2)}\,
(\vesigma\cdot{\bf p})(\vesigma^{\prime}\cdot{\bf p})
\vetau\cdot\vetau^{\prime}
=\\
&\displaystyle
-\frac{f^2}{m^2_\pi}\left[
\frac{S_{T}}{m^2_\pi+p^2}+ {1\over 3}\vesigma\cdot\vesigma^{\prime}
- \frac{m^2_\pi}{3\,(m^2_\pi+p^2)}\,\vesigma\cdot\vesigma^{\prime}\right]
\,\vetau\cdot\vetau^{\prime}\nonumber \, ,
\end{eqnarray}
where $S_{T}=(\vesigma\cdot {\bf p})(\vesigma^{\prime}\cdot {\bf p})-
(\vesigma\cdot\vesigma^{\prime})\, p^2/3 $.
The potential arising from $\rho$-exchange is
\begin{eqnarray}\label{vrho}
&\displaystyle
V_{\rho}(\vp)=-\frac{f^2_{\rho}}{m^2_{\rho}(m^2_{\rho}+p^2)}\,
[\vesigma\cdot\vesigma^{\prime}\, p^2-
(\vesigma\cdot{\bf p})(\vesigma^{\prime}\cdot{\bf p})]\,
\vetau\cdot\vetau^{\prime}
=\\
&\displaystyle
-\frac{f^2_{\rho}}{m^2_{\rho}}\left[
-\frac{S_{T}}{m^2_{\rho}+p^2}+ {2\over 3}\vesigma\cdot\vesigma^{\prime}
- \frac{2m^2_{\rho}}{3\,(m^2_{\rho}+p^2)}
\,\vesigma\cdot\vesigma^{\prime}\right]
\,\vetau\cdot\vetau^{\prime}\nonumber\, .
\end{eqnarray}
Taking into account $\omega $- and $\sigma $-exchange potentials one is
lead to replacing $f^2_{\rho}$ in  (\ref{vrho}) by
$\tilde f^2_{\rho}=0.4\,f^2_{\rho}$ \cite{BR80}.
Equations (\ref{vpi}) and (\ref{vrho}) correspond to the contributions
to the nuclear exchange potential given in Eq.\ (\ref{ExchangePotentials}),
except that here a separation into
central and tensor parts is given.

Brown and Rho \cite{BR80} arrived at a quantitative derivation of the
effective mass $M^*$  making use of such a
model, where the isospin dependent interactions between nucleons
consist of $\pi$ and $\rho$ exchanges.
The effective mass $M^*$ is obtained by evaluating the self
energy stemming from the Fock term. The results of this investigation are
summarized in the following.
Writing $\epsilon(p)$ as
\begin{equation}\label{sigma}
\epsilon(p)=\frac{p^2}{M}+\Sigma(p,\epsilon(p))\, ,
\end{equation}
where $\Sigma(p,\epsilon(p))$ is the quasiparticle self-energy.
Then using the definition
of the effective mass  (\ref{mass}) one has \cite{BR80}:
\begin{equation}
{1\over M^*}= {1\over M}+{1\over p_F}
\left.\frac{\partial\Sigma(p,\epsilon(p)}{\partial p}\right |_{p=p_F}
\end{equation}
Substituting this relation into  (\ref{mass1}) and calculating the
one-pion-exchange contribution to the self-energy, one obtains an
expression for the pion
contribution to $F_1$  \cite{BR80}:
\begin{equation}\label{fpi}
F_{1\,\pi}=-\frac{9f^2M^*}{8\pi^2p_F}\left[
\frac{m^2_\pi+2p_F^2}{2p_F^2}\ln\frac{m^2_\pi+4p_F^2}{m^2_\pi}-2\right]=
-0.46{M^*\over M}\, .
\end{equation}
The contribution of $\rho$-exchange to $F_1$ reads  \cite{BR80}:
\begin{equation}\label{frho}
F_{1\,\rho}=-0.56{M^*\over M}\, .
\end{equation}
Taking a sum of  (\ref{fpi}) and (\ref{frho}) and then using (\ref{mass1}) one
has:
\begin{equation}\label{value1}
{M^*\over M}=0.75\quad ,\quad F_1=-0.76\quad .
\end{equation}
The numerical value for $M^*/M$ is in a good agreement with the experimental
value given in
Table \ref{EffectiveMasses} for the mass region $A = 197$--209 of
spherical nuclei.
It is interesting to note that the $A$-dependence of the effective mass $M^*$
also finds a satisfactory explanation in the work of Brown and Rho
\cite{BR80}. The explanation which can be extracted from that work
is as follows. First of all we have to realize that the excellent agreement
between theory and
experiment found in (\ref{value1}) and the $A=197$--209 mass range
discussed in Table (\ref{EffectiveMasses}), respectively,
partly
comes as a surprise because in its theoretical derivation
it was assumed \cite{BR80}
that most of the difference $M-M^*$ comes from the Fock terms for $\pi-$
and $\rho-$ exchange, whereas it was discussed in that paper that
quasiparticle-phonon couplings also have an effect on the difference  $M-M^*$
but with the opposite sign.
The mathematical procedure of treating the two effects
is very much the same. The only difference is that the pion propagator
has to be replaced by the appropriate energy denominator for the
intermediate state involving quasiparticle plus vibration. The essential
point is that the contributions from couplings to vibrations
leads to effects close to the nuclear surface only. High-$l$ orbitals
are confined by the centrifugal barrier  more to the interior of nuclei
than those of low-$l$ ones  \cite{BR80}.
Since states with high angular momentum dominate at the $^{208}$Pb shell
closures
this property is exactly what we need to understand the $A$-dependence of
$M^*/M$ with a minimum in the $^{208}$Pb mass range.

The operators of the pion-exchange and $\rho$-exchange
potentials are proportional to
$$
\vetau\cdot\vetau^{\prime}=2(\tau_{+}\tau^{\prime}_{-}+
\tau_{-}\tau^{\prime}_{+})+\tau_{3}\tau^{\prime}_{3}\quad ,
$$
where $\tau_{+}$ and $\tau_{-}$ are isospin raising and lowering operators,
respectively. Therefore, the following relation between $F_1^{pp}$ and
$F_1^{pn}$ is obtained  \cite{BR80}:
\begin{equation}
\label{fpn}
F_1^{pn}=2F_1^{pp}
\end{equation}

Starting from Eqs.\  (\ref{vpi}) and (\ref{vrho}) one can obtain the
contributions of pion-exchange and $\rho$-exchange
to the quasiparticle-quasiparticle interaction  (\ref{fermi5}).
For this purpose  one can calculate the forward scattering amplitude
of a quasiparticle with $p > p_F$ and a hole with  momentum $p < p_F$.
After the Fourier transform to configuration space
the momentum independent terms in  (\ref{vpi}) and (\ref{vrho}) are
proportional to $\delta({\bf r})$ . Due to the strong short-range
repulsion from the $\omega$-meson, the wave function vanishes at zero
distance between the nucleons. Therefore, the momentum-independent
terms in  (\ref{vpi}) and (\ref{vrho}) can be omitted.
Neglecting also the tensor part of the interaction
(proportional to $S_{T}$), we have  \cite{29} :
\begin{eqnarray}\label{pi}
&\displaystyle
\hat f_{\pi}({\bf p},{\bf p}^{\prime})=
\vesigma\cdot\vesigma^{\prime}\vetau\cdot\vetau^{\prime}\,
\frac{f^2}{3\,m^2_\pi}+\\
&\displaystyle
(-9+3\vesigma\cdot\vesigma^{\prime}+
3\vetau\cdot\vetau^{\prime}-
\vesigma\cdot\vesigma^{\prime}\vetau\cdot\vetau^{\prime})\,
\frac{f^2}{12\,(m^2_\pi+({\bf p}-{\bf p}^{\prime})^2)}\, .\nonumber
\end{eqnarray}
Here the first term corresponds to  particle-hole annihilation and
the second one to  particle-hole exchange. In the last case the
following identities have been used:
$$\vesigma_{\alpha\beta}\vesigma_{\delta\gamma}=
{3\over 2}\delta_{\alpha\gamma}\delta_{\delta\beta}-
{1\over 2}\vesigma_{\alpha\gamma}\vesigma_{\delta\beta}\quad
$$
and the same for isospin matrices.
The contribution of $\rho$-exchange has the form:
\begin{eqnarray}\label{rho}
&\displaystyle
\hat f_{\rho}({\bf p},{\bf p}^{\prime})=
\vesigma\cdot\vesigma^{\prime}\vetau\cdot\vetau^{\prime}\,
\frac{2\tilde f^2_{\rho}}{3\,m^2_\rho}+\\
&\displaystyle
(-9+3\vesigma\cdot\vesigma^{\prime}+
3\vetau\cdot\vetau^{\prime}-
\vesigma\cdot\vesigma^{\prime}\vetau\cdot\vetau^{\prime})\,
\frac{\tilde f^2_{\rho}}{6\,(m^2_\rho+({\bf p}-{\bf p}^{\prime})^2)}\,
   .\nonumber
\end{eqnarray}
Multiplying both sides of  (\ref{pi}) by $P_1(x)=x=
{\bf p}\cdot{\bf p}^{\prime}/p^2$ and taking the integral
over $x$ from minus one to one, we obtain expressions for the
Landau parameters at $l=1$. As it should be, the result for
$F_{1\,\pi}$ agrees with (\ref{fpi}). It is clear from  (\ref{pi}) and
 (\ref{rho}) that
\begin{equation}\label{value2}
F_1^{\prime}=\, G_1=\,-{1\over 3}\,F_1=\, 0.25\quad ,
\quad G_1^{\prime}=\, {1\over 9}\,F_1=\,-0.08\quad .
\end{equation}
Even though $F'_1$ is not known with great precision, a 10--20 per cent
uncertainty in it \cite{BR95} only has a 1--2 per cent effect
in the quantities of interest here.
The discussion of the Landau parameters at $l=0$ as well as
the contribution of tensor interaction can be found in the review \cite{29}.

Let us now pass to the consideration of the enhancement constant $\ko$
and its connection with the parameters of the Fermi liquid. It was shown
in \cite{22} that without taking into account tensor correlations
the following relation is valid for the modified TRK sum rule:
\begin{eqnarray}\label{sumrul}
&\displaystyle
\int\limits_0^\infty  {\,\unret{\sigma }{GDR} (\omega )}\,d\omega =
2\pi ^2{{e^2} \over M}\,{{NZ} \over A} (1 + \ko)=\nonumber\\
&\displaystyle
2\pi ^2{{e^2} \over M^*}\,{{NZ} \over A}
\left(1 + {1\over 3}F_1^{\prime}\right)=
2\pi ^2{{e^2} \over M}\,{{NZ} \over A}
\frac{\left(1 + {1\over 3}F_1^{\prime}\right)}
{\left(1 + {1\over 3}F_1\right)}\quad ,
\end{eqnarray}
where $\unret{\sigma }{GDR} (\omega )$ is the unretarded
electric giant-dipole cross section for nuclear photoabsorption and
$\ko$ is the unretarded  enhancement constant.
It is seen from  (\ref{sumrul}) that $ \ko $
is related not only to the effective mass $M^*$ but also to
the constant $F_1^{\prime}$. Besides, if $F_1^{pn}=0$ then
$\ko=0$ (in this case  $F_1^{\prime}=F_1$).
Equation (\ref{sumrul}) gives a   relation for   $\unret{\kappa }{GDR}$
in the form:
\begin{equation}
\label{kappagdr}
\unret{\kappa }{GDR}=\frac{{1\over 3}F_1^{\prime}- {1\over 3}F_1}
{1 +{1\over 3}F_1} \quad .
\end{equation}
Alternatively, one may also express the ratio $M^*/M$ of the effective mass and
the free mass via $\ko$. Using Eqs.\ (\ref{mass1}), (\ref{fpn})
and (\ref{kappagdr}) one obtains
\begin{equation}
	{{M^*} \over M}=\left( {1+{{3}
	\over 4} \ko } \right)^{-1}\quad .
	\label{mstar}
\end{equation}
Using the relations given above it is easy to see that the theoretical results
for $M^*/M$, $F_1$ and $F'_1$ are nicely consistent with the experimental
quantity $\ko=0.46$  (see also Table \ref{EffectiveMasses}).

We now return to tensor correlations.
The calculation of the total enhancement constant
$\kappa$ with the use of its representation as a double commutator
\cite{arima73}  gives
$\kappa \sim 1$, if the tensor correlations in the ground state
are taken into account. In this paper  \cite{arima73} it was noted that
the total enhancement constant $\kappa$ consists of two pieces,
$ \kappa = \kappa' + \Delta \kappa$, where only $\kappa'$ has a relation
to the Sachs magnetic moment.
Therefore, we have to conclude that the quantities
$\ko$ and $\kappa'$ may be identified with each other. One consequence
of this identification is that the short-range tensor correlations
giving rise to $\Delta \kappa$
are not important for  giant resonances as anticipated above.
In fact  short-range tensor correlations give the main contribution
to the  quasi-deuteron part of the total enhancement constant
(see \cite{arima73}).

We present here a simple derivation of  (\ref{sumrul}).
The cross section  $\unret{\sigma }{GDR}$ is of the form:
\begin{equation}\label{n1}
\unret{\sigma }{GDR}(\omega)=\frac{4\pi^2}{3}\omega\sum_{n}
|\left\langle n \right|\,
{\bf d}\,\left| 0 \right\rangle |^2\,\delta(E_0+\omega-E_n)\, ,
\end{equation}
where ${\bf d}$ is the dipole moment operator with respect to the center
of mass of the nucleus,
\begin{equation}\label{dip}
{\bf d}=e{N\over A}\sum_i {\bf r}_{i\,p}
 - e{Z\over A}\sum_i {\bf r}_{i\,n}\, .
\end{equation}
The additional indices $p$ and $n$ indicate that in that term the sum is
taken over only protons and only neutrons, respectively.
Using the relation ${\bf j}_{em}=i[H,{\bf d}]$ , we get
\begin{equation}\label{crossint}
\int\limits_0^\infty  {\,\unret{\sigma }{GDR} (\omega )}\,d\omega =
\frac{2\pi ^2}{3i}\left\langle 0
\right|[{\bf d},{\bf j}_{em}]\left| 0 \right\rangle \quad .
\end{equation}
It follows from (\ref{emcurrent}) that in the c.m. frame the
electromagnetic current is given by
\begin{eqnarray}\label{emcurrent1}
&\displaystyle
{\bf j}_{em}=\frac{e}{2M}\left[(2+\ko)
\sum_i\left( {\bf p}_{i\,p}-
\sum_j\frac{ {\bf p}_{j\,p}+ {\bf p}_{j\,n}}{A}\right)-\ko
\sum_i\left( {\bf p}_{i\,n}-
\sum_j\frac{ {\bf p}_{j\,p}+ {\bf p}_{j\,n}}{A}\right)\right]\, \nonumber\\
&\displaystyle
=\,\frac{e(1+\ko)}{M}\left[\frac{N}{A}\sum_i {\bf p}_{i\,p}-
\frac{Z}{A}\sum_i {\bf p}_{i\,n}\right] \quad .
\end{eqnarray}
Substituting (\ref{dip}) and (\ref{emcurrent1}) into (\ref{crossint})
and evaluating the commutator, we get  (\ref{sumrul}).

Next we discuss the low-energy limit of the Compton scattering
amplitude. Within the framework of second-order perturbation theory
the resonance amplitude in the long-wavelength approximation
is given by:
\begin{equation}
	R(\omega ,\theta )=-\sum\limits_\nu
	{\left\{ {{{\left\langle f \right|\vj_{em}\cdot
	\veceps _2\left| \nu  \right\rangle \left\langle \nu
	\right|\vj_{em}\cdot \veceps _1\left| i \right\rangle }
	\over {E_0-E_\nu +\omega }}+{{\left\langle f \right|\vj_{em}\cdot
	\veceps _1\left| \nu  \right\rangle \left\langle \nu
	\right|\vj_{em}\cdot \veceps _2\left| i \right\rangle }
	\over {E_0-E_\nu -\omega }}} \right\}}
	\label{pertres}
\end{equation}
Again expressing the current via the commutator of the electric dipole
moment with the Hamiltonian it is easy to extract the low-energy limit of
the resonance amplitude:
\begin{equation}\label{compres}
R(0,\theta )=\frac{1}{2i}\left\{ \left\langle 0 \right|[\veceps _1\cdot {\bf d},
\veceps _2\cdot {\bf j}_{em}]\left| 0 \right\rangle
+ \left\langle 0 \right|[\veceps _2\cdot {\bf d},
\veceps _1\cdot {\bf j}_{em}]\left| 0 \right\rangle \right\} \quad .
\end{equation}
Then it follows from  (\ref{dip}) and
(\ref{emcurrent1}) that
\begin{equation}\label{compres1}
R(0,\theta )=\veceps _1\cdot \veceps _2\frac{e^2}{M}\frac{ZN}{A}
(1+\ko)\, .
\end{equation}
Using the low-energy theorem we may also obtain the low-energy limit of
the relevant seagull amplitude
\begin{equation}
\label{lowseag}
B(0,\theta )+S_{GR}(0,\theta )=
-\veceps _1\cdot \veceps _2\frac{Z^2e^2}{MA}-R(0,\theta )=
-\veceps _1\cdot \veceps _2\frac{Ze^2}{M}\left(1+\frac{N}{A}\ko \right)\, .
\end{equation}
It is possible to derive Eq.\ (\ref{lowseag}) directly, when the
explicit expression, Eq.\ (\ref{seaglimit2}), for the total
seagull amplitude is translated into the language of Fermi liquid theory.
One has
\begin{equation}
B(0,\theta )+S_{GR}(0,\theta ) =
{i \over 2}\left\{ {\left\langle f \right|[\veceps _1\cdot
        \vD\,,\;\veceps _2\cdot \vJ]+[\veceps _2\cdot
        \vD\,,\;\veceps _1\cdot \vJ]\left| i \right\rangle }
	\right\}
	\label{lowseag2}
\end{equation}
with
\begin{equation}
	\vD=\vd+{{eZ} \over A}\sum\limits_i
	{(\vr_{i\,p}+}\vr_{i\,n})\;,\quad \vJ=\vj_{em}+{{eZ}
	\over {AM}}\sum\limits_i {(\vp_{i\,p}+}\vp_{i\,n})  \, .
	\label{dj}
\end{equation}
Calculating the commutator we obtain again Eq.\ (\ref{lowseag}).

When nucleons inside a nucleus have an angular momentum $l $ they also
produce an angular (Sachs) magnetic moment. Meson exchange currents modify
this angular magnetic moment and it is easy to derive from Eq.\
(\ref{emcurrent1}) that for the proton this modification is given by
\begin{equation}
\label{DeltaGlP}
\delta g^{meson}_l(p) = \frac{N}{A} \ko .
\end{equation}
For the neutron one has
\begin{equation}
\label{DeltaGlN}
\delta g^{meson}_l(n) = - \frac{Z}{A} \ko  .
\end{equation}
As a result, we obtain that
the isovector part of the angular momentum correction is given by
\begin{equation}
\label{DeltaIsovector}
\delta g_l^{(iv)} = \frac12 \left( \delta g^{meson}_l(p) -\delta
g^{meson}_l(n) \right) =\frac12 \ko.
\end{equation}
This relations between $\delta g_l^{(iv)}$ and $\ko $ is known as the
Fujita-Hirata relation
\cite{BR80,arima73,EricsonWeise88,fujita71,fujita76,fujita79}.
A detailed investigation of it has been performed in \cite{arima73} with
special attention on the role of tensor correlations.

The Fujita-Hirata relation opens the possibility to compare different
measurable quantities related to meson exchange currents with each other
as there are the strength of the giant-dipole resonance and the magnetic
moments of nucleons in a nucleus. Taking the result for nuclei
with a $^{208}$Pb double-magic core, ({\it cf.}  (\ref{NumericalKappa}))
$\ko=0.46\pm0.05$, we arrive at
\begin{equation}
\delta g_l^{meson}(p)=0.28\pm0.02, \hspace{20mm}
\delta g_l^{meson}(n)= -0.18\pm 0.02
\label{p-n-dgl-a}
\end{equation}
corresponding to
\begin{equation}
\label{numerical-dgl}
\delta g_l^{(iv)} = 0.23 \pm 0.03.
\end{equation}
Given the validity of the Fujita-Hirata relation we may consider
these quantities in (\ref{p-n-dgl-a}) and (\ref{numerical-dgl})
as experimental  results which may be compared with
predictions.
We find good agreement with the prediction of
Brown and Rho \cite{BR80}, {\it viz.}
\begin{equation}
\delta g_l^{meson}(p)=0.27, \hspace{20mm}
\delta g_l^{meson}(n)= -0.17
\label{p-n-Brown}
\end{equation}
and with the prediction of Hyuga et al.\  \cite{hyuga80}, {\it viz.}
\begin{equation}
\delta g_l^{meson}(p)=0.27, \hspace{20mm}
\delta g_l^{meson}(n)= -0.14.
\label{p-n-dgl-b}
\end{equation}

The quantities $\delta g_l^{meson}$ are not directly
observable in measurements of magnetic moments because of the effects
of higher-order configuration mixing.
For a certain fraction of time protons and neutrons are correlated to
each other producing an additional angular momentum, and related to this
a correction $\delta g_l^{high}$ to the orbital g-factor. An analysis of
mesonic effects on nuclear magnetic moments has been carried out by Yamazaki
\cite{yamazaki79}. After correcting for first-order configuration mixing
he arrives at anomalous orbital g-factors of
\begin{equation}
\delta g_l^{obs}(p)=0.15\pm 0.02, \hspace{20mm}
\delta g_l^{obs}(n)=-0.05\pm 0.02,
\label{gobs}
\end{equation}
consistent for all measured g-factors of nuclear states having a
$^{208}$Pb core. Comparing these quantities with our results of
({\ref{p-n-dgl-a}})
we find
\begin{equation}
\delta g_l^{obs}  - \delta g_l^{meson}= - (0.13\pm0.03) \, \tau_3,
\label{dgl-obs}
\end{equation}
[$\tau_3(p) =+1$, $\tau_3(n) =-1$]. This result (\ref{dgl-obs})
may be compared with the predicted
\cite{hyuga80,arima79} effects of higher-order configuration mixing,
{\it viz.},
\begin{equation}
\delta g_l^{high}(p)=-0.15, \hspace{20mm}
\delta g_l^{high}(n)= +0.10.
\label{p-n-high}
\end{equation}
We notice that the ``experimental" result (\ref{dgl-obs})  is purely
isovector, whereas the predictions  (\ref{p-n-high}) suggests some isoscalar
component. However, in view of the error given  (\ref{dgl-obs}) it is
not possible to draw any conclusion from this difference.

A summary of these results is given  in Fig.\
\ref{FujitaHirata}. The first column  shows the result of the
measurement of magnetic moments carried out by Yamazaki
\cite{yamazaki79}. The second column shows the quantities
$\delta_l^{meson}$ obtained from the experimental $\ko$ on the basis of the
Fijita-Hirata relation together with higher-order correction $\delta
g_l^{high}$ obtained as a difference of $\delta_l^{meson}$ in column II
and $\delta g_l^{obs}$ in column I. Column III shows the predictions of
Hyuga et al.\ \cite{hyuga80} and column 4 the prediction of Brown and Rho
\cite{BR80}.
\begin{figure}[htb]
\caption{ \em Summary of results on the Fujita-Hirata relation. Column I:
Experimental result of Yamazaki \cite{yamazaki79}. Column II:
$\delta g_L^{meson}$ calculated from the experimental $\protect\ko$ and
$\delta g_L^{high}$ obtained as a difference between $\delta_l^{meson}$
of column II
and $\delta g_L^{obs}$ of column I. Column III: Predictions of Hyuga et al.\
\cite{hyuga80}. Column IV: Predictions of Brown and Rho \cite{BR80}.}
\label{FujitaHirata}
\end{figure}


The quasiparticle aspect of nucleons in a nucleus  may be discussed in
different forms.
Instead of considering the dynamics of nucleon interactions
explicitly,
as done in Landau's Fermi liquid theory and its interpretation by Brown
and Rho \cite{BR80} there are attempts to view the role of
the nuclear environment as a modification of the vacuum
\cite{NambuJonaLasinio61}.
The advantages of this approach are that it
is possible to make use of different models like the chiral mean field
and the Walecka model of nuclear matter \cite{walecka86} and to arrive at new
predictions which are not
that easy to obtain within the framework of the original Fermi
liquid theory. One disadvantage is that these new vacua are less familiar than
the building blocks of Fermi liquid theory,
thus making it desirable to build bridges between
these different aspects of nuclear matter. The most recent works going into
this direction are those of G.E. Brown \cite{BR95}, Friman and
Rho \cite{FR96}, Brown and Rho \cite{BR96} and Rho \cite{mrho98}.

In \cite{BR95}  essentially the same effective mass $M^*$, which is
related to the
Migdal parameter $F_1$ as in Eq.\ (\ref{mass1}),
is also related to a (negative) scalar mean field $\Phi$ of the Walecka type
\cite{walecka86}    via
\begin{equation}
\label{Walecka}
M^* = M + g_{\sigma NN} \Phi  .
\end{equation}
In terms of
(\ref{Walecka}) the primary process is the lowering of the mass from
$M$ to $M^*$. The physical reason for this lowering of the mass
is  that  the scalar mean field with the coupling constant
$g_{\sigma NN}$  gives an attractive scalar potential.
Both descriptions, {\it viz.}\ the velocity dependence
of quasi-particle interactions in the Fermi liquid and the lowering
of the {\it in-medium} mass lead to the same increase of the quasiparticle
velocity $v=p/M^*$ as compared to $p/M$ and these two interpretations
are equivalent.
This equivalence is seen in a relativistic formulation, where Lorentz
invariance provides a connection between the two interpretations.
These matters are reviewed in Brown et al.\ \cite{BWBS87}.

A further aspect of the in-medium modification of
the nucleon-nucleon interaction
has been investigated Brown and Rho in \cite{BR90}.
They discuss the enhancement of the $\rho $ meson tensor coupling  to
the nucleon. The $\rho$-meson is assumed to acquire an effective
mass inside the nuclear medium with
\begin{equation}
\label{RhoMass}
\frac{m^*_{\rho}}{m_{\rho}} \simeq  \frac{M^*}{M}  \simeq  \frac34
\end{equation}
as suggested by the general concept of Brown and Rho scaling \cite{BR91}
and by the Fermi liquid theory
together with  the experimental value of $\ko$ = 0.46.
It is shown \cite{BR90} that this implies that in
the nuclear medium the $\rho$-meson tensor force is enhanced by a
factor $16/9$.

\clearpage

\section{Dispersion relations at fixed momentum transfer for nuclear
Compton scattering}

In the previous sections phenomenological approaches have been applied in
order to obtain the angular dependence of the Compton scattering amplitude.
One of the main guidelines has been the fulfilment of a dispersion relation
in forward direction, together with the requirement that the resonance
parts of the amplitude should vanish in the high-energy limit at
arbitrary scattering angle. Another ingredient in this construction of
the Compton amplitude has been the correct low-energy behaviour of each
(electric and magnetic) multipole. In this section we investigate the
angular dependence of the nuclear Compton amplitude within a different
approach based on a dispersion theory for the invariant amplitudes
at fixed momentum transfer.
This approach gives a correct form of dispersion relations
for the resonance and partial-wave amplitudes
which sometimes are written too naively.

\subsection{General structure of the Compton amplitude}

In the case of spin-0 nuclei, which are considered in this review, the
general form (see {\it e.g.} \cite{petrunkin81})
of the Compton scattering amplitude $T$ is
\begin{equation}
        T(\omega ,t)=\omega ^2 \left\{
      \Big(A_1(\omega ,t)+A_2(\omega ,t)\Big)
 \,\veceps_1\cdot\veceps_2 +
      \Big(A_1(\omega ,t)-A_2(\omega ,t)\Big)
 \,\vs_1\cdot\vs_2 \right\},
	\label{dispy1}
\end{equation}
where $A_1(\omega ,t)$ and $A_2(\omega ,t)$ are invariant
amplitudes and
\begin{equation}
	t=-{\bf q}^2=-2\omega ^2(1-\cos \theta )
	\label{dispy2}
\end{equation}
is the square of momentum transfer. Here all recoil corrections are neglected.
The form (\ref{dispy1}) for the amplitude is most convenient, when a
discussion of
low-energy properties, such as sum rules and the behaviour of multipole
amplitude below resonance energies, is intended.
The invariant amplitudes $A_i$, which appear in a Lorentz-invariant
form of the Compton amplitude as soon as unphysical (kinematical)
singularities and zeroes are eliminated,
are related to
the helicity non-flip amplitude
$T_{1,1}$ and the helicity flip amplitude $T_{1,-1}$
for a circularly polarized photon via
\begin{equation}
	A_1={{T_{1,1}(\omega ,t)} \over {\omega ^2(1+\cos \theta )}}=
	2{{T_{1,1}(\omega ,t)} \over {(4\omega ^2+t)}}\;,\quad A_2={{T_{1,-1}
   (\omega ,t)} \over {\omega ^2(1-\cos \theta )}}=-2{{T_{1,-1}(\omega ,t)}
	\over t}  .
	\label{dispersion1}
\end{equation}
Due to the conservation of angular momentum only
$T_{1,1}$ differs from zero in forward direction.
For linearly polarized photons the two non-vanishing
amplitudes are
\begin{equation}
	T_{|\,|}=T_{1,1}-T_{1,-1}\;,\quad T_\bot =T_{1,1}+T_{1,-1}\;  ,
	\label{dispersion2}
\end{equation}
where $T_{|\,|}$ ($T_\bot$) corresponds to the polarization of
both photons being parallel (perpendicular) to
the scattering plane.

Furthermore, we will apply the usual partial wave expansion of the helicity
amplitudes (see {\it e.g.} \cite{jacob59,hara}),
\begin{equation}
	T_{1,\pm 1}(\omega ,\theta )=\sum\limits_{j=1}^\infty
	{T_{1,\pm 1}^j(\omega )\;d_{1,\pm 1}^j(\theta )}  ,
	\label{dispy7}
\end{equation}
where the $d$-functions are given by the formulae \cite{hara,erdely}
\begin{eqnarray}
   d^{\,j}_{1,1}(\theta) &=& \cos^2 \frac{\theta}{2} \,
         F(-j+1,j+2,1,\sin^2 \frac{\theta}{2}),
     \quad  \nonumber \\
   d^{\,j}_{1,-1}(\theta) &=&  \frac{j(j+1)}{2}
      \sin^2 \frac{\theta}{2} \, F(-j+1,j+2,3,\sin^2 \frac{\theta}{2}).
      \label{dfun}
\end{eqnarray}
Here $F(a,b,c,x) = 1 + (ab/c)(x/1!) + \ldots$
is the hypergeometric function.
Equations (\ref{dfun}) can be rewritten as
\begin{eqnarray}
      \label{dfun-Legendre}
  d_{1,\pm 1}^j(\theta ) &=&
     \frac{1 \pm z}{j(j+1)} \Big(P'_j(z)
    + (z \mp 1) P^{\prime\prime}_j(z) \Big)
\nonumber \\
   &=& \pm P_j(z)  +  {{1 \mp z}
     \over {j(j+1)}}\,P'_j(z),  \quad  z=\cos \theta,
\\
     \quad d_{1,-1}^j(\theta ) &=&
	(-1)^{j-1}d_{1,+1}^j(\pi -\theta )   .
     \nonumber
\end{eqnarray}
The functions $d^{\,j}_{\lambda\lambda'}$ satisfy an orthogonality relation:
\begin{equation}
  \int_{-1}^1 d^{\,j }_{\lambda\lambda'}(\theta)
       d^{\,j'}_{\lambda\lambda'}(\theta) \, d(\cos\theta)
   = \frac{2}{2j+1} \delta_{jj'}.
\label{orthog}
\end{equation}
The partial waves $T_{1,\pm 1}^j(\omega )$ contain electric
($T^{Ej}$) and magnetic ($T^{Mj}$) multipoles:
\begin{equation}
	T_{1,\pm 1}^j(\omega )=
      {T^{Ej}\pm T^{Mj}}  ,
	\label{dispy8}
\end{equation}
what can be easily inferred from (\ref{MultipoleDecomp}).
Due to the optical theorem
the imaginary parts of the partial amplitudes $T^{Ej}$ and $T^{Mj}$ give
rise to partial absorption cross sections $\sigma ^{Ej}$ and
$\sigma ^{Mj}$ via
\begin{equation}
	{\omega \over {4\pi } }\,\sigma ^{\lambda j}(\omega )=
	\mbox{Im }T^{\lambda j}(\omega )  ,
	\label{sigmapart}
\end{equation}
where $\lambda = E,M$.

In order to obtain a definite multipole amplitude, let us start from
a dispersion relation at fixed momentum transfer $t$ for the
invariant amplitudes $A_i$. From Eq.\ (\ref{dispersion1}) one can
expect that at fixed $t$ the amplitude $A_1$ tends to zero as $\omega$
tends to infinity and hence satisfies an ordinary unsubtracted
dispersion relation.
At the same time the amplitude $A_2$ does not generally tend to zero
and therefore a dispersion relation for $A_2$ should include either a
subtraction or a safe regularization.
The regularization can be implemented by cutting the dispersion
integral at some high maximal energy $\omega_{\rm max}$
and by adding an asymptotic contribution $a(t)$
which is energy independent (but is momentum-transfer dependent)
at low and medium energies $\omega^2 \ll \omega^2_{\rm max}$
--- see a parallel discussion of the nucleon case in
\cite{petrunkin81,lvov97}.
For the sake of simplicity we disregard here such a regularization
which can be easily restored. Since the asymptotic piece $a(t)$
does not depend on the energy, it does not affect the resonance part
$R(\omega,t)$ of the Compton scattering amplitude and
enters only into the seagull $S(\omega,t)$, to its helicity-flip
energy-dependent part.

So, we write the dispersion relation for either amplitude $A_i$ in
the same form with the dispersion integral taken up to infinity and
without the explicit addition $a(t)$ to the amplitude $A_2$:
\begin{equation}
   A_i(\omega ,t)=-{{Z^2e^2} \over {2AM\omega ^2}}
     +{2 \over \pi }\int\limits_0^\infty
      {\mbox{Im} A_i(\omega ',t)\;{{\omega '\;d\omega '} \over
      {\omega '^2-\omega ^2-i0}}} .
  \label{dispy3}
\end{equation}
Here the pole contribution ({\it viz.}\ the first term on the r.h.s.)
ensures the Thomson limit for the total amplitude $T(\omega,t)$.

In the following we apply the partial-wave expansion (\ref{dispy7})
and write the imaginary parts of $A_i$ through the partial absorption
cross sections $\sigma^{Ej}$ and $\sigma^{Mj}$.  Such a procedure may
be invalid at high $|t|$ exceeding a binding threshold for
disintegrating the nucleus into separate parts.  In this case we
consider the resulting formulas in the sense of an analytical
continuation. Nevertheless, this procedure allows us to investigate
the deviation of the exact partial scattering amplitudes from the
naive assumption of a purely Lorentzian form for its resonance part.

The use of Eq.\ (\ref{dispersion1}) leads to dispersion relations
for the helicity amplitudes. With the help of Eq.\ (\ref{dispy7}) and
Eq.\ (\ref{sigmapart})
for the r.h.s.\ of Eq.\ (\ref{dispy3}), one obtains
\begin{equation}
	\label{dispy10}
	 T_{1,\pm 1}(\omega ,\theta )=
	 (1\pm \cos \theta )
	 \Biggl\{ -{{Z^2e^2} \over {2AM}}  +
	 {\omega ^2 \over {2\pi ^2}}\;\sum\limits_{j=1}^\infty
	 {\int\limits_0^\infty
	{{{d_{1,\pm 1}^j(\theta ')} \over
        {1\pm \cos \theta '}}\;{{\sigma ^{Ej}(\omega ')
      \pm \sigma ^{Mj}(\omega ')}
	\over {\omega '^2-\omega ^2-i0}}\;d\omega '}} \Biggr\}
\end{equation}
where $\theta '$ is a function of $\theta $, $\omega $ and
$\omega '$, due to the condition that the square of the momentum transfer
$q ^2=-t$ is fixed,
\begin{equation}
        \omega '^2\left( {1-\cos \theta '} \right)
    =\omega ^2\left( {1-\cos \theta }
	\right)=-{t \over 2}  .
	\label{momentumfixed}
\end{equation}
Here the $\theta'$-dependent factors in the braces of Eq.\ (\ref{dispy10})
are polynomials in $\cos\theta'$ and hence polynomials in
$x=\omega ^2/\omega '^2$.

Let us multiply both sides of Eq.\ (\ref{dispy10}) by
$d_{\lambda ,\lambda '}^{J}$ and take the integral with respect to $\theta$
using Eqs.\ (\ref{dispy7}) and (\ref{orthog}).
As a result we obtain
\begin{equation}
	\label{dispersion11}
	T^{EJ}(\omega )\pm T^{MJ}(\omega )
	 =-{{Z^2e^2} \over {AM}}\,\delta _{J,1}
	+ {{\omega ^2}
	\over {2\pi ^2}}\,\sum\limits_{L=1}^\infty  {\int\limits_0^\infty
	{\Phi ^{(\pm )}_{JL}(\omega ,\omega ')\,{{\sigma ^{EL}(\omega ')\pm
        \sigma ^{ML}(\omega ')} \over {\omega '^2-\omega ^2-i0}}}}\,d\omega'.
	\nonumber
\end{equation}
The function $\Phi ^{(\pm )}_{JL}$ has the following integral
representation:
\begin{equation}
	\Phi _{JL}^{(\pm )}(\omega ,\omega ')=
	{{2J+1} \over 2}\int\limits_{-1}^1
	{{{1\pm \cos \theta } \over
	{1\pm \cos \theta '}}\;d_{1,\pm 1}^J(\theta )
	d_{1,\pm 1}^L(\theta ')\,d(\cos \theta )}   .
	\label{dispersion12}
\end{equation}
In all cases the function $\Phi ^{(\pm )}_{JL}$ is a polynomial
in $x$, whoes properties follows from the orthogonality relation
(\ref{orthog}) and can be summarized in the following way:
\begin{equation}
\label{polynom}
   \Phi ^{(\pm )}_{JL}(\omega,\omega')= \cases{
     0,    &$J>L$,  \cr
     x^{J-1},  &$J=L$,  \cr
     x^{J-1}(1-x) \tilde \Phi ^{(\pm )}_{JL}(x),   &$J<L$,}
\end{equation}
where $\tilde \Phi ^{(\pm )}_{JL}(x)$ is a polynomial
of the $(L-J-1)$th power in $x$.
By applying Eq.\ (\ref{polynom}) to the dispersion relation
(\ref{dispersion11}) via  (\ref{dispersion12}) we obtain
dispersion representations for the
multipole amplitudes of nuclear Compton scattering:
\begin{eqnarray}
   && T^{EJ}(\omega) = -
   \frac{(Ze)^2}{AM}\delta_{J1}
   +\frac{1}{2\pi^2}
    \int_0^\infty
   \left(\frac{\omega}{\omega'}\right)^{2J} \bigg\{
    \frac{ {\omega'}^2 \sigma^{EJ}(\omega')}
          { {\omega'}^2 - \omega^2 -i0}   +
   \nonumber\\ && \qquad \displaystyle
     \sum_{L>J}
       \left[ \sigma^{EL}(\omega')
         A_{JL}\left(x\right) +
       \sigma^{ML}(\omega')
         B_{JL}\left(x\right)\right]\bigg\}
     d\omega'  ,
\label{T-EJ}
\end{eqnarray}
and
\begin{eqnarray}
   && T^{MJ}(\omega) =
   \frac{1}{2\pi^2}
    \int_0^\infty
   \left(\frac{\omega}{\omega'}\right)^{2J} \bigg\{
     \frac{ {\omega'}^2 \sigma^{MJ}(\omega')}
          { {\omega'}^2 - \omega^2 -i0}  +
   \nonumber\\ && \qquad \displaystyle
    \sum_{L>J}
     \left[ \sigma^{ML}(\omega')
         A_{JL}\left(x\right) +
            \sigma^{EL}(\omega')
         B_{JL}\left(x\right)\right]\bigg\}
     d\omega'  .
\label{T-MJ}
\end{eqnarray}
Here $A_{JL}(x)$ and $B_{JL}(x)$ are polynomials
of $(L-J-1)$th power in $x$, with $x=(\omega /\omega ')^2$:
\begin{equation}
  A_{JL}(x)=\frac12 [ \tilde \Phi ^{(+)} _{JL}(x)
        + \tilde \Phi ^{(-)} _{JL}(x)],
  \quad
  B_{JL}(x)=\frac12 [ \tilde \Phi ^{(+)} _{JL}(x)
        - \tilde \Phi ^{(-)} _{JL}(x)].
\end{equation}
The first few of them are given by
\begin{equation}\begin{array}[b]{ll}
  A_{12}(x)= 2,  &
  B_{12}(x)=-1,  \\[2mm]
  A_{13}(x)= \frac{ 7}{2}-\frac{21}{4}x, &
  B_{13}(x)=-\frac{ 5}{2}+\frac{15}{4}x,  \\[2mm]
  A_{14}(x)= \frac{11}{2}-\frac{77}{4}x+\frac{77}{5}x^2, &
  B_{14}(x)=-\frac{ 9}{2}+\frac{63}{4}x-\frac{63}{5}x^2,  \\[2mm]
  A_{23}(x)= \frac{15}{4}, &
  B_{23}(x)=-\frac{ 5}{4},  \\[2mm]
  A_{24}(x)= \frac{39}{4}-13x, &
  B_{24}(x)=-\frac{21}{4}+ 7x,  \\[2mm]
  A_{34}(x)= \frac{28}{5},  &
  B_{34}(x)=-\frac{ 7}{5} .
\end{array}\end{equation}
Since each partial absorption cross section is nonzero only above an
energy threshold corresponding
to the first excitation level of the nucleus, no divergence at
$\omega '=0$ appears in Eqs.\ (\ref{T-EJ}) and (\ref{T-MJ}).
Similar multipole dispersion relations are used in the analysis of pion
photoproduction data \cite{donnachie}.
The partial-wave series in (\ref{T-EJ}), (\ref{T-MJ})
may become divergent at high $\omega$ and should be understand
in the sense of an analytical continuation in such a case.

One can see from the relations (\ref{T-EJ}) and (\ref{T-MJ})
that at $\omega \rightarrow 0$ the partial
photoabsorption cross sections $\sigma ^{EJ}$ and $\sigma ^{MJ}$ give
contributions to the partial amplitudes $T ^{EJ}$ and $T ^{MJ}$,
which are proportional to $\omega ^{2J}$.
The asymptotic contribution $a(t)$ which is disregarded in the
above consideration and which makes equal but opposite additions
to $T ^{EJ}$ and $T ^{MJ}$ does not change this threshold behavior.
The nonzero contributions of $\sigma ^{\lambda ' L}$ to $T^{\lambda J}$
with $L>J$ is another important property of the partial scattering
amplitudes, which
follows from (\ref{T-EJ}) and (\ref{T-MJ}).
These properties may be used as a starting point to investigate the
deviation of multipole amplitudes from a purely Lorentzian form.

\subsection{Resonance and seagull amplitudes at fixed momentum transfer}

In the previous section we obtained a representation for the total
Compton scattering amplitude in non-forward direction, Eq.\ (\ref{dispy10}).
Starting from this expression we now want to construct the $t$-dependence
of the giant resonance, the quasi-deuteron and the seagull parts, which
enter into the Compton amplitude $T_A$. The additional contribution
$T_N$, which contains the electromagnetic polarizabilities of the
nucleon, is not considered here.
This decomposition should be consistent with the definitions (\ref{mesonGR1}),
(\ref{mesonGR2}) and (\ref{PolCorr}) given for the case of forward direction.
Here our dispersion-theoretical method (as
opposed to the operator-based techniques found {\it e.g.} in
\cite{fri76,arenh85,stich}) can be efficiently used to
reconstruct these contributions to the Compton amplitude
on the base of an experimental information.
The $t$-dependence can
consistently be constructed due to the use of dispersion relations at
fixed momentum transfer.

Let us define the contributions of each of the three  parts to the total
(nuclear) helicity amplitudes $T^A_{1,\pm 1}$ by
\begin{equation}
	T^A_{1,\pm 1}(\omega ,t)=R_{1,\pm 1}^{GR}(\omega ,t)
	+R_{1,\pm 1}^{QD}(\omega ,t)+S^{tot}_{1,\pm 1}(\omega ,t)  .
	\label{parthel}
\end{equation}

The representations of the two resonant parts should fulfill certain
requirements. In the dispersion integral contributing to the
amplitude $R_{1,\pm 1}^{GR}(\omega ,t)$ instead of the full partial wave
cross section $\sigma ^{\lambda j}$
only $\sigma ^{\lambda j}_{GR}$ should appear.
In all other aspects it has the same form as the one in Eq.\
(\ref{dispy10}).
Furthermore, some function of $t$,
which does not depend on $\omega $
and ensures vanishing $R^{GR}$ at high $\omega$, has to be added.
The amplitude $R_{1,\pm 1}^{QD}(\omega ,t)$ should have similar properties.
It also is represented as a sum of a dispersion integral containing
$\sigma ^{\lambda j}_{QD}$ and some function of $t$.
As for the seagull amplitude $S^{tot}_{1,\pm 1}(\omega ,t)$, the integration in
the dispersion contribution should start from pion mass
$m_\pi$ with the integrand
depending on the difference
$\sigma ^{\lambda j} - \sigma ^{\lambda j}_{QD} $.
The seagull amplitude has no resonance structure and corresponds to the
scattering by an object of small size
in comparison with the nuclear radius $R$. More specifically,
it can be argued (see {\it e.g.} \cite{christillin4,C78} and the
discussion in sections 3.1 and 3.2) that two
distinct processes contribute to $S_{tot}$, namely the Thomson scattering by
individual nucleons (kinetic seagull) and the scattering by correlated
nucleon pairs (mesonic seagull). For the latter, the characteristic size
is of the order of $1/m_\pi$.
The comparison of the nuclear radius $R$ with
the relevant interaction range for the processes
contributing to the seagull amplitude
leads to the conclusion that the $t$-dependence of the total seagull
amplitude is similar to the nuclear form factor.
We will investigate this point in more detail in section 9.
Finally, using Eq.\ (\ref{mesonS2}) and taking the above-mentioned properties
into account, we get
\begin{eqnarray}
	 & &S_{1,\pm 1}(\omega ,\theta )=\left( {1\pm \cos \theta }
	 \right)\left\{ {\matrix{{}\cr
{}\cr
}} \right.-{{Ze^2} \over {2M}}\left( {F_1(t)+\kappa {N
\over A}\,F_2(t)} \right)+
	\label{dispseagull} \\
	& &{{\omega ^2} \over {2\pi ^2}}\,\sum\limits_{j=1}^\infty
     {\int\limits_{m_\pi}^\infty  {{{d_{1,\pm 1}^j(\theta ')}
	\over {1\pm \cos \theta '}}}}\,\;{{\Delta \sigma ^{Ej}(\omega ')\pm
        \Delta \sigma ^{Mj}(\omega ')} \over {\omega '^2-\omega ^2}}\,d\omega'
	\left. {\matrix{{}\cr
{}\cr
}} \right\}  ,
	\nonumber
\end{eqnarray}
with $\Delta \sigma (\omega )=
\sigma _{tot}(\omega )-\sigma _{QD}(\omega )
-A\sigma _{N}^{bound}(\omega )$,
the function $F_1(t)$ being the nuclear one-body form factor
and $F_2(t)$ corresponding also to a
nuclear form factor but taking into account the finite size of a
correlated nucleon pair   ({\it cf.} sections 3.5 and 9).

Let us discuss the behaviour of Eq.\ (\ref{dispseagull})
in the case $\omega \ll m_\pi$. Then $(-t) \ll m^2_\pi$
due to (\ref{momentumfixed}).
Since the integral with respect to $\omega '$ starts from $m_\pi$,
the same relation shows that also $\theta ' \ll 1$.
We can omit $\omega ^2$ in the denominator of the
integrand in  Eq.\ (\ref{dispseagull}). However, it is not possible also
to neglect $\theta '$, which would correspond to $t=0$, because
as described above
the $t$-dependence of any contribution to the seagull amplitude should
approximately be given by a nuclear form factor.
Since quasi-deuteron production occures from small regions of size
$\sim 1/m_\pi \ll R$ scattered over the whole volume of the
nucleus, many multipoles contribute to
the integrand in (\ref{dispseagull}), thus making
the partial-wave expansion almost useless.
It is physically clear, however, that the $t$-dependence
of the integrand reflects a distribution of participating
$pn$ pairs in the nucleus, so that the integral is proportional to
the two-body form factor $F_2(t)$.
The coefficient of the proportionality can be identified with
modifications of the electric and magnetic polarizabilities of the
nucleon, $\delta \alpha $ and $\delta \beta $,
caused by the mesonic seagull amplitude:
\begin{equation}
	\label{dispseagull2}
	S_{1,\pm 1}(\omega ,\theta )={{\left( {1\pm \cos \theta } \right)}
        \over 2}\left\{ -{{Ze^2} \over M}
  \left( {F_1(t)+\kappa {N \over A}\,F_2(t)} \right)+
	A\omega ^2\,
	\left( {\delta \alpha \pm \delta \beta }
	\right)\,F_2(t)
	 \right\}  .
	\nonumber
\end{equation}
The important quantities $\delta \alpha $ and $\delta \beta $
will be discussed in section 9 within a specific model,
where it will also be shown that form factors accompaning
$\kappa$, $\delta \alpha $ and $\delta \beta $
in (\ref{dispseagull}) are actually not identical, as a consequence
of a different configuration size of the correlated $pn$ pairs
determining these values.
In this section, however, we disregard this nontrivial feature and
use the same form factor $F_2(t)$ everywhere.
It is also possible to discuss this particular form (\ref{dispseagull2}) of
the seagull amplitude on the level of nuclear matrix elements, as was
done in section 3.

Now, taking into account Eqs.\ (\ref{resonance}),
(\ref{mesonQD2}) and (\ref{dispseagull}), we obtain an explicit
representation of the resonance amplitudes from Eq.\
(\ref{dispy10}). For $R_{1,\pm 1}^{GR}(\omega ,t)$ we have
\begin{eqnarray}
        & &R_{1,\pm 1}^{GR}(\omega ,\theta )
   ={{\left( {1\pm \cos \theta } \right)}
	\over 2}\left\{ {\matrix{{}\cr
{}\cr
}} \right.{{Ze^2} \over M}\left( {-{Z \over A}+F_1(t)+\k \,{N
\over A}\,F_2(t)} \right)+
    \label{dispGR} \\
    &  & {\omega ^2 \over {\pi ^2}}\;\sum\limits_{j=1}^\infty
	 {\int\limits_0^\infty
	{{{d_{1,\pm 1}^j(\theta ')} \over
	{1\pm \cos \theta '}}\;{{\sigma _{GR}^{Ej}(\omega ')\pm
	\sigma _{GR}^{Mj}(\omega ')}
	\over {\omega '^2-\omega ^2-i0}}\;d\omega '}} \Biggr\} \, .
	\nonumber
\end{eqnarray}
This representation of the giant resonance amplitude, in which form factors
appear explicitly, is convenient, as in the dispersive part each
multipole contains the correct $\omega $-dependence at low energies.
Similarly, for $R_{1,\pm 1}^{QD}(\omega ,t)$  one has
\begin{eqnarray}
	& &R_{1,\pm 1}^{QD}(\omega ,\theta )={{\left( {1\pm \cos \theta }
	\right)} \over 2}\left\{ {\matrix{{}\cr
{}\cr
}} \right.\kappa ^{QD}\,{{e^2} \over M}{{ZN} \over A}\,F_2(t)+
    \label{dispQD} \\
    &  & {\omega ^2 \over {\pi ^2}}\;\sum\limits_{j=1}^\infty
	 {\int\limits_0^\infty
	{{{d_{1,\pm 1}^j(\theta ')} \over
	{1\pm \cos \theta '}}\;{{\sigma _{QD}^{Ej}(\omega ')\pm
	\sigma _{QD}^{Mj}(\omega ')}
	\over {\omega '^2-\omega ^2-i0}}\;d\omega '}} \Biggr\} \, .
	\nonumber
\end{eqnarray}
In contrast to (\ref{dispGR}), the integral here involves
many multipoles and therefore is less useful. It's physically
clear again that the $t$-dependence of the integral mainly follows
the two-body form factor $F_2$.

Now we should fulfill the requirement that the resonance amplitudes tend
to zero at high $\omega $ and fixed $t$. Since in
both amplitudes $R_{1,-1}^{GR}$ and $R_{1,-1}^{QD}$
the external factor $(1-\cos \theta )/2=-t/(4\omega ^2)$ tends to zero
in this limit, they vanish independently of the explicit shape of the
form factors.
Moreover, the asymptotic piece $a(t)$ can also affect
the helicity-flip amplitudes.
Therefore, model-independent restrictions appear only from
the asymptotic behaviour of $R_{1,1}^{GR}$ and $R_{1,1}^{QD}$.
In the limit $\omega \rightarrow \infty $ the requirement of vanishing
$R_{1,1}^{GR}$ leads to
\begin{equation}
        {{Ze^2} \over M}\left( {-{Z \over A}+F_1(t)+\k \,{N
\over A}\,F_2(t)} \right) =
     {1 \over {\pi ^2}}\;\sum\limits_{j=1}^\infty
	 {\int\limits_0^\infty
	{{{d_{1,1}^j(\theta ')} \over
	{1+\cos \theta '}}\;[\sigma _{GR}^{Ej}(\omega ')+
	\sigma _{GR}^{Mj}(\omega ') ]\;d\omega '}}  \, ,
	\label{sumrule1}
\end{equation}
while the same condition for $R_{1,1}^{QD}$ gives
\begin{equation}
	\kappa ^{QD}\,{{e^2} \over M}\,{NZ
     \over A}\,F_2(t)  =
     {1 \over {\pi ^2}}\;\sum\limits_{j=1}^\infty
	 {\int\limits_0^\infty
	{{{d_{1,1}^j(\theta ')} \over
	{1+\cos \theta '}}\;[\sigma _{QD}^{Ej}(\omega ')+
	\sigma _{QD}^{Mj}(\omega ') ]\;d\omega '}} \, .
	\label{sumrule2}
\end{equation}
Using Eq.\ (\ref{sumrule1}) we can express the giant resonance amplitude
$R_{1,\pm 1}^{GR}$ of Eq.\ (\ref{dispGR}) via the absorption cross sections
$\sigma ^{\lambda j}_{GR}$ only, without the explicit appearance of form
factors.
Putting $t$ equal to zero in Eqs.\ (\ref{sumrule1}) and
(\ref{sumrule2}) we obtain again the sum rules (\ref{mesonGR3}) and
(\ref{mesonQD3}), since
\begin{displaymath}
	\sigma _{GR}(\omega )=\sum\limits_{j=1}^\infty
      {\left( {\sigma _{GR}^{Ej}(\omega )+\sigma _{GR}^{Mj}(\omega )}
     \right)}\;,\quad \sigma _{QD}(\omega )=\sum\limits_{j=1}^\infty
   {\left( {\sigma _{QD}^{Ej}(\omega )+\sigma _{QD}^{Mj}(\omega )} \right)} .
\end{displaymath}
Expanding both sides of (\ref{sumrule1}) with respect to $t$ and
comparing the coefficients, we obtain additional sum rules
\begin{eqnarray}
	 & &{{Ze^2} \over M}\,{{\pi ^2(n!)^22^{2n+1}}
	 \over {(2n+1)!}}\,\left[ {\left\langle {r^{2n}}
         \right\rangle _1+\k \,{N \over A}\left\langle {r^{2n}}
      \right\rangle_2} \right]=
	\label{sumrule3} \\
	 &  & \sum\limits_{j=n+1}^\infty  {{{(j+n+1)!}
	 \over {j(j+1)(j-n-1)!}}}\int\limits_0^\infty  {{{d\omega }
	 \over {\omega ^{2n}}}\left( {\sigma _{GR}^{Ej}(\omega )+
      \sigma _{GR}^{Mj}(\omega )} \right)}  ,    \quad , n > 0
	\nonumber
\end{eqnarray}
where the ordinary definition of the average values of $r^{2n}$ is used.
In the particular case of $n=1$ Eq.\ (\ref{sumrule3}) gives
\begin{equation}
	{{Ze^2} \over M}\,{{4\pi ^2} \over 3}\,\left[
        {\left\langle {r^2} \right\rangle _1+\k \,{N \over A}\left\langle
     {r^2} \right\rangle_2} \right]=\sum\limits_{j=2}^\infty
	{(j+2)(j-1)}\int\limits_0^\infty  {{{d\omega }
	\over {\omega ^2}}\left( {\sigma _{GR}^{Ej}(\omega )+
	\sigma _{GR}^{Mj}(\omega )} \right)}   .
	\label{sumrule4}
\end{equation}
Restricting ourselves to the lowest electric multipole on the r.h.s.\ of
Eq.\ (\ref{sumrule3}) we reclaim the usual energy-weighted sum rules
\cite{connell,hayward}.
In reality, only the lowest multipoles $E1$,$M1$,$E2$
are experimentally observed and taken into account in the analysis of
scattering data \cite{W5,pb-lund,H15}.
For this case Eq.\ (\ref{dispGR}) is of the following form:
\begin{eqnarray}
         &  & R_{1,\pm 1}^{GR}(\omega ,\theta )
   ={{\left( {1\pm \cos \theta } \right)}
	\over 2}\left\{ {\matrix{{}\cr
{}\cr
}} \right.{{Ze^2} \over M}\left( {-{Z \over A}+F_1(t)+\k \,{N
\over A}\,F_2^{GR}(t)} \right)+
	\label{threemult} \\
	 &  & {{\omega ^2} \over {2\pi ^2}}\int\limits_0^\infty  {{{d\omega '}
         \over {\omega '^2-\omega ^2-i0}}}
    \left[ {\sigma _{GR}^{E1}(\omega ')\pm
	 \sigma _{GR}^{M1}(\omega ')+\sigma _{GR}^{E2}(\omega ')\left( {2-
	 {{2\omega ^2} \over {\omega '^2}}(1-\cos \theta )\mp 1} \right)}
	 \right]  .
	\nonumber
\end{eqnarray}
At very low energies it agrees with the standard formalism,
as is discussed below. However, this approximation fails
at high energies where higher multipoles are not negligible
--- see section 10.

If as before we assume that the quasi-deuteron contribution $R_{1,\pm 1}^{QD}$
to the resonance amplitude corresponds to the scattering by correlated
nucleon pairs, its dependence on $t$
is described by the form factor $F_2(t)$.
From Eq.\ (\ref{dispQD}), together with (\ref{mesonQD3}), one has approximately
\begin{equation}
	R_{1,\pm 1}^{QD}={{1\pm \cos \theta } \over 2}\,F_2(t)\,{1
	\over {2\pi ^2}}\int\limits_0^\infty
	{{{\omega '^2\,d\omega '\,\sigma _{QD}(\omega ')}
	\over {\omega '^2-\omega ^2-i0}}}
	\label{rQD}
\end{equation}
For practical purposes the absorption cross section
$\sigma _{QD}(\omega )$ may then be expressed using a phenomenological
description as discussed in section 5.
It is also possible to represent $\sigma _{QD}$ as a Lorentzian with a
very big width \cite{S8,L12}.

A straightforward application of the fixed-$t$ dispersion relations
discussed in this section is the dispersion representation of the nuclear
polarizabilities
$\bar \alpha _{A}$ and $\bar \beta _{A}$. Expanding the
amplitudes $T^{E1}$ and $T^{M1}$ from Eqs.\ (\ref{T-EJ})
and (\ref{T-MJ}) with respect to $\omega $ up to the order $\omega ^2$
we obtain \cite{petrunkin81,lvov79}
\begin{eqnarray}
     \bar \alpha _{A}&=&{1 \over {2\pi ^2}}\int_0^\infty {{{d\omega }
	\over {\omega ^2}}}\,\left\{ {\sigma ^{E1}(\omega )+{1
	\over 2}\sum\limits_{L>1} {\left[ {\sigma ^{EL}(\omega )\left(
	{1+{{L(L+1)} \over 2}} \right)} \right.}} \right.
	\nonumber \\
	&&\left. {\left. {+\,\sigma ^{ML}(\omega )\left( {1-{{L(L+1)}
	\over 2}} \right)} \right]} \right\}\;,\quad
	\nonumber \\
     \bar \beta _{A}&=&{1 \over {2\pi ^2}}\int_0^\infty {{{d\omega }
	\over {\omega ^2}}}\,\left\{ {\sigma ^{M1}(\omega )+{1
        \over 2}\sum\limits_{L>1}
     {\left[ {\sigma ^{EL}(\omega )\left( {1-{{L(L+1)}
	\over 2}} \right)} \right.}} \right.
	\nonumber \\
     &&\left. {\left. {+\,\sigma ^{ML}(\omega )\left( {1+{{L(L+1)} \over 2}}
	\right)} \right]} \right\}  .
	\label{disppol}
\end{eqnarray}
One can see that not only $\sigma ^{E1}$ and $\sigma ^{M1}$ give a
contribution to the electromagnetic polarizabilities, but also higher
multipoles of the photoabsorption cross section.
In general, the asymptotic piece $a(t)$ also contributes and gives
equal but opposite additions $\pm a(0)$
to $\bar\alpha_A$ and $\bar\beta_A$, respectively.
There are also recoil corrections $\sim 1/AM$
\cite{petrunkin81,lvov79} which are again
equal but opposite for $\bar\alpha_A$ and $\bar\beta_A$.
When in Eq.\ (\ref{disppol})
the sum ${\bar \alpha} _{A} + {\bar \beta}_{A}$ is
taken, one recovers the usual Baldin-Lapidus sum rule.
If only the lowest three multipoles are taken into account, one has
\begin{eqnarray}
     \bar \alpha _{A} &=& {1 \over {2\pi ^2}}\int_0^\infty {{{d\omega }
        \over {\omega ^2}}}\,\left( \sigma ^{E1}(\omega )+
     2\sigma ^{E2}(\omega )
       + \frac{\omega}{AM} \sigma ^{E1}(\omega )   \right) + a(0),
\nonumber \\
     \bar \beta _{A} &=& {1 \over {2\pi ^2}}\int_0^\infty {{{d\omega }
     \over {\omega ^2}}}\,\left(
        {\sigma ^{M1}(\omega )-\sigma ^{E2}(\omega )}
       - \frac{\omega}{AM} \sigma ^{E1}(\omega )   \right) - a(0),
\label{disppol2}
\end{eqnarray}
where the asymptotic contributions and recoil corrections are
explicitly written for the sake of completeness
though the latter are negligible for nuclei with $A \gg 1$.
Evaluating these formulas
with a nonrelativistic sum rules technique
and using a closure, one can reproduce known answers for the
polarizabilities of nonrelativistic systems \cite{lvov93,friar77}.
Clearly, when in Eqs.\ (\ref{disppol}) and (\ref{disppol2})
one passes from the total cross sections
$\sigma ^{\lambda J}$ to the giant resonance contributions
$\sigma ^{\lambda J}_{GR}$ the quantities $\bar \alpha _{GR}$ and
$\bar \beta _{GR}$ are obtained. In particular,
the diamagnetic correction (\ref{PolCorr1}) is derived using
the sum rule (\ref{sumrule4}).

Furthermore, it is instructive to have a closer look at the interplay of
the seagull and resonance parts in the total amplitude. To this end we
again restrict ourselves to the lowest three multipoles.
We make use of the previously defined quantities $\tilde R^{\lambda j}$
(see Eqs.\ (\ref{ResGR}) and (\ref{DispersionRelation})), i.e.
\begin{equation}
	\tilde R^{\lambda j}(\omega ,\theta )={{\omega ^2}
	\over {2\pi ^2}}\,\int\limits_0^\infty
	{{{\sigma ^{\lambda j}(\omega ')\,d\omega '}
	\over {\omega '^2-\omega ^2-i0}}}\,g^{\lambda j}(\theta )
	\label{rtilde2}
\end{equation}
to express the giant resonance amplitude for this special case.
From Eqs.\ (\ref{threemult}) and (\ref{sumrule1}) we obtain
\begin{eqnarray}
	R_{GR}(\omega ,\theta )&=&\tilde R^{E1}(\omega ,\theta )+
	\tilde R^{E2}(\omega ,\theta )+\tilde R^{M1}(\omega ,\theta )
        \nonumber \\ && {}
        + {1 \over {2\pi ^2}}\int\limits_0^\infty  {d\omega '}\,\left(
	{\sigma _{GR}^{E1}(\omega ')+\sigma _{GR}^{M1}(\omega ')+
	\sigma _{GR}^{E2}(\omega ')\,} \right)\,g^{E1}(\theta )  .
	\label{grspecial}
\end{eqnarray}
Similarly, the seagull amplitude has the form
\begin{eqnarray}
	 &&B(\omega ,\theta )+S_{GR}(\omega ,\theta )=
        \nonumber \\ && \qquad {}
        -g^{E1}(\theta )\,\left[ {{{Z^2e^2} \over {AM}}+{1
	\over {2\pi ^2}}\int\limits_0^\infty  {d\omega '}\,\left(
	{\sigma _{GR}^{E1}(\omega ')+\sigma _{GR}^{M1}(\omega ')+
	\sigma _{GR}^{E2}(\omega ')\,} \right)\,} \right]
        \nonumber \\ && \qquad {}
    +{{\omega ^2} \over {2\pi ^2}}\,\int\limits_0^\infty  {{{\,d\omega '}
     \over {\omega '^2}}}\,\sigma _{GR}^{E2}(\omega ')
     \left( {2g^{E1}(\theta )-
        g^{M1}(\theta )-g^{E2}(\theta )} \right) .
	\label{seagsspecial}
\end{eqnarray}
If one takes into account the modified TRK sum rule,
Eq.\ (\ref{mesonGR3}), for the
last term on the r.h.s.\ of Eq.\ (\ref{grspecial}),
this expression is in agreement with the
formulation of the giant resonance amplitude given in Eq.\ (\ref{ResGR}).
As before, when considered at fixed momentum transfer, the amplitude
$R_{GR}(\omega ,\theta )$ vanishes in the high-energy limit. However, the
individual multipole contributions from $E2$ and $M1$ do not vanish in
this limit.
Note that the applicability of fixed-$t$ dispersion relations is
restricted to small momentum transfer, since
due to the relation (\ref{momentumfixed}) for some $\omega '$ the angle
$\theta '$ becomes imaginary at high $-t$ and the cosine $\cos\theta'$
runs away the range of convergence
of partial-wave series ({\it viz.}\ the Lehmann ellipse).
Thus, for large $\omega $ and $t$, but fixed $\theta $
the sum with respect to $j$ in
Eq.\ (\ref{dispy10}) is not necessarily convergent.
Unfortunately, this is precisely
the regime, where the effects of retardation as discussed in
section 4.4 become important.
Therefore, at this point one has to get back
to the phenomenological discussion of retardation given in section 4.4.

\clearpage

\section{The mesonic seagull amplitude}

It was pointed out above that the seagull amplitude has no imaginary part
below pion threshold. It contains contributions from two fundamentally
different physical sources, Thomson scattering on individual nucleons
inside the nucleus
and the scattering by correlated nucleon pairs.
Such correlations occur as a result of the nucleon-nucleon interaction,
which can be described in terms of mesonic exchange between the
nucleons.
As discussed before, the amplitude for the first process contains the
form factor $F_1(t)$, while
the second contribution is proportional to $F_2(t)$.
Since the correlation radius is of the order $1/m_\pi$, i.e.\ much smaller
than the nuclear radius $R$, one can expect that the form factor $F_2$
does not differ tremendously from $F_1$. Nevertheless, this difference
produces an observable effect.  We will confirm this statement within a
model calculation. The same model will be used to investigate the
different contributions to the enhancement constant $\kappa $, as well as
the energy-dependence of the mesonic seagull amplitude, which is
expressed in terms of polarizability modifications $\delta \alpha$ and
$\delta \beta$ and the influence of the finite nuclear size on all these
quantities.

\subsection{Construction of the mesonic seagull amplitude for nuclear
matter}

In this section the contribution of correlated nucleon pairs
to the seagull amplitude, Eq.\ (\ref{dispseagull2}), will be discussed on
the basis of the results obtained in \cite{huett96}.
There the contribution of mesonic exchange currents to the nuclear Compton
scattering amplitude was investigated within the framework of a
(modified) Fermi gas model
of nuclear matter in the non-relativistic limit.
Such an
approach was suggested in \cite{fri76,christillin5} and for the case of
forward direction used in \cite{R85,ME87}.
There it was also shown that the nuclear degrees of
freedom can be taken into account by
representing the amplitude $S$ as a convolution of a two-body
spin-isospin correlation function with matrix elements corresponding to
the amputated irreducible Feynman diagrams for meson exchange.
Such a correlation function consists of a central and a tensor part. In a pure
Fermi gas model the tensor correlator is zero, but the importance of tensor
correlations was pointed out in \cite{ME87,BR80,arima73}.
In its effect on
the enhancement constant $\kappa $, for example, the central correlator
makes a comparatively small contribution due to the strong compensation
between the different Feynman diagrams taken into account.
In \cite{ME87} the results of
numerical calculations for the correlators in nuclear matter were used and
compared to a similar calculation based on the variational principle
\cite{fantoni}.

In \cite{huett96} corrections to the pure Fermi gas wave function
were calculated up to first order in a non-covariant perturbation
theory.  As a result of this model calculation the values of the
enhancement constant $\kappa $ and the polarizability modifications
$\delta \alpha $ and $\delta \beta $ were obtained and the central
and tensor part of the correlation function due to pion exchange
between the nucleons were represented in an analytical form.
In section 9.3 we will use this representation as a starting point to
investigate the difference between the form factors $F_1$ and $F_2$.
In the case of the present section, for the nuclear radius tending to
infinity, the correlation function is proportional to $F_1$.  Since
in \cite{huett96} a Fermi gas model has been used, these results are
only applicable for heavy nuclei, where it can be expected that the
quasi-classical approximation is valid.  The effects of a finite
nuclear size as well as of a realistic nuclear density have been
investigated in \cite{huett97} and will also be discussed in section 9.3.

Let us start our discussion from the static Hamiltonian for the
interaction of the nucleon with the pion field $\vephi $ (see {\it e.g.}\
\cite{EricsonWeise88}),
\begin{equation}
	H_{\pi NN}=-{f \over {m_\pi }}\left(
        {\vecsigma \cdot \vecnab}
	\right)\left( {\vetau \cdot \vephi } \right)   ,
	\label{s2-7}
\end{equation}
together with a minimal coupling to the photon field.
For all coupling constants we use the same notation and values as given in
\cite{EricsonWeise88}, in particular $f^2/(4\pi )$=0.08.
The mesonic seagull amplitude $S$ can be written in the following
form \cite{fri76,R85,ME87,huett96}:
\begin{equation}
        S=\int {{{d\vQ} \over
        {(2\pi )^3}}}\;{\cal F}^{ij}(\vQ)\;T_{ij}(\vQ) .
	\label{s2-8}
\end{equation}
The contribution ${\cal T}^{ij}_{(\pi )}$
 of $\pi $-meson exchange to ${\cal T}_{ij}$, corresponding to the diagrams
shown in Fig.\ \ref{pr22}, is given by:
\begin{eqnarray}
        {\cal T}_{(\pi )}^{ij} &=& {{2e^2f^2} \over {m^2_\pi }}\;
\left\{ {\matrix{{}\cr
{}\cr
{}\cr
}} \right.{{\epsilon _1^i\epsilon _2^j} \over {D_1}}
-2\;{{\epsilon _1^iQ_2^j\;\veceps _2\cdot \vQ_2}
\over {D_1d_2}}-2\;{{\epsilon _2^iQ_2^j\;\veceps _1\cdot \vQ_2}
	\over {D_2d_2}}   \nonumber  \\
        &+& 4\;{{Q_2^jQ_1^i\;\veceps _1\cdot \vQ_1\;\veceps _2\cdot
        \vQ_2} \over {D_1d_1d_2}}
	-\;{{Q_2^jQ_1^i\;\veceps _1\cdot \veceps _2}
	\over {d_1d_2}}
	+\left( {\matrix{{i\leftrightarrow j}\cr
{\vQ\leftrightarrow -\vQ}\cr
}} \right)
	\left. {\matrix{{}\cr
{}\cr
{}\cr
}} \right\}   ,
	\label{b1}
\end{eqnarray}
where the following abbreviations have been used:
\begin{displaymath}
        D_{1,2}=\left( {\vQ\pm \vK} \right)^2+m^2_\pi
     -\omega ^2  \;,\quad
                \vQ_{1,2}=\vQ\pm {{\vq} \over 2}\;,\quad
        d_i=\vQ_i^2+m^2_\pi \;,\quad
        \vK = (\vk_1 + \vk_2)/2    .
\end{displaymath}
The correlator ${\cal F}^{ij}$ entering Eq.\ (\ref{s2-8}) has the following
general form:
\begin{equation}
	{\cal F}^{ij}=\sum\limits_{a\ne b} {\left\langle 0
        \right|\tau _a^{(-)}\tau _b^{(+)}\sigma _a^i
      \sigma _b^j\;}e^{i\;\vQ\cdot
        (\vx_b-\vx_a)}\;e^{-i\;\vq
        \cdot (\vx_a+\vx_b)/2}\left| 0
	\right\rangle   .
	\label{s2-9}
\end{equation}
Here the summation with respect to $a$ and $b$ is performed over all
nucleons, $\tau _a^{(\pm )} = (\tau _a^1 \pm i \tau _a^2)/2$ are the isospin
raising and lowering operators, while $\sigma _a ^i/2$ denotes the $i$-th
component of the spin operator for the $a$-th nucleon.

\begin{figure}[htb]
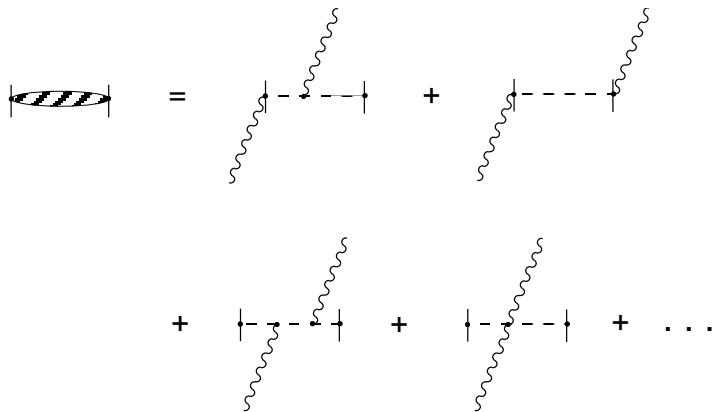

\caption{\it Typical diagrams contributing to ${\cal T}_{(\pi )}^{ij} $.
The wavy lines denote photons and dashed lines denote pions. The
amputation indicates that $T^{ij}$ contains only the nucleon vertices, but
not its wave functions.}
\label{pr22}
\end{figure}

For the simplest case of a pure
Fermi gas model the correlator (\ref{s2-9}) only has a central part:
${\cal F}^{ij} = {\cal F}_C \delta ^{ij}$,
where
\begin{equation}
        {\cal F}_C= -2 \int {{{d\vp_1d\vp_2}
	\over {(2\pi )^6}}}\int
        {d\vx_1d\vx_2}\,e^{-i(\vx_1+\vx_2)\vq/2}
        \,e^{i(\vx_1-\vx_2)((\vp_1-\vQ-\vp_2)}   .
	\label{a1-1}
\end{equation}
The ranges of integration for the nucleon
momenta $\vp_1$ and $\vp_2$ are the proton and neutron
Fermi spheres, with the radii
\begin{displaymath}
     p_F^{(p)}=2.27\,m_\pi\,\left( {{Z \over A}}
     \right)^{1/3}\;,\quad p_F^{(n)}=2.27\,m_\pi\,\left( {{N \over A}}
	\right)^{1/3}\;,\quad A=N+Z .
\end{displaymath}
This form is easily obtained in a usual Fermi gas model, where the proton
($i = p$) and neutron ($i = n$) densities are written as
\begin{displaymath}
	\rho ^{(i)}=2\int {{{d\vp}
        \over {(2\pi )^3}}}\,n^{(i)}(\vp)\;,\quad n^{(i)}(\vp)=
        \theta (p_F^{(i)}-\left| {\vp} \right|)\;,\;\;\mbox{i.e.}\quad
	\rho ^{(i)}={{\left( {p_F^{(i)}} \right)^3} \over {3\pi ^2}} .
\end{displaymath}
Here $\theta (x)$ is the step function.
Together with the assumption of a homogeneous distribution of protons and
neutrons inside the nuclear volume one has
\begin{displaymath}
	{{4\left( {p_F^{(p)}R} \right)^3}
	\over {9\pi }}=Z\;,\quad {{4\left( {p_F^{(n)}R} \right)^3}
	\over {9\pi }}=N  .
\end{displaymath}
Finally, by writing $R = 1.2\,\,A^{1/3}$ fm one finds the above relations for
$p_F^{(i)}$.
In Eq.\ (\ref{a1-1}) the integrations with respect to $\vx_i$ are taken
over a sphere with the radius $R$.

The dependence of the correlation function (\ref{a1-1}) on the nuclear
radius $R$, on $N$ and $Z$ and on the momentum transfer $\vq$
were investigated in \cite{huett96}.
In the following we will put $N = Z = A/2$.
Let us pass in Eq.\ (\ref{a1-1}) to the variables
$\vX = (\vx_1+\vx_2)/2 $ and $\vr = \vx_1-\vx_2$,
which denote the c.m. of the nucleon pair and the relative distance
between the two nucleons, respectively.
Taking the integral with respect to $\vX$ over the region
$|\vX| < R$ and with respect to $\vr$  over an infinite
range, one obtains as a first approximation for the correlator
\begin{equation}
	{\cal F}_C={\cal F}_C^{(0)}(q)\,F_1(t)  ,
	\label{correlappr}
\end{equation}
where $q=Q/p_F$, $\,\,p_F = p_F^{(p)} =  p_F^{(n)} = 1.8\,m_\pi\,\,$ and
\begin{equation}
	{\cal F}_C^{(0)}(q)=-{A \over 2}\left( {1-{q \over 2}}
	\right)^2\left( {1+{q \over 4}} \right)\;\theta (2-q)     ,
	\label{a1}
\end{equation}
which is a well-known result (see {\it e.g.} \cite{molin80}) of the Fermi model.
In the model under consideration the form factor
$F_1(q)$ is equal to $F(Rq )$, where
\begin{equation}
	F(x)={3 \over {x^2}}\left( {{{\sin x} \over x}-\cos x} \right)   .
	\label{formfac}
\end{equation}
Thus, the use of the approximate result (\ref{correlappr})
for the correlator leads to a proportionality of the mesonic seagull
amplitude (\ref{s2-8}) to a form factor $F_1(q)$.

Although this conclusion was obtained for the pure Fermi gas model, where
only the central part of the correlator contributes, the same is valid
for the general case, if one integrates in Eq.\ (\ref{s2-9})
with respect to $\vx_b-\vx_a$ over an infinite range.
It is important to understand the difference between
the mesonic seagull amplitude arising from the approximate
correlator and the exact amplitude, which is obtained by
integrating in (\ref{s2-9}) over a finite volume with
respect to both, $\vx_a$ and $\vx_b$.
In the amplitude this difference manifests itself as a difference between
the form factors $F_1(q)$ and $F_2(q)$.
At low energies the dependence of the amplitude $S$
on momentum transfer
$\vq$ is determined
by the distribution of nucleon pairs inside the nucleus.
In the case of heavy nuclei
the scale of nucleon correlations is essentially smaller than the nuclear
radius $R$. Thus, one can expect that the
$\vq$-dependence of $S$
is similar to the nuclear charge form factor
$F_1(q )$.
However, experimental data clearly indicate
\cite{S8,pb-lund,SC94} that this
$\vq$-dependence cannot fully be identified
with the form factor $F_1$
In \cite{C78,C75} it was suggested to use
for the amplitude $S$
another form factor
$F_2$ instead of $F_1$.
It has been proposed to apply
$F_2(q ) = F_1^2(q /2)$, which corresponds to
the distribution of uncorrelated
nucleon pairs \cite{Riska,W5}.
A first attempt to quantitatively discuss the function $F_2$ within
a model calculation has been made in \cite{A82}.
As a first step, we will consider this effect
calculating the $q$-dependence of the term in $S$ proportional to
the enhancement constant $\kappa $ ({\it cf.} Eq.\ (\ref{dispseagull2})).
Putting $\vK$ and $\vq$ equal to zero in
Eq.\ (\ref{b1}), substituting (\ref{a1-1}) into (\ref{s2-8}) and
integrating first with respect to $\vQ$ and then with respect to the
variables $\vp_i$ and $\vx_i$ we get the following representation
of $F_2$:
\begin{equation}
	F_2(q)={{J(Rq )} \over {J(0)}}    ,
	\label{f2}
\end{equation}
where
\begin{eqnarray}
        & &J(\zeta )=\int\limits_0^2 {dx}\,x^3
   \exp(-m_\pi  R\,x)\,F^2(p_FR\,x)\times
	\label{analytf2} \\
	& &\left[ {{1 \over 3}\left( {1-{x \over 2}}
	\right)^3F(\zeta (1-x/2))+{1
	\over {\zeta x}}\,\,\int\limits_{1-{x \over 2}}^{\sqrt {1-{{x^2}
        \over 4}}} {dy\,\,\sin (\zeta y)
 \left( {1-y^2-{{x^2} \over 4}} \right)}}
	\right]
	\nonumber
\end{eqnarray}
and $F(x)$ is given in Eq.\ (\ref{formfac}). The integral with respect to $y$
in (\ref{analytf2}) can easily be taken analytically, but we represent the
result in the form (\ref{analytf2}) for the sake of brevity. In Fig.\
\ref{pr17} the two form factors $F_1(q)$ and $F_2(q)$
as given in Eqs.\ (\ref{formfac}) and (\ref{f2}), respectively, are compared
for $A$=40. As expected, the difference is not very big. In Fig.\
\ref{pr17} we also show the widely used approximation $F_1^2(q/2)$.
As one sees, this approximation
is not valid.
Note that in Fig.\ \ref{pr17} the form factor $F(x)$ is not the experimental
(charge) form factor. For the general discussion of exchange form
factors here this difference is not important. However, experimental form
factors $F_1$ will be used as a reference in section 9.3 and appendix B,
where exchange
form factors will be discussed in a more quantitative form.

\begin{figure}[ht]
\caption{\it Comparison of the form factors from Eqs.\
(\ref{formfac}) (dotted curve) and (\ref{f2}) (full curve) for the case
of $^{40}$Ca. The
approximation $F_1^2(q/2)$ is shown as a dashed curve.}
\label{pr17}
\end{figure}

A statement better suited for practical purposes than the analytical
result (\ref{analytf2}) is obtained by considering
the slope of the form factor.
We represent the form factors $F_i$ at small momentum transfer in the form:
\begin{equation}
	F_i(q)=1+{q^2 \over 6}\left\langle {r^2} \right\rangle _i   .
	\label{formexpan}
\end{equation}
As known, $\left\langle {r^2} \right\rangle _1=(3/5)\;R^2$. In order to
get the value of $\left\langle {r^2} \right\rangle _2$, it is necessary to
expand the r.h.s.\ of Eq.\ (\ref{analytf2}) up to the order  $(Rq )^2$.
The resulting ratio $\chi (A)=\left\langle {r^2}
\right\rangle _2/\left\langle {r^2} \right\rangle _1$
is shown in Fig.\ \ref{pr18}.
We checked numerically that a good approximation for the function
$F_2$ in a wide range of $t$ is given by $F_2(q)\approx F_1(\chi t)$.
In a range for $A$ between 4 and 200 the relation
\begin{equation}
	\chi \approx \left( {1-{{0.32\,\mbox{fm}}
	\over {\sqrt {\left\langle {r^2} \right\rangle _1}}}} \right)^2
	\label{chi}
\end{equation}
accounts for the dependence of $\chi $ on $A$ within a few percent.
This corresponds to a finite shift
of the mean charge radius:
\begin{equation}
	\sqrt {\left\langle {r^2} \right\rangle _2}\approx
\sqrt {\left\langle {r^2} \right\rangle _1}-0.32\,\mbox{fm}   .
	\label{shift}
\end{equation}
In the following section it will become evident that tensor correlations
influence mostly the short-range behavior of the correlation function.
We can therefore expect that the difference between $F_1$ and $F_2$
is mainly governed by central correlations. The
more refined discussion in the next section will confirm this.

\begin{figure}[ht]
\caption{\it Ratio $\chi (A)=\left\langle {r^2}
\right\rangle _2/\left\langle {r^2} \right\rangle _1$ as a function of
the nuclear mass number $A$}
\label{pr18}
\end{figure}

\subsection{Corrections to the correlation function}

In order to obtain the correct behavior of the mesonic seagull
amplitude, it is necessary to take into account not only the central
part of the correlator, but also its tensor part.
To this end the correction to the nuclear wave function has been
calculated in \cite{huett96} using perturbation theory
with respect to $\pi $- and $\rho $-meson exchange.
These corrections are important, because they determine an essential
part of the nuclear correlation functions.
In \cite{huett96} all diagrams up to the order $f^4$ have been taken into
account, where $f$ is the meson-nucleon coupling constant.
In the course of that calculation it has been realized that three-body
corrections to the correlator are as important as two-body corrections.
A word of caution is necessary concerning the gauge invariance of the
amplitude obtained in such a model. In \cite{huett96} it was shown that the
modifications of electromagnetic polarizabilities are, indeed, gauge
invariant quantities. However, in higher order with respect to $\omega
^2$ terms violating gauge invariance appear, which by definition have to
cancel with some corresponding terms in the resonance part of the
amplitude. Such cancellations either can be incorporated implicitly in
the phenomenology of the resonance amplitude or may give rise to
additional photonuclear sum rules.

As pointed out in the last section all dependence of the
correlator on the momentum transfer is given by the form factor $F_2(q)$.
So, putting $\vq=0$ in Eq.\ (\ref{s2-9}) we represent
without change of notation the correlator ${\cal F}^{ij}$
in the following form:
\begin{equation}
	{\cal F}^{ij}={\cal F}_C\;\delta ^{ij}+{\cal F}_T\;t^{ij}\,,\quad
	t^{ij}={{3Q^iQ^j} \over {Q^2}}-\delta ^{ij}  .
	\label{s2-10}
\end{equation}
The diagrams, which correspond to the two-body part of the correlator
correction, are shown in Fig.\ \ref{pr23}.
Explicit evaluation of these diagrams leads to the following result:
\begin{eqnarray}
        {\cal F}_{(1)}^{ij}&=&-{{4Mf^2} \over {m^2_\pi }}V
   \int {{{d\vp_1d\vp_2}
	\over {(2\pi )^6}}}\;\left\{ {{{4Q^iQ^j}
        \over {\vQ^2+m^2_\pi }}+{{2r^ir^j-\vr\,^2\delta ^{ij}}
        \over {\vr\,^2+m^2_\pi }}} \right\} \times
	\nonumber \\
        & &{{n(\vp_1)n(\vp_2)\left[ {1-n(\vp_2+\vQ)}
        \right]\left[ {1-n(\vp_1-\vQ)} \right]\,} \over {\vQ\cdot
        \vr}} ,
	\label{a4}
\end{eqnarray}
where $\vr = \vp_1-\vp_2-\vQ$
and $V$ is the nuclear volume.
The function $n(\vp) = \theta (p_F - |\vp\,|)$ is the
occupation number in the Fermi gas model.

\begin{figure}[ht]
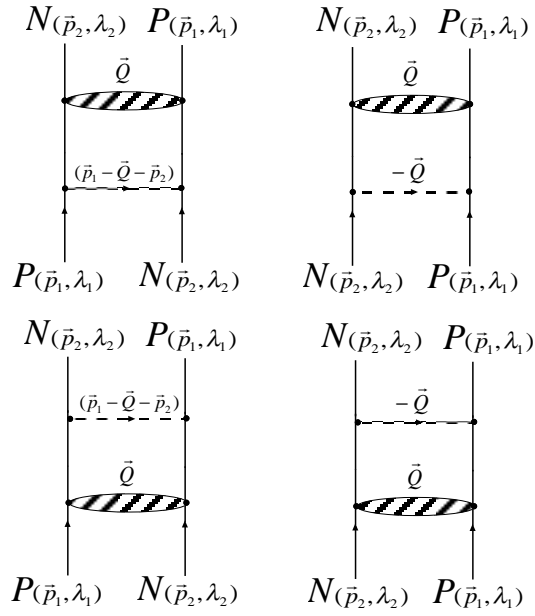

\caption{\it Set of diagrams,
     from which the two-body correction to the correlator
     is extracted. Here the same symbolic abbreviation is used as in Fig.\
     \ref{pr22}. The nucleon spin projections are denoted by $\lambda _i$.}
\label{pr23}
\end{figure}

\begin{figure}[ht]
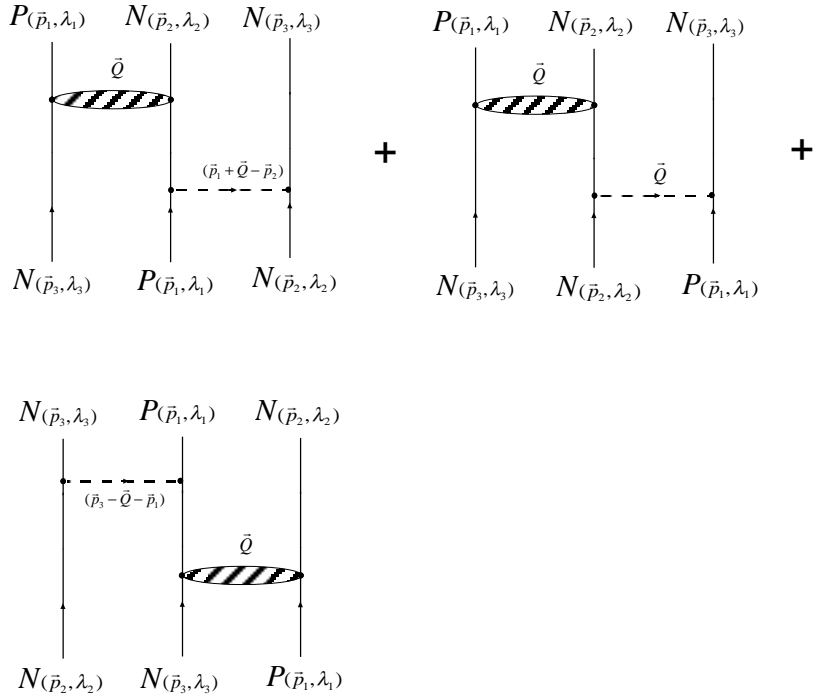

\caption{\it Three-body diagrams yielding an additional correction
to the correlator, which is of the same order in $f/m_{\pi}$ as the
two-body diagrams shown in Fig.\ \ref{pr23}. The notations are the
same as in Fig.\ \ref{pr22} and  Fig.\ \ref{pr23}.}
\label{pr24}
\end{figure}

It is interesting to note that the three-body diagrams displayed
in Fig.\ \ref{pr24} also give a significant contribution.
The corresponding general
expression for the three-body contribution to the correlator reads
\begin{eqnarray}
        {\cal F}_{(2)}^{ij}&=&{{16Mf^2} \over {m^2_\pi }}V
 \int {{{d\vp_1d\vp_2}
	\over {(2\pi )^6}}}\;\left\{ {{{2Q^iQ^j}
        \over {\vQ^2+m^2_\pi }}+{{2r^ir^j-\vr\,^2\delta ^{ij}}
        \over {\vr\,^2+m^2_\pi }}} \right\}
\times
	\nonumber \\
        & &{{n(\vp_1)n(\vp_2)n(\vp_1+\vQ)\,}
        \over {\vQ\cdot \vr}} .
	\label{threebody}
\end{eqnarray}
The result of the explicit analytical integration for Eqs.\,(\ref{threebody})
and (\ref{a4}) can be found in \cite{huett96}.
As before, the correction (\ref{threebody}) contributes to the central
part of the correlator, as well as to its tensor part.
Thus, at this stage of our discussion the correlation function
${\cal F}^{ij}$ has the following form:
\begin{equation}
	{\cal F}^{ij}={\cal F}_C^{(0)}\delta ^{ij}+
	{\cal F}_{(1)}^{ij}+{\cal F}_{(2)}^{ij}  .
	\label{fullcorr}
\end{equation}

Note that central and tensor parts of the corrections can be extracted
by using the definitions in (\ref{s2-10}).
In Fig.\ \ref{pr25} the full central part of the correlator
as a function of $Q/p_F$ is shown, together with the
zeroth order approximation ${\cal F}_C^{(0)}$ and the one, where only two-body
correlations have been taken into account.
It can be seen that almost over the whole range of $Q$ the
asymptotic form of the central correlator is modified,
although the overall shape remains the same.
Figure \ref{pr26} shows the tensor correlator with and without three-body
effects.
Here a strong modification due to three-body correlations occurs at
comparatively low $Q$, where we found a strong damping that is in fact
essential for obtaining reasonable physical results.
One can see from Fig.\ \ref{pr26} that the contribution of the high-$Q$
region is not negligible. It is known that taking into account only the
one-pion exchange leads to a wrong behavior in the tensor part
of the nucleon-nucleon potential at small
distances (i.e.\ at large $Q$) (see {\it e.g.} \cite{EricsonWeise88}).
Thus, it is necessary to include $\rho $-meson exchange in the
calculation of the tensor part of the correlator.
This can be done \cite{castel,BR80} by replacing ${\cal F}_T$ by:
\begin{equation}
   \tilde {\cal F}_T={\cal F}_T\,\left[ {1-2\,{{q^2+m^2_\pi  / p_F^2}
   \over {q^2+m_\rho ^2 / p_F^2}}} \right]  ,
   	\label{s3-5}
\end{equation}
where $m_\rho $ is the $\rho $-meson mass.
This inclusion of $\rho $-meson exchange improves essentially the behaviour
of the correlator at $Q\sim p_F$ and higher ({\it cf.} Fig.\ \ref{pr27}).

\begin{figure}[ht]
\caption{\it Central part of the correlator with three-body
corrections (full curve), without
	    three-body corrections (dotted curve) and Fermi correlator
	    without any corrections (dashed curve).}
\label{pr25}
\end{figure}

\begin{figure}[ht]
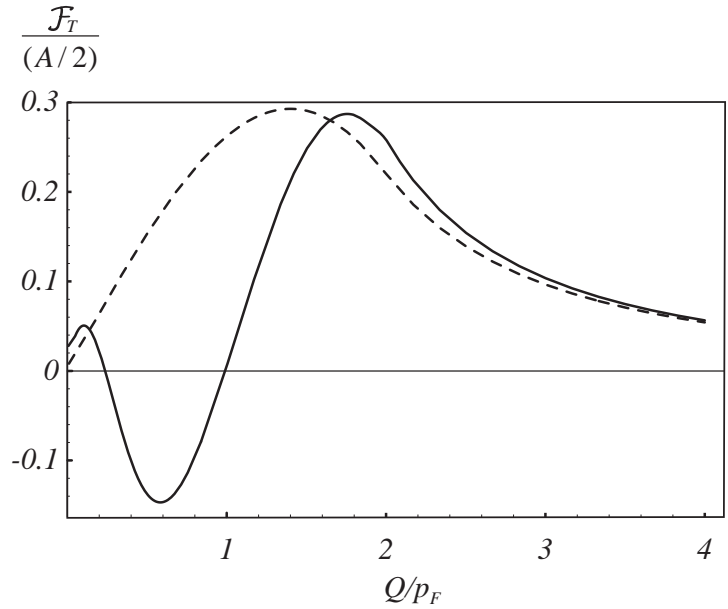

\caption{\it Tensor part of the correlator with three-body
corrections (full curve) and without
	    three-body corrections (dashed curve).}
\label{pr26}
\end{figure}

\begin{figure}[ht]
\caption{\it Comparison of the tensor part of the correlator with
(full curve) and without (dashed curve) the $\rho
	    $-meson contribution.}
\label{pr27}
\end{figure}

Even below pion threshold the contribution of a virtual $\Delta $-isobar
excitation to the mesonic part of the seagull amplitude is not negligible.
General properties of such an effect were discussed in
\cite{AR80}. In \cite{huett96} this contribution was investigated
quantitatively using the same diagrammatic approach as for the
amplitude $T_{(\pi )}^{ij}$.
In the static limit the corresponding Hamiltonians, which determine the
interaction, are of the form  (see {\it e.g.} \cite{EricsonWeise88})
\begin{equation}
        H_{\gamma N\Delta }= - {e f_{\gamma N\Delta } \over m_\pi }
        \,{\bf S}^+ \cdot ( \vecnab \times \vA) ~T_3^+
	\label{s5-1}
\end{equation}
and
\begin{equation}
        H_{\pi N\Delta }= - { f_\Delta \over m_\pi }
      \,({\bf S}^+\cdot \vecnab) ({\bf T}^+ \cdot\vephi),
	\label{s5-2}
\end{equation}
with the hermitian conjugate to be added in both cases. Here
${\bf S}$ and $\bf T$ are the 1/2-to-3/2 transition operators in
spin space and isospin space, respectively.
Evaluating the corresponding diagrams on this basis, one
obtains the following expression for the isobar contribution to the
tensor $T^{ij}$ entering into Eq.\ (\ref{s2-8}):
\begin{equation}
	T_{(\Delta )}^{ij}=\eta \,\left[ {{{h_1^il_2^j+l_1^ih_2^j}
        \over {D_1}}+{{Q_1^i} \over {d_1}}\,
 \left[ {\veceps_1\times (\vk_2\times
        \veceps_2)-\,\veceps_2\times (\vk_1\times
	\veceps_1)} \right]^j} \right]   ,
	\label{deltatij}
\end{equation}
with
\begin{displaymath}
     {\bf h}_a=\veceps_a-{{2(\vQ_a\cdot \veceps_a)\vQ_a}
     \over {d_a}}\;,\quad {\bf l}_a=(\vQ+\vK)\times (\vk_a\times
     \veceps_a)
\end{displaymath}
and the coefficient
\begin{displaymath}
     \eta =-{{8\Omega e^2f_\Delta f_{\gamma N\Delta }f}
     \over {9(\omega ^2-\Omega ^2)m_\pi ^3 }}   .
\end{displaymath}
In (\ref{deltatij}) the same abbreviations have been used as in
Eq.\ (\ref{b1}) and, in addition, $\Omega = M_{\Delta } - M$ is the
mass difference between the $\Delta $-isobar and the nucleon.
Also, the symmetry with respect to the substitution
$\vQ \rightarrow -\vQ$, which is due to the integration in Eq.\
(\ref{s2-8}), has been used to bring $T^{ij}_{(\Delta )}$ into the form
(\ref{deltatij}).
Comparison of this mesonic seagull amplitude for nuclear matter with Eq.\
(\ref{dispseagull2}) allows us to give the numerical results in terms
of explicit
values for $\kappa $, $\delta \alpha $ and $\delta \beta $. These
numerical values are given in appendix A.

\subsection{The mesonic seagull amplitude in finite nuclei}

In order to discuss the properties of the mesonic seagull amplitude for
the case of specific nuclei rather than for nuclear matter, it is
convenient to consider the Fourier
transform of the correlation function discussed in section 9.2. Then the
function ${\cal F }^{ij}$ can be written as
\begin{eqnarray}
         & &{\cal F }^{ij}=\int {d\vx_1d\vx_2}\,e^{-i\vq
      \cdot (\vx_1+\vx_2)/2}e^{-i\vQ\cdot (\vx_1-\vx_2)}
	 \times
	\label{mwithg} \\
        & &\left[ {g_C(\vx_1-\vx_2)\;\,\delta ^{ij}+
        g_T(\vx_1-\vx_2)\;t^{ij}} \right] ,
	\nonumber
\end{eqnarray}
with $t^{ij}$ from (\ref{s2-10}).
In Eq.\ (\ref{mwithg}) the functions $g_C$ and $g_T$ describe the
central and tensor correlations of two
nucleons, while the exponential function depending on $\vq$
is responsible for the distribution of such nucleon pairs inside the
nucleus.
The functions $g_C$ and $g_T$ are related to the momentum space
correlation functions
${\cal F }_C$ and ${\cal F }_T$ via
\begin{equation}
        g_{C,T}(\rho )= \frac{1}{V} \int {{{d\vQ}
	\over {(2\pi )^3}}}\,{\cal F}_{C,T}(Q)\,e^{i\,
        \vecrho \cdot \vQ}  .
	\label{gct}
\end{equation}
Note that in the functions ${\cal F}_{C}$ and ${\cal F}_{T}$ the
contribution from $\rho $-meson exchange has been taken into account.
In the case of a pure Fermi gas model we have
\begin{equation}
	g_C^{(0)}(\rho )=-2\left( {{{p_F^3}
	\over {6\pi ^3}}} \right)^2 F^2(p_F\rho )\;,\quad
	g_T^{(0)}(\rho )=0
	\label{g0f}
\end{equation}
with the function $F$ from Eq.\ (\ref{formfac}).
Next, we expand $T^{ij}_{(\pi )}$ in Eq.\ (\ref{s2-8})
with respect to $\vk_1$ and $\vk_2$ up to
$O(\omega ^2)$, pass to the variables $\vecrho = \vx_2 - \vx_1$
and  $\vecxi = (\vx_1 + \vx_2)/2$.
Then, taking the integral with respect to $\vQ$ and the angles
of $\vecrho $ and $\vecxi $ we obtain the contribution of
$T^{ij}_{(\pi )}$ to the mesonic seagull amplitude:
\begin{eqnarray}
	S_{(\pi )}&=&{{Ae^2} \over {4M}}\,\left\{ {\matrix{{}\cr
{}\cr
}} \right.\Phi _1(q )\,\veceps _1\cdot \veceps _2+
{{\omega ^2} \over {m^2_\pi }}\,\Phi _2(q )\,\veceps _1\cdot
\veceps _2+
	\nonumber \\
	& &{1 \over {m^2_\pi }}\Phi _3(q )\,(\veceps _1\times
        \vk_1)\cdot (\veceps _2\times \vk_2)\left. {\matrix{{}\cr
{}\cr
}} \right\}  ,
	\label{mphi}
\end{eqnarray}
where
\begin{eqnarray}
     & &\Phi _i={{2Mf^2} \over {3m_\pi \pi ^2}}(2Rp_F)^3
	\int\limits_0^1 {dx}\,\left[ {G_i^C(\rho _1)\,\tilde g_C(\rho _2)+}
	\right.\left. {G_i^T(\rho _1)\,\tilde g_T(\rho _2)} \right]\times
	\nonumber \\
        & &x^2 e^{-\rho _1}\left( {\int\limits_0^{1-x}
  {d\xi }\;\xi \,{{\sin \xi Rq }
	\over {Rq }}+\int\limits_{1-x}^{\sqrt {1-x^2}}
	{d\xi }\;{{\sin \xi Rq } \over {Rq }}\;{{1-x^2-\xi ^2}
	\over {2x}}} \right)  .
	\label{phi}
\end{eqnarray}
In Eq.\ (\ref{phi}) the following abbreviations have been used:
\begin{displaymath}
	\tilde g_{C,T}(\rho _2 )={1 \over 2}\left( {{{6\pi ^2}
	\over {p_F^3}}} \right)^2
    g_{C,T}(\rho _2/p_F )\;,\quad \rho _1=2Rm_\pi x\;,\quad
	\rho _2=2Rp_Fx\quad .
\end{displaymath}
The functions $G_i^{C,T}$ are of the form
\begin{eqnarray}
	 &  & G_1^C(\rho _1 )=\rho _1 \;,\quad G_1^T(\rho _1 )=
	 {{2\rho _1 ^2-12} \over \rho _1 }\;,
	\nonumber \\
	 &  & G_2^C(\rho _1 )={{150+30\rho _1 -\rho _1 ^3} \over {60}}\;,
	 \quad G_2^T(\rho _1 )=
	 {{24+12\rho _1 -\rho _1 ^3} \over {30}}
	\nonumber \\
	 &  & G_3^C(\rho _1 )={{3-21\rho _1 +2\rho _1 ^2}
         \over {12}}\;,\quad G_3^T(\rho _1 )=
  {{2\rho _1 ^2-3\rho _1 -15} \over 6}
	\nonumber
\end{eqnarray}
The integral with respect to $\xi $
in (\ref{phi}) can easily be taken analytically, but we represent the
result in this form for the sake of brevity.
By comparing Eq.\ (\ref{mphi}) with the corresponding terms in
Eq.\ (\ref{dispseagull2}) one sees that
the parameters appearing in the mesonic seagull amplitude
are given by the functions $\Phi _i(q )$ at $q $=0:
\begin{equation}
	\kappa =-\Phi _1(0)\;,\quad \delta \alpha ={{e^2}
	\over {4Mm^2_\pi }}\Phi _2(0)\;,\quad \delta \beta ={{e^2}
	\over {4Mm^2_\pi }}\Phi _3(0)  .
	\label{phinorm}
\end{equation}
It is evident from Eq.\ (\ref{phi}) that three different form factors
\begin{equation}
	F_2^{(i)}(q )={{\Phi _i(q )}
	\over {\Phi _i(0)}}
	\label{formphi}
\end{equation}
appear instead of only $F_2$.

In order to account for the contribution of $\rho $-meson exchange
to $T^{ij}$ one may follow the prescription of \cite{BR80} and
make the substitution
$f^2\rightarrow 2\tilde f^2_\rho$ for the central part of each
quantity and $f^2\rightarrow -\tilde f^2_\rho$ for the tensor
part. The pion mass $m$ is substituted in all cases by the
$\rho $-meson mass $m_\rho $.

In the case of the magnetic polarizability $\delta \beta $ the contribution
of the $\Delta $-isobar excitation to the mesonic seagull amplitude
should also be taken into account \cite{AR80,huett96}.
In our notation this corresponds to an additional contribution
to the function $\Phi _3$, which is of the following form:
\begin{eqnarray}
	& &\delta \Phi _3={{8Mf_\Delta f_{\gamma N\Delta}f(2Rp_F)^3}
	\over {81(M_\Delta -M)\pi ^2}}
        \int\limits_0^1 {dx}\,\left[ {G_\Delta ^C(\rho _1)\,\tilde
    g_C(\rho _2)+} \right.\left.
  {G_\Delta ^T(\rho _1)\,\tilde g_T(\rho _2)} \right]\times
	\nonumber \\
	& &x^2 e^{-\rho _1}\left( {\int\limits_0^{1-x}
	{d\xi }\;\xi \,{{\sin \xi Rq }
	\over {Rq }}+\int\limits_{1-x}^{\sqrt {1-x^2}}
	{d\xi }\;{{\sin \xi Rq } \over {Rq }}\;{{1-x^2-\xi ^2}
	\over {2x}}} \right)  ,
	\label{phidelta}
\end{eqnarray}
where again $M_\Delta $ is the $\Delta $-isobar mass and
\begin{displaymath}
     G_\Delta ^C(\rho _1)={{6-12\rho _1}
	\over {\rho _1}}\;,\quad G_\Delta ^T(\rho _1)={{12-6\rho _1}
	\over {\rho _1}} .
\end{displaymath}
The coupling constants appearing in Eq.\ (\ref{phidelta}) are taken to
be $f_\Delta = 2f$ and $f_{\gamma N\Delta}=0.35$.

Up to now we have considered a constant
nucleon density $n_0=p_F^3/3\pi ^2$
inside the nucleus, which is normalized as $n_0 V=Z$.
Using a local-density approximation we will extend our
consideration to realistic nuclear densities $n(r)$.
With the help of the usual plane-wave expansion via Legendre polynomials
$P_l(x)$ we obtain the realistic-density ($rd$) form
\begin{eqnarray}
	& &\Phi _i^{(rd)}={{64\pi Mf^2}
     \over {3m_\pi Z}}\,\int\limits_0^\infty
      {dx}\,x^2e^{-2m_\pi x}\int\limits_x^\infty
	{dr}\,r^2n^2(r)\,\;\times
	\label{phireal} \\
     & &\left[ {G_i^C(2xm_\pi)\,\tilde g_C(2xp(r))+
     G_i^T(2xm_\pi)\,\tilde g_T(2xp(r))} \right]\;\times
	\nonumber \\
	& &\sum\limits_{l=0}^\infty  {j_l(r\Delta )\,j_l(x\Delta )
	\left( {P_{l-1}(x/r)-P_{l+1}(x/r)} \right)}  ,
	\nonumber
\end{eqnarray}
where $p(r)=(3\pi ^2 n(r))^{1/3}$ is the local Fermi momentum.
In numerical calculations a three-parameter Fermi parameterization
of the densities $n(r)$ has been used with values
for the different nuclei taken from \cite{datatab}.

Equations (\ref{mphi})--(\ref{phireal}) form
the starting point of our numerical investigation of the mesonic seagull
amplitude.
Note that due to the use of either the Fermi gas model or a local
density approximation the accuracy of our results decreases with
decreasing $Z$.

First we discuss
the numerical results for the different contributions to the enhancement
constant $\kappa $.
Note that, as it was argued in \cite{arima73} and is also
discussed in \cite{EricsonWeise88},
the main contribution to $\kappa ^{GR}$ comes from
central correlations, while  $\kappa ^{QD}$ is mainly determined by
tensor correlations. Obviously, it is impossible to give a precise
separation of $\kappa $ into a GR part and a QD part, but this
approximation seems reasonable, as the GR
contribution should be associated with a small momentum transfer between
the two nucleons in comparison with the QD part.
In the pure Fermi gas model, where $g_C$= $g_C^{(0)}$ and
$g_T$=0, the contribution to $\kappa $ from pion exchange $\kappa ^\pi$
is approximately equal to the $\rho $-meson contribution $\kappa ^\rho $.
For nuclear matter (infinite nuclear radius) one has
$\kappa ^\pi=\kappa ^\rho$=0.2, which is in agreement with a variety of
model calculations, {\it e.g.} \cite{BR80,arima73,brown4}.
For finite nuclei, again
in the pure Fermi gas, the value for  $\kappa ^\pi$ decreases slightly
with decreasing $Z$, whereas $\kappa ^\rho$ remains the same.
With inclusion of the full correlation functions $g_C$ and $g_T$
the situation for $\kappa $ changes drastically.
Now the main contribution to $\kappa $ comes from tensor correlations
related to pion exchange. The pionic central contribution is still
of the same order as before, while $\kappa _C^\rho $ becomes
negligible. The only significant contribution from $\rho $-meson
exchange is now due to tensor correlations and has a negative value.
All these relations between the different ingredients to $\kappa $
remain valid, when realistic nuclear densities are considered.
In Fig.\ \ref{pr39} the different contributions to $\kappa $ are shown
as a function of $Z$ for the modified Fermi gas model.
In addition, the realistic-density result $\kappa ^{(rd)}$,
which is obtained from
Eq.\ (\ref{phireal}), is shown in the same figure.

\begin{figure}[ht]
\caption{\it Dependence of enhancement constant $\kappa $ on proton number $Z$.
The dashed curve corresponds to the pionic tensor contribution
$\kappa _T^\pi $, the dash-dotted curve includes also
the central contribution $\kappa _C^\pi $ and the dotted
curve gives the total $\kappa $, including the
contribution from $\rho $-meson exchange.
The realistic-density result $\kappa ^{(rd)}$ for the full
enhancement constant  ({\it cf.} Eq.\ (\ref{phireal}))
is shown as a full curve.}
\label{pr39}
\end{figure}

In the case of electric and magnetic polarizabilities
$\delta \alpha $ and $\delta \beta $ the contributions from
$\rho $-meson exchange are suppressed by a factor of $m^2_\pi /m_\rho ^2$
in comparison with the pion contributions and, therefore, are negligible.
The values of $\delta \alpha $ and $\delta \beta $ are determined
mainly by pionic central correlations, as can be seen in Figs.\
\ref{pr40} and \ref{pr41}.
In the case of $\delta \beta $ the inclusion of
the $\Delta $-isobar intermediate state produces a noticeable effect
({\it cf.} Fig.\ \ref{pr41}).
In this contribution the values due to central and tensor
correlations are of the same order.
Note that $\kappa $, $\delta \alpha $ and
$\delta \beta $ get close to their asymptotic (nuclear matter)
values calculated in \cite{huett96} only at extremely high $Z$.
The size of both, $\delta \alpha $ and $\delta \beta $ becomes
noticeably smaller, when a realistic density is taken into account.
This effect gains importance with decreasing $Z$.
The ratio of central and tensor contributions to $\delta \alpha $
is approximately the same for a realistic density as in a modified
Fermi gas model. For $\delta \beta $ the influence of tensor
correlations in the realistic-density case is slightly stronger than
for homogeneous nuclear density.

We consider now the dependence of the mesonic seagull amplitude
$S$ on momentum transfer $q $, which is determined
by the form factors $F_2^{(i)}(q )$ ({\it cf.} Eq.\ (\ref{formphi})).
Here these form factors are given for $^{40}$Ca. A wider range in $A$ is
covered in Appendix B, where explicit results are given for
$^{208}$Pb, $^{16}$O and $^{12}$C.
The results from \cite{huett97}
indicate that $F_2^{(2)}$ for the term proportional to
$\delta \alpha $ and $F_2^{(3)}$ (for $\delta \beta $) are
equal with high accuracy, but differ significantly from
the form factor $F_2^{(1)}$ for the
term containing $\kappa $.
All three functions  $F_2^{(i)}$ differ noticeably from $F_1$.
Figure \ref{pr13} shows the
corresponding curves for $^{40}$Ca,
for the case of a realistic density.
One can see that the frequently used phenomenological approximation
$F_2(q )=F_1^2(q /2)$, which is also shown in
Fig.\ \ref{pr13}, is not in agreement
with the $q $-dependence
of the amplitude $S$ obtained here.
For very small $q $ it is convenient to represent the form
factors as $F_2^{(i)} = 1 - q ^2 r^2_i/6$.
Then for calcium we find $r_1$=3.0\,fm and $r_2$=2.5\,fm. For
other nuclei the corresponding numbers are  given in Appendix B.
In the case of $^{208}$Pb, the result from \cite{A82} coincides
within good accuracy with the corresponding form factor $F_2^{(2)}$
from \cite{huett97}.

\begin{figure}[ht]
\caption{\it Pion-exchange contribution to electric polarizability
	    $\delta \alpha $ as a function of $Z$.
	    The dashed curve corresponds to the central contribution
	    $\delta \alpha _C$ and the dotted
	    curve gives the total $\delta \alpha
            =\delta \alpha _C + \delta \alpha _T$.
	    The use of a realistic density leads to the full curve.}
\label{pr40}
\end{figure}

\begin{figure}[ht]
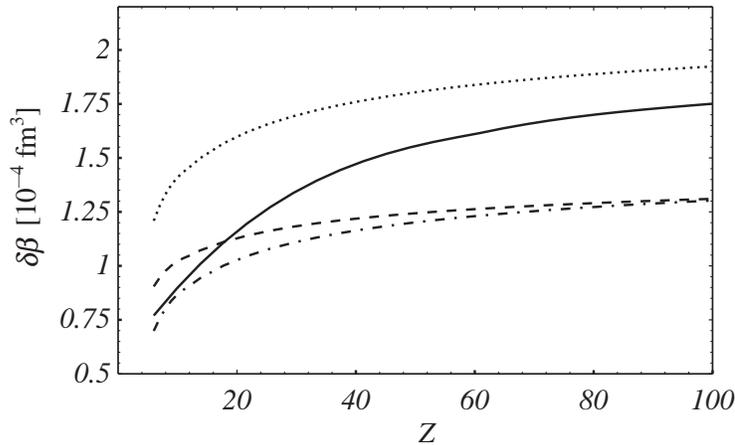

\caption{\it Pion-exchange contribution to magnetic polarizability
	    $\delta \beta $ as a function of $Z$.
	    The dashed curve corresponds to the central contribution
	    $\delta \beta _C$ and the dash-dotted
            curve gives the sum $\delta \beta _C + \delta \beta _T$.
	    Adding the contribution of the $\Delta $-isobar excitation
            as given in Eq.\ (\ref{phidelta}) leads to the total
	    value of $\delta \beta $ given as the dotted curve.
	    The use of a realistic density leads to the full curve.}
\label{pr41}
\end{figure}

\begin{figure}[ht]
\caption{\it Form factors $F_2^{(i)}(\Delta )$ for $^{40}$Ca. The dashed
	    curve is $F_2^{(1)}$ and the full curve is $F_2^{(2)}$. For
	    comparison the (experimental) charge form factor $F_1$ is
	    also shown (dash-dotted curve), as well as the
	    function $F_1^2(\Delta /2)$ (dotted curve).}
\label{pr13}
\end{figure}

\clearpage

\section{Retardation effects in Compton scattering. A model study with
a relativistic oscillator}

\label{Section-osc}

\subsection{Motivation and aims}

In many earlier works \cite{D3,W5,ziegl85,martin1}, a simple ansatz
for the Compton scattering amplitude through the GR region
was widely used. It is formulated in terms of the lowest partial-wave
amplitudes,
\begin{equation}
  R_{GR}(\omega,\theta) = \sum_{\lambda L= E1,M1,E2}
    R_{GR}^{\lambda L}(\omega) g^{\lambda L}(\theta),
\end{equation}
whose imaginary parts, $\mbox{Im} R_{GR}^{\lambda L}(\omega) =
(\omega/4\pi) \sigma_{GR}^{\lambda L}(\omega)$, exhaust
photoabsorption in the GR region.  The real parts of these partial
amplitudes are assumed to be given by the dispersion integrals
\begin{equation}
\label{naive-DR}
  \mbox{Re} R_{GR}^{\lambda L}(\omega) = R_{GR}^{E1}(0)
   \delta_{\lambda L,\,E1}
   + \frac{\omega^2}{2\pi^2} {\cal P} \int_0^\infty
     \frac{\sigma_{GR}^{\lambda L}(\omega')\,d\omega'}
     {\omega'^2-\omega^2},
\end{equation}
where
\begin{equation}
  R_{ GR}^{E1}(0)  = \frac{1}{2\pi^2} \int_0^\infty
     \sigma_{GR}(\omega)\,d\omega.
\end{equation}
This ansatz was constructed in a way that ensured the fulfillment of
the Gell-Mann--Goldberger--Thirring dispersion relation in forward
direction, when all $g^{\lambda L}=1$:
\begin{equation}
  R_{GR}(\omega,0) =  \frac{1}{2\pi^2} \int_0^\infty
   \frac{\omega'^2}{\omega'^2-\omega^2-i0}
    \sigma_{GR}(\omega')\,d\omega'.
\end{equation}
In particular, when $\omega\to\infty$, the amplitude $R_{GR}
(\omega,0)$ vanishes, and the total Compton scattering amplitude
$T_{GR}$ is determined by the seagull contribution $S^{GR}$.

This nice feature is destroyed by the $E2$ and $M1$
contributions when the scattering angle $\theta$ is not zero.  For
example, when $\theta=\pi$, the asymptotically nonvanishing part
of the resonance amplitude reads
\begin{equation}
        R_{GR} (\infty,\pi) =  \frac{1}{\pi^2}
          \int \left( \sigma_{GR}^{E2}(\omega)+
     \sigma_{GR}^{M1}(\omega) \right)\,d\omega \ne 0 .
\end{equation}
However, the vanishing of $R_{GR}$ for $\omega=\infty$ and for
all angles is physically necessary. Such a vanishing was used in
section 8.2  to analyze the seagull contribution emerging from the
fixed-$t$ dispersion relations and to derive a set of related sum
rules.

Phenomenologically, the dispersion relations (\ref{naive-DR}) for the
partial amplitudes could be corrected through introducing special
$\omega$-dependent form factors $F(\omega)$
\cite{SC94,baron91} which erase the unphysical asymptotic
pieces.  Though well motivated by retardation effects calculated
in the frame of nuclear models \cite{baron91}, 
such a procedure may however need a further theoretical
justification because the form factors $F(\omega)$ generally destroy
the asymptotic behavior of the amplitude at high complex energies and
violate the validity of dispersion relations.  Only taken consistently
by taking into account the 
contributions of higher multipoles and relativistic
corrections, they lead to a correct result.  Accordingly, the present
section is aimed to illustrate an interplay between the retardation
effects, higher multipoles, relativistic corrections, asymptotic
behavior and the dispersion relations by using a model which contains
the Compton scattering amplitude with all correct analytical
properties.  Such a model has to be relativistic, and therefore we
consider a relativistic harmonic oscillator, which allows an analytical
treatment of this problem.
In addition to its theoretical transparency, this model has the advantage
of having a very close relation to the physical problem we are
considering here, {\it i.e.} to giant resonances.
The $1\hbar\omega_0$ transition from the ground state to the
$n=1$ oscillator state corresponds to the electric giant-dipole
resonance, the $2\hbar\omega_0$ transition to the $d$-component
of the $n=2$ oscillator states corresponds to the electric giant-quadrupole
resonance, {\it etc}.
This interpretation shows that each electric giant-multipole resonance
is accompanied by higher harmonics which are closely related to the
retardation problem. Magnetic giant resonances are not taken into
consideration.

In the following subsections we explicitly formulate the model. We find
an analytical representation of
the Compton scattering amplitude\footnote{We drop here and in the following
the subscript GR}
$T=S+R$, study the
high-energy behavior of $T$ and show that the amplitude satisfies the
fixed-$t$ dispersion relation. We discuss why relativistic effects are
important for the validity of the dispersion relation.  Finally, we
study the retardation effects on the amplitude at low and high energies
and on the $E1$ and $E2$ sum rules and discuss how the retardation form
factors can appear in the Compton scattering amplitude and in the
dispersion relations.

\subsection{Basics of the model. The seagull and resonance amplitudes}

We consider a scalar particle with the electric charge $e$ described by
the Klein-Gordan equation with an oscillator Lorentz-scalar potential,
\begin{equation}
\label{KG}
  \Big[\Big( i\frac{\partial}{\partial t} - eA_0\Big)^2
   - (i\vecnab + e\vA)^2 - (\mu^2+\gamma^4r^2)\Big]
  \psi(\vr,t)=0.
\end{equation}
When the electromagnetic potential $A_\mu$ is absent, the normalized
solutions $\psi_{n\alpha}^\pm$ of the above Klein-Gordan equation with
positive and negative energies read
\begin{equation}
\label{psi_n}
  \psi_{n\alpha}^\pm(\vr,t) = \frac 1{\sqrt{2E_n}}
    \exp(\mp iE_n t)\phi_{n\alpha}(\vr),
\end{equation}
where $n=2n_r+j$ is the quantum number which determines the energies
$\pm E_n$ of degenerated levels with different radial quantum numbers
$n_r$ and angular momenta $j$, and the generic parameter $\alpha=(jm)$
describing the angular momentum $j$ and its projection $m$. The eigen
functions $\phi_{n\alpha}$ satisfy the equation
\begin{equation}
\label{phi_n}
  -\vecnab^2 \phi_{n\alpha} (\vr)
  + \gamma^4 r^2 \phi_{n\alpha}(\vr) =
  (E_n^2 - \mu^2)\phi_{n\alpha}(\vr),
  \qquad \int |\phi_{n\alpha}(\vr)|^2 d\vr =1,
\end{equation}
and the eigen energies $E_n$ are
\begin{equation}
  E_n = \sqrt{\mu^2 + (2n+3)\gamma^2}
      = \sqrt{E_0^2 + 2n\gamma^2}, \quad n=0,1,2,\ldots .
\end{equation}

The wave functions and the energy spectrum are characterized by two
parameters, $\gamma$ and $E_0$.  The parameter $\gamma$ determines the
scale for the wave function of the ground state,
\begin{equation}
   \phi_0(r)=N_0 \exp\left(-\frac12\gamma^2 r^2\right),
   \quad  |N_0|^2=  \gamma^3 \pi^{-3/2},
\end{equation}
the corresponding r.m.s.\ electric radius,
$\langle r^2_E \rangle ^{1/2}= \sqrt{\frac32} \gamma^{-1}$,
and the form factor of the charge distribution in the ground state,
\begin{equation}
\label{FF}
  F(q)= \int \exp(i\vq\cdot\vr) \, |\phi_0(\vr)|^2 d\vr
    =\exp\left( -\frac{q^2}{4\gamma^2}\right).
\end{equation}
The parameter $2E_0$ determines the energy gap between positive and
negative parts of the spectrum.  In the nonrelativistic limit, $E_0$
becomes the mass of the particle, and the quantity
\begin{equation}
  \omega_0 \equiv \frac{\gamma^2}{E_0}
\end{equation}
becomes the oscillator frequency.  The scale of relativistic
effects is determined by the dimensionless ratio
\begin{equation}
\label{damping}
  \eta = \frac{\gamma^2}{E_0^2} = \frac{\omega_0}{E_0}
    = \frac{\omega_0^2}{\gamma^2} , \quad 0 < \eta < \infty.
\end{equation}
In the nonrelativistic regime, $\eta\ll 1$.  When $\eta\simge 1$, the
energies which are necessary to excite the oscillator or create a
particle-antiparticle pair become of the same order.


\begin{figure}[htb]
\caption{\it Diagrams of Compton scattering off the relativistic
oscillator (crossed terms are not shown)}
\label{fig-osc}
\end{figure}

The total amplitude of photon scattering on the particle confined in
the potential wall consists of two parts.  The first one, $T_D$, is the
so-called Delbr\"uck scattering amplitude which describes scattering
by the empty potential wall via virtual particle-antiparticle pairs
(see the recent review \cite{milstein94} and references therein).  It is
usually small except for a region of near-forward scattering angles.  For
the case of the oscillator potential, the Delbr\"uck amplitude was
calculated in \cite{LM94}.  The second part $T$, which solely will be
discussed in the following, is the amplitude for Compton scattering by the
particle itself.  It is given by diagrams of the noncovariant
perturbation theory which are shown in Fig.~\ref{fig-osc}.
There, the
diagram $a$ gives the seagull contribution $S$, and the diagrams $b$
and $c$ correspond to the resonance part $R$
of the amplitude $T=S+R$.  The seagull contribution
reads
\begin{equation}
\label{S-ampl}
  S= -\frac{e^2}{E_0} (\veceps_1\cdot\veceps_2)
   \langle \phi_0 | \exp(i\vq\cdot\vr) | \phi_0 \rangle
   = -\frac{e^2}{E_0} (\veceps_1\cdot\veceps_2) F(q)
\end{equation}
with $\vq=\vk_1-\vk_2$, whereas the resonance amplitude $R$
has the form
\begin{eqnarray}
\label{R-ampl}
   R &=& \sum_{n\alpha}  \frac{e^2}{E_0 E_n}
  \frac{\langle \phi_0 |
    (\veceps_2\cdot\vp)\exp(-i\vk_2\cdot\vr)
   | \phi_{n\alpha} \rangle   \langle \phi_{n\alpha} |
    (\veceps_1\cdot\vp) \exp(i\vk_1\cdot\vr)
   | \phi_0 \rangle}    {E_n-\omega-E_0 - i0} +
\nonumber \\  &&
        \sum_{n\alpha} \frac{e^2}{E_0 E_n}
  \frac{\langle \phi_0 |
    (\veceps_2\cdot\vp)\exp(-i\vk_2\cdot\vr)
   | \phi_{n\alpha} \rangle   \langle \phi_{n\alpha} |
    (\veceps_1\cdot\vp) \exp(i\vk_1\cdot\vr)
   | \phi_0 \rangle}    {E_n+\omega+E_0 - i0}
  + \nonumber \\ && \hspace{10em}
  (\omega \to -\omega,~ \vk_1 \leftrightarrow -\vk_2,~
    \veceps_1 \leftrightarrow \veceps_2 ).
\end{eqnarray}
The factors $1/E_0$ in (\ref{S-ampl}) and $1/E_0E_n$ in (\ref{R-ampl})
are related with the proper normalization of the wave functions, {\it cf.}\
Eq.~(\ref{psi_n}).  The first sum in (\ref{R-ampl}) corresponds to the
contribution of positive-energy intermediate states (the diagram $b$),
whereas the second one accounts for negative-energy states, or virtual
pair production (the diagram $c$).%
\footnote{In terms of physical states having positive energies and
consisting of particles and antiparticles (i.e.\ $p$ and $\bar p$), 
the two sums
in (\ref{R-ampl}) have a different meaning.  The first sum is related
with transitions of the particle $p$ from the ground state to an excited
state ($p_0 \leftrightarrow p_{n\alpha}$).  The second sum is related
with pair production from the vacuum.  Most of the pair production is
included through the Delbr\"uck scattering amplitude $T_D$ which takes
into account all possible transitions $vac \leftrightarrow p_i \bar p_j$.
However, due to Bose correlations, the presence of the particle $p_0$
in the initial state doubles the probability of pair production
to the occupied state $i=0$.
Respectively, the second sum in Eq.\ (\ref{R-ampl}) takes into account
these additional transitions $vac \leftrightarrow p_0 \bar p_j$
caused by the initial particle $p_0$.
Note that in the case of a spin-1/2 particle the second sum would change
its sign. It would describe Fermi correlations with $p_0$, i.e.\
Pauli blocking of pair production to the occupied state $p_0$.}
Taken together, they give the
resonance amplitude in the Feynman approach with the Feynman propagator
$(E_n^2-(\omega+E_0)^2 -i0)^{-1}$. The crossed term has a similar
structure and interpretation.  The $k$-dependent exponents in
(\ref{R-ampl}) describe the retardation effects. Keeping in the
operators $(\veceps\cdot\vp) \exp(\pm i\vk\cdot\vr)$ the
leading-in-$k$ terms of a given total angular momentum $j$ ({\it cf.}\
section 3.4), we can obtain unretarded contributions to different
multipole amplitudes $R^{\lambda j}$.  The unretarded amplitudes can
also be extracted from leading terms in an expansion of the total
amplitude (\ref{R-ampl}) in powers of photon momenta, $k_1$ and $k_2$,
considered at arbitrary absolute values of $k_1\ne\omega$ and $k_2\ne
\omega$. The momenta are still assumed to be orthogonal to the photon
polarizations, $\veceps_1\cdot\vk_1 = \veceps_2\cdot\vk_2 = 0$.
Since it is equally easy, in our simple model, to calculate
the amplitude $R$ for such a general case of arbitrary $k_1$ and $k_2$,
we follow this way to make the retardation effects manifestly evident.
In other words, we calculate the amplitude of virtual Compton
scattering with transversely polarized photons.

A powerful tool to calculate the resonance amplitude (\ref{R-ampl}) is
provided by the Greens function which is explicitly known for the
oscillator potential. Introducing
\begin{equation}
\label{eq:GFdef}
  G(\vr_2,\vr_1 | E) = \sum_{n\alpha} \frac
 {\langle\vr_2 |\phi_{n\alpha} \rangle
  \langle \phi_{n\alpha} |\vr_1\rangle} {E^2 - E_n^2 + i0},
\end{equation}
we rewrite the resonance amplitude like
\begin{eqnarray}
  && R=-\frac{2e^2}{E_0}\int
  \exp(i\vk_1\cdot\vr_1 -i\vk_2\cdot\vr_2)
  \Big(\veceps_2\cdot\vecnab\phi_0(\vr_2)\Big)^*
  \, G(\vr_2,\vr_1 | \omega+E_0) \times
\nonumber \\ && \hspace{10em}
  \Big(\veceps_1\cdot\vecnab\phi_0(\vr_1)\Big)
  \,d\vr_1 \,d\vr_2  + \mbox{crossed term}.
\end{eqnarray}
Substituting the Greens function for the relativistic oscillator
\cite{LM94,FH65},
\begin{eqnarray}
\label{eq:GF}
  && G(\vr_2,\vr_1 | E) = -\frac{i\gamma}2
  \int_0^\infty
  \frac{ds}{(2\pi i\sin s)^{3/2}}\times \nonumber\\ && \qquad
  \qquad \exp\left\{ \frac{is}{2\gamma^2}(E^2-\mu^2+i0)
  +\frac{i\gamma^2}{2\sin s}
  \left[(r_1^2+r_2^2)\cos s - 2\vr_1\cdot\vr_2\right] \right\}
\end{eqnarray}
and taking simple Gaussian integrals with respect to
$\vr_1$ and $\vr_2$, we finally get
\begin{equation}
\label{R-answ1}
  R=\frac{e^2 \,F(q)}{2E_0}
   \left\{g(a,x)(\veceps_1\cdot\veceps_2)
    + g(a-1,x)\frac{
   (\veceps_1\cdot\vk_2)(\veceps_2\cdot\vk_1)}
    {2\gamma^2}\right\} +(\omega\to -\omega).
\end{equation}
Here the function $g$, a close relative of the standard incomplete
gamma-function, reads
\begin{equation}
  g(a,x) = \frac{i}{e^{-2\pi ia}-1} \int_0^{2\pi}
    \exp\Big( {-ias} + x(e^{is}-1)\Big)\, ds,
\label{g-function}
\end{equation}
and the quantities $x$ and $a$ are defined as
\begin{equation}
\label{a}
   x=\frac{\vk_1\cdot\vk_2}{2\gamma^2},  \qquad
   a=\frac{\omega^2 + 2\omega E_0}{2\gamma^2} -1+i0 .
\end{equation}
In this form, Eqs.\ (\ref{S-ampl}) and (\ref{R-answ1})  for the
seagull and the resonance amplitude, respectively, are valid for any $k_1$
and $k_2$.  We have to set $k_1=k_2=\omega$ whenever we consider Compton
scattering with real photons, including the retardation effects.

For further references note that
\begin{equation}
\label{g-series}
 g(a,x) = e^{-x} \sum_{n=0}^\infty \frac{x^n}{n!(n-a)}
  = - \sum_{n=0}^\infty \frac{x^n}{a(a-1) \ldots (a-n)}
\end{equation}
and
\begin{equation}
\label{g(a-1)}
   xg(a-1,x)=ag(a,x)+1.
\end{equation}
It is seen that $g(a,x)$ is a regular function of $x$ and $a$ except for
simple poles at $a=0,1,2,\ldots$\,.  Asymptotic expansions of $g$ at low
energies (when $x\ll a\approx -1$) or at ``nonrelativistically high
energies" ({\it i.e.} when $\omega\sim \gamma \gg \omega_0$ 
and $\omega \ll E_0$, or, in other words,
when $a\gg x\sim 1$; such a regime is possible, provided $\eta \ll 1$)
directly follow from Eq.~(\ref{g-series}).  The important case of
$a\approx x \sim (a-x)^2 \to\infty$, which arises for real Compton
scattering at ultra-relativistic energies, is described by the
saddle-point asymptotics
\begin{eqnarray}
\label{g-as}
   && g(a,x) =
  -\sqrt{\frac{\pi}{2x}}
  \exp\Big(-\frac{(a-x)^2}{2x}\Big)
  {\,\rm ctg\,}(\pi a) +
 \nonumber \\ && \hspace{10em}
   \frac{i}{\sqrt x} \int_0^\infty
 \sin\Big(\frac{a-x}{\sqrt x}s\Big) \, e^{-s^2/2} \,ds + O(\frac 1x).
\end{eqnarray}
Here the sign of $\sqrt x$ must be chosen in such a way that
$\mbox{Re}\sqrt x >0$.

Given (\ref{g-series}), the amplitude $R$ is recast to the form%
\footnote{This form can be easily derived from (\ref{R-ampl}) by using,
instead of the Greens function method, the well-known technique of the
creation and annihilation operators for the harmonic oscillator.}
\begin{eqnarray}
\label{R-answ2}
  && R=\frac{e^2}{2E_0} \sum_{n=1}^\infty
    \frac{ \omega_0 \,F(k_1) \, F(k_2)}
  {n\omega_0-\omega-  \omega^2/2E_0 -i0} \times
  \nonumber \\ && \qquad\qquad
   \left\{ \frac{x^{n-1}}{(n-1)!}(\veceps_1\cdot\veceps_2)
          +\frac{x^{n-2}}{(n-2)!}
   \frac{(\veceps_1\cdot\vk_2)
         (\veceps_2\cdot\vk_1)}
    {2\gamma^2}\right\} +(\omega\to -\omega),
\end{eqnarray}
which is useful to single out the individual contributions of different
intermediate states.  In particular, from (\ref{R-answ2})
one can directly read off individual contributions of oscillator excitations
and pair production to the amplitude (\ref{R-ampl}). These contributions
correspond to the pole parts of the $\omega$-dependent propagator in
(\ref{R-answ2}),
\begin{equation}
  \frac{\omega_0/E_0}{n\omega_0 - \omega - \omega^2/2E_0} =
  \frac{\gamma^2}{E_0 E_n} \left(
  \frac1{E_n-E_0-\omega} + \frac1{E_n+E_0+\omega}\right).
\end{equation}
The sum over $n$ in (\ref{R-answ2}) starts from $n=1$ because the
matrix elements in (\ref{R-ampl}) vanish at $n=0$. Also,
in (\ref{R-answ2}) it should be taken into account that
$(-1)!=\infty$. For a weakly relativistic oscillator
(i.e.\ $\eta \ll 1$) and low energies $\omega \sim \omega_0$, the
resonance amplitude $R$ is dominated by the dipole term ($n=1$),
whereas the higher-level corrections are generally suppressed as
$\eta^{n-1}$.

Equation (\ref{R-answ2}) is convenient for obtaining the multipole
structure of the resonance amplitudes.  All the angular dependence of
$R$ is contained in the $x$-dependent braces in (\ref{R-answ2}), where
$x\propto \cos\theta \equiv z$.  Applying the same technique of
helicity amplitudes as was used in section 8.1, we find the partial
$Ej$-components of the resonance amplitude as a sum over radial
excitations,%
\footnote{The partial amplitudes here are normalized as
$R(\omega,\theta) = \sum_j R^{Ej} (\omega) \,
g^{Ej} (\theta)$.}
\begin{equation}
\label{REj-answ}
   R^{Ej} = \frac{e^2}{2E_0} \sum_{n=j,j+2,j+4,\ldots}
    \frac{ \omega_0 \,F(k_1) \, F(k_2)}
  {n\omega_0-\omega-  \omega^2/2E_0 -i0}
  \left(\frac{k_1 k_2}{2\gamma^2}\right)^{n-1}
  \frac{C_{jn}}{(n-1)!} + (\omega\to -\omega).
\end{equation}
Here the coefficients $C_{jn}$ determine the multipole expansion of the
braces in (\ref{R-answ2}),
\begin{equation}
   z^{n-1} (\veceps_1\cdot\veceps_2)
   + (n-1) z^{n-2}
   (\veceps_1\cdot\hat \vk_2) (\veceps_2\cdot\hat \vk_1)
   = \sum_j C_{jn}\, g_{Ej}.
\end{equation}
They are normalized as $\sum_j C_{jn}=1$ and are equal to
\begin{eqnarray}
   && C_{jn}=\frac{2j+1}4 \int_{-1}^1
    \Big( n z^{n-1} - (n-1) z^{n-2} \Big)
     (1+z)\, d^{\,j}_{1,1}(\theta)\,dz
    \nonumber\\ &&  \qquad = \frac 14 \,\Big(1+(-1)^{n-j}\Big) \,
   \frac{j(j+1)(2j+1)}{(n+j+1)(n+j-1)}
     \prod_{i=1}^{[(j-1)/2]} \frac{n-j+2i}{n+j-2i-1}~,
\end{eqnarray}
provided $n\ge j$ (otherwise $C_{jn}=0$).  In particular,
\begin{equation}
  C_{1n}=\frac{3}{n(n+2)},\quad C_{2n}=\frac{15}{(n+1)(n+3)},
  \quad\mbox{\it etc.}
 \qquad (\mbox{for}\quad n-j=\mbox{even}).
\end{equation}
The (magnetic) $Mj$-components of $R$ are identically zero in our
model due to parity conservation.

Keeping in (\ref{REj-answ}) the leading-in-$k$ term, i.e.\ omitting the
form factors $F(k_1)$ and $F(k_2)$ and retaining only the lowest $n=j$
contribution, we obtain the unretarded multipoles,
\begin{equation}
\label{REj-unret}
   \unret{R}{Ej} = \frac{e^2}{E_0}\,
    \frac{ \omega_0 \,(j\omega_0 - \omega^2/2E_0)}
  {(j\omega_0-\omega^2/2E_0)^2 - \omega^2 -i0}
  \left(\frac{k_1 k_2}{2\gamma^2}\right)^{j-1}
  \frac{C_{jj}}{(j-1)!} ~.
\end{equation}
For a weakly relativistic oscillator ($\eta \ll 1$), the main difference
between the unretarded and retarded multipoles $R^{Ej}$ at low energies
$\omega \simle j\omega_0$ is described by the $k$-dependent form factor
squared of the ground state. Also, the form factor damps the amplitude
at the resonance peak, {\it e.g.} to the amount of
$F^2(\omega_0)=e^{-\eta/2}$ in the case of the $E1$-multipole at
$\omega\simeq \omega_0$.

In order to understand what happens at somewhat higher energies
$\omega$ (but still a few $\omega_0$), note that all the coefficients
$C_{jn}$ are positive.
They determine relative contributions of different radial
excitations to $R^{Ej}$ and are proportional to the matrix element
squared of the transition operator $(\veceps_1\cdot\vp)
\,\exp(i\vk_1\cdot\vr)$ between the ground state $|\phi_0\rangle$
and a state $|\phi_{n(jm)}\rangle$ of the energy $E_n$ and the angular
momentum $j$. So, this feature does not depend on a specific shape of
the potential.
Then, in the region
between the energies $\omega_0$ and $3\omega_0$, the
positive contributions $O(\eta^2)$ of the radial excitations to
$R^{E1}$ tend to cancel the leading negative contribution from the
first level $n=1$ and thus slightly intensify the damping effect
$O(\eta)$ caused by the $k$-dependent form factors alone.  Similar
small local strengthening of the damping effects just above the first
peak holds for higher multipoles as well.

At even higher energies, $\omega \sim \gamma \gg \omega_0$, which we
would call intermediate energies, when the $k$-dependent form factors
produce essential damping of individual transitions, the contributions
of radial excitations are not negligible at all.  They are enhanced by
powers of the same parameter $k_1 k_2 /2 \gamma^2$ which determines the
magnitude of the form factors.  As a result, many multipoles $R^{Ej}$
become as large as $R^{E1}$, which dominates at low energies.  They act
coherently at small angles and tend to compensate the effect of the
form factors.  Below we explicitly demonstrate this important feature.

\subsection{Asymptotic behavior and fixed-$t$ dispersion relations}

The resonance amplitude (\ref{R-answ2}) at $k_1 = k_2 = \omega$ is
given by a convergent series in $x=(2\omega^2-q^2)/4\gamma^2$ and has
no singularities at finite energies $\omega$ and momentum transfer
$q^2=-t$, except for the poles at $\omega=\omega_n^\pm$ and
$\omega=-\omega_n^\pm$, where the quantities
\begin{equation}
\label{omega+-}
  \omega_n^\pm=E_n \mp E_0
\end{equation}
describe the energies of the oscillator excitation and pair production,
respectively.  Moreover, the seagull amplitude
(\ref{S-ampl}), considered as a
function of the energy, is a ($t$-dependent) constant.  Therefore, the
validity of the fixed-$t$ dispersion relation for the Compton
scattering amplitude $T$ and for the appropriate invariant amplitudes depends
on whether or not they vanish at high complex $\omega$.

Using (\ref{S-ampl}) and (\ref{R-answ1}), we find coefficients of the
structures $(\veceps_1\cdot\veceps_2)$ and $(\vs_1\cdot\vs_2)$
in the total amplitude $T$ and then obtain the
invariant amplitudes $A_{1,2}(\omega,t)$. They read
\begin{eqnarray}
\label{A1-answ}
  \omega^2\,(A_1+A_2) &=& \frac{e^2}{2E_0} F(q)
    \Big\{ (1+a)g(a,x) + (\omega\to-\omega)\Big\},
  \\
\label{A2-answ}
  (A_1-A_2) &=& -\frac{e^2}{4\gamma^2 E_0} F(q)
    \Big\{ g(a-1,x) + (\omega\to-\omega) \Big\}.
\end{eqnarray}
Note that, at low energies $\omega\to 0$ and fixed $t$, the parameter
$a\to -1$ and $x\to t/4\gamma^2$. In this limit, the r.h.s.\ of Eq.\
(\ref{A1-answ}) vanishes.  Respectively, both the amplitudes
$A_{1,2}(\omega,t)$ are finite at $\omega=0$ and thus are free from the
$1/\omega^2$ pole.  This perfectly agrees with the low-energy theorem,
since the system of ``the particle plus the potential wall" has an
infinite mass and, therefore, its Thomson amplitude has to vanish.

At high energies, $\omega\to\infty$, and fixed $t$, both $a$ and $x$
go to infinity like $\omega^2$, and $(a-x)^2 =O(\omega^2)$ as well.
Using Eq.~(\ref{g-as}), we find that both amplitudes
$A_{1,2}(\omega,t)$ vanish like $O(1/\omega)$ in the complex plane of
$\omega$ and, therefore, satisfy unsubtracted fixed-$t$ dispersion
relations (\ref{T-EJ}).  The ultra-relativistic
asymptotics of the resonance amplitude reads
\begin{equation}
  R\simeq -\frac{e^2}{2E_0} F(q) \,e^{-1/\eta}
   \sqrt{\frac{\pi}{2x}}
  \Big(\mbox{ctg\,}(\pi a) + (\omega\to -\omega)\Big)
   \left\{ (\veceps_1\cdot\veceps_2)
   +\frac{(\veceps_1\cdot\vk_2)(\veceps_2\cdot\vk_1)}
  {2\gamma^2} \right\},
\end{equation}
where $\mbox{Re}\sqrt x >0$, like in (\ref{g-as}).  Beyond the real axis of
$\omega$, the amplitude $R$ is predominately imaginary,
\begin{equation}
\label{R-ultra}
  R \simeq  i \mbox{\,sign(Im\,$\omega$)\,} \frac{e^2}{\omega}
  F(q) \sqrt{\pi\eta} \,e^{-1/\eta}
   \left\{ (\veceps_1\cdot\veceps_2)
   +\frac{(\veceps_1\cdot\vk_2)(\veceps_2\cdot\vk_1)}
  {2\gamma^2} \right\}.
\end{equation}
When the relativistic parameter $\eta$ is small, the coefficient of the
leading $1/\omega$ term in this asymptotics is exponentially
suppressed.  Then the asymptotic regime (\ref{R-ultra}) holds only at
extremely high energies, whereas at intermediate energies of
$\omega\sim \gamma \gg\omega_0$ the resonance amplitude is
predominately real and follows the $1/\omega^2$ dependence,
\begin{equation}
\label{R-as}
  R \simeq  - \frac{e^2}{E_0} \, \frac{\omega_0^2}{\omega^2}  F(q)
   \left\{
  \Big(1-\frac{q^2}{4\gamma^2}\Big)
   (\veceps_1\cdot\veceps_2)
   +
  \Big(2-\frac{q^2}{4\gamma^2}\Big)
  \frac{(\veceps_1\cdot\vk_2)
        (\veceps_2\cdot\vk_1)} {2\gamma^2} \right\}.
\end{equation}
The remarkable feature of these asymptotics is that the retardation
$k$-dependent form factors $F(k_1)$ and $F(k_2)$, which accompany all
individual contributions of the transitions $|0\rangle \to
|n\alpha\rangle$ and exponentially grow at high complex energies, are
canceled when many transitions come into play.  In the asymptotic
regime, the resulting amplitude depends on the form factor only through
the momentum transfer $q$ rather than the energy $\omega$ and,
therefore, is free from the exponential growth at $\omega\to i\infty$.
This is why the fixed-$t$ dispersion relation holds.

As an important particular case, note that the asymptotics (\ref{R-as})
of the exact amplitude $R$ at $\theta=0$ and at intermediate energies
behaves like  the asymptotics of the naive unretarded electric
dipole amplitude
\begin{equation}
   \unret{R}{E1,{\rm naive}} = \frac{e^2}{E_0} \,
    \frac{\omega_0^2}{\omega_0^2  - \omega^2 - i0} ~.
\end{equation}
This feature is connected with a validity of the dispersion relation
and its consequence, the $E1$-sum rule, is discussed in the next
subsection.

It is instructive to compare the high-energy behavior of the Compton
scattering amplitude off the relativistic and nonrelativistic (nonrel)
oscillator and to see where a difference comes from. The
nonrelativistic oscillator has the energy spectrum
\begin{equation}
  E_n^{\rm nonrel}=E_0^{\rm nonrel} + n\omega_0
\end{equation}
(only with positive energies) and is described by the same eigen
functions $\phi_{n\alpha}$ as in (\ref{phi_n}), with
the substitution $E_n^2-\mu^2\to (2n+3)\gamma^2$. The nonrelativistic
oscillator frequency $\omega_0=\gamma^2/M$ is determined by the size of
the charge distribution in the ground state, $1/\gamma$, and the mass
of the particle, $M$. The nonrelativistic resonance amplitude $R^{\rm
nonrel}$ is given by expressions similar to (\ref{R-ampl}) and
(\ref{R-answ2}),
\begin{eqnarray}
   R^{\rm nonrel} &=& \sum_{n\alpha}  \frac{e^2}{M^2}
  \frac{\langle \phi_0 |
    (\veceps_2\cdot\vp)\exp(-i\vk_2\cdot\vr)
   | \phi_{n\alpha} \rangle   \langle \phi_{n\alpha} |
    (\veceps_1\cdot\vp) \exp(i\vk_1\cdot\vr)
   | \phi_0 \rangle}    {E_n-\omega-E_0 - i0}
\nonumber \\ && \hspace{5em} {} +
  (\omega \to -\omega,~ \vk_1 \leftrightarrow -\vk_2,~
    \veceps_1 \leftrightarrow \veceps_2 )
\label{R-nonrel}
  \\[1em] &=& \sum_{n=1}^\infty
    \frac{ e^2\gamma^2 \,F(k_1) \, F(k_2)}
  {2M^2 (n\omega_0-\omega- i0)}
   \left\{ \frac{x^{n-1}}{(n-1)!}(\veceps_1\cdot\veceps_2)
          +\frac{x^{n-2}}{(n-2)!}
   \frac{(\veceps_1\cdot\vk_2)
         (\veceps_2\cdot\vk_1)}
    {2\gamma^2}\right\}
  \nonumber \\ && \hspace{5em} {}+ (\omega\to -\omega)
\label{R-nonrel2}
  \\[1em]  &=&  \frac{e^2 \,F(q)}{2M}
   \left\{g(a_0,x)(\veceps_1\cdot\veceps_2)
    + g(a_0-1,x)\frac{
   (\veceps_1\cdot\vk_2)(\veceps_2\cdot\vk_1)}
    {2\gamma^2}\right\}
  \nonumber \\ && \hspace{5em} {}+ (\omega\to -\omega),
     \qquad  a_0=\omega/\omega_0-1+i0.
\label{R-nonrel1}
\end{eqnarray}
The absence of $\omega^2$-pieces in the denominator in
(\ref{R-nonrel2}) and in $a_0$ drastically changes the behavior of the
amplitude $R^{\rm nonrel}$ at very high complex energies. When
$\omega\to i\infty$, the function $g(a_0,x)$ grows exponentially like
$e^{-x}$ with $x$ given by Eq.\ (\ref{a})..  
Respectively, the resonance amplitude $R^{\rm nonrel}$ grows
like the retardation form factors $F^2(\omega)$ and does not obey the
fixed-$t$ dispersion relation.
\footnote{The violation of the dispersion relation for the
nonrelativistic Compton scattering amplitude was discussed in Ref.\
\cite{matsuura73,friar75,arenh77} where it was connected with
additional poles or cuts at $\omega\simeq 2M$.
In the nonrelativistic oscillator model such singularities are absent,
but the exponential growth at $\omega\to i\infty$ does make
standard dispersion relations inapplicable.
It is still possible to write a
"practical" dispersion relation with a Cauchy loop of a finite size
$\omega_{\rm max}$ with $\gamma \ll \omega_{\rm max} \ll M$ and use the
asymptotics discussed in the text for allowing for high-$\omega$
contributions.}

A difference between the asymptotics of the relativistic and
nonrelativistic amplitudes appears not only at extremely high energies.
The retardation corrections become essential already at intermediate
energies, $\omega \sim \gamma$, and the amplitude $R^{\rm nonrel}$
behaves like
\begin{equation}
\label{R-as-nonrel}
  R^{\rm nonrel} \simeq  - \frac{e^2}{M} \,
\frac{\omega_0^2}{\omega^2} F(q) \left\{
  (1+x) (\veceps_1\cdot\veceps_2)
   + (2+x)  \frac{(\veceps_1\cdot\vk_2)
        (\veceps_2\cdot\vk_1)} {2\gamma^2} \right\}
\end{equation}
at such $\omega$, with $x=(-q^2+2\omega^2)/4\gamma^2$. Therefore, it
shows no tendency to decrease, in contrast to (\ref{R-as}).

It is again instructive to obtain the nonrelativistic asymptotics
(\ref{R-as-nonrel}) directly from Eq.~(\ref{R-nonrel}) and to trace a
difference with the relativistic consideration.  Since the sum in
(\ref{R-nonrel}) is saturated by intermediate states with the typical
momentum $\sim k$ and, hence, with the excitation energies $E_n-E_0\sim
k^2/(2M) \ll \omega$, the denominator in (\ref{R-nonrel}) can be expanded
in the inverse powers of $\omega$,
\begin{equation}
\label{1/omega}
  \frac{1}{E_n-E_0-\omega} = -\frac{1}{\omega} -
     \frac{E_n-E_0}{\omega^2} - \cdots.
\end{equation}
Here the leading term is odd and is canceled by the crossed term.
The asymptotics is determined by the next term.  Using closure,
one can recast the result in terms of the double commutator of the
electromagnetic currents and the nonrelativistic Hamiltonian
$H=p^2/(2M) + M\omega_0^2r^2/2$,
\begin{equation}
  R^{\rm nonrel} \simeq  -\frac{1}{\omega^2}
   \langle \phi_0 |\, [j_2,[H,j_1] \, | \phi_0 \rangle .
\end{equation}
Here $j_1=(e/M)(\veceps_1\cdot\vp) \exp(i\vk_1\cdot\vr)$
and $j_2=(e/M)(\veceps_2\cdot\vp)\exp(-i\vk_2\cdot\vr)$.
Calculating the above matrix element, we arrive at
Eq.~(\ref{R-as-nonrel}). The derivatives of the $k$-dependent exponents in
the operators $j_1$ and $j_2$, coming from the commutator with the
kinetic energy, lead to the $\omega^2$-terms in the braces
of (\ref{R-as-nonrel}).  With similar steps for the relativistic
oscillator, we would get a difference due to the negative-energy
intermediate states which results in an $\omega^2$-correction in
the energy denominator.
So, instead of Eq.~(\ref{1/omega}), we get the
terms
\begin{equation}
  \frac{1}{E_n-E_0-\omega-\omega^2/(2M)} = -\frac{1}{\omega} -
     \frac{E_n-E_0-\omega^2/(2M)}{\omega^2} - \cdots
\end{equation}
in this case. Here, the relativistic correction $\omega^2/(2M)$ is of the
same order of magnitude as $E_n-E_0$ itself . It gives an additional
$\omega$-independent contribution to the asymptotic resonance
amplitude,
\begin{equation}
  R_{-} = \frac{1}{2M}
   \langle \phi_0 | j_1 j_2 + j_2 j_1 | \phi_0 \rangle .
\label{Rminus}
\end{equation}
This $R_{-}$, which is the contribution of negative energies, compensates the
$\omega$-independent term, which appears in (\ref{R-as-nonrel}) through
the terms containing $x$, and results in a relativistic
$1/\omega^2$-asymptotics, Eq.~(\ref{R-as}), consistent with the
fixed-$t$ dispersion relation.

\subsection{Retarded and unretarded sum rules}

Since the amplitude $R$ vanishes at $\omega\to\infty$ and fixed $t$, it
obeys the unsubtracted dispersion relation.  That means that both the
coefficients of $(\veceps_1\cdot\veceps_2)$ and
$(\veceps_1\cdot\vk_2) (\veceps_2\cdot\vk_1)$
obey the relation:
\begin{eqnarray}
\label{R-disp}
    R(\omega,t) &=& \frac{2}{\pi} \int_0^\infty
    \mbox{Im}\,R(\omega',t) \,  \frac{\omega' d\omega'}
    {\omega'^2-\omega^2-i0}
   \nonumber \\
  &=&  \sum_\pm \sum_{n=1}^\infty \frac{e^2}{E_0 E_n}\,
  \frac{\gamma^2 \omega_n^\pm \,F^2(\omega_n^\pm)}
   {(\omega_n^\pm)^2 - \omega^2 -i0} \times
  \nonumber \\ && \qquad\qquad
   \left\{ \frac{(x_n^\pm)^{n-1}}{(n-1)!}
   (\veceps_1\cdot\veceps_2)
          +\frac{(x_n^\pm)^{n-2}}{(n-2)!}
   \frac{(\veceps_1\cdot\vk_2)
         (\veceps_2\cdot\vk_1)}
   {2\gamma^2}\right\}.
\end{eqnarray}
Here the sign of $\pm$ refers to the contributions of positive and
negative energies, and $x_n^\pm = (2(\omega_n^\pm)^2 + t)/4\gamma^2$.
Matching the r.h.s.\ of Eq.~(\ref{R-disp}) with the relation
\begin{equation}
\label{LET-for-R}
   R(0,t) = -S(t)
   = \frac{e^2}{E_0} (\veceps_1\cdot\veceps_2) F(t),
\end{equation}
which strictly follows from the low-energy theorem for the total
amplitude, which reads $T(\omega=0,t)=0$ in our case, we arrive at the
series of sum rules as discussed in section 8.2.  In the following we
consider two of them which are closely related with the $E1$ and $E2$
photoabsorption, respectively.

(i) The first one, the Gell-Mann--Goldberger--Thirring (GGT) sum rule,
corresponds to the particular case of $t=0$ in (\ref{R-disp}).  It
reads
\begin{equation}
\label{GGT2}
    \Sigma \equiv
    \int_0^\infty \sigma(\omega)\,d\omega
    = 2\pi^2 R(0,0)= \frac{2\pi^2 e^2}{E_0},
\end{equation}
where
\begin{equation}
\label{optic}
  \sigma(\omega) =\frac{4\pi}{\omega}\, \mbox{Im}\,R(\omega,0)
\end{equation}
is the total photoabsorption cross section. In accordance with
Eqs.~(\ref{R-answ2}) and (\ref{R-disp}), this cross section consists of
the contribution of particle excitation, $\sigma_+$, and that of pair
production, $\sigma_-$,
\begin{equation}
  \sigma=\sigma_+ + \sigma_-, \qquad
  \sigma_\pm(\omega)= \sum_{n=1}^\infty
  \frac{2\pi^2 e^2\gamma^2}{\omega E_0 E_n} e^{-A} \frac{A^{n-1}}{(n-1)!}
  \delta(\omega-E_n \pm E_0), \quad A=\frac{\omega^2}{2\gamma^2}.
\end{equation}
Respectively, the integral in (\ref{GGT2}) consists of two parts as well,
\begin{equation}
\label{Sigma}
  \Sigma =  \Sigma_+  +   \Sigma_-,   \quad
     \Sigma_\pm =  \sum_{n=1}^\infty
   \frac{2\pi^2 e^2\gamma^2}{\omega_n^\pm E_0 E_n}
    \frac{(A_n^\pm)^{n-1}} {(n-1)!} \exp(-A_n^\pm) ,
   \qquad A_n^\pm =\frac{(\omega_n^\pm)^2}{2\gamma^2}.
\end{equation}
Both of them are needed to strictly fulfil the GGT sum rule.  However,
in the weakly relativistic case, the pair contribution is exponentially
suppressed by the retardation form factors, $\Sigma_- \propto
\exp(-1/\eta)$, and the sum rule is saturated with the positive-energy
contribution $\Sigma_+$.  Even when $\eta=1$, $\Sigma_-$ contributes
still only 5\% to the total magnitude of $\Sigma$.  Only in the extreme
case of $\eta\gg1$ both parts of $\Sigma$ become compatible. $\Sigma_-
= \Sigma_+ = \frac12 \Sigma$ in this limit.

When the retardation in Eq.~(\ref{GGT2}) is switched off, i.e.\ when
$k_1$ and $k_2$ are set to zero, we find the unretarded
$E1$-amplitude,
\begin{equation}
  \unret{R}{E1}(\omega)
  = \frac{e^2\gamma^2}{E_0} \frac{1}{2\gamma^2-\omega^2-2\omega E_0}
   + (\omega\to -\omega).
\label{unretarded-RE1}
\end{equation}
It also satisfies the unsubtracted dispersion relation, and, being
matched with the low-energy theorem (\ref{LET-for-R}), gives the
unretarded partner of the GGT sum rule, which is the
Thomas--Reiche--Kuhn (TRK) sum rule:
\begin{equation}
\label{TRK}
    \unret{\Sigma}{} \equiv
  \int_{\omega_{\rm thr}}^\infty \unret{\sigma}{E1}(\omega)
   \,d\omega = 2\pi^2 \unret{R}{E1}(0)= \frac{2\pi^2 e^2}{E_0}.
\end{equation}
Here the unretarded photoabsorption cross section, $\unret{\sigma}{E1}
=(4\pi/\omega)\mbox{Im} \unret{R}{E1}$, also consists of the
contributions of excitation and pair production,
\begin{equation}
  \unret{\sigma}{E1}=
  \unret{\sigma}{E1,+} + \unret{\sigma}{E1,-}, \qquad
  \unret{\sigma}{E1,\pm} =
  \frac{2\pi^2 e^2\gamma^2}{\omega E_0 E_1} \delta(\omega-E_1 \pm E_0).
\end{equation}
Corresponding parts of $\unret{\Sigma}{}$ read
\begin{equation}
\label{Sigma-unret+-}
   \unret{\Sigma}{}_\pm =
   \frac{2\pi^2 e^2\gamma^2}{\omega_1^\pm E_0 E_1} ,
\end{equation}
and their sum, $\unret{\Sigma}{} =  \unret{\Sigma}{}_+  +
\unret{\Sigma}{}_-$, coincides with the retarded integral in Eq.\
(\ref{GGT2}),
\begin{equation}
\label{Gerasimov}
  \unret{\Sigma}{} = \Sigma.
\end{equation}
The last equation was discussed in literature many times after
S.~Gerasimov
\cite{gerasimov64,matsuura73,friar75,arenh77,arenh79,schmitt85,SC94}.
It strictly holds in our model.
Moreover, as is seen from the above derivation, it
is valid in a more general case, with a general form of the current
operators or the energy spectrum, provided the unsubtracted dispersion
relation is valid for the retarded amplitude $R(\omega,0)$.  Other
ingredients of the derivation, which are the unsubtracted dispersion
relation for $\unret{R}{E1}(\omega)$ and the identity
$R(0,0)=\unret{R}{E1}(0)$, are model independent.%
\footnote{These requirements are not fulfilled for a Dirac particle,
for which the seagull contribution is absent and
$R(\infty,0) = T(\infty,0) \ne 0$.
In that case a sensical comparison of the GGT and TRK sum rule
needs a consideration of positive-energy components
\cite{levinger57,arenh85}.}

However, Gerasimov's relation (\ref{Gerasimov}) is valid only when both
positive- and negative-energy parts are included. It is not valid for
each of these parts separately.  For a weakly relativistic oscillator, the
contribution of pair production is exponentially small, $\Sigma_-
\propto \exp(-1/\eta)$, whereas it is of order ${\cal O}(\eta)$ for the
unretarded counterpart. Respectively,
\begin{equation}
\label{GGT+-}
  \unret{\Sigma}{}_+ - \Sigma_+ =
  - (\unret{\Sigma}{}_- - \Sigma_-) \simeq -\unret{\Sigma}{}_-
   \simeq -\frac12\eta\Sigma
\end{equation}
in this case. In other words, the positive-energy TRK sum rule
differs from the GGT sum rule.
Saturation of the full TRK sum rule needs the consideration of
relativistically high energies, whereas such energies are irrelevant
for the GGT sum rule.  When the GGT and TRK integrals are truncated
below the pair production threshold (that is what is usually done in
most nuclear applications),
they become different in order ${\cal O}(\eta)$.
Such a ``practical" violation of the Gerasimov's relation was studied
and exemplified in \cite{matsuura73,friar75,arenh77,schmitt85}.
As was said, another
reason for the violation of Eq.~(\ref{Gerasimov}) can be caused by
a nonvanishing amplitude $R$ at high real or complex energies, which
makes the {\em unsubtracted} dispersion relation inapplicable for $R$
(see an example in \cite{arenh79}).

Note that the retardation form factor in $\Sigma_+$ reduces the
dominating $E1$ contribution by $F^2(\omega_1) \simeq 1-\frac12\eta$.
Since the $E2$ contribution also increases $\Sigma_+$ by $\eta$, the
resulting GGT magnitude is bigger than the nonrelativistic TRK by
$\frac12\eta$. Only after adding the negative-energy part of TRK,
$\unret{\Sigma}{}_- \simeq \frac12\eta \Sigma$, the balance between
GGT and TRK is achieved.%
\footnote{Since the relativistic corrections affect the energy $E_0$ of
the bound state, they change also the high energy limit of the
Compton scattering amplitude $T$ and the magnitude of the GGT and TRK
sum rule, {\it cf.} \cite{goldberger68,friar75}.
However, this in itself does not destroy Gerasimov's theorem
(\ref{Gerasimov}). What does is a projection to positive-energy states.}
Equation (\ref{GGT+-}) perfectly agrees with the evaluation of the
difference between positive-energy GGT and TRK sum rules for a bound
particle done by Friar and Fallieros \cite{friar75} who showed that
this difference is equal to $-2\pi^2 e^2 \langle p^2/3M^3 \rangle$
(note that for the oscillator model
$\langle p^2 \rangle = \frac32 \gamma^2$).
The origin of the difference (\ref{GGT+-}) can be traced to
the asymptotic behavior of the unretarded amplitude (\ref{unretarded-RE1})
at intermediate energies $\omega_0 \ll \omega \ll M$. This asymptotics is
determined by the energy-independent contribution
$\unret{R}{E1}_-$ of negative-energy states
(see Eq.\ (\ref{Rminus}), in which one has to use $j_{1,2}$
with $\vk_{1,2}=0$) and reads
\begin{equation}
  \unret{R}{E1}(\omega) \simeq \unret{R}{E1}_-
  \simeq  \frac{e^2\gamma^2}{2M^3} = \frac{e^2 \langle p^2 \rangle}{3M^3}.
\label{R-unret-asympt}
\end{equation}
Accordingly, the dispersion integral (\ref{TRK}) truncated at an
intermediate energy $\omega_m$ becomes
\begin{equation}
    \unret{\Sigma}{}(\omega_m) \equiv
  \int_{\omega_{\rm thr}}^{\omega_m} \unret{\sigma}{E1}(\omega)
   \,d\omega = 2\pi^2 \Big(\unret{R}{E1}(0) - \unret{R}{E1}(\omega_m) \Big)
   \simeq \Big( 1- \frac{\gamma^2}{2M^2} \Big) \,\Sigma,
\end{equation}
from which Eq.\ (\ref{GGT+-}) follows.

Usually in low-energy applications, all contributions of
negative-energy states which do not depend on the energy $\omega$
and thus survive at high $\omega$ are included
into an effective seagull amplitude.
Doing so, we conclude from Eq.\ (\ref{R-unret-asympt}) that
the effective (i.e.\ vanishing at high $\omega$) unretarded
resonance amplitude is exactly the positive-energy contribution
$\unret{R}{E1}_+(\omega)$.
Due to $\unret{R}{E1}_- \ne 0$, it is {\em different}
from the retarded (i.e.\ dispersion) amplitude $R(\omega,0)$ at zero energy,
and, correspondingly, the effective unretarded seagull
$\unret{S}{}(\omega,\theta) = S(\omega,\theta) +
\unret{R}{E1}_- g^{E1}(\theta)$
is different from the retarded seagull $S(\omega,\theta)$ as well.

Considering unretarded and retarded amplitudes which vanish
at intermediate energies, we find a difference in that
these vanishing amplitudes are exactly
the positive-energy part $\unret{R}{E1}_+$
and the total amplitude $R = R_+ + R_-$, respectively.
The latter includes both positive- and negative-energy states,
because the positive-energy part $R_+$ alone does not vanish,
as was discussed in section 10.3.
Therefore, using dispersion relations with an intermediate-energy cut
$\omega_m$, we obtain through the retarded cross section
$\sigma(\omega)$ the amplitude $R$ which includes {\em both} positive- and
negative-energy states.
At the same time, the dispersion integral of the unretarded cross section
$\unret{\sigma}{E1}(\omega)$
gives {\em only} the positive-energy part of $\unret{R}{E1}$.
That is why the obtained amplitudes are not exactly the same at zero energy
and that is why Gerasimov's argument, which is a combination
of the (retarded) dispersion relation and of the equation
$R(0,0)=\unret{R}{E1}(0)$, Eq.\ (\ref{R-gerasimov}), has a flaw
when applied to photonuclear reactions at intermediate energies.

(ii) The second sum rule, which is briefly discussed below along the same
line, appears as a derivative $d/dt$ of the relation (\ref{R-disp})
taken at the point $t=0$. More specifically, (i) we consider the case
of equal photon helicities, $\lambda_1=\lambda_2$, so that
$(\veceps_1\cdot\vk_2) (\veceps_2\cdot\vk_1) =
(t/2) (\veceps_1 \cdot \veceps_2) $, and (ii) we do not
apply the derivative to the product $(\veceps_1 \cdot
\veceps_2)$.%
\footnote{In other words, we consider the $t$-derivative of the
invariant amplitude $A_1$. {\it Cf.} Eq.~(\ref{sumrule4}).}
We therefore have
\begin{equation}
\label{GGT-E2}
    \Sigma_t \equiv
    \int_0^\infty \sigma_t(\omega)\,d\omega
    = 2\pi^2 \frac{\partial}{\partial t} R_1(0,0)
   =\frac{2\pi^2 e^2}{E_0} \frac16 \langle r_E^2 \rangle
   =\frac{\pi^2 e^2}{2\gamma^2 E_0},
\end{equation}
where $R_1(\omega,t) = R(\omega,t;~ \lambda_1=\lambda_2,~
   \veceps_1 \cdot \veceps_2 =1)$
and the integrand
\begin{equation}
  \sigma_t(\omega)=\frac{4\pi}{\omega}
   \frac{\partial}{\partial t} \mbox{Im} R_1(\omega,0)
\end{equation}
is generally determined by the partial $Ej$ and $Mj$ cross sections,
\begin{equation}
  \sigma_t = \sum_{j=2}^\infty
    \frac{(j+2)(j-1)}{4\omega^2} (\sigma^{Ej} + \sigma^{Mj})
  = \frac{1}{\omega^2}(\sigma^{E2} + \sigma^{M2}) + \mbox{higher $j$}.
\end{equation}
(In fact, the magnetic cross sections are absent in our model.)
The derivative cross section $\sigma_t$ consists of contributions of
the excitation and the pair production,
\begin{equation}
  \sigma_t = \sigma_{t,+} + \sigma_{t,-}, \qquad
  \sigma_{t,\pm}(\omega)= \sum_{n=2}^\infty
  \frac{\pi^2 e^2}{\omega E_0 E_n} e^{-A} \frac{A^{n-2}}{(n-2)!}
  \delta(\omega-E_n \pm E_0), \quad A=\frac{\omega^2}{2\gamma^2},
\end{equation}
and so does the corresponding integral,
\begin{equation}
\label{Sigma-t}
  \Sigma_t =  \Sigma_{t,+}  +   \Sigma_{t,-},   \quad
     \Sigma_{t,\pm} =  \sum_{n=2}^\infty
   \frac{\pi^2 e^2}{\omega_n^\pm E_0 E_n}
    \frac{(A_n^\pm)^{n-2}} {(n-2)!} \exp(-A_n^\pm) ,
   \qquad A_n^\pm = \frac{(\omega_n^\pm)^2}{2\gamma^2}.
\end{equation}
With the relativistic parameter $\eta\ll 1$, the pair contribution
is exponentially suppressed, $\Sigma_{t,-}\propto \exp(-1/\eta)$.

Again, we can consider the unretarded analogue of the sum rule
(\ref{GGT-E2}) which involves the unretarded $t$-derivative of the
amplitude $R(\omega,t)$,
\begin{equation}
  \unret{R}{}_t(\omega) = \frac{1}{\omega^2} \unret{R}{E2}
    = \frac{e^2}{2E_0} \, \frac{1}{4\gamma^2-\omega^2-2\omega E_0}
             + (\omega\to-\omega),
\end{equation}
and the unretarded cross section $\unret{\sigma}{}_t = (1/\omega^2)
\unret{\sigma}{E2}$.  Its positive- and negative-energy components
read
\begin{equation}
  \unret{\sigma}{E2} = \unret{\sigma}{E2}_+ +
       \unret{\sigma}{E2}_-, \qquad
  \unret{\sigma}{E2}_\pm
    = \frac{\pi^2 e^2 \omega}{E_0 E_2} \delta(\omega - E_2 \pm E_0).
\end{equation}
The corresponding integral
\begin{equation}
  \unret{\Sigma}{}_t = \int_0^\infty \unret{\sigma}{E2}(\omega)
   \frac{d\omega}{\omega^2} =
  \unret{\Sigma}{}_{t,+} + \unret{\Sigma}{}_{t,-}, \qquad
  \unret{\Sigma}{}_{t,\pm} = \frac{\pi^2 e^2} {\omega_2^\pm E_0 E_2}
\end{equation}
exactly satisfies the Gerasimov-like relation
\begin{equation}
\label{Gerasimov-E2}
  \Sigma_t = \unret{\Sigma}{}_t.
\end{equation}
Again, this relation does not hold separately for the $+$ and $-$
components. In essence, the relation (\ref{Gerasimov-E2}) is based only
on the assumption that the amplitude $R_t(\omega)$ satisfies an
unsubtracted dispersion relation. Other ingredients of the derivation,
{\it viz.}\ an unsubtracted dispersion relation for $\unret{R}{}_t(\omega)$
and the identity $R_t(0)=\unret{R}{}_t(0)$ are model independent.

Since $\unret{\Sigma}{}_{t,-}\simeq \eta \Sigma_t$, when $\eta$ is
small, the ``practical" violation of the Gerasimov $E2$-theorem
(\ref{Gerasimov-E2}) happens in first order in $\eta$ whenever the
integrals are truncated at nonrelativistic energies $\omega\ll E_0$.
Details of the balance between the retarded and full unretarded sum
rules are as follows.  The retardation form factor reduces the
dominating $E2$ contribution to $\Sigma_t$ by $F^2(\omega_2) \simeq
1-2\eta$, but the $E3$ contribution (additional $+3\eta\Sigma_t$) increases
$\Sigma_t$ with respect to $\unret{\Sigma}{}_{t,+}$ by $\eta\Sigma_t$.
After taking into account $\unret{\Sigma}{}_{t,-}$ the balance is
restored.

The important conclusion, which holds in a more general situation than the
oscillator model suggests, is that relativistic corrections to the
TRK-like sum rules are of the same scale as the retardation effects and
effects of higher multipoles in the GGT-like sum rules.  Taken
together, they lead to an exact balance between retarded and unretarded
sum rules. Incomplete saturation of the unretarded sum rules with
nonrelativistically low energies results in an error of the same scale
as the relativistic and retardation corrections themselves.

Relativistic TRK sum rule matches the GGT sum rule, nonrelativistic TRK
does not.

\subsection{Comparison with the phenomenological approach}

The amplitude $R \equiv R_{\rm osc}$ suggested by our oscillator model
is too simple to include all details of nuclear Compton scattering in
the giant-resonance region.  For example, it cannot describe the ensemble
of isoscalar giant resonances which begins with the quadrupole
excitation. (The dipole isoscalar mode is a translation.)  The
oscillator model better works for the isovector modes, for which the
$E2/E1$ ratio of excitation energies and widths is close to the
oscillator value of 2.  (We can include the widths into the oscillator
model by using a complex parameter $\omega_0$, although this gives an
imaginary part to a few other parameters such as the oscillator radius or the
ground-state energy $E_0$.)  An other restriction of the oscillator model,
which is caused by a shape of the oscillator potential, is that it
keeps a firm relation between the nuclear form factor and the
quadrupole photoabsorption strength which may not work well for all
nuclei.  In terms of the oscillator frequency $\omega_0$ and of the
small relativistic parameter $\eta$, this relation reads
\begin{equation}
   F^2(\omega_0) \simeq 1-\frac{\eta}{2}, \quad
   F^2(2\omega_0) \simeq 1-2\eta, \quad
   \int_0^\infty \sigma^{E2}(\omega)\,d\omega \simeq
   \eta \times \int_0^\infty \sigma^{E1}(\omega)\,d\omega.
\end{equation}
So, we cannot directly use the oscillator-model amplitude for the
comparison with or the interpretation of experimental data on Compton
scattering. But we observe that the phenomenological amplitude
$R_{GR}$ from  Eq.~(\ref{Retardation2}), which is used for practical
data analyses, indeed incorporates some gross features of the amplitude
$R_{\rm osc}$ at energies in the GR region.  These include (i) the
fulfillment of the GGT sum rule, (ii) retardation form factors
accompanying the lowest resonance peaks, and (iii) the asymptotic tail
${\cal O}(\omega^{-2})$. One difference which does exist between these two
amplitudes is that the asymptotics of $R_{\rm osc}$ contains a
$q$-dependent form factor which erases the amplitude at large angles.
This feature is absent in the phenomenological amplitude $R_{GR}$
where the vanishing in the high-energy limit and at all angles is caused by
an exact cancellation of the high-energy limit of the $\omega$-dependent
part of $R_{GR}$ and the $\omega$-independent subtraction.

We can check how the prescription (\ref{Retardation2}), being
applied to the oscillator-model cross sections, matches the amplitude
$R_{\rm osc}$ which is known exactly. For a numerical illustration we
take the parameters $\omega_0$ and $\eta$ as determined by the
isovector giant dipole and quadrupole resonances in $^{208}$Pb \cite{SC94}
(see Table 4.1):
\begin{equation}
   \omega_0 = (13.4 - 2.1i) {\rm ~MeV}, \quad \eta=0.06 \, .
\end{equation}
With these parameters, the r.m.s.\ radius of the oscillator is $\langle
r^2 \rangle^{1/2} = \omega_0^{-1}\sqrt{\frac32\eta} \simeq 4.4$ fm.
The energy region of our interest will be $\omega \le 100$ MeV, that is
well below the threshold of pair production $2E_0=2\omega_0/\eta \simeq
450$ MeV.

Applying the prescription (\ref{Retardation2}), we act in line with
the phenomenological procedure and keep in the dispersion integrals
(\ref{naive-DR}) only the lowest $E1$ and $E2$ peaks. Thus, we
omit radial excitations, which give corrections of order ${\cal O}(\eta^2)$,
and the pair contribution. Then we get
\begin{equation}
\label{R-ph}
  R_{GR}(\omega,\theta) = g^{E1}(\theta) R_{GR}(\omega,\theta=0)
     + (g^{E2}(\theta)-g^{E1}(\theta)) \hat R^{E2}_{GR}(\omega) \, ,
\end{equation}
where
\begin{equation}
\label{R-GR-osc}
  R_{GR}(\omega,\theta=0) = \frac{e^2\gamma^2}{E_0} \left\{
     \frac{\omega_1^+ F^2(\omega_1^+)} {((\omega_1^+)^2-\omega^2) E_1}
    +\frac{\omega_2^+ F^2(\omega_2^+)} {((\omega_2^+)^2-\omega^2) E_2} \,
     \frac{(\omega_2^+)^2}{2\gamma^2} \right\}
\end{equation}
and
\begin{equation}
\label{R-GR-E2-osc}
  \hat R^{E2}_{GR}(\omega) = \frac{e^2\gamma^2}{E_0 E_2} \,
     \frac{\omega_2^+ F^2(\omega)} {(\omega_2^+)^2-\omega^2} \,
      \frac{\omega^2}{2\gamma^2}.
\end{equation}
By construction, $R_{\rm osc}$ and the amplitudes from (\ref{R-ph}) are very
close at zero energy (the difference there is ${\cal O}(\eta^2)$, see the
previous discussion of the GGT sum rule), near the $E1$ and $E2$ peaks,
and at forward angle. They remain very close at other energies and
angles, see Fig.~\ref{fig:phen-vs-osc}.

This would not be the case if the retardation form factors were removed
in Eqs.\ (\ref{Retardation2}), (\ref{R-GR-osc}), (\ref{R-GR-E2-osc}).
Then, in particular, the backward scattering amplitude
$R_{GR}(\omega,180^\circ)$ at energies above the GR region
would behave as $R_{GR}(\infty,180^\circ) \simeq  2\eta e^2/E_0$
and would show a wrong asymptotic tail
$2\eta = 0.12$ in Fig.~\ref{fig:phen-vs-osc}.

Thus, the oscillator model
suggests that the phenomenological prescription
(\ref{Retardation2}) provides a very good approximation to the
exact amplitude.

\begin{figure}[htb]
\caption{\it Phenomenological, Eq.~(\ref{R-ph}),  (dashed lines)
and exact (solid lines)
amplitudes $R_{GR}$ in the relativistic oscillator model at
$\omega_0=(13.4-2.1i)$ MeV and $\eta=0.06$. Units are $e^2/{E_0}$. }
\label{fig:phen-vs-osc}
\end{figure}

\appendix
\clearpage

\setcounter{table}{0}
\section{Parameters of the mesonic seagull amplitude for different nuclei}

In section 9 the parameters  $\kappa $, $\delta \alpha $ and $\delta
\beta $, which constitute  the mesonic seagull amplitude, are shown as a
function of the proton number $Z$. In this appendix the corresponding
numerical values are given, including all values for the various
contributions to each of these parameters. In the following tables this
has been done for both, a pure Fermi gas model and a realistic nuclear
density based upon the parameterizations in \cite{datatab}. The
results for the mesonic seagull amplitude in nuclear matter, which has
been discussed in sections 9.1 and 9.2 are also included in the tables
(denoted ``$A\rightarrow\infty $"), as
well as the case of an asymptotic correlation function
${\cal F}_0^{ij}$ as in section 9.1
(superscript ``0"). The polarizability modifications
$\delta \alpha $ and $\delta \beta $ are given in the usual units
10$^{-4}$fm$^3$.

\begin{table}[htb]
\caption{\it
    Numerical values for the various contributions to the enhancement
    constant $\kappa $ shown for nuclear matter ($A\rightarrow
  \infty$) and different finite nuclei. The quantities $\kappa _{\pi}^C$
    and $\kappa _{\rho}^C$ denote the pion and $\rho $-meson
    contributions due to central correlations, respectively,
    while $\kappa _{\pi}^T$
    and $\kappa _{\rho}^T$ denote the contributions due to tensor
    correlations. When only the asymptotic correlation function has
    been used, the contributions are termed
    $\kappa _{\pi}^0$ and $\kappa _{\rho}^0$.
    The right-hand half of the table shows the values for realistic
    nuclear densities based on the Fermi parameterizations from
    \protect\cite{datatab}.}
\label{tab1}
\begin{center}
\begin{tabular}{|c||c|c|c|c||c|c|c|c||}
\hline
&\multicolumn{4}{c||}{constant density}&\multicolumn{4}{c||}{realistic
density}\\
\hline
&$A \to \infty$& $^{208}$Pb & $^{40}$Ca & $^{12}$C& $^{208}$Pb&$^{40}$Ca&
$^{12}$C&$^4$He\\
\hline
\hline
$\kappa^0_{\pi}$&0.19&0.16&0.13&0.11&0.15&0.11&0.09&0.05\\
$\kappa^C_{\pi}$&0.15&0.13&0.11&0.09&0.12&0.09&0.07&0.03\\
$\kappa^T_{\pi}$&0.86&0.85&0.84&0.83&0.79&0.68&0.67&0.60\\
$\kappa^0_{\rho}$&0.19&0.17&0.16&0.15&0.16&0.11&0.10&0.08\\
$\kappa^C_{\rho}$&0.01&0&0&0&-0.03&-0.01&-0.01&-0.18\\
$\kappa^T_{\rho}$&-0.11&-0.11&-0.11&-0.10&-0.06&-0.07&-0.07&-0.07\\
$\kappa$&0.91&0.87&0.84&0.84&0.82&0.69&0.66&0.52\\
\hline
\end{tabular}
\end{center}
\end{table}

\begin{table}[htb]
\caption{\it
Same as Table \protect\ref{tab1},
but for the electric polarizability
modification $\delta \alpha $. The $\rho$ contribution is negligible
in comparison with the $\pi$ contribution.}
\label{tab3}
\begin{center}
\begin{tabular}{|c||c|c|c|c||c|c|c|c||}
\hline
&\multicolumn{4}{c||}{constant density}&\multicolumn{4}{c||}{realistic
density}\\
\hline
&$A \to \infty$& $^{208}$Pb & $^{40}$Ca & $^{12}$C& $^{208}$Pb&$^{40}$Ca&
$^{12}$C&$^4$He\\
\hline
\hline
${\delta \alpha}^0$&-4.17&-3.58&-3.11&-2.59&-3.26&-2.44&-1.94&-1.22\\
${\delta \alpha}^C$&-2.87&-2.42&-2.05&-1.62&-2.21&-1.51&-1.07&-0.41\\
${\delta \alpha}^T$&-0.15&-0.09&-0.04&-0.02&-0.06&0.02&0.08&0.13\\
${\delta \alpha}$&-3.02&-2.51&-2.09&-1.64&-2.27&-1.49&-0.99&-0.28\\
\hline
\end{tabular}
\end{center}
\end{table}

\begin{table}[htb]
\caption{\it
Same as Table \protect\ref{tab1},
but for the magnetic polarizability
modification $\delta \beta $.  The $\rho$ contribution is negligible
in comparison with the $\pi$ contribution.
The subscript $\Delta $ denotes contributions to $\delta \beta $
from diagrams with a $\Delta $-isobar as an intermediate state.}
\label{tab2}
\begin{center}
\begin{tabular}{|c||c|c|c|c||c|c|c|c||}
\hline
&\multicolumn{4}{c||}{constant density}&\multicolumn{4}{c||}{realistic
density}\\
\hline
&$A \to \infty$& $^{208}$Pb & $^{40}$Ca & $^{12}$C& $^{208}$Pb&$^{40}$Ca&
$^{12}$C&$^4$He\\
\hline
\hline
$\delta \beta^0$&1.86&1.55&1.30&1.03&1.42&1.10&0.83&0.47\\
$\delta \beta^C$&1.54&1.31&1.13&0.91&1.25&0.93&0.69&0.30\\
$\delta \beta^T$&0.11&0&-0.10&-0.20&-0.06&-0.21&-0.30&-0.38\\
$\delta \beta^0_{\Delta}$&0.29&0.24&0.19&0.19&0.22&0.17&0.12&0.05\\
$\delta \beta^C_{\Delta}$&0.38&0.33&0.29&0.29&0.30&0.22&0.17&0.09\\
$\delta \beta^T_{\Delta}$&0.30&0.29&0.28&0.28&0.27&0.22&0.21&0.18\\
$\delta \beta$&2.33&1.93&1.60&1.28&1.76&1.16&0.77&0.19\\
\hline
\end{tabular}
\end{center}
\end{table}

\newpage
\setcounter{table}{0}
\section{Exchange form factors}

At low energies the dependence of the mesonic seagull amplitude on
momentum transfer is given by exchange form factors. In this appendix the
numerical results for such form factors are presented on the
basis of the model calculation discussed in section 9. Notations for the
different types of exchange form factors are given in Table \ref{exchange1}.

\begin{table}[hb]
\caption{\it
Notations for the different form factors and their formulation
within a specific model.}
\label{exchange1}
\begin{center}
\begin{tabular}{|c|c|c|}
	\hline
	{\bf notation} & {\bf description} & {\bf model realization}  \\
	\hline
        $F_1$ & one-nucleon form factor & taken from ref.\ \cite{datatab}  \\
	\hline
	$F_2^{(1)}$ & exchange form factor for the static &
	$F_2^{(1)}(q) = \Phi _1(q)/\Phi _1(0)$  \\
	 & part of the mesonic
	seagull amplitude &   \\
	\hline
	$F_2^{(2)}$ & same as $F_2^{(1)}$, but for  &
	$F_2^{(2)}(q) = \Phi _2(q)/\Phi _2(0)$  \\
	 & the  energy-dependent part &   \\
	\hline
	$F_2^{(GR)}$ & same as $F_2^{(1)}$, but for  &
	$F_2^{(GR)}(q)=\left. {\left( {{\Phi _1(q) /
      \Phi _1(0)}} \right)} \right|_{g_T=0}$  \\
	 & the giant resonance
	contribution only &   \\
	\hline
	$F_2^{(QD)}$ & same as $F_2^{(1)}$, but for  &
	$F_2^{(QD)}(q)=\left. {\left( {{\Phi _1(q) /
    \Phi _1(0)}} \right)} \right|_{g_C=0}$   \\
	 & the quasideuteron
	contribution only &   \\
	\hline
\end{tabular}
\end{center}
\end{table}

Figure \ref{pr12} (for $^{208}$Pb) and Fig.\ \ref{pr14} (for $^{12}$C)
show the form factors in comparison
with both, the nuclear charge form factor and the form factor for
uncorrelated nucleon pairs.
Furthermore, our model prediction for the
giant resonance and quasideuteron exchange form factors are shown for
for $^{40}$Ca (Fig.\ \ref{pr15}) and $^{12}$C (Fig.\ \ref{pr16}).
Again in
these two figures the
nuclear charge form factor and the approximation of uncorrelated nucleon
pairs is shown for the sake of comparison.

Table  \ref{exchange2} gives the radii corresponding to the exchange form
factors shown in Figs.\ \ref{pr12}, \ref{pr14} and
\ref{pr13}, which are obtained by
representing the functions $F_2^{(i)}$ in the form
$F_2^{(i)} = 1 - q ^2 r^2_i/6$.

\begin{table}[htb]
\caption{\it
Numerical values for the radii appearing in the expansion of the
exchange form factors from Figs.
(\protect\ref{pr12}), (\protect\ref{pr14}) and
(\protect\ref{pr13}).}
\label{exchange2}
\begin{center}
	\begin{tabular}{|c|c|c|}
	\hline
        $A$ & $r_1$ [fm] & $r_2$ [fm]
                        \rule[-1ex]{0ex}{2.5ex} \\  \hline
        208 & 5.0 & 4.7 \rule[-1ex]{0ex}{2.5ex} \\  \hline
        40 & 3.0 & 2.5  \rule[-1ex]{0ex}{2.5ex} \\  \hline
        12 & 1.9 & 1.4  \rule[-1ex]{0ex}{2.5ex} \\  \hline
\end{tabular}
\end{center}
\end{table}

\begin{figure}[htb]
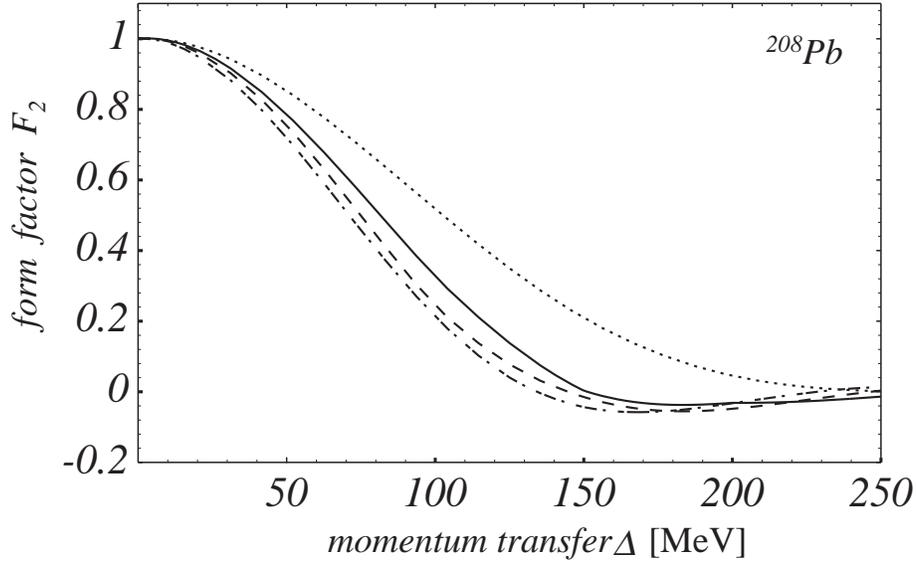

\caption{\it
Exchange form factors $F_2^{(i)}(q)$ for the case of $^{208}$Pb.
The dashed curve represents $F_2^{(1)}$, while the full curve
corresponds to $F_2^{(2)}$. For comparison the (experimental) charge
form factor $F_1$  (dash-dotted curve) and the function
$F_1^2(q/2)$ (dotted curve) are also shown. The parameters for the
form factor $F_1$ were taken from ref.\ \protect\cite{datatab}.}
\label{pr12}
\end{figure}

\begin{figure}[htb]
\caption{\it
Same as Fig.\ \protect\ref{pr12}, but for $^{12}$C.}
\label{pr14}
\end{figure}

\begin{figure}[htb]
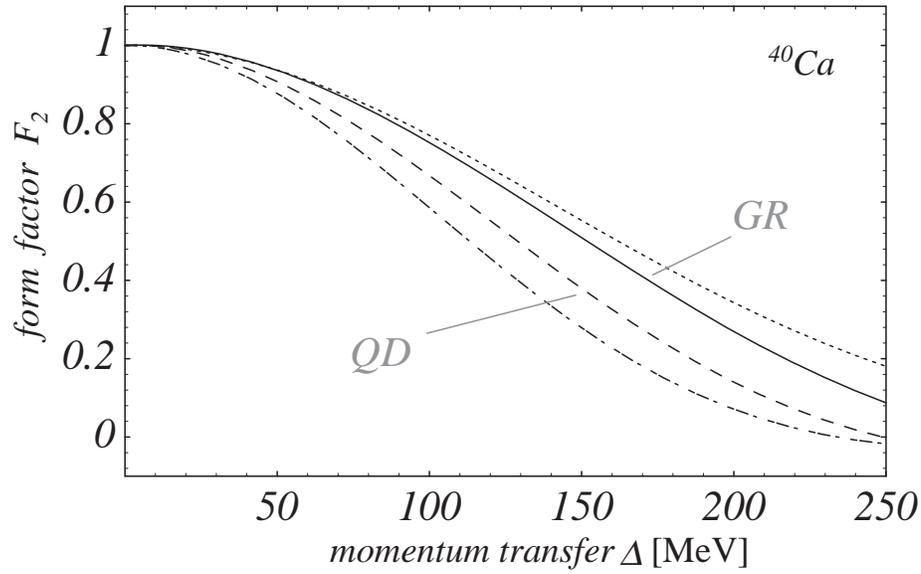

\caption{\it
Form factors $F_2^{(GR)}(\Delta )$ and
$F_2^{(QD)}(\Delta )$for $^{40}$Ca. The dashed
curve is $F_2^{(QD)}$ and the full curve is $F_2^{(GR)}$. In
addition, the (experimental) charge form factor $F_1$
(dash-dotted curve) and the
function $F_1^2(\Delta /2)$ (dotted curve) are shown for the
sake of comparison.}
\label{pr15}
\end{figure}

\begin{figure}[htb]
\caption{\it
Same as Fig.\ \protect\ref{pr15}, but for $^{12}$C.}
\label{pr16}
\end{figure}

\vfill


\clearpage
\section{Multipole angular distribution functions}

Here the partial-wave expansion of the seagull
amplitude (\ref{FormMultipoles})
given in Section 3 in terms of the functions
$g^{\lambda L}$ is derived. When both photons have definite
helicities $\lambda_1,\lambda_2=\pm1$ and the scattering plane
is $xz$, these functions read
$$
   g^{El}(\theta) = d^l_{\lambda_1,\lambda_2}(\theta), \quad
   g^{Ml}(\theta) = \lambda_1\lambda_2 \,
     d^l_{\lambda_1,\lambda_2}(\theta).
$$
To expand the product $\veceps_1\cdot \veceps_2^*
\,P_l(\cos\theta)$, we use $\veceps_1\cdot \veceps_2^* =
(1+\lambda_1\lambda_2 \cos\theta)/2 = d^1_{\lambda_1,\lambda_2}(\theta)$
and $P_l(\cos\theta)=d^l_{0,0}(\theta)$ and apply the general
rule of adding angular momenta for the $d$-functions:
$$
   d^{j_1}_{m_1,n_1}(\theta) \,d^{j_2}_{m_2,n_2}(\theta)
    = \sum_{jmn}   C^{jm}_{j_1 m_1, j_2 m_2}
       C^{jn}_{j_1 n_1, j_2 n_2} d^j_{m,n}(\theta).
$$
Using specific values for the Clebsch-Gordan coefficients
$C^{j\lambda}_{1\lambda, l0}$ with $\lambda=\pm 1$, we get
$$
  d^1_{\lambda_1,\lambda_2}(\theta)\, d^l_{0,0}(\theta) =
       \frac{l+2}{2(2l+1)} d^{l+1}_{\lambda_1,\lambda_2}(\theta)
     + \frac{\lambda_1\lambda_2}{2} d^l_{\lambda_1,\lambda_2}(\theta)
     + \frac{l-1}{2(2l+1)} d^{l-1}_{\lambda_1,\lambda_2}(\theta).
$$
Coming back to the functions $g^{\lambda L}$, we find
$$
   G_l(\theta) \equiv (2l+1)\veceps_1\cdot \veceps_2^* \,P_l(\cos\theta)
      = \frac{l+2}{2} g^{E(l+1)}(\theta)
     + \frac{2l+1}{2} g^{Ml}(\theta)
     + \frac{l-1}{2} g^{E(l-1)}(\theta)
$$
which is Eq.\ (\ref{GlMultipoles}).

\section*{Acknowledgements}

The authors are indebted to Prof.\ G.E.~Brown for his continuing
interest to this work. They thank Deutsche Forschungsgemeinschaft
for support of several visits of A.I.M.\ and A.I.L.\ in
G\"ottingen and Deutscher Akademischer Austauschdienst for
support of visits of M.T.H.\ in Novosibirsk. One of us (M.S.)
thanks B.~Schr\"oder (MAX--LAB, Lund) and B.~Ziegler, J.~Ahrens
and Th.~Walcher (MAMI, Mainz) for fruitful cooperations during
the experiments and Deutsche Forschungsgemeinschaft for the 
support of the experimental work.

\clearpage

\clearpage
\def\thesection{\arabic{section}}


\setcounter{section}{2}
\setcounter{figure}{0}

\begin{figure}[htb]
\vspace*{-3cm}
\centering\epsfig{figure=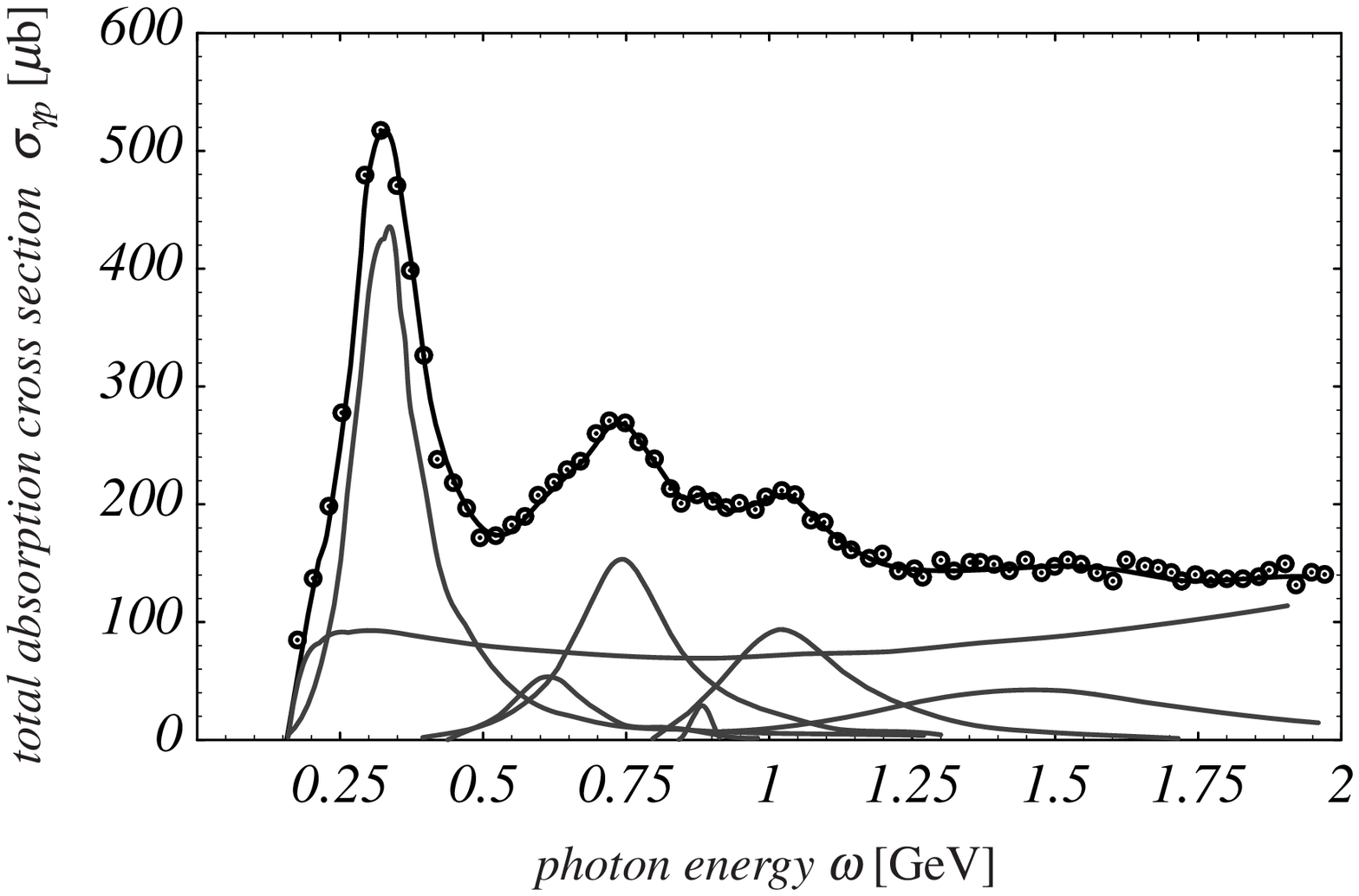,height=12cm}
\vspace*{-2cm}
\caption{\em Photoabsorption cross section of the proton schematically
separated into partial cross sections \cite{armstrong72}  containing
nucleon resonances and a nonresonant component.}
%
%
\vspace*{-1cm}
\centering\epsfig{figure=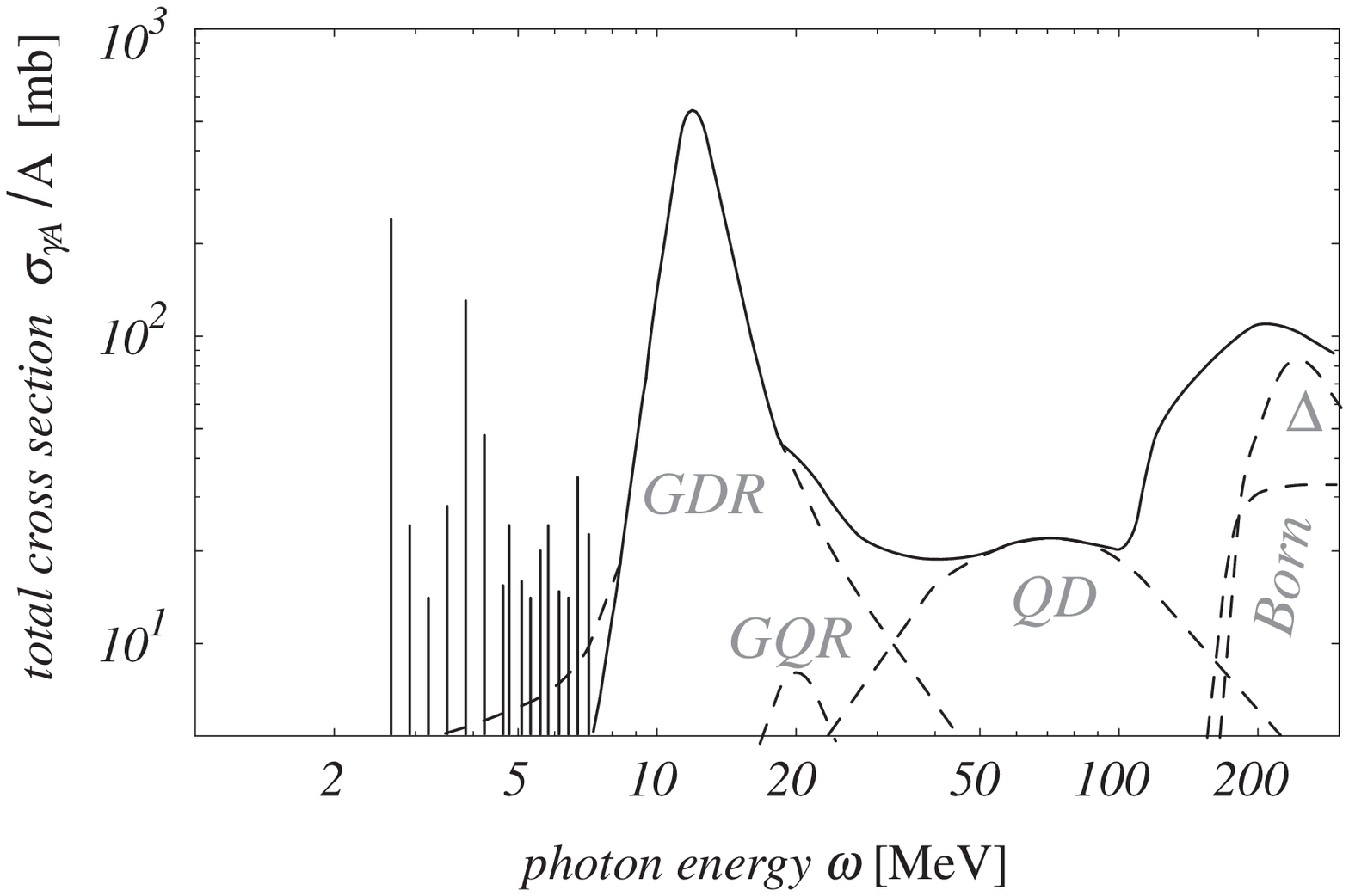,height=12cm}
\vspace*{-2cm}
\caption{\em Schematic view of the nuclear photoabsorption cross section
of $^{208}$Pb
from the giant resonance to the $\Delta$ resonance range. GDR:
giant-dipole resonance, GQR: isovector giant-quadrupole resonance,
QD: quasideuteron mode, $\Delta$: $Delta$ resonance of nucleons in the
nucleus, Born: nonresonant photoexcitation of nucleons in the nucleus
through the Born terms of the photo-pion amplitudes  \cite{schumacher88Conf}.}
\end{figure}

\begin{figure}[htb]    
\vspace*{-4cm}
\centering\epsfig{figure=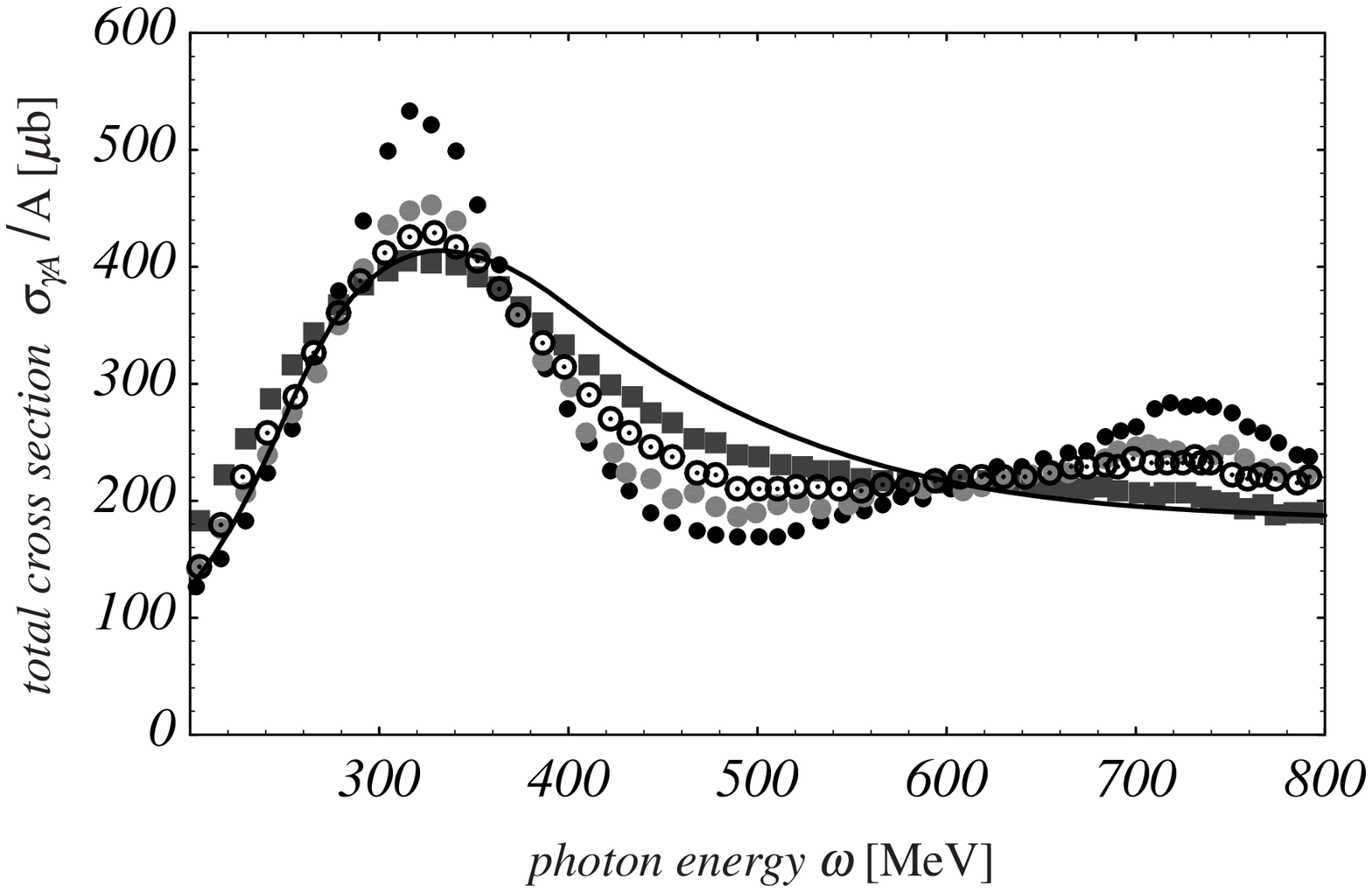,height=12cm}
\vspace*{-2cm}
\caption{\em Total photoabsorption cross sections per nucleon. Data are
shown for $^1$H (black circles), $^2$H (gray circles), $^3$He (open circles)
and $^4$He (grey squares) \cite{MacCormic97}.
The solid line represents the universal curve for complex nuclei.}
%
%
\vspace*{-1cm}
\centering\epsfig{figure=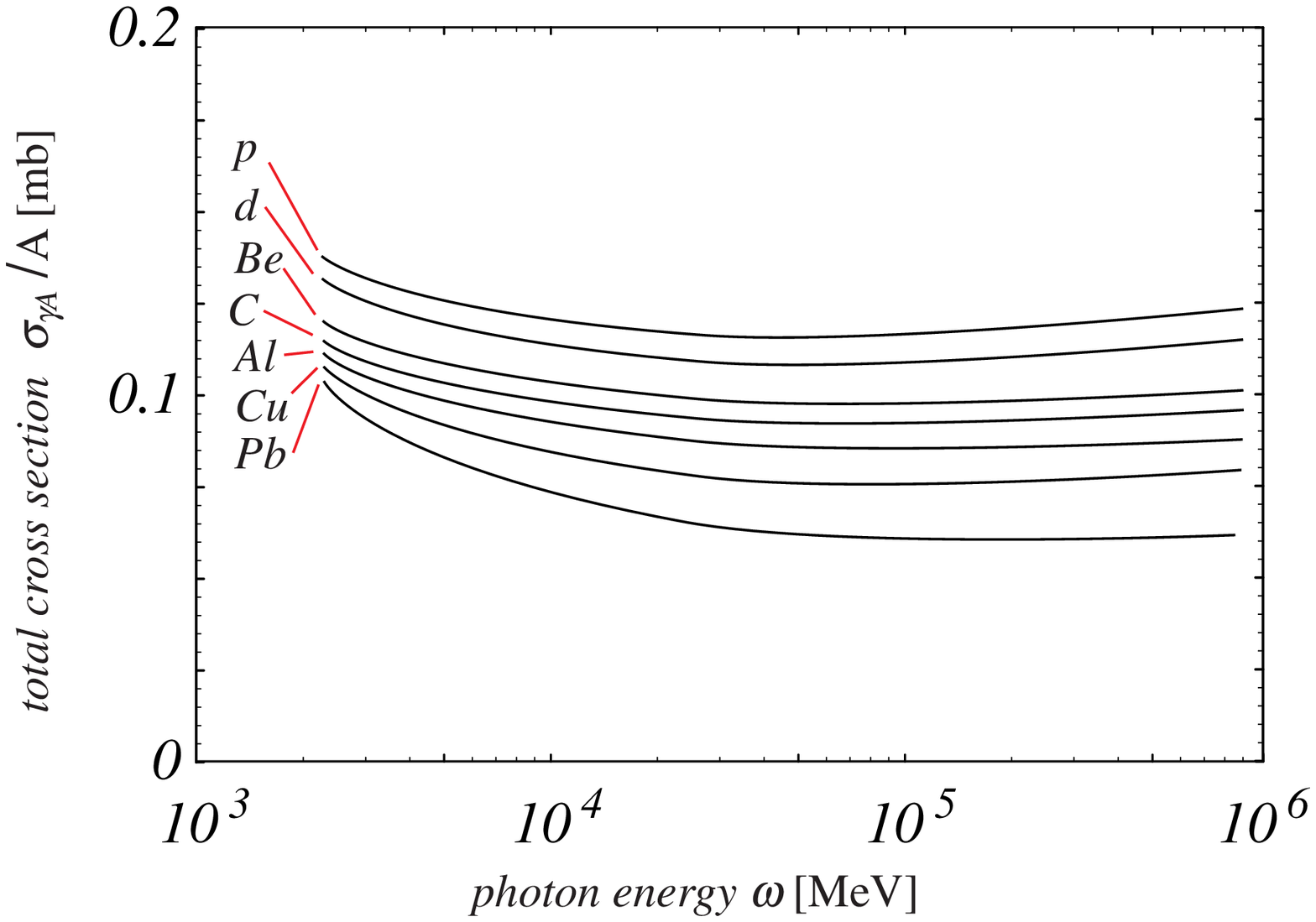,height=12cm}
\vspace*{-2cm}
\caption{\em Fits to the total nuclear photoabsorption cross sections
per nucleon as an example of shadowing in the asymptotic region
\cite{ahrens85c}.}
\end{figure}

\begin{figure}[htb]  
\vspace*{-4cm}
\centering\epsfig{figure=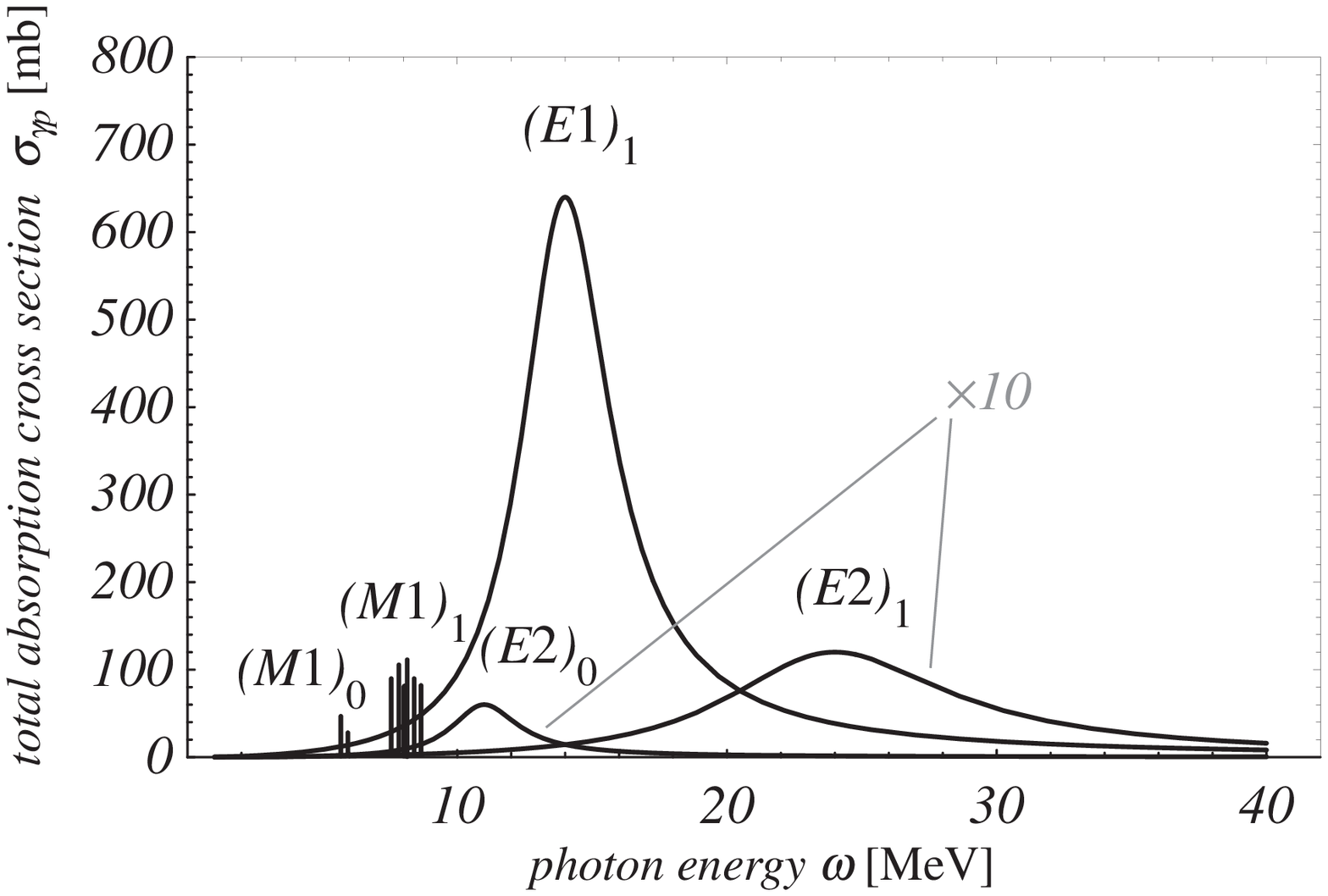,height=12cm}
\vspace*{-2cm}
\caption{\it Schematic representation of  giant resonance
multipoles for $^{208}$Pb. (E1)$_1$: isovector giant electric-dipole
resonance (GDR), (M1)$_0$: isoscalar  giant magnetic-dipole
resonance, (M1)$_1$: isovector giant magnetic-dipole
resonance,(E2)$_0$: isoscalar giant  electric-quadrupole
resonance, (E2)$_1$: isovector giant  electric-quadrupole
resonance \cite{SC94}.}
%
%
\vspace*{-1cm}
\centering\epsfig{figure=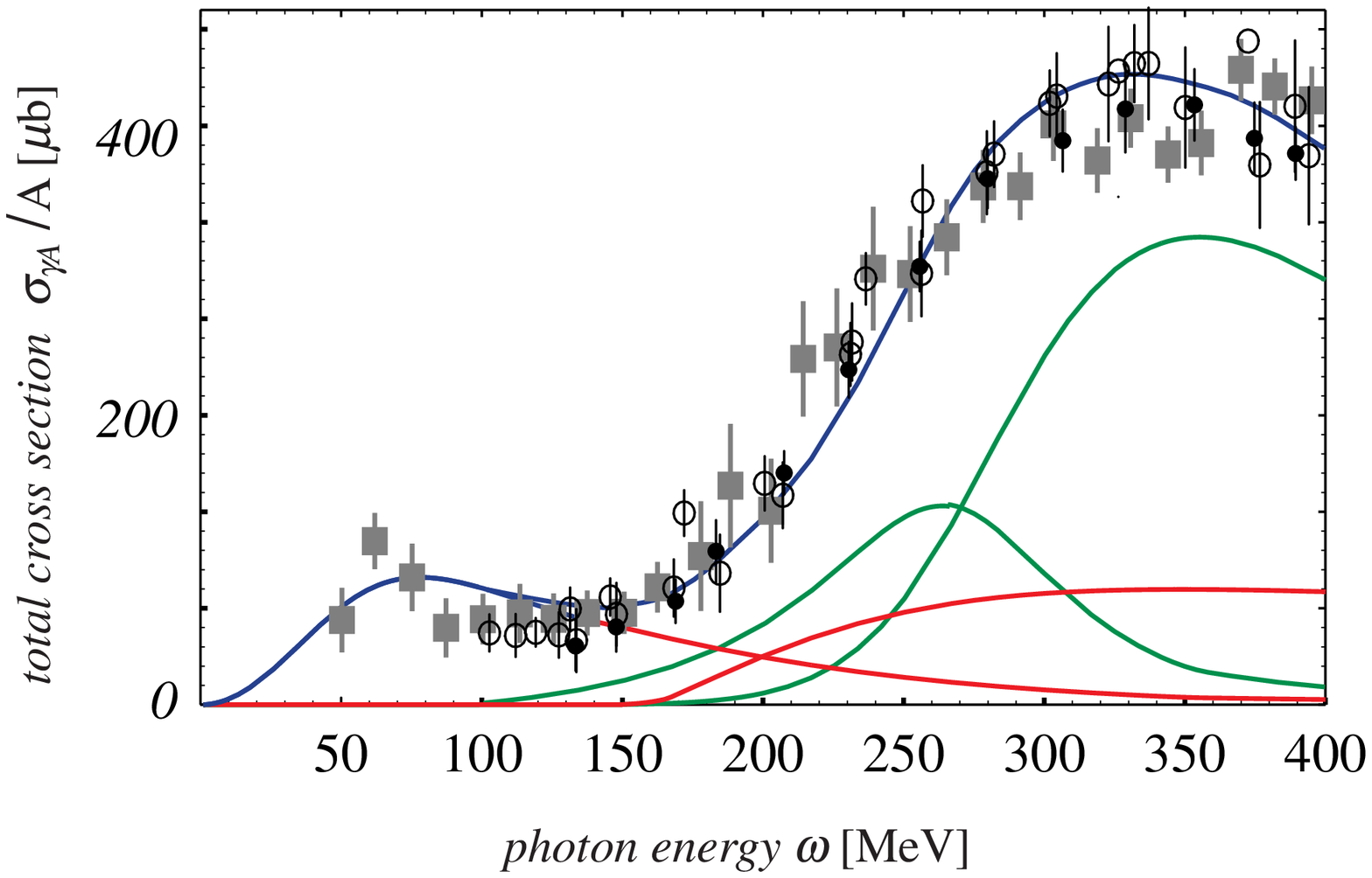,height=12cm}
\vspace*{-2cm}
\caption{\em Total nuclear photoabsorption cross sections per nucleon.
Data are shown for nuclei between $^4$He and $^{238}$U
\cite{ahrens85c}. In addition the result of a model
calculation for the universal photoabsorption curve is shown
\cite{HuettMiSchu97}.
The left curve extending underneath the $\Delta$ resonance is the
QD cross section. The $\Delta$-resonance cross section is partitioned
into  two curves, into a non-mesonic part (left resonant curve) and a
mesonic part (right resonant curve). The nonresonant curve
in the $\Delta$ range corresponds to
the Born term of meson photoproduction.}
\end{figure}


\setcounter{section}{4}
\setcounter{figure}{0}

\begin{figure}[htb]
\vspace*{-4cm}
\centering\epsfig{figure=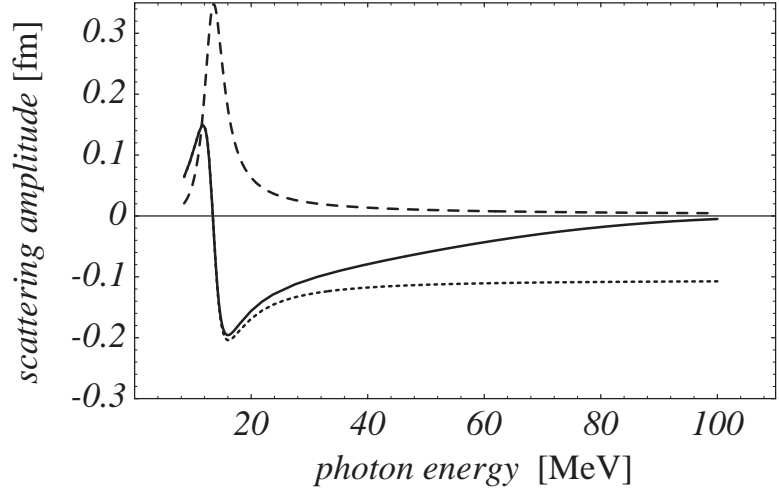,height=10cm}
\centering\epsfig{figure=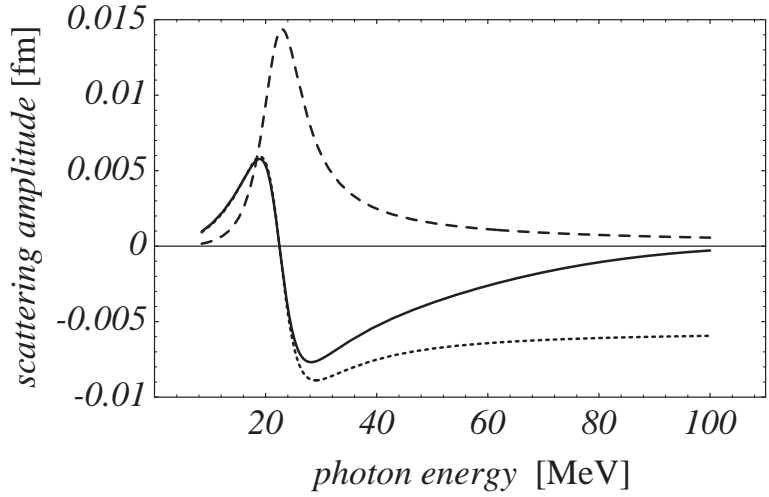,height=10cm}
\vspace*{-1cm}
\caption{\it Scattering amplitudes for the giant-dipole resonances
(upper figure) and isovector giant-quadrupole resonance (lower figure)
of $^{208}$Pb in the forward direction: Im~$R^{\lambda L}(\omega,\theta=0)$
(dashed),  Re~$\tilde R^{\lambda L}(\omega,\theta=0)$ (dotted) and
Re~${\hat  R}^{\lambda L}(\omega,\theta=0)$ (solid).
}
\end{figure}

\begin{figure}[htb] 
\vspace*{-8cm}
\centering\epsfig{figure=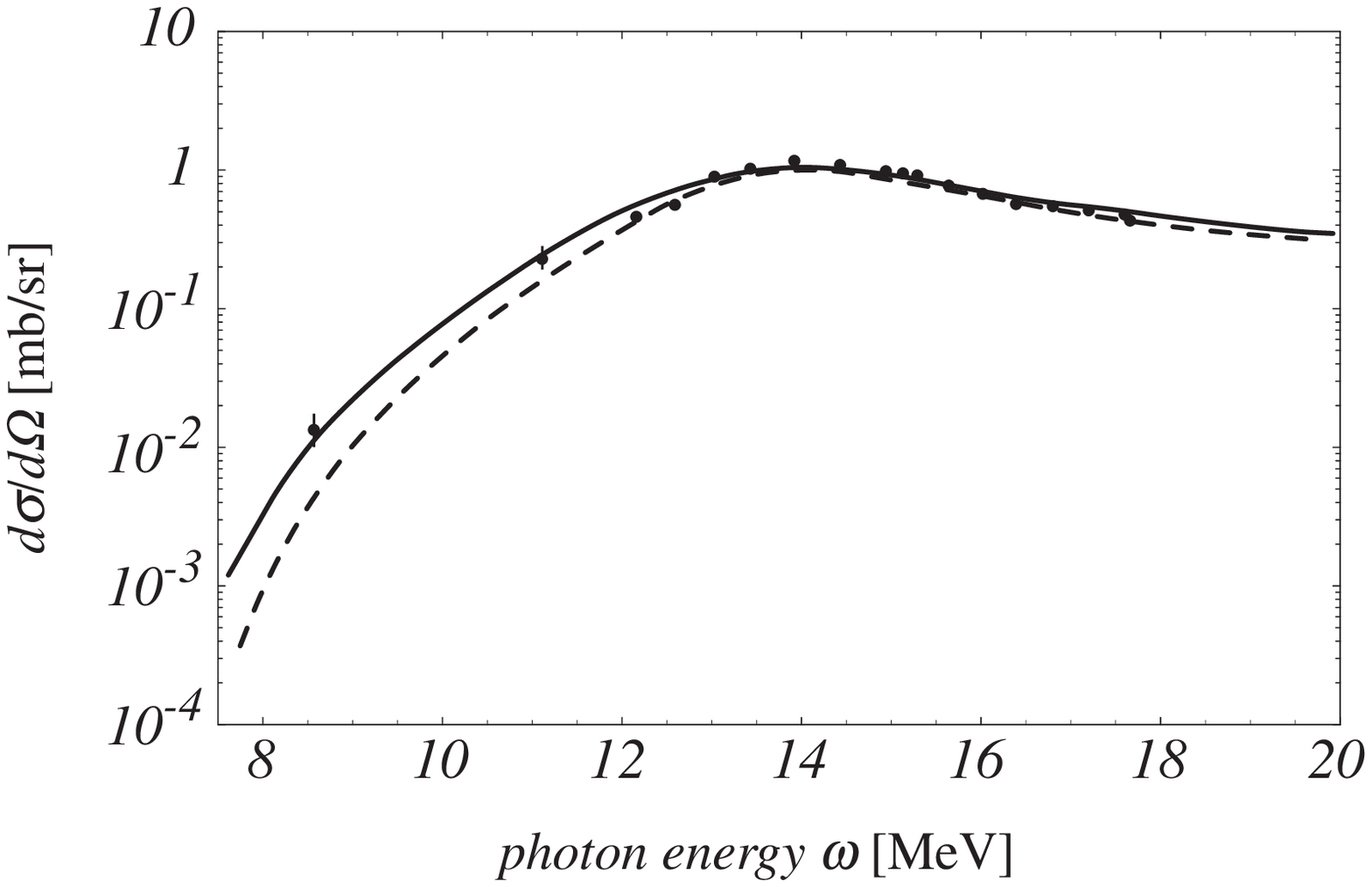,height=12cm}
\vspace*{-2cm}
\caption{\em Differential cross sections for Compton scattering by
$^{209}$Bi. Scattering angle $\theta$ = 135$^{\circ}$. Data points at
9.0, 11.4 and 17.74 MeV are
measured by \cite{nolte86}. Other data  points are measured by  \cite{dale88}.
Solid curve: calculated using the GDR parameters of \cite{nolte86}.
Dashed curve: calculated using the GDR parameters of \cite{dale88}.}
%
%
\vspace*{-2cm}
\centering\epsfig{figure=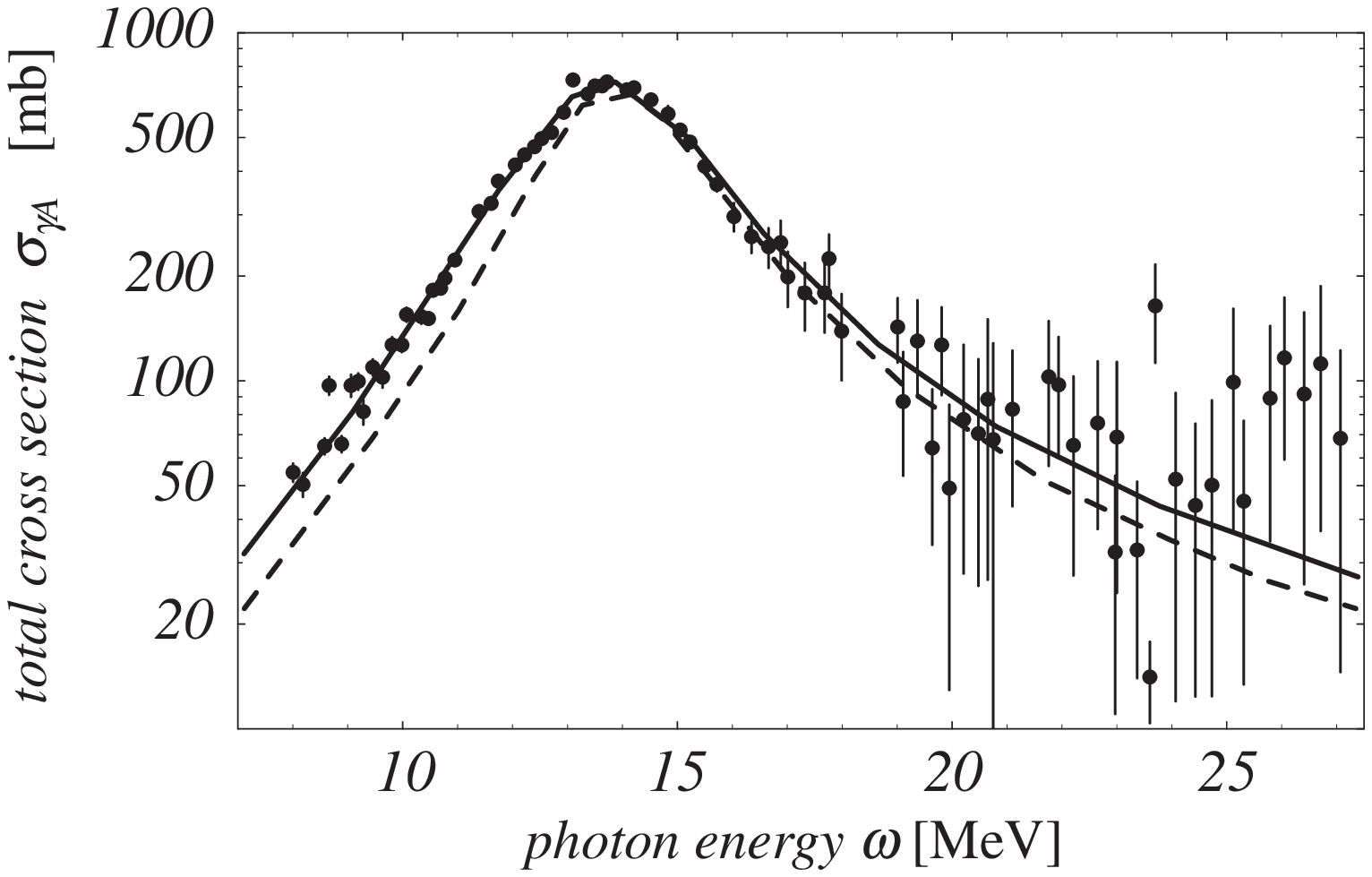,height=12cm}
\vspace*{-1cm}
\caption{\em Photoabsorption cross sections for  $^{209}$Bi
measured by \cite{harvey64}  multiplied by a scaling factor of 1.35.
Solid curve:
calculated using the GDR parameters of \cite{nolte86}. Dashed curve: calculated
using the GDR parameters of \cite{dale88}.}
\end{figure}

\begin{figure}[htb]  
\vspace*{-8cm}
\centering\epsfig{figure=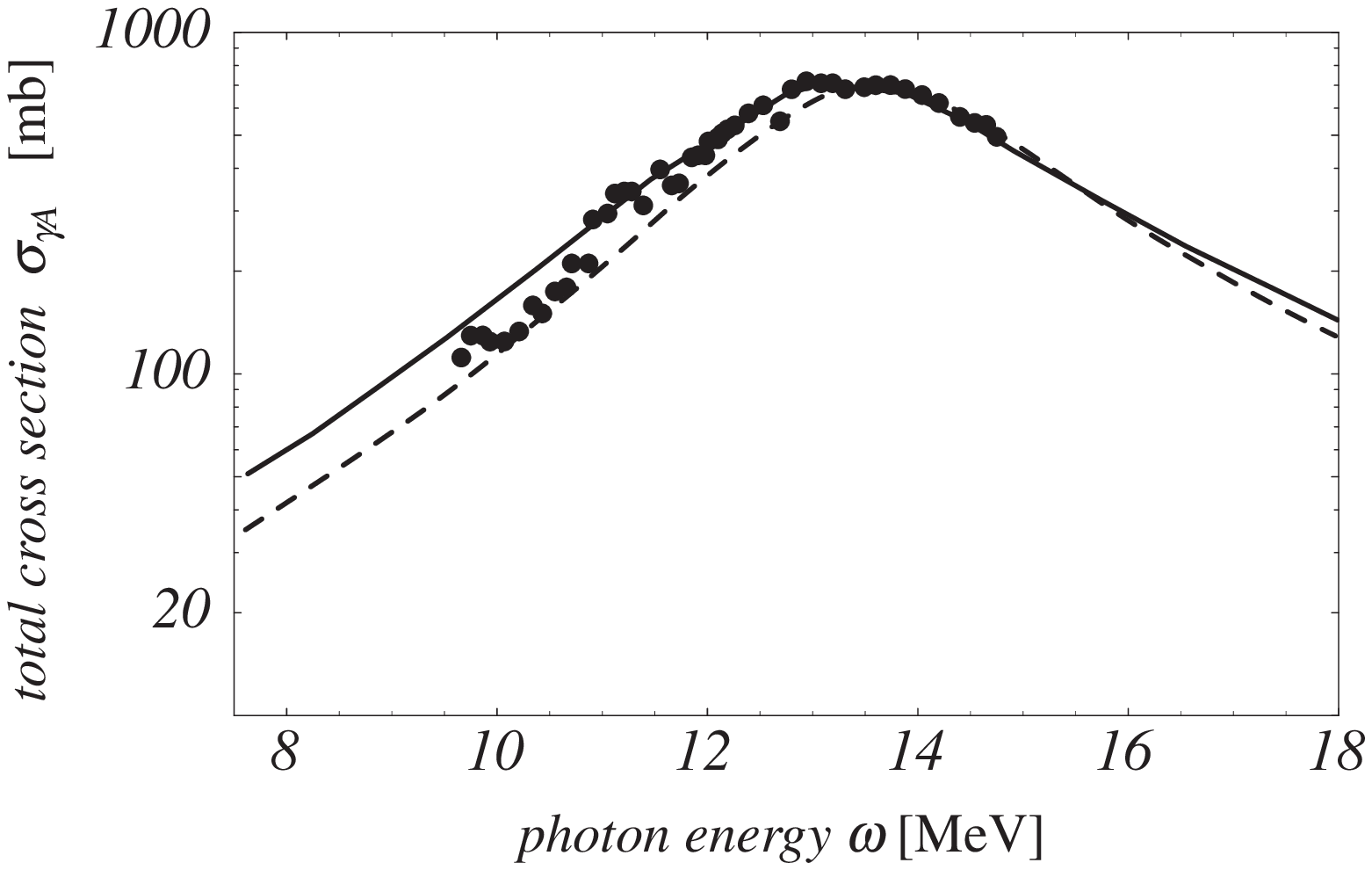,height=14cm}
\vspace*{-2cm}
\caption{\em Photoabsorption cross sections for $^{209}$Bi
measured by \cite{young72} multiplied by a scaling factor of 1.09. Solid curve:
calculated using the GDR parameters of \cite{nolte86}.
Dashed curve: calculated using the GDR parameters of \cite{dale88}.}
%
%
\vspace*{-1cm}
\centering\epsfig{figure=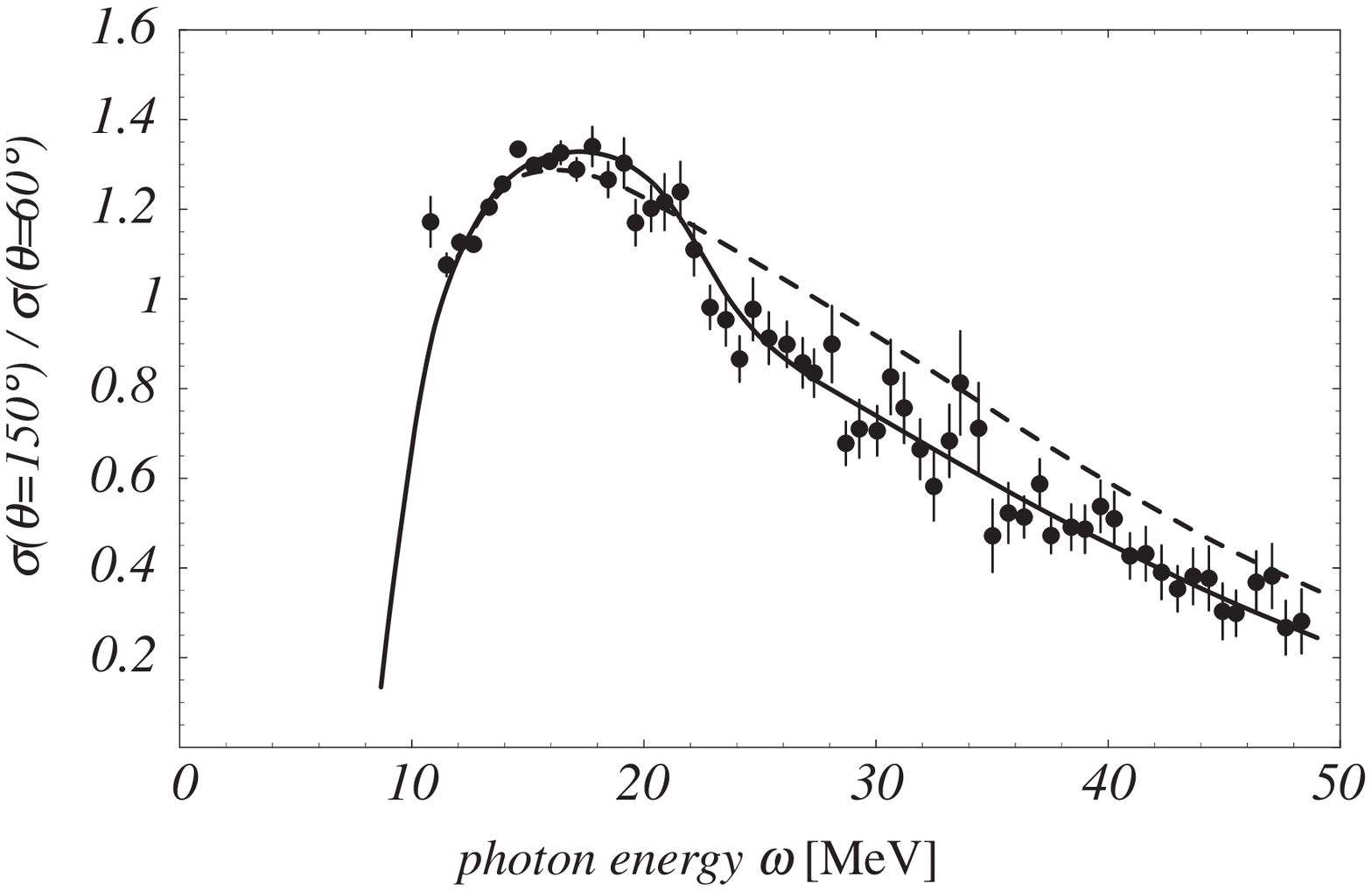,height=12cm}
\vspace*{-2cm}
\caption{\em  Cross-section ration
$\sigma(\theta=150^{\circ})/\sigma(\theta=60^{\circ})$
for Compton scattering of unpolarized photons by $^{208}$Pb \cite{S8}.
Solid line: Calculated including the IVGQR. Dashed line: Calculated not
including the IVGQR.
The IVGQR shows up as
interference of the $E2$ amplitude with the predominant $E1$ amplitude from
the GDR.}
\end{figure}

\setcounter{section}{5}
\setcounter{figure}{0}

\begin{figure}[htb]
\vspace*{-5cm}
\centering\epsfig{figure=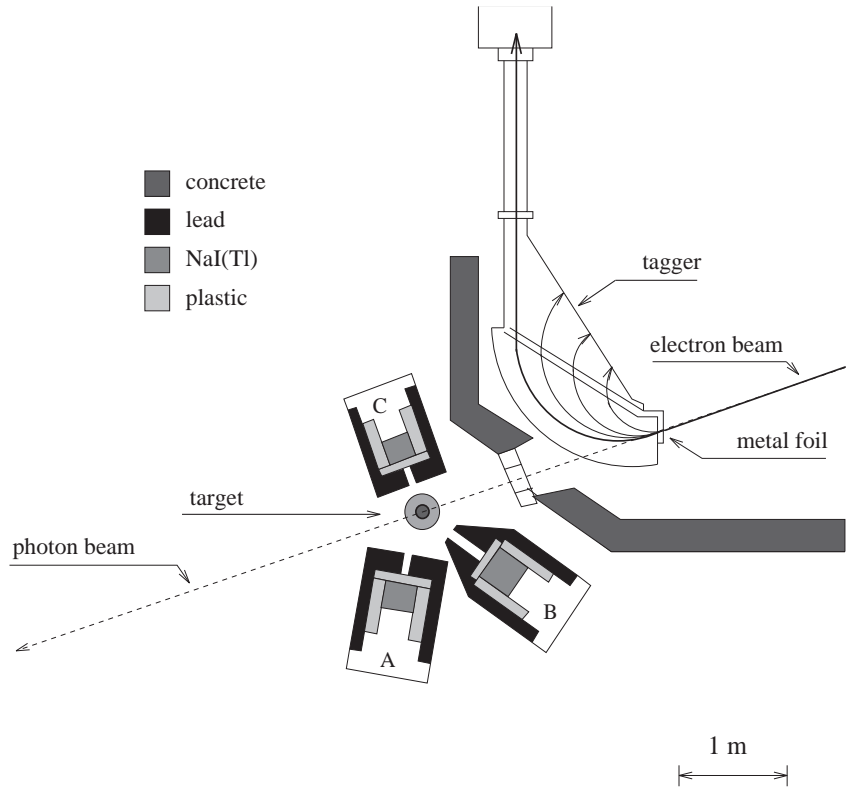,width=20cm}
\caption{\em Typical experimental arrangement as used at the MAX laboratory
of the university of Lund (Sweden) \cite{adler90,adler97}.
The photon beam having
an energy of about 100 MeV hits a thin metal foil serving as a
bremsstrahlung radiator. Quasi monochromatic photons are produced
through a coincidence condition  between an event in one of the
NaI(Tl) detectors (A, B, C) and an event in the tagger.}
\end{figure}

\begin{figure}[t]
\vspace*{-3cm}
\centering\epsfig{figure=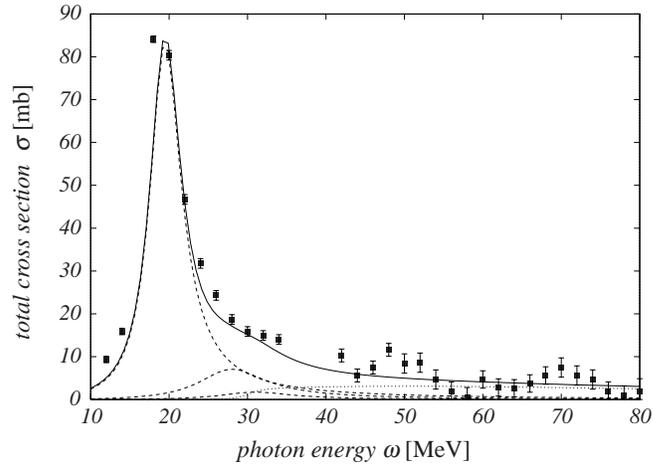,height=8cm}
\centering\epsfig{figure=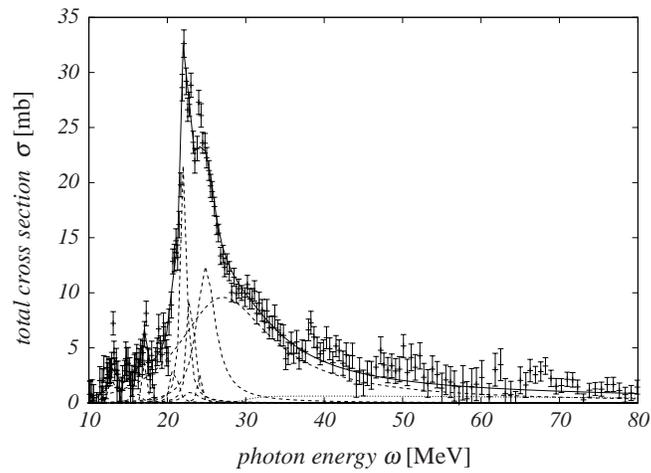,height=8cm}
\caption{\em
Total photoabsorption cross sections \cite{ahrens75}
for $^{40}Ca$ (upper figure) and
$^{16}O$ (lower figure)
partitioned into components having Lorentzian shapes. Dashed curves: GR
components. Dotted curves: QD components.
}
\end{figure}

\begin{figure}[htb]
\vspace*{-3cm}
\centering\epsfig{figure=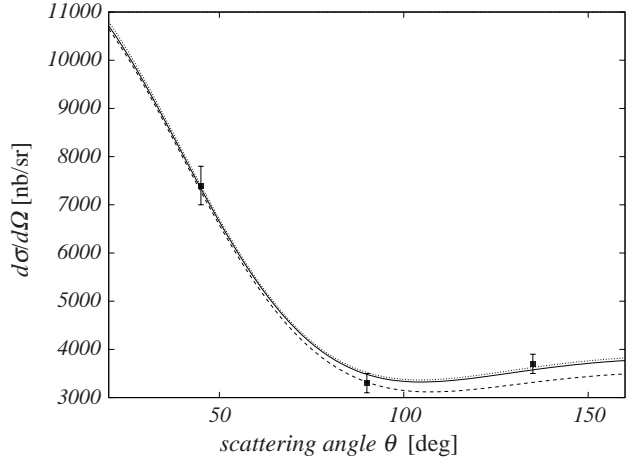,height=8cm}
\centering\epsfig{figure=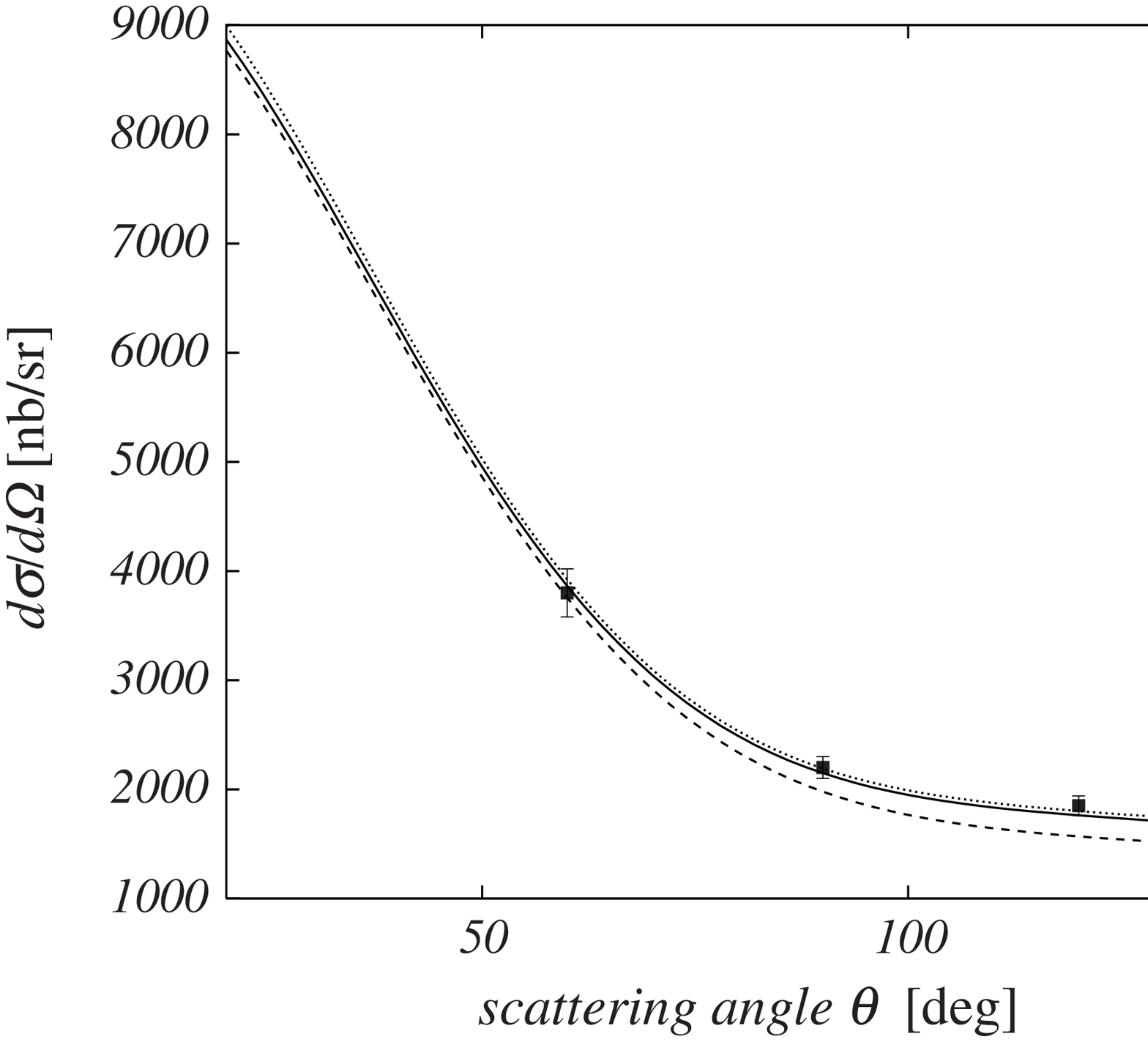,height=8cm}
\caption{ \em Experimental elastic differential cross
sections \cite{P16,P18,P18a}  for $^{40}$Ca
versus scattering angle compared with predictions.
E$_{\gamma}$ = 58 MeV (upper Figure),
E$_{\gamma}$ = 75 MeV (lower Figure).
The curves are calculated for  (i) the free-nucleon electromagnetic
polarizabilities (dashed), (ii) the free-nucleon electromagnetic
polarizabilities supplemented by meson exchange corrections predicted for
the finite nucleus (solid), and (iii) the free-nucleon electromagnetic
polarizabilities supplemented by meson exchange corrections predicted for
nuclear matter (dotted).
}
\end{figure}

\begin{figure}[htb]
\vspace*{-2cm}
\centering\epsfig{figure=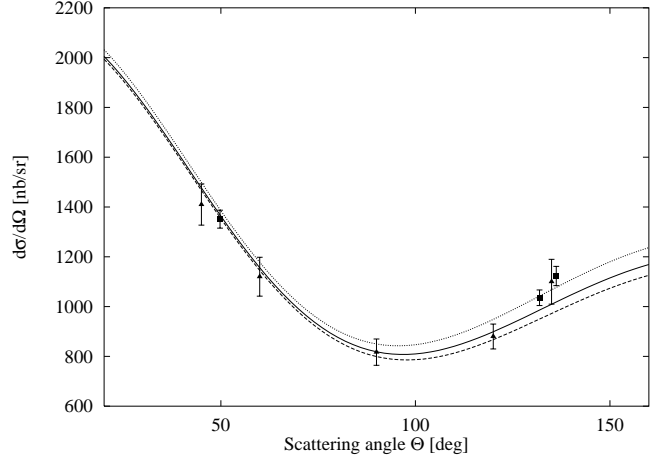,angle=-90,width=9cm}
\centering\epsfig{figure=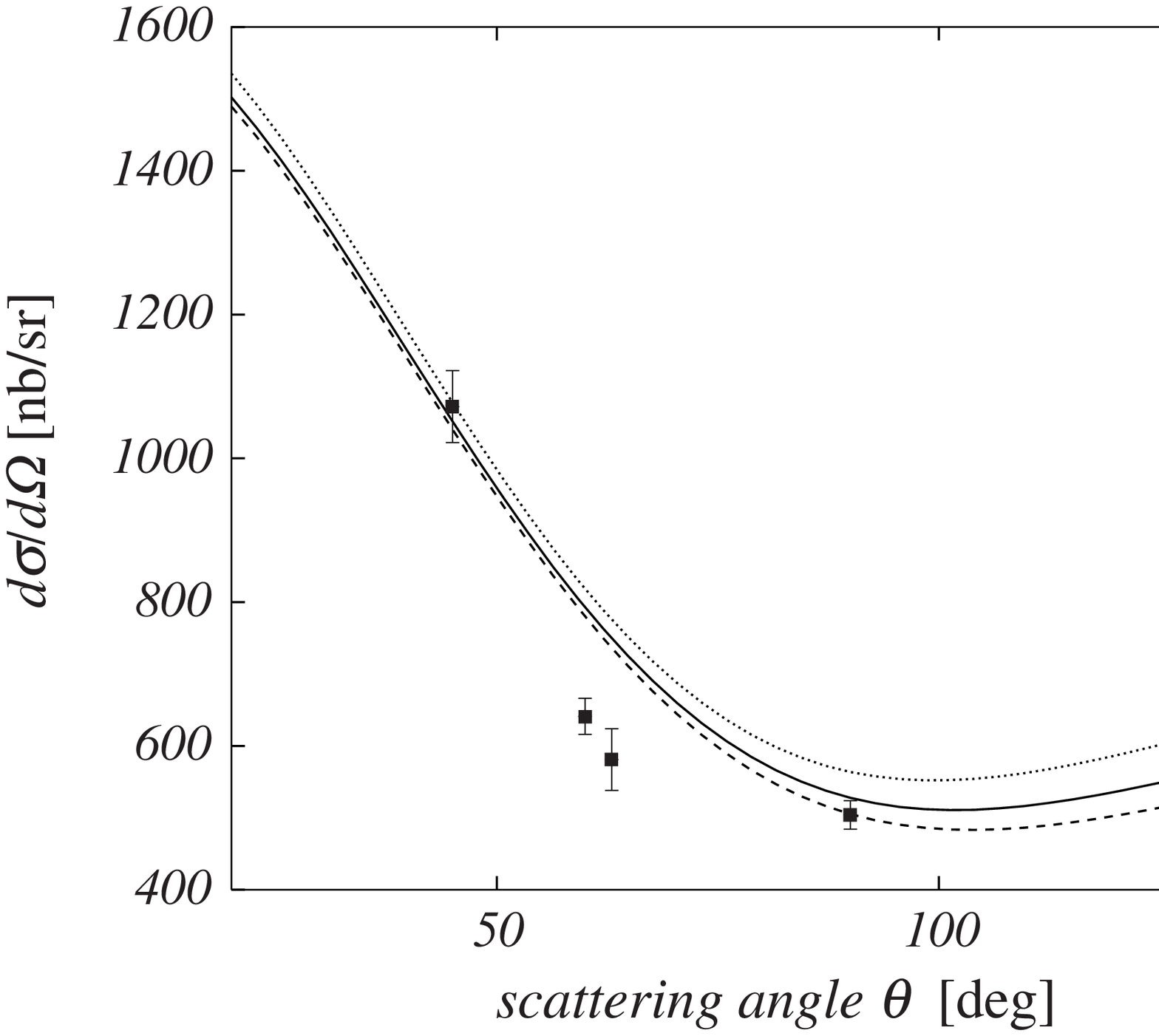,height=8cm}
\caption{\em Differential cross sections \cite{H15,P18,P18a}
for Compton scattering by $^{16}O$
versus scattering angle compared with predictions. $E_{\gamma}$ = 58 MeV
(upper figure); $E_{\gamma}$ = 75 MeV (lower figure).
The curves are calculated for (i) the free-nucleon electromagnetic
polarizabilities (dashed), (ii) the free-nucleon electromagnetic
polarizabilities supplemented by meson exchange corrections predicted for
the finite nucleus (solid), and (iii) the free-nucleon electromagnetic
polarizabilities supplemented by meson exchange corrections predicted for
nuclear matter (dotted).
}
\end{figure}

\setcounter{section}{6}
\setcounter{figure}{0}

\begin{figure}[htb]
\vspace*{-1cm}
\centering\epsfig{figure=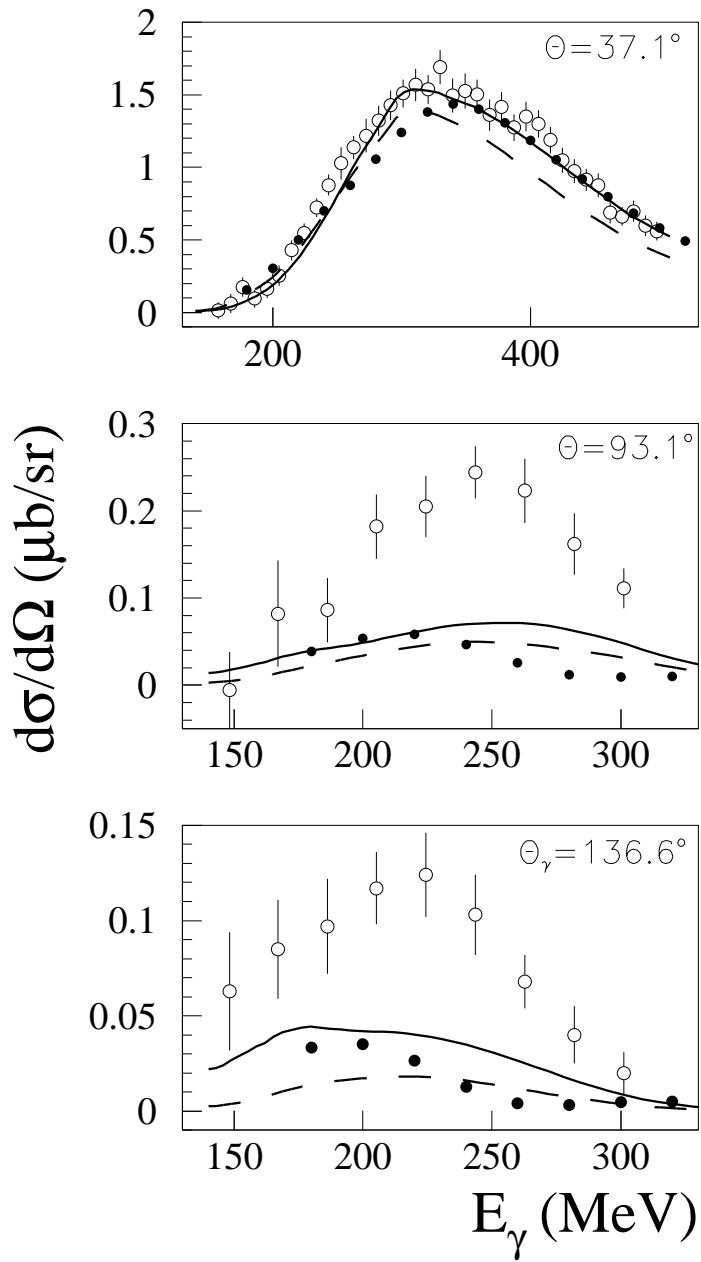,height=18cm}
\caption{\em Differential cross sections $(d\sigma/d\Omega)_{LAB}$
for Compton scattering by  $^4$He \cite{K18} compared with predictions.
The full circles are calculated within the $\Delta$-hole model
including the seagull amplitudes \cite{K18}.
The results of the schematic model (dashed)
and the extended schematic model (solid) are also shown \cite{K18}.}
\end{figure}

\begin{figure}[htb]
\vspace*{-4cm}
\centering\epsfig{figure=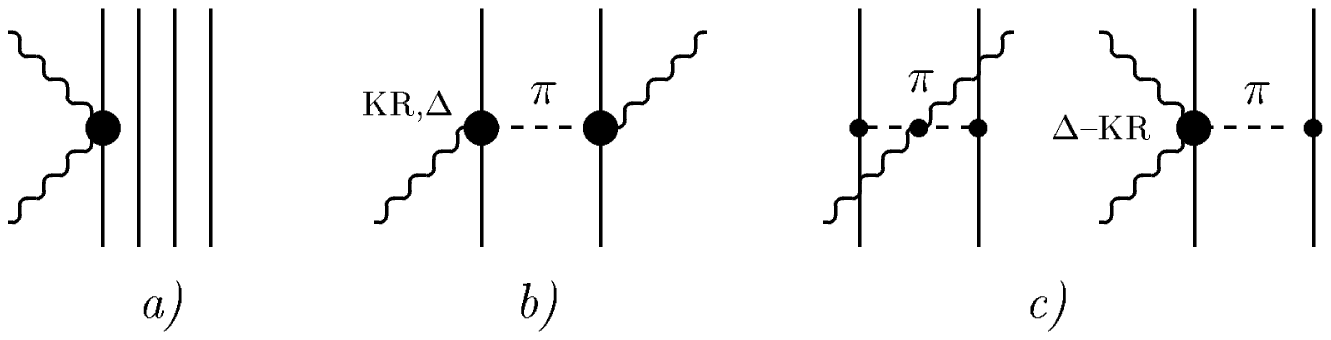,
bbllx=35,bblly=635,bburx=470,bbury=760,height=4cm}
\caption{\em
One- and two-body contributions to nuclear Compton scattering.
Diagram {\it a} is the impulse approximation.
Diagram {\it b} describes real pion production followed by
pion absorption.
Diagrams {\it c} are required by the gauge invariance; in particular,
the last diagram contains the contact $\gamma\pi N\Delta$ vertex.}
%
\vspace*{2cm}
\centering\epsfig{figure=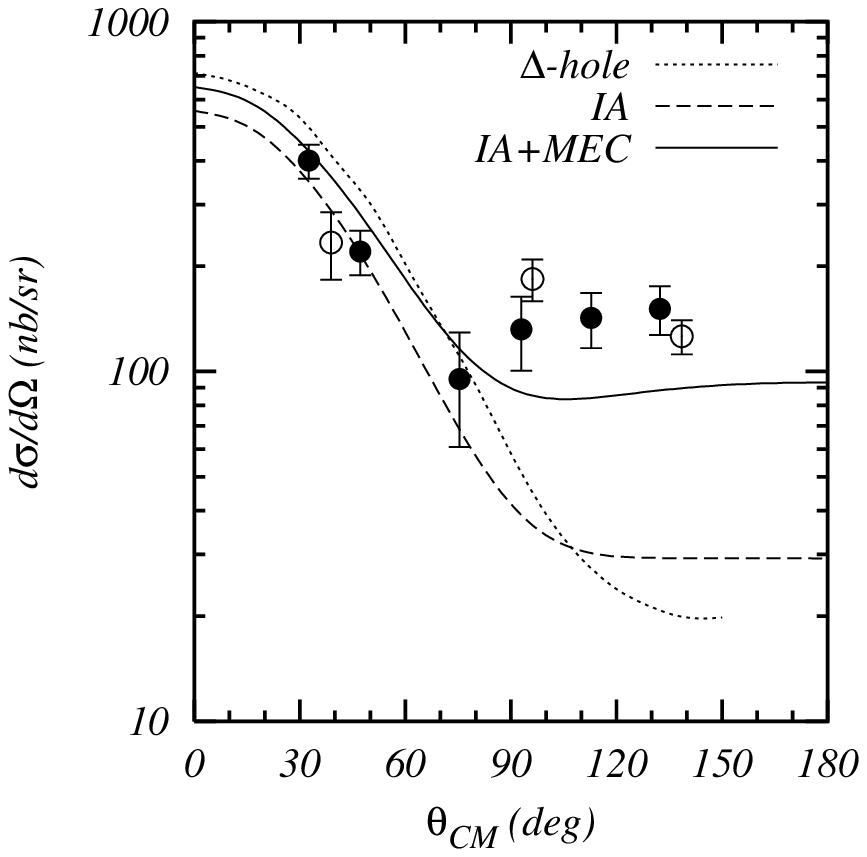,height=10cm}
\vspace*{1cm}
\caption{\em
 Differential cross sections $(d\sigma/d\Omega)_{CM}$
for Compton scattering by $^4$He at $E_\gamma = 206$ MeV.
Data are from \protect\cite{B17} (solid circles) and from
\protect\cite{K18} (open circles). Shown are predictions
of the $\Delta$-hole model in the local-density approximation
\protect\cite{pas96}, impulse approximation, Eq.\ (\protect\ref{IA}),
and the total contribution including the two-body MEC,
Eq.\ (\protect\ref{MEC}).}
\end{figure}

\clearpage

\setcounter{section}{7}
\setcounter{figure}{0}

\begin{figure}[htb]
\vspace*{-2cm}
\centering\epsfig{figure=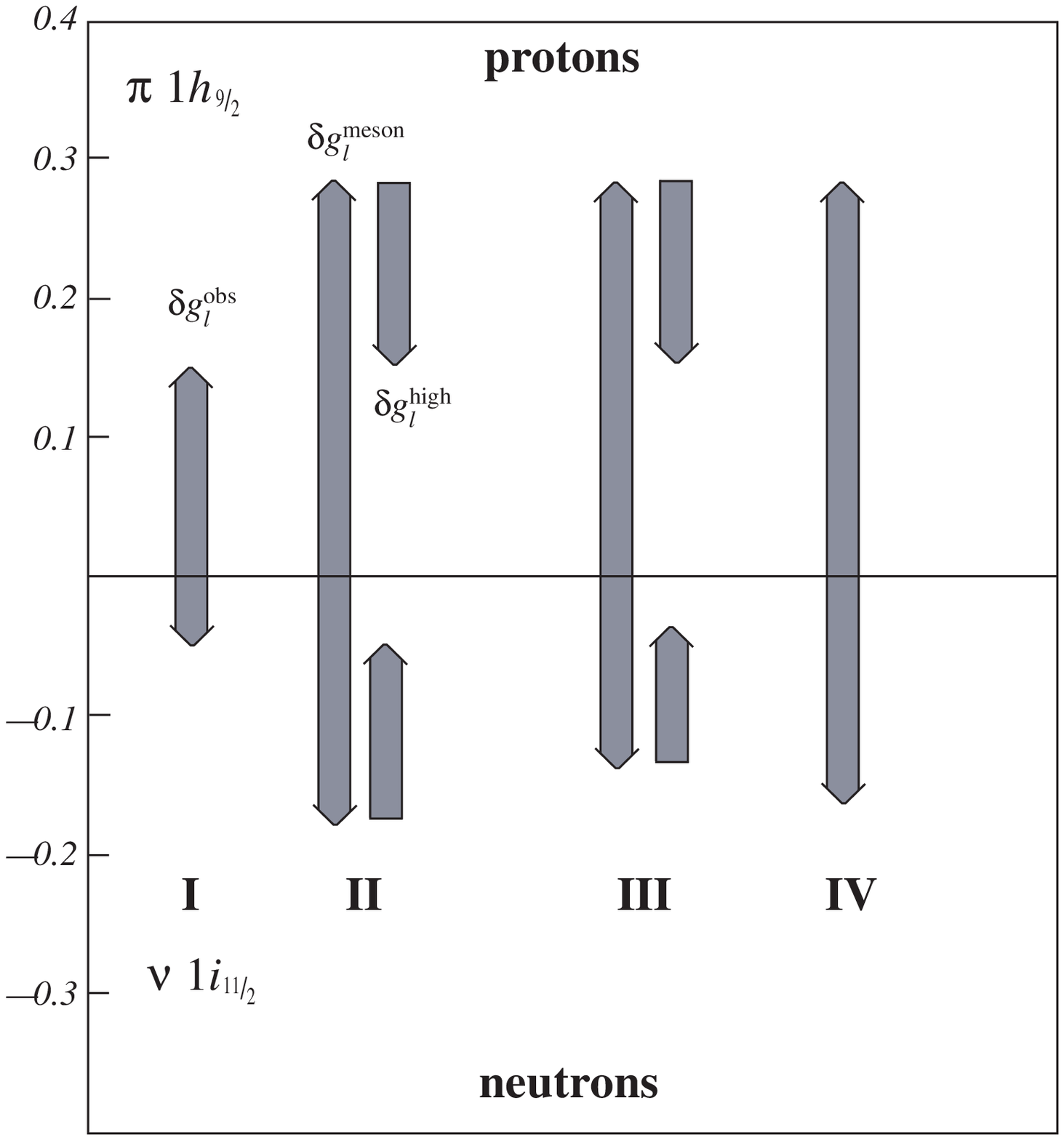,height=10cm}
\caption{ \em Summary of results on the Fujita-Hirata relation. Column I:
Experimental result of Yamazaki \cite{yamazaki79}. Column II:
$\delta g_L^{meson}$ calculated from the experimental $\protect\ko$ and
$\delta g_L^{high}$ obtained as a difference between $\delta_l^{meson}$
of column II
and $\delta g_L^{obs}$ of column I. Column III: Predictions of Hyuga et al.\
\cite{hyuga80}. Column IV: Predictions of Brown and Rho \cite{BR80}.}
%
%
%
\setcounter{section}{9}
\setcounter{figure}{0}
%
%
\centering\epsfig{figure=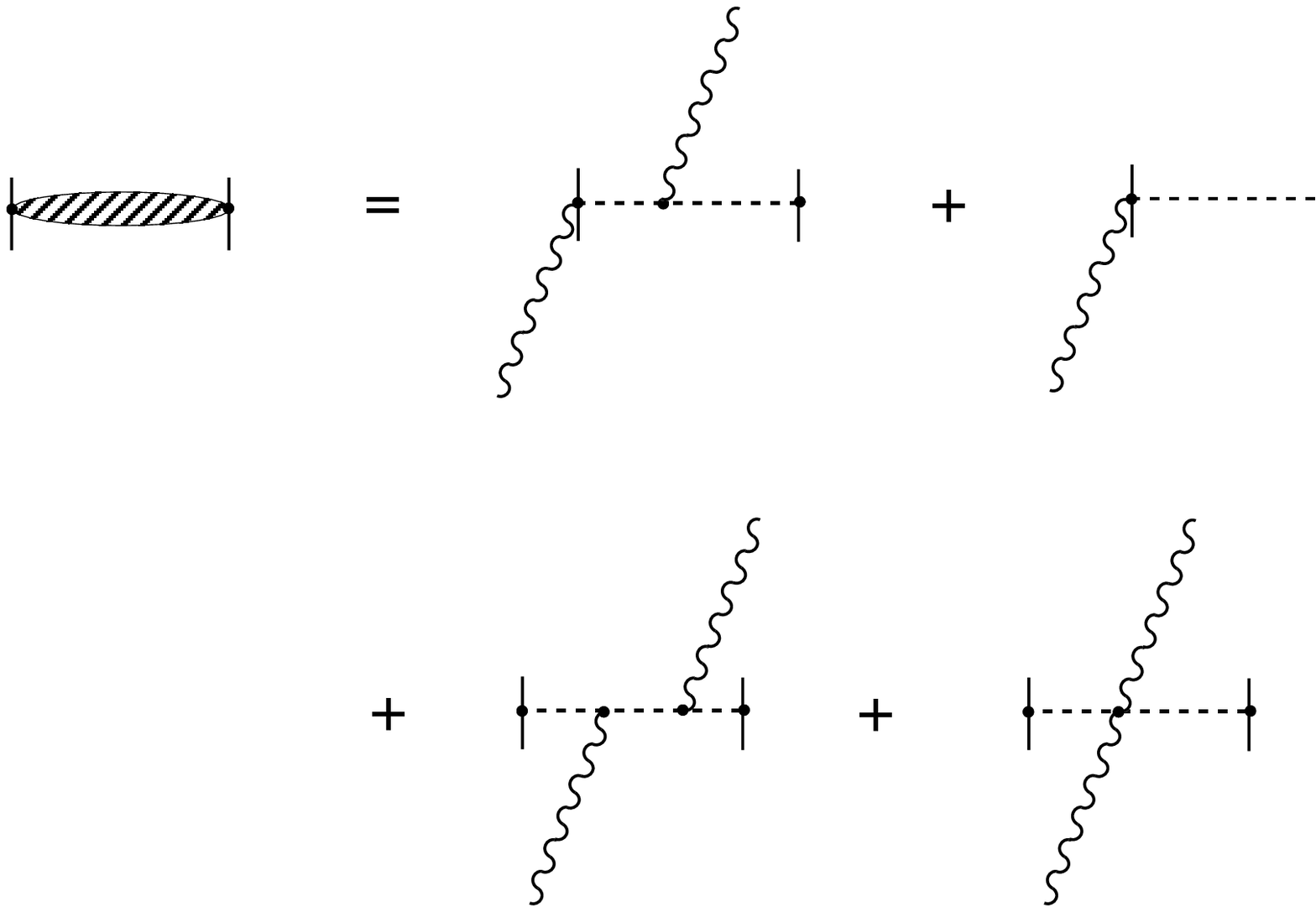,height=10cm}
\vspace*{-2cm}
\caption{\it Typical diagrams contributing to ${\cal T}_{(\pi )}^{ij} $.
The wavy lines denote photons and dashed lines denote pions. The
amputation indicates that $T^{ij}$ contains only the nucleon vertices, but
not its wave functions.}
\end{figure}

\begin{figure}[ht]
\vspace*{-2cm}
\centering\epsfig{figure=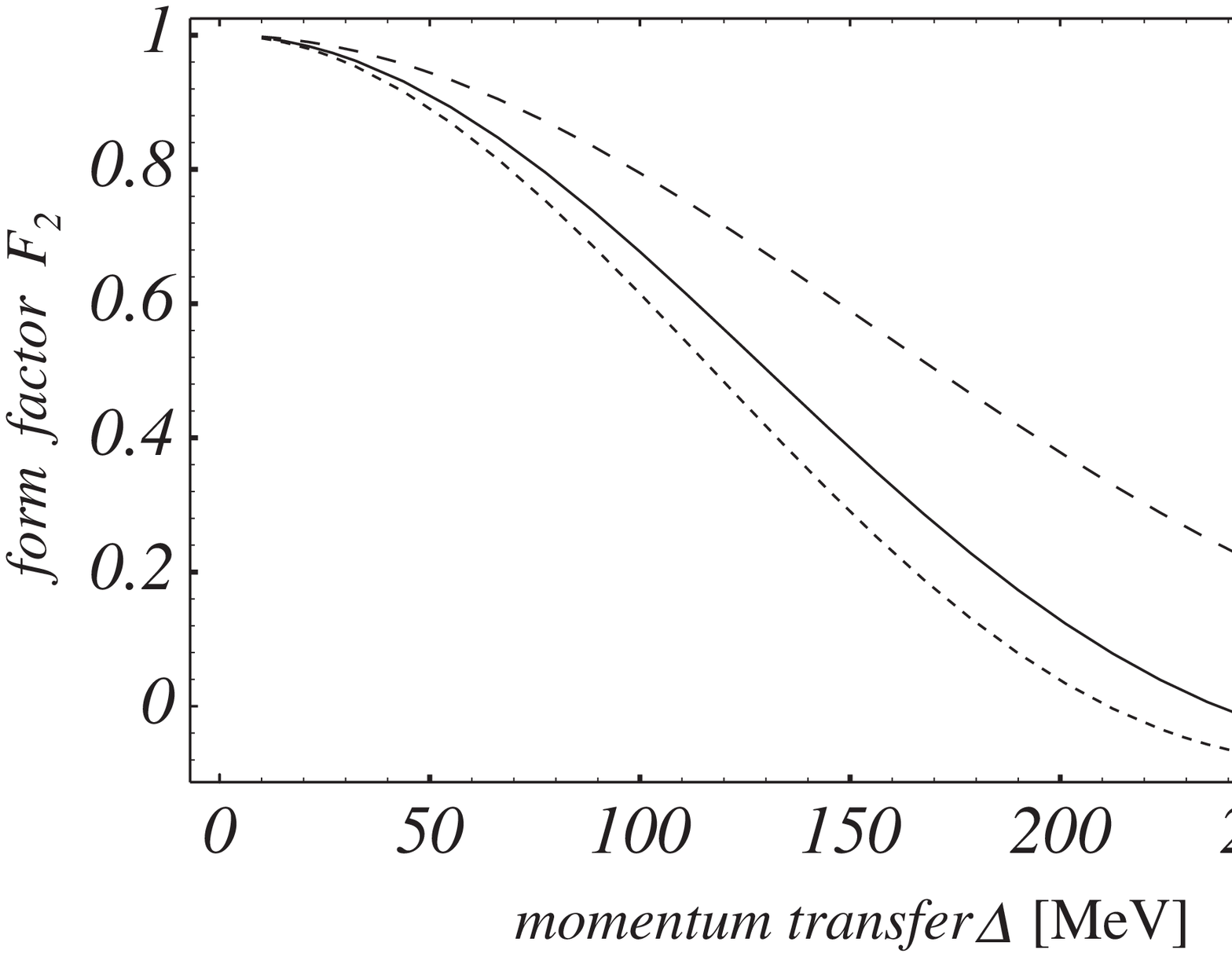,height=12cm}
\vspace*{-2cm}
\caption{\it Comparison of the form factors from Eqs.\
(\ref{formfac}) (dotted curve) and (\ref{f2}) (full curve) for the case
of $^{40}$Ca. The
approximation $F_1^2(q/2)$ is shown as a dashed curve.}
%
%
\centering\epsfig{figure=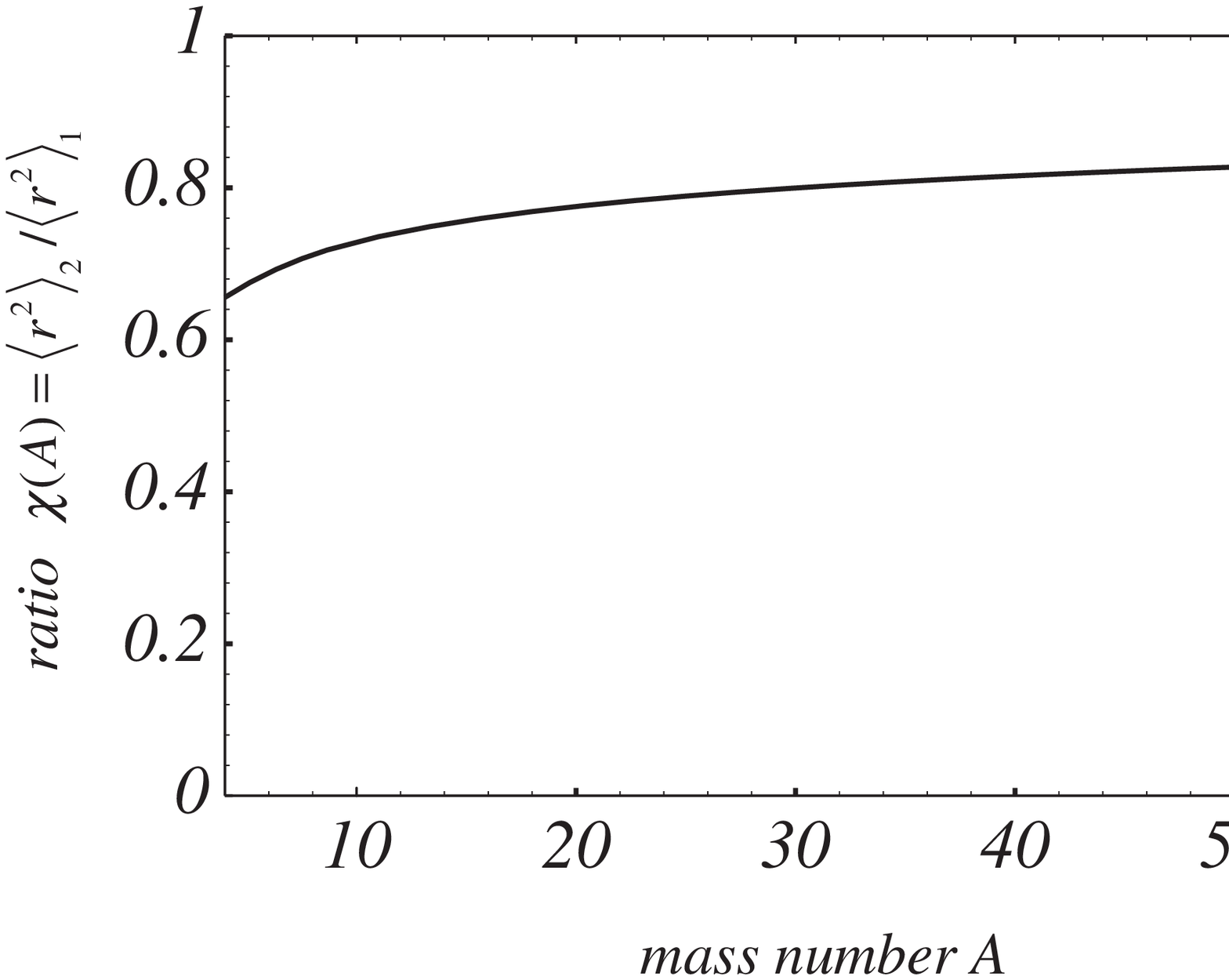,height=12cm}
\vspace*{-2cm}
\caption{\it Ratio $\chi (A)=\left\langle {r^2}
\right\rangle _2/\left\langle {r^2} \right\rangle _1$ as a function of
the nuclear mass number $A$}
\end{figure}

\begin{figure}[ht]
\vspace*{-3cm}
\centering\epsfig{figure=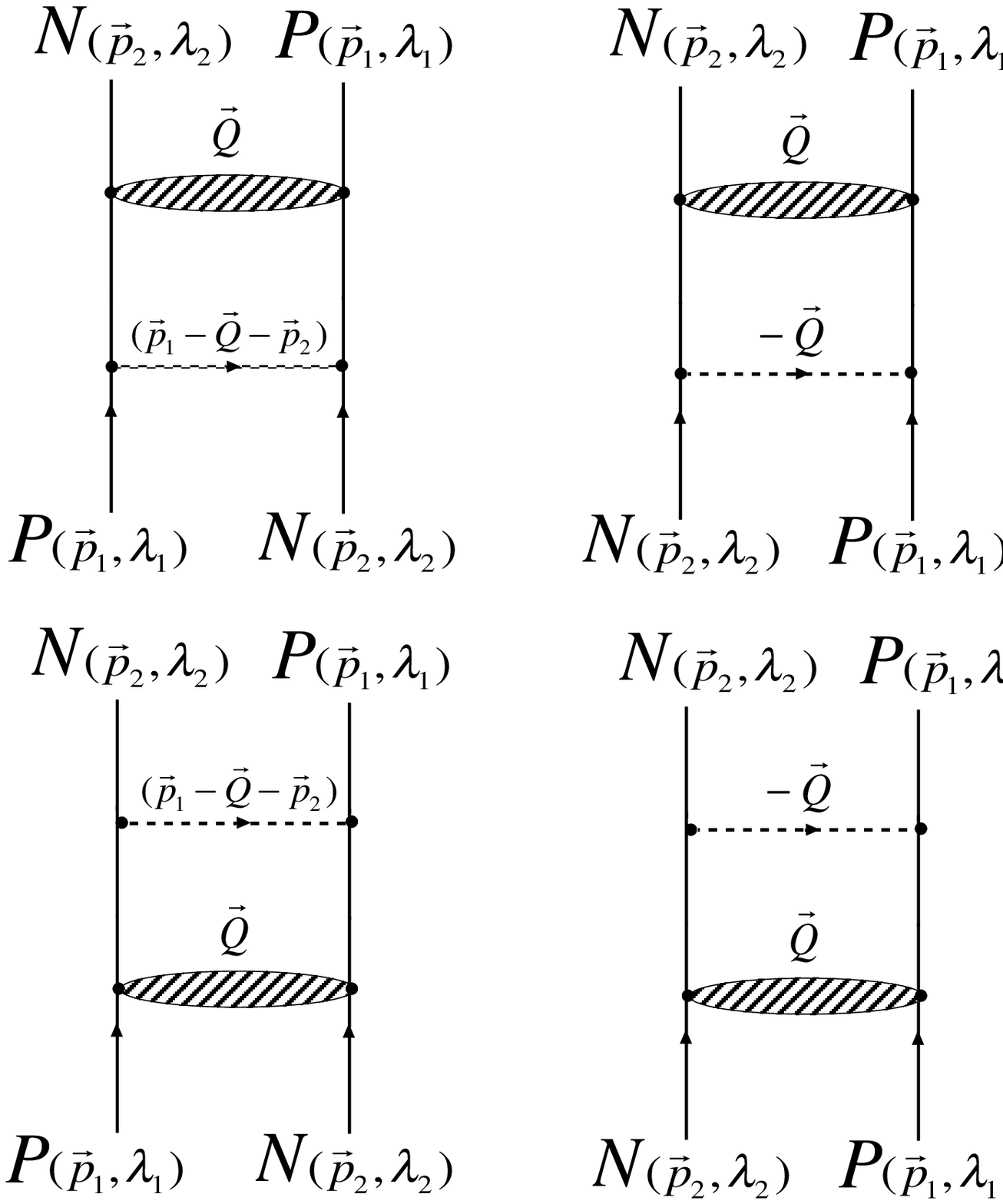,height=10cm}
\caption{\it Set of diagrams,
     from which the two-body correction to the correlator
     is extracted. Here the same symbolic abbreviation is used as in Fig.\
     \ref{pr22}. The nucleon spin projections are denoted by $\lambda _i$.}
%
%
\centering\epsfig{figure=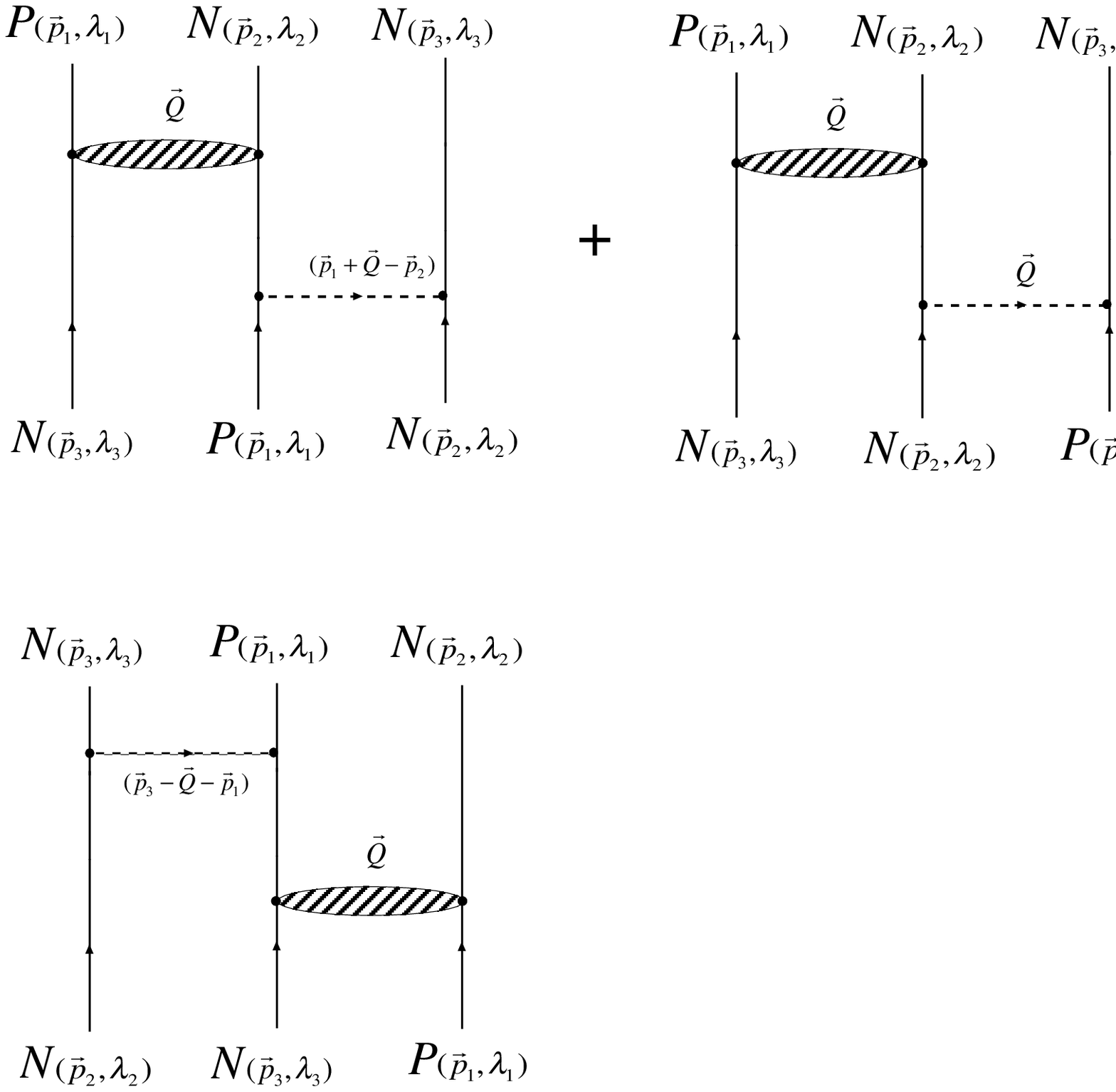,height=12cm}
\vspace*{-1cm}
\caption{\it Three-body diagrams yielding an additional correction
to the correlator, which is of the same order in $f/m_{\pi}$ as the
two-body diagrams shown in Fig.\ \ref{pr23}. The notations are the
same as in Fig.\ \ref{pr22} and  Fig.\ \ref{pr23}.}
\end{figure}

\begin{figure}[ht]
\vspace*{-4cm}
\centering\epsfig{figure=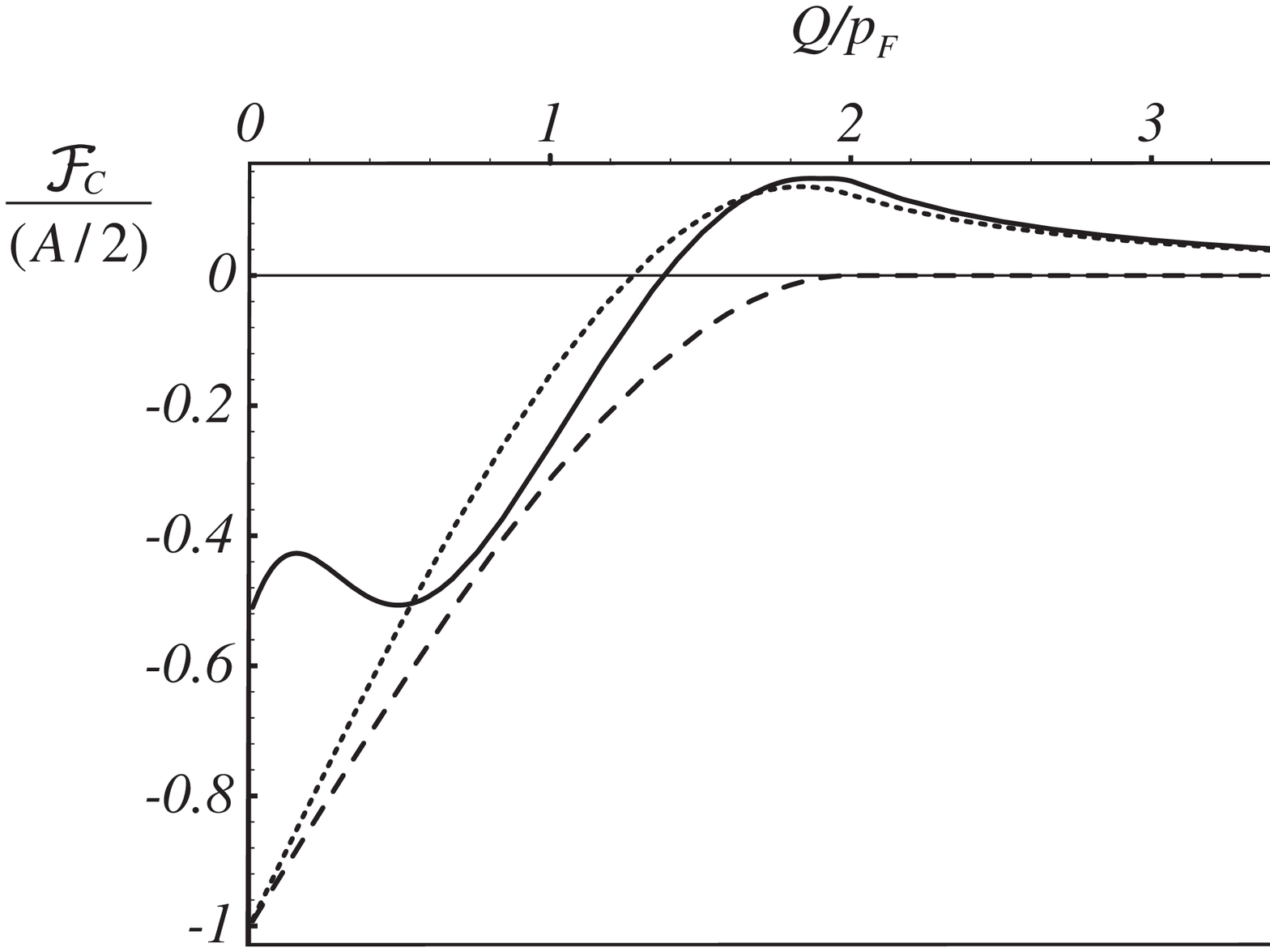,height=10cm}
\vspace*{-1cm}
\caption{\it Central part of the correlator with three-body
corrections (full curve), without
	    three-body corrections (dotted curve) and Fermi correlator
	    without any corrections (dashed curve).}
%
%
\vspace*{1cm}
\centering\epsfig{figure=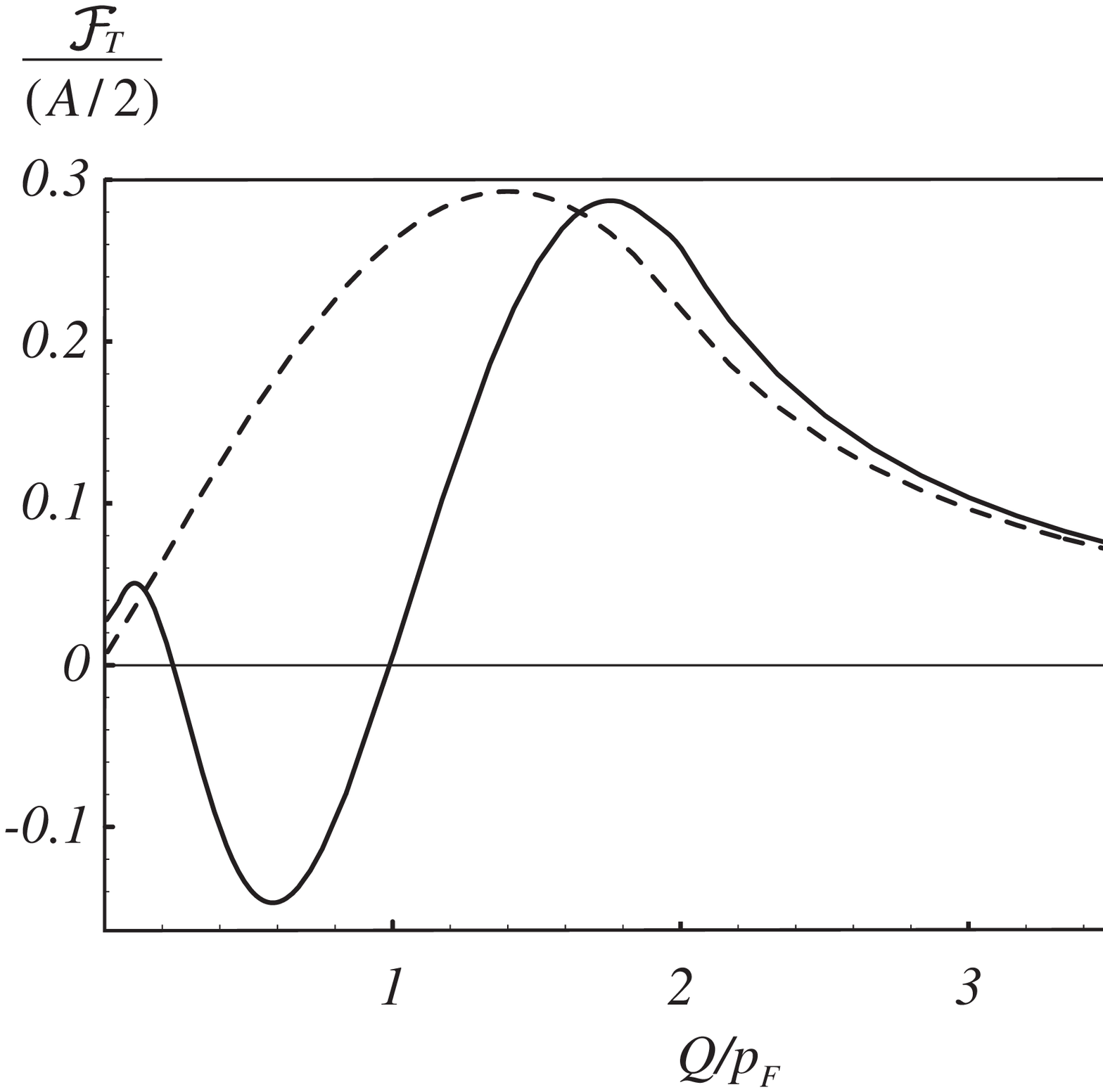,height=10cm}
\vspace*{-1cm}
\caption{\it Tensor part of the correlator with three-body
corrections (full curve) and without
	    three-body corrections (dashed curve).}
\end{figure}

\begin{figure}[ht]
\vspace*{-3cm}
\centering\epsfig{figure=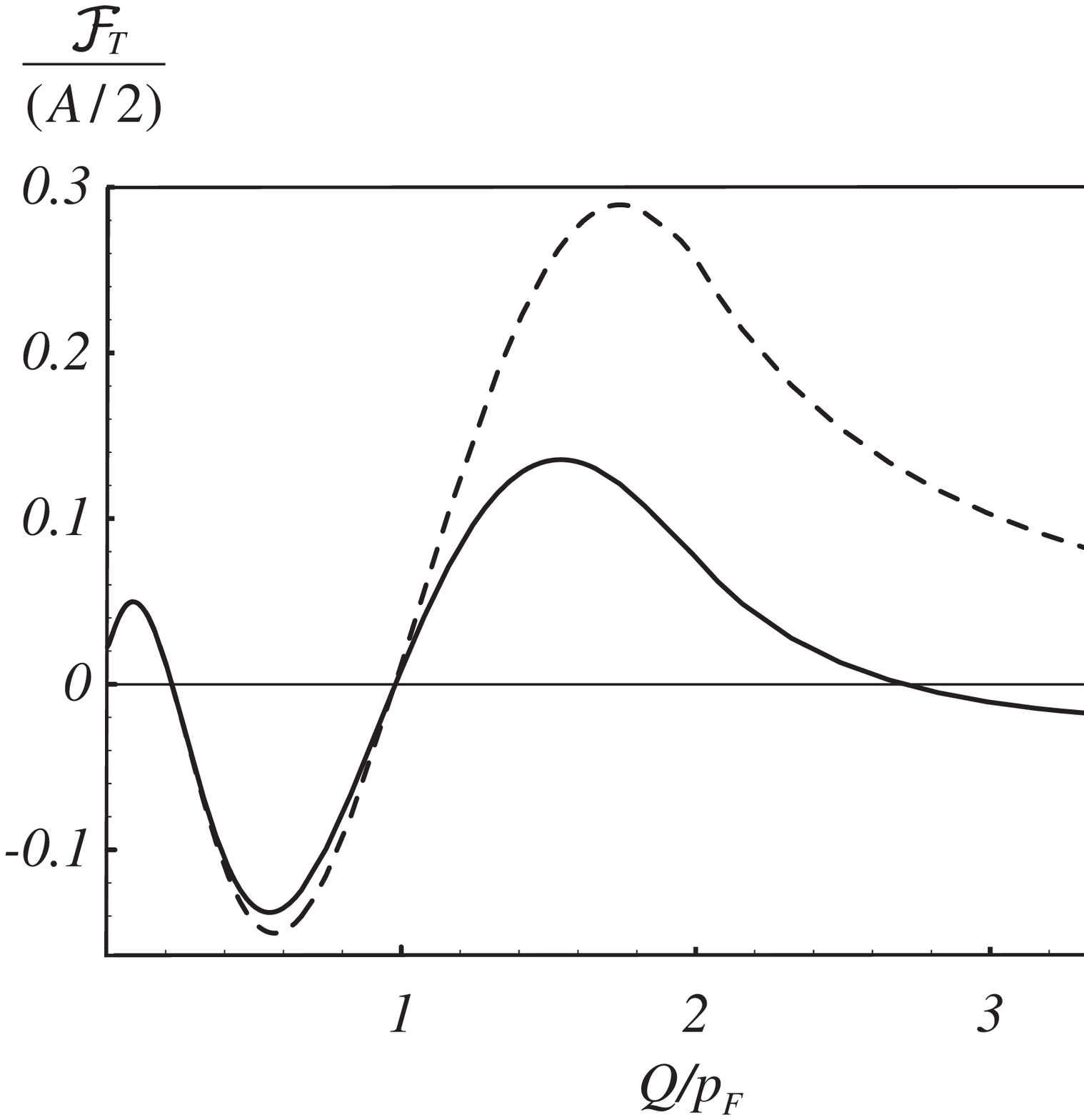,height=10cm}
\vspace*{-1cm}
\caption{\it Comparison of the tensor part of the correlator with
(full curve) and without (dashed curve) the $\rho
	    $-meson contribution.}
%
%
\vspace*{-3cm}
\centering\epsfig{figure=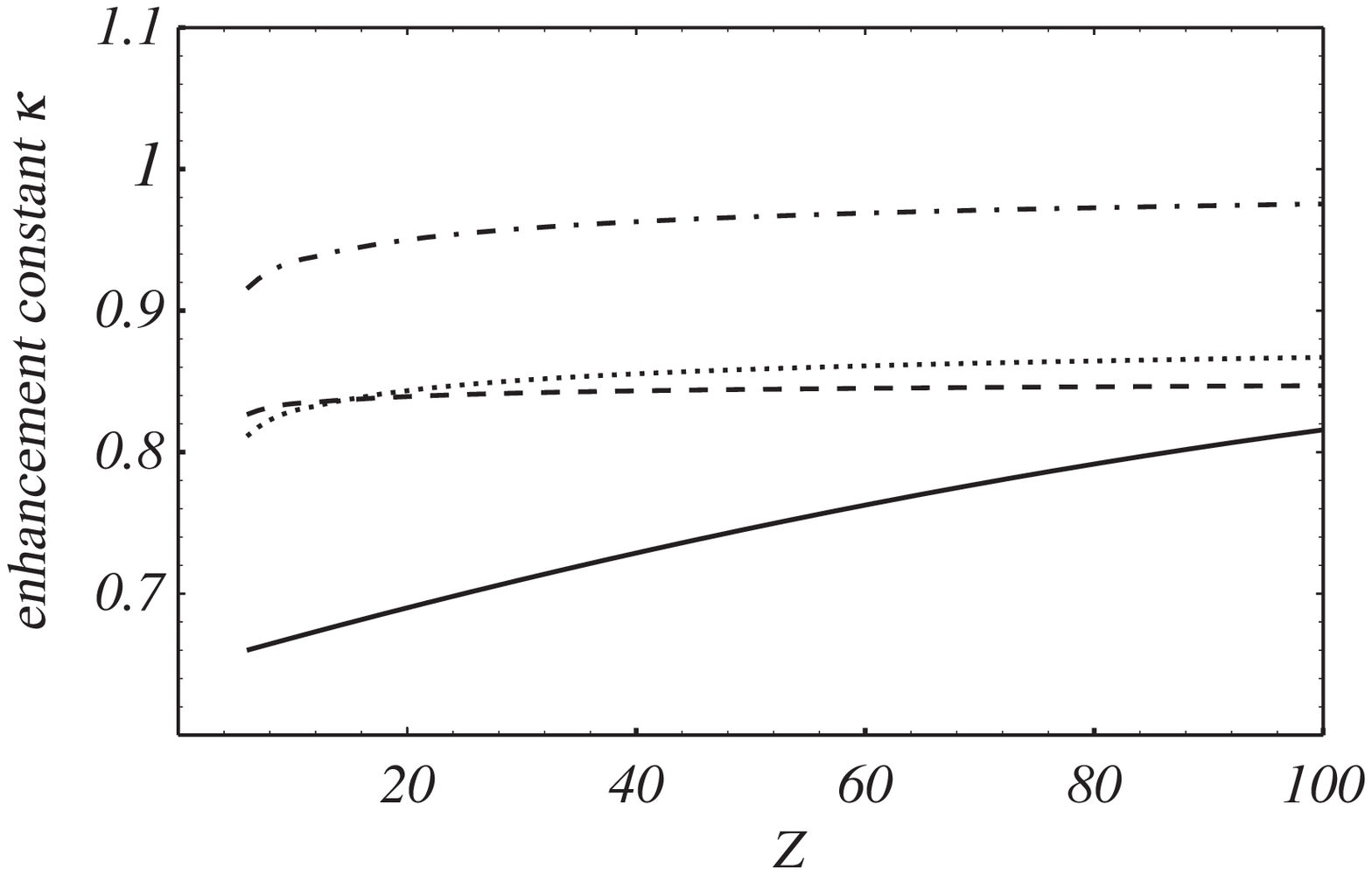,height=15cm}
\vspace*{-2cm}
\caption{\it Dependence of enhancement constant $\kappa $ on proton number $Z$.
The dashed curve corresponds to the pionic tensor contribution
$\kappa _T^\pi $, the dash-dotted curve includes also
the central contribution $\kappa _C^\pi $ and the dotted
curve gives the total $\kappa $, including the
contribution from $\rho $-meson exchange.
The realistic-density result $\kappa ^{(rd)}$ for the full
enhancement constant  ({\it cf.} Eq.\ (\ref{phireal}))
is shown as a full curve.}
\end{figure}

\begin{figure}[ht]
\vspace*{-3cm}
\centering\epsfig{figure=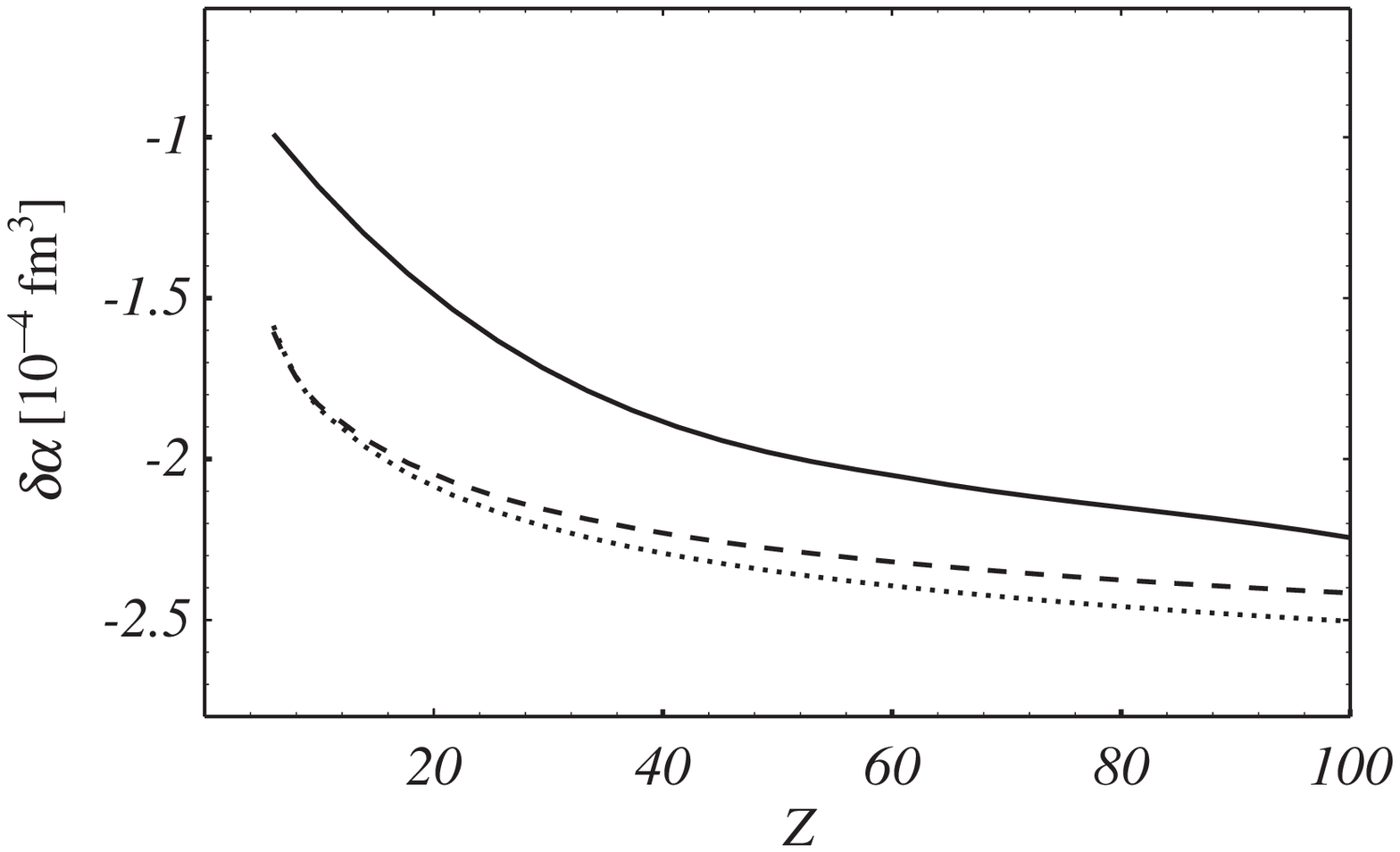,height=12cm}
\vspace*{-2cm}
\caption{\it Pion-exchange contribution to electric polarizability
	    $\delta \alpha $ as a function of $Z$.
	    The dashed curve corresponds to the central contribution
	    $\delta \alpha _C$ and the dotted
	    curve gives the total $\delta \alpha
            =\delta \alpha _C + \delta \alpha _T$.
	    The use of a realistic density leads to the full curve.}
%
%
\vspace*{-2cm}
\centering\epsfig{figure=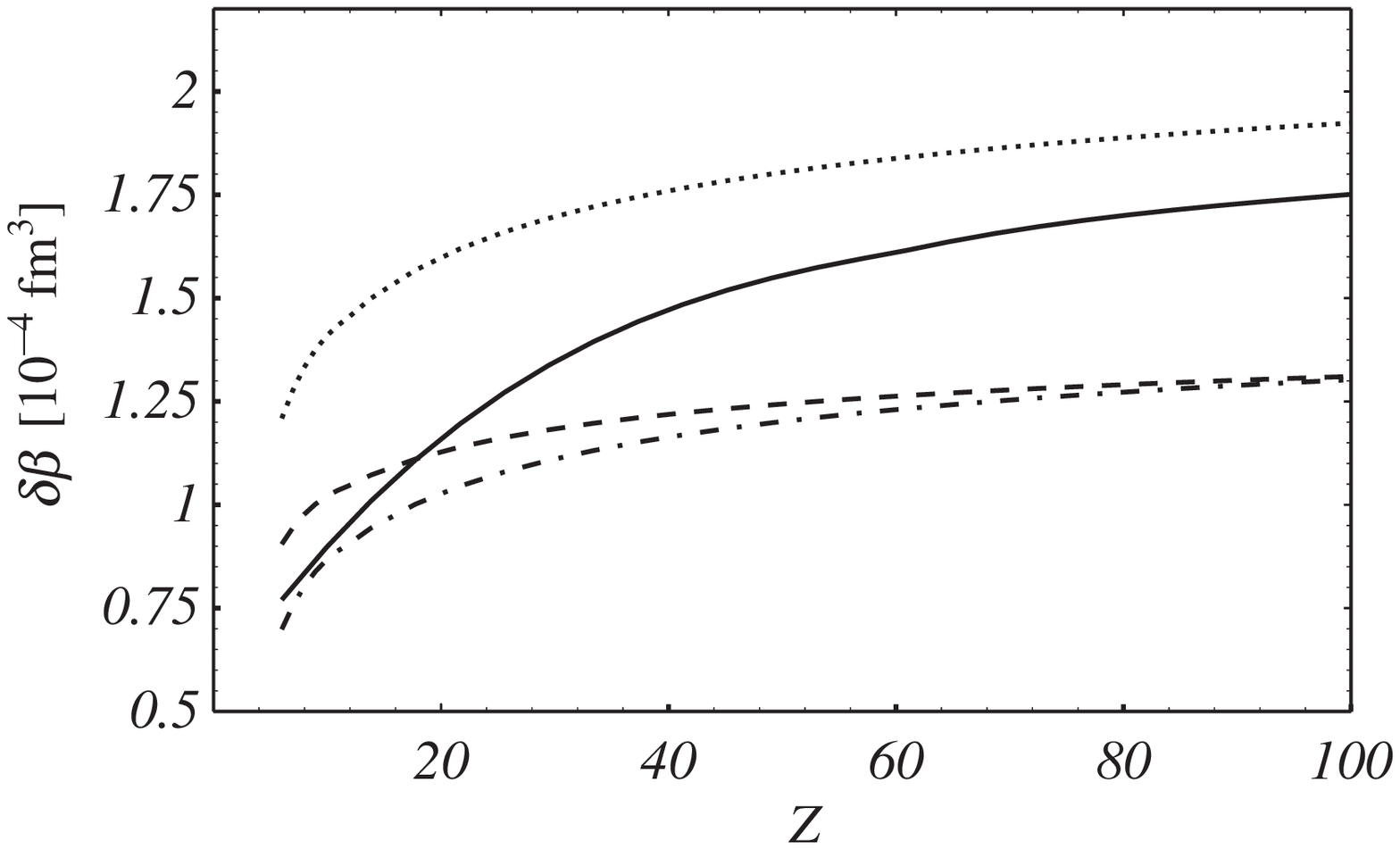,height=12cm}
\vspace*{-2cm}
\caption{\it Pion-exchange contribution to magnetic polarizability
	    $\delta \beta $ as a function of $Z$.
	    The dashed curve corresponds to the central contribution
	    $\delta \beta _C$ and the dash-dotted
            curve gives the sum $\delta \beta _C + \delta \beta _T$.
	    Adding the contribution of the $\Delta $-isobar excitation
            as given in Eq.\ (\ref{phidelta}) leads to the total
	    value of $\delta \beta $ given as the dotted curve.
	    The use of a realistic density leads to the full curve.}
\end{figure}

\begin{figure}[ht]
\vspace*{-3cm}
\centering\epsfig{figure=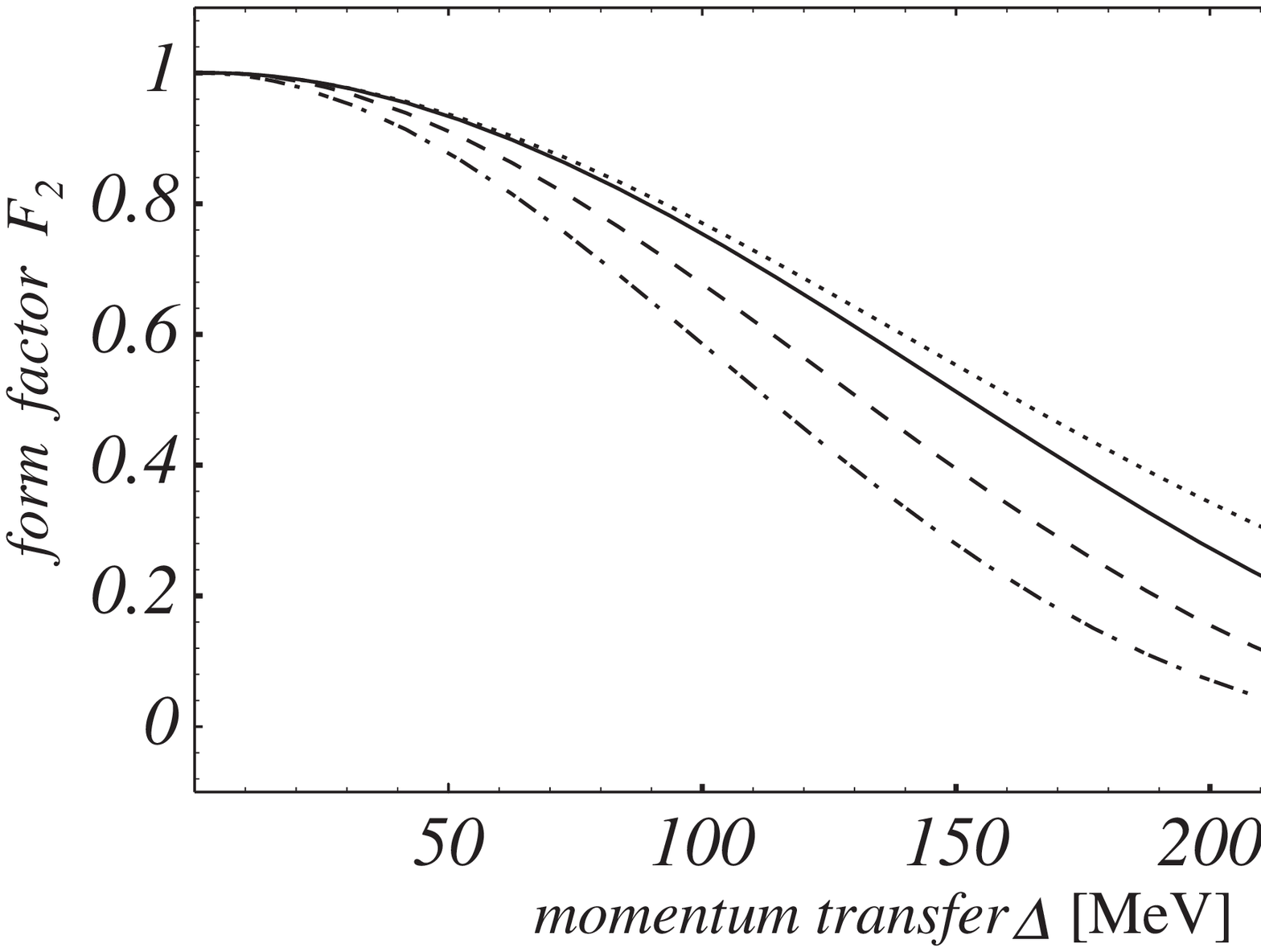,height=10cm}
\vspace*{-1cm}
\caption{\it Form factors $F_2^{(i)}(\Delta )$ for $^{40}$Ca. The dashed
	    curve is $F_2^{(1)}$ and the full curve is $F_2^{(2)}$. For
	    comparison the (experimental) charge form factor $F_1$ is
	    also shown (dash-dotted curve), as well as the
	    function $F_1^2(\Delta /2)$ (dotted curve).}
%
%
%
\setcounter{section}{10}
\setcounter{figure}{0}
%
\centering\epsfig{figure=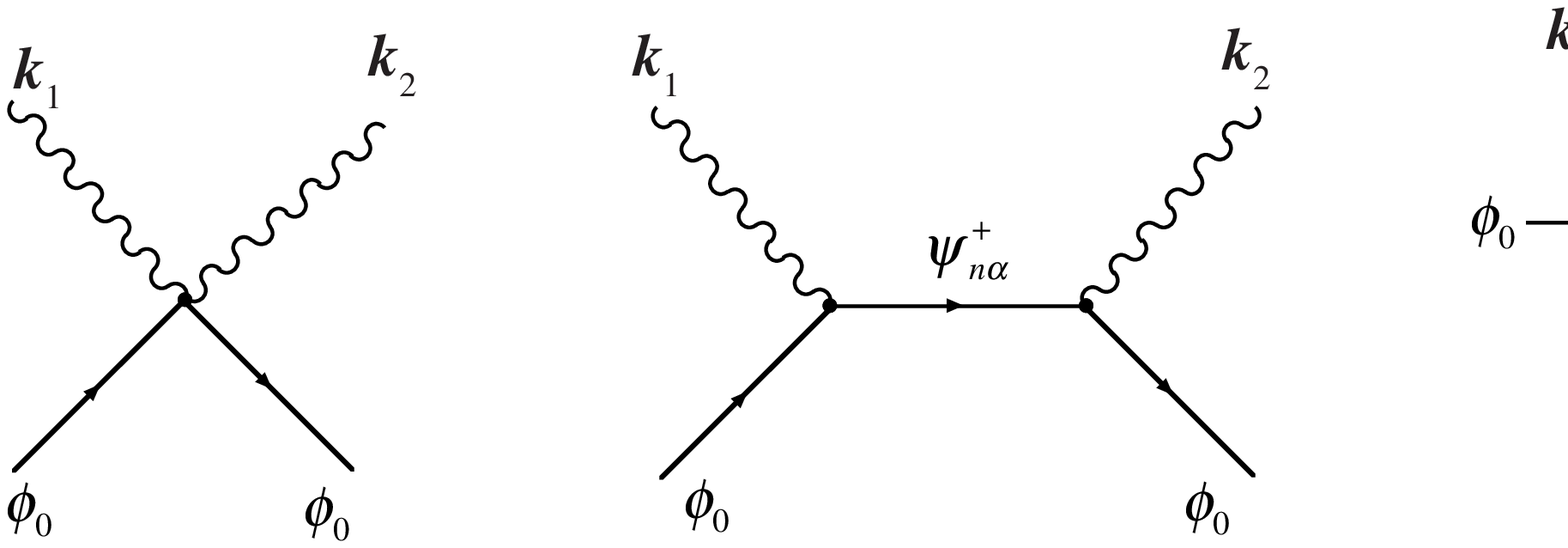,height=12cm}
\vspace*{-5cm}
\caption{\it Diagrams of Compton scattering off the relativistic
oscillator (crossed terms are not shown)}
\end{figure}

\begin{figure}[htb]
\vspace*{-15cm}
\centering\epsfig{figure=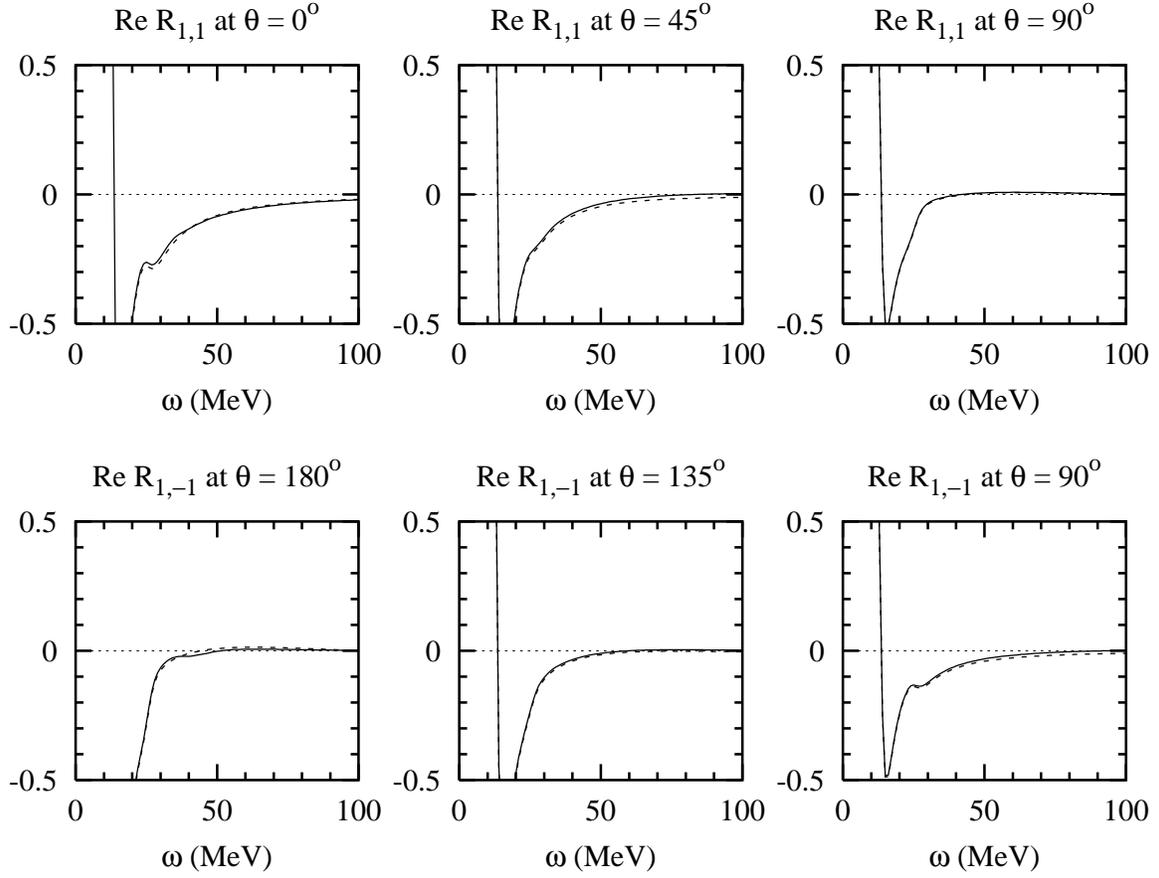,width=17cm}
\caption{\it Phenomenological, Eq.~(\ref{R-ph}),  (dashed lines)
and exact (solid lines)
amplitudes $R_{GR}$ in the relativistic oscillator model at
$\omega_0=(13.4-2.1i)$ MeV and $\eta=0.06$. Units are $e^2/{E_0}$. }
\end{figure}

\appendix
\setcounter{section}{2}
\setcounter{figure}{0}

\begin{figure}[htb]
\vspace*{-2cm}
\centering\epsfig{figure=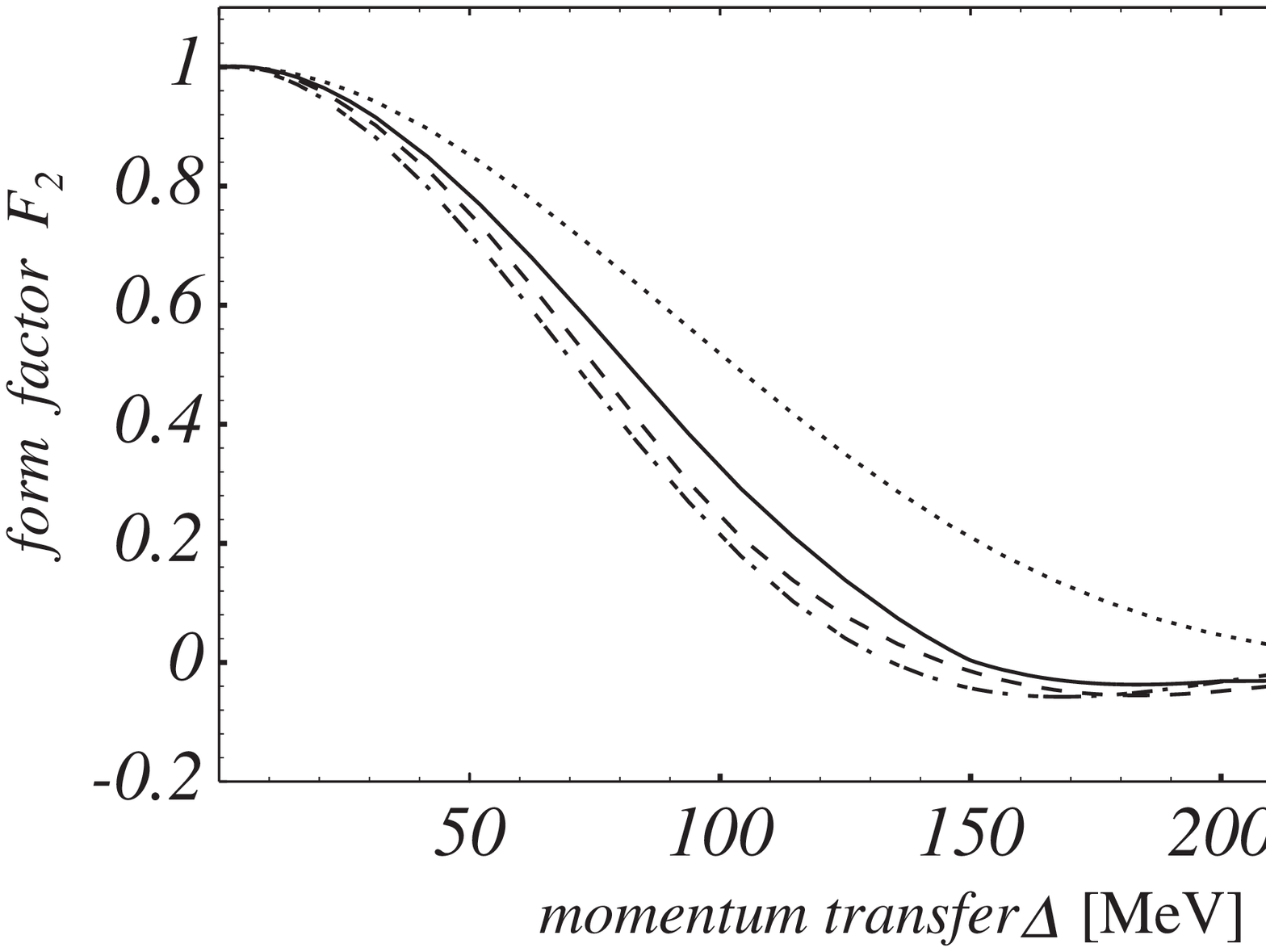,height=12cm}
\vspace*{-2cm}
\caption{\it
Exchange form factors $F_2^{(i)}(q)$ for the case of $^{208}$Pb.
The dashed curve represents $F_2^{(1)}$, while the full curve
corresponds to $F_2^{(2)}$. For comparison the (experimental) charge
form factor $F_1$  (dash-dotted curve) and the function
$F_1^2(q/2)$ (dotted curve) are also shown. The parameters for the
form factor $F_1$ were taken from ref.\ \protect\cite{datatab}.}
%
%
\centering\epsfig{figure=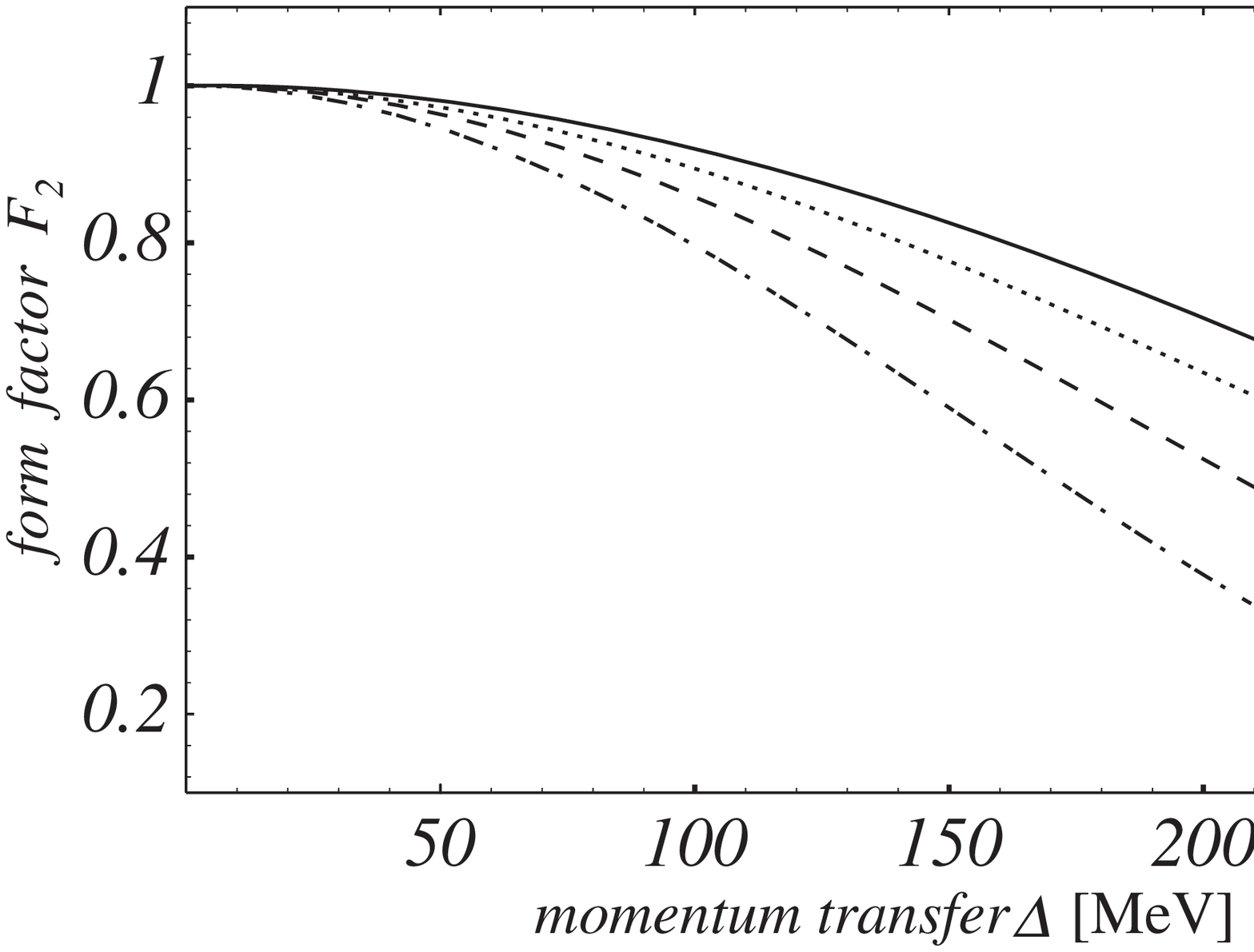,height=12cm}
\vspace*{-2cm}
\caption{\it
Same as Fig.\ \protect\ref{pr12}, but for $^{12}$C.}
\end{figure}

\begin{figure}[htb]
\vspace*{-2cm}
\centering\epsfig{figure=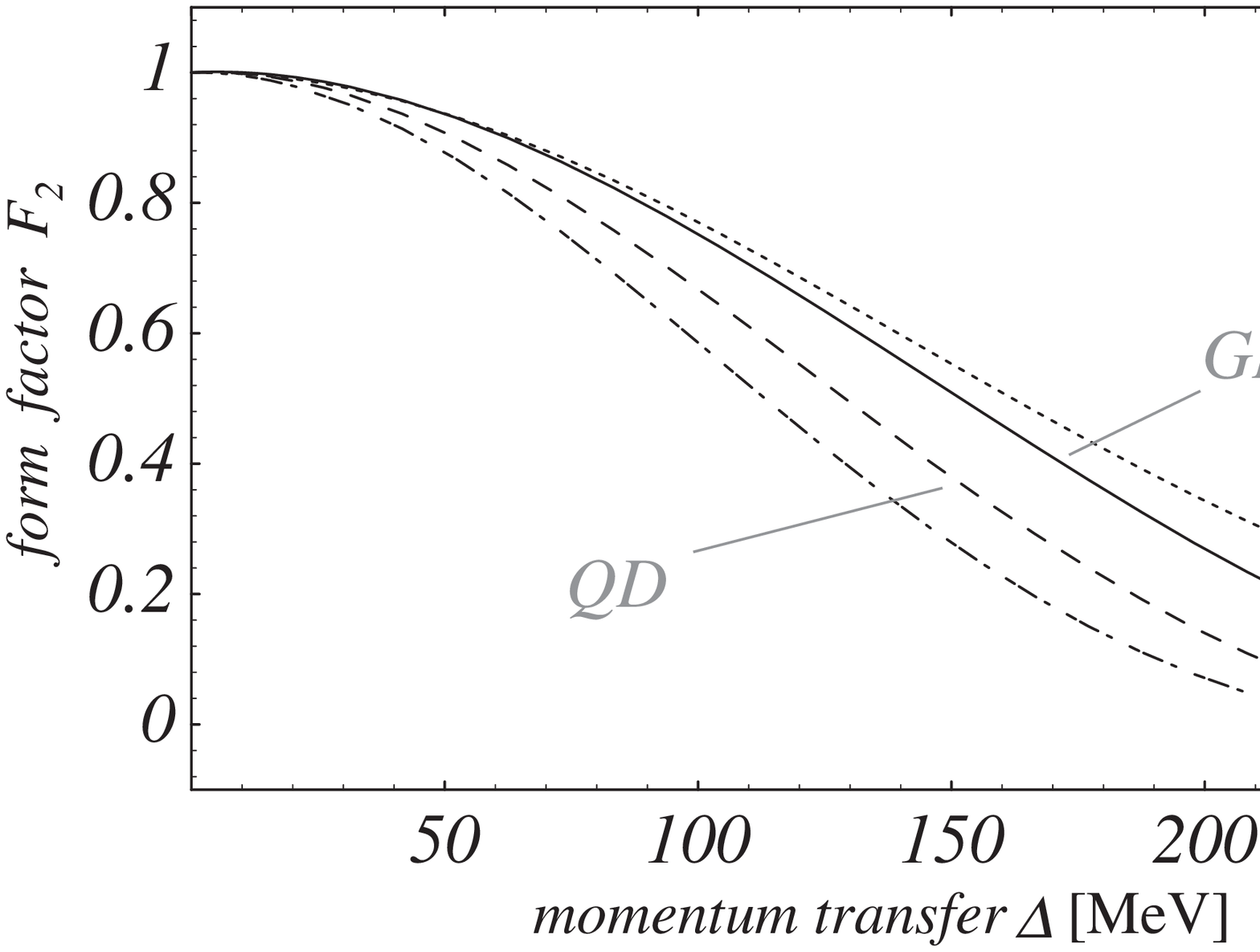,height=12cm}
\vspace*{-2cm}
\caption{\it
Form factors $F_2^{(GR)}(\Delta )$ and
$F_2^{(QD)}(\Delta )$for $^{40}$Ca. The dashed
curve is $F_2^{(QD)}$ and the full curve is $F_2^{(GR)}$. In
addition, the (experimental) charge form factor $F_1$
(dash-dotted curve) and the
function $F_1^2(\Delta /2)$ (dotted curve) are shown for the
sake of comparison.}
%
%
\centering\epsfig{figure=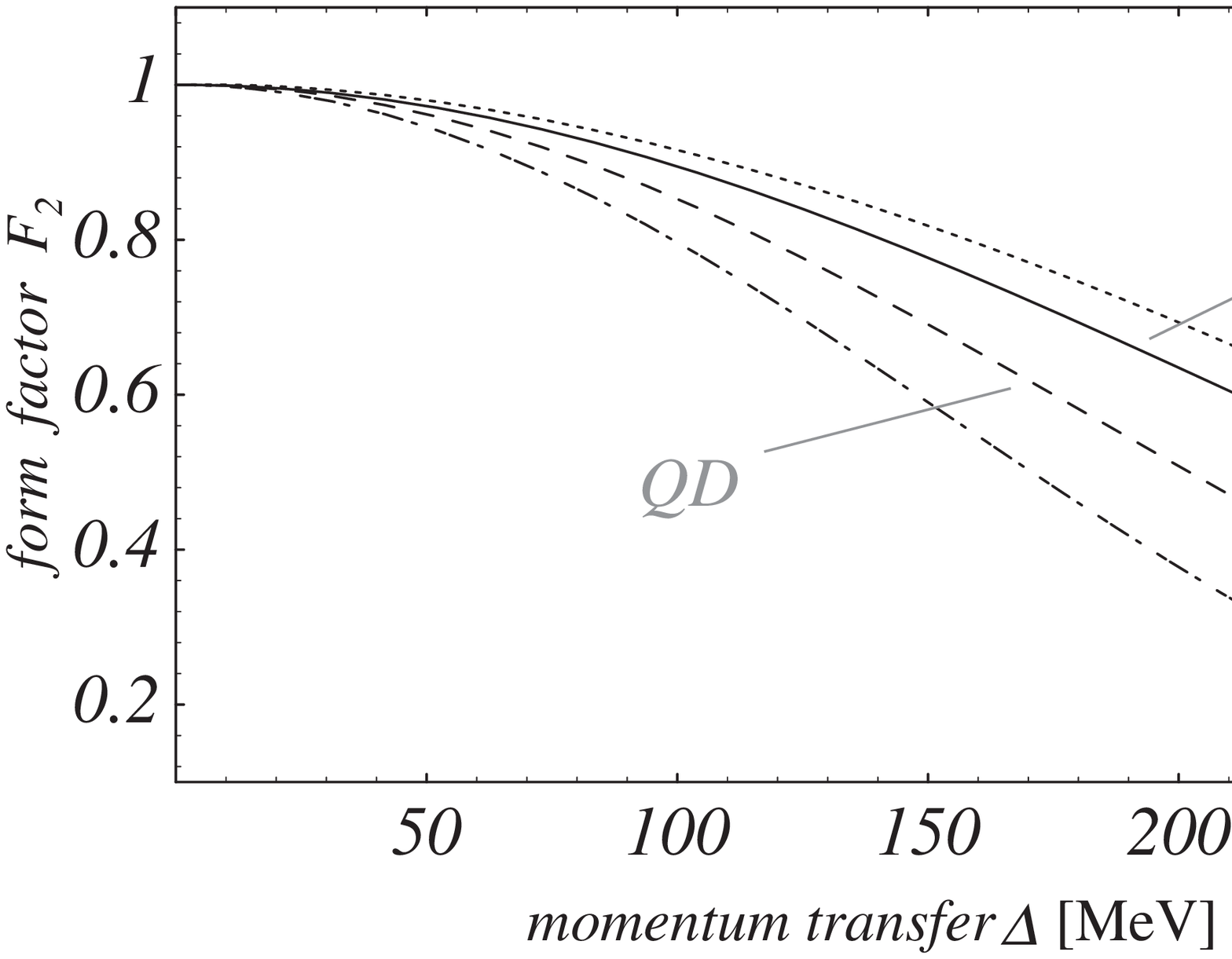,height=12cm}
\vspace*{-2cm}
\caption{\it
Same as Fig.\ \protect\ref{pr15}, but for $^{12}$C.}
\end{figure}

\end{document}